\documentclass{aa}  

\usepackage{graphicx}
\usepackage{txfonts}
\usepackage{multicol}
\usepackage{subfig}
\usepackage{hyperref}
\usepackage{gensymb}
\usepackage{amsmath,amssymb}
\usepackage{float}
\usepackage{longtable}
\usepackage[title]{appendix}
\usepackage{color}

\begin{document}

\title{The distribution of globular clusters in kinematic spaces does not trace the accretion history of the host galaxy}

   \author{G. Pagnini\inst{1},
          P. Di Matteo\inst{1},
          S. Khoperskov\inst{2,1},
          A. Mastrobuono-Battisti\inst{1},
          M. Haywood\inst{1},
          F. Renaud\inst{3},
          F. Combes\inst{4}
          }

   \institute{GEPI, Observatoire de Paris, PSL Research University, CNRS, Place Jules Janssen, 92195 Meudon, France
         \and
             Leibniz-Institut für Astrophysik Potsdam (AIP), Potsdam, Germany
             \and Lund Observatory, Department of Astronomy and Theoretical Physics, Box 43, SE-221 00 Lund, Sweden
             \and Observatoire de Paris, LERMA, Collège de France, CNRS, PSL University, Sorbonne University, Paris, France\\
             }
  \authorrunning{Pagnini et al.}

   \date{Received ; accepted }

 
  \abstract
   {Reconstructing how all the stellar components of the Galaxy formed and assembled over time, by studying the properties of the stars which make it, is the aim of Galactic archeology. In these last years, thanks to the launch of the ESA Gaia astrometric mission, and the development of many spectroscopic surveys, we are for the first time in the position to delve into the layers of the past of our galaxy. Globular clusters play a fundamental role in this research field since they are among the oldest stellar systems in the Milky Way and so bear witness of its entire past.}
   {As a natural result of galaxy formation, globular clusters did not necessarily all formed in the Galaxy itself: a fraction of them can indeed have been formed in satellite galaxies accreted by the Milky Way over time. In the recent years, there have been several attempts to constrain the nature of clusters (accreted or formed in the Milky Way itself) through the analysis of kinematic spaces - such as the $E - L_z$, $L_{perp} - L_z$, $eccentricity - L_z$, and action space - and to reconstruct from this the properties of the accretions events experienced by the Milky Way through time. This work aims to test a widely-used assumption about the clustering of the accreted populations of globular clusters in the integrals of motions space.}
   {In this paper we analyse a set of dissipation-less 
   $N$-body simulations that reproduce the accretion of one or two satellites with their globular cluster population on a Milky Way-type galaxy.}
   {Our results demonstrate that a significant overlap between accreted and ``kinematically-heated'' in-situ globular clusters is expected in kinematic spaces, for mergers with mass ratios of 1:10. In contrast with standard assumptions made in the literature so far, we find that accreted globular clusters do not show dynamical coherence, that is they do not cluster in kinematic spaces. In addition, we show that globular clusters can also be found in regions dominated by stars which have a different origin (i.e. different progenitor). This casts doubt on the association between clusters and field stars that is generally made in the literature to assign them to a common origin. By means of Gaussian Mixture Models, we demonstrate that the overlap of clusters is not only a projection effect on specific planes but it is found also when the whole set of kinematic properties (i.e. $E, L_z, L_{perp}$, eccentricity, radial and vertical actions) is taken into account.
   Overall, our findings severely question the recovered accretion history of the Milky Way based on the phase-space clustering of the globular cluster population.}
   {}

   \keywords{Galaxy: formation -- Galaxy: evolution -- Galaxy: kinematics and dynamics -- Methods: numerical}

   \maketitle

%

\section{Introduction}

Our Galaxy, the Milky Way, is a collection of hundreds of billions stars mostly distributed in a disc which extends over distances of at least 20 kiloparsecs from the Galactic centre. The disc is surrounded by a diffuse stellar, oblate-shaped halo, which contains only few percent of the total stellar mass of the Galaxy \citep{hartwick1987structure, deason11, belokurov2018co, deason2019total}. A stellar bulge-bar sits at the centre of the Galaxy and dominates the mass, and light distribution, in the inner kiloparsecs \citep[e.g.][]{wegg2013milky, ness2016x}. All these components are made of stars which have, on average, different properties, in term of ages, chemical abundances, and kinematics, thus suggesting different scenarios for their formation. Reconstructing how all these stellar components formed and assembled over time, by studying the properties of the stars which make it, is the aim of Galactic archaeology.

In these last years, this field of research has seen a succession of fascinating new discoveries, thanks to the launch of the ESA Gaia astrometric mission\footnote{\href{1}{https://sci.esa.int/web/gaia}} and the delivery of its catalogues comprising full 6D positions and motions for millions of stars \citep[33 million stars have full 6D phase-space information in the Gaia Data Release~3, see][]{gaiaDR3cat}. In addition, many spectroscopic surveys such as the APOGEE survey with SDSS \citep{prieto2008apogee}, and the soon operational WEAVE@WHT \citep{dalton2012weave}, MOONS@VLT \citep{cirasuolo2014moons, cirasuolo20} and 4MOST \citep{de20124most} surveys aim at complementing Gaia by providing detailed chemical abundances for millions of stars, each. We are, for the first time, in the position to delve into the layers of the far and recent past of our Galaxy, bringing to light the timeline of events which helped make the Milky Way the galaxy that we observe today.
Having survived billions of years to Milky Way's changes, globular clusters (hereafter GCs) observed today in the Galaxy bear witness of this entire past. GCs are dense, gravitationally bound stellar systems distributed from the Galactic centre to the most remote regions \citep[up to 150 kpc from the galactic centre, see][]{harris1979globular, harris96, meylan97}. They are very luminous objects, with typical masses of few $10^5\,M_{\odot}$, and very compact, having sizes of few pc only \citep{harris2013catalog, baumgardt18,  gratton2019globular}. Being very old, typically older than 10 Gyr \citep{marin09, carretta2010properties, vandenberg13}, GCs are tracers of the early epochs of the Galaxy and, as such, they keep precious information about its formation and evolution. At present, more than 160 GCs are known in the Milky Way \citep[see, for example, ][for a recent derivation of their distances]{baumgardt21}.\\

According to the $\Lambda$CDM cosmological paradigm, galaxy formation proceeds in a bottom-up scenario, as small structures merge together to build up the larger galaxies we observe today \citep{white1978core}. The Milky Way is a prime example of this formation mechanism, as demonstrated, for example, by the discovery of the Sagittarius dwarf spheroidal galaxy in the process of being accreted by our Galaxy and disrupted by its gravitational tidal field \citep{ibata1994dwarf, newberg02, majewski03}. 
As a natural result of galaxy formation, not only field stars but also globular clusters may have been accreted \citep{penarrubia2009tidal}: a fraction of them can indeed have been formed in satellite galaxies accreted by the Milky Way over time, while the remaining fraction would have formed in-situ, that is in the Galaxy itself. This double nature of the Galactic globular cluster system is now fully accepted, and evidence of this duality come from the analysis of observational data \citep{marin09, forbes10, dotter10, leaman2013bifurcated, vandenberg13} 
and of models, as well \citep{renaud2016origin, kruijssen2019formation, kruijssen2020kraken, pfeffer20, chen_gnedin2022}. In the recent years, there have been several attempts to constrain the origin of clusters (accreted or formed in the Milky Way itself), and to reconstruct the properties (mass, numbers) of the accretions events experienced by the Milky Way through time. Concerning globular clusters originally belonging to satellite galaxies that are still being accreted (e.g. the Sagittarius dwarf galaxy), it is possible to reconstruct their origin from the similarity between their positions, velocities and orbits, and those of stars belonging to the tidal stream tracing the on-going accretion \citep{bellazzini2020globular}. For globular clusters accreted in the past instead, it is not feasible to proceed in this way since systems accreted long ago are expected to be spatially mixed with the in-situ population and, as a consequence, they should have lost all spatial coherence.
Since the publications of the Gaia~DR2 catalogue \citep{brown18gaia}, and the access to phase-space information for almost the entire clusters population \citep{helmi2018gaia, vasiliev2019proper}, different works have tried to establish the accreted or in-situ nature of Galactic globular clusters using kinematic spaces, such as the energy-angular momentum space (hereafter $E - L_{z}$, $E$ being the total orbital energy and $L_z$ being the $z$ component of the angular momentum space in a reference frame with the Galactic disc in the $x-y$ plane). Under the hypothesis that accreted clusters conserve a dynamical coherence even several billion years after their accretion onto the Galaxy -- i.e. that they cluster in kinematic spaces -- and that different groups are related to different accretion events \citep{helmi2000mapping}, Galactic globular clusters located in different regions of the $E-L_z$ space have been associated to different progenitors \citep[see for example][]{massari2019origin}. Since the models in \citet{helmi2000mapping} assume an analytic and static Galactic potential and no dynamical friction is taken into account, the energies and angular momenta of the accreted satellites are overall conserved by definition. Thus, the fact that accreted globular clusters stand out as clumps in the $E-L_z$ space corresponding to different satellites, is, generally, a direct consequence of the assumptions made in the models, rather than an intrinsic feature of the accretion event. In addition to the $E - L_z$ space, the $L_{perp} - L_z$  space ($L_{perp}$ being the projection of the total angular momentum onto the Galactic plane) has been used several times as a natural space to search for the presence of stellar currents, ever since the initial work of \citet{helmi2000mapping}. This has been done by assuming that $L_{perp}$, although not strictly conserved, varies only marginally during the merging.
In particular, the Helmi stream was identified in the $L_{perp} - L_z$ space as a localised overdensity in the prograde region ($L_z > 0$) with a high orbital inclination (high $L_{perp}$) \citep[see, for example][]{helmi99,kepley2007halo, koppelman2019}. As a consequence, this space has been applied in literature also to retrieve the origin of globular cluster together with the $E - L_z$ space \citep[see for instance][]{massari2019origin}. Eccentricity is another important parameter in describing an orbit which has been used for instance to separate the Gaia-Sausage-Enceladus debris from the rest of the stellar halo \citep[e.g.][]{belokurov2018co, mackereth2019, naidu20}. In particular the $eccentricity - L_z$ plane has been suggested by \citet{lane2022kinematic} as another tool to identify accreted systems \citep[see also][]{cordoni2021exploring}. Another kinematic space that has been considered in literature is the action space where the horizontal axis is the (normalized) azimuthal action ($J_{\phi}/J_{tot} \equiv L_z/J_{tot}$, where $J_{tot}=\sqrt{J_R^2 + J_z^2 + J_{\phi}^2}$), while the vertical axis is the (normalized) difference between the vertical and radial actions ($(J_z - J_R )/J_{tot}$). The action space has been suggested as the ideal plane to identify and study possible substructures and debris from accretion events \citep[e.g.][]{myeong2018milky, vasiliev2019proper, myeong19, lane2022kinematic} since actions are nearly conserved under the hypothesis that the potential is slowly evolving \citep{binney1982spectral}.
Dynamical coherence is the common thread that has guided the interpretation of the kinematic spaces just listed, leading to several tentative reconstructions of the merger history of the Milky Way. For instance, by making use of the globular clusters classification proposed by \citet{massari2019origin}, \citet{kruijssen2020kraken} inferred the stellar masses and accretion redshifts of five satellite galaxies accreted by our Galaxy over time, namely Kraken \citep{kruijssen2019formation}, the Helmi stream \citep{helmi99}, Gaia-Sausage Enceladus \citep{nissen10, belokurov2018co, haywood18, helmi18}, Sequoia \citep{myeong19}, and the still accreting Sagittarius galaxy \citep{ibata1994dwarf}.  \citet{malhan2022global} recovered other past mergers  as Cetus \citep{newberg09}, Arjuna/Sequoia/I'itoi \citep{naidu20}, LMS-1/Wukong \citep{yuan20}, and discovered a possible new merger called Pontus \citep[see also][]{malhan22pontus}. The classification by \citet{massari2019origin}, slightly revised, has been also used by \citet{forbes20} who combined it with ages, metallicities and [$\alpha$/Fe] abundances of globular clusters to reconstruct the properties of their progenitor galaxies, while the classifications by \citet{kruijssen2019formation} and \citet{malhan2022global} have been used by \citet{hammer2023} to estimate their infall time.

However, the assumptions used in the previous cited works -- namely the hypothesis that accreted clusters should show a dynamical coherence -- has not been proven even if several works have started doubting it. For example, various simulations showed that even a single merger debris span over the large range in the $E-L_z$ space~\citep{ jean2017kinematic, grand2019effects,simpson2019simulating, koppelman2020messy}, where, e.g. chaotic mixing may further affect the coherence of the tidal debris~\citep{vogelsberger2008fine, price2016chaotic}. Moreover, a single merger could result in a number of features or overdensities in the $E-L_z$ ~\citep{jean2017kinematic, grand2019effects, koppelman2020messy}; finally, both $E$ and $L_z$ are not conserved quantities in case of evolving and non-axisymmetric galaxy, because of the substantial mass growth of the galaxy over time \citep{panithanpaisal2021galaxy} and pericentric passages of massive satellites~\citep{erkal2021detection, conroy2021all} perturbing both accreted and in-situ components of the halo.\\

In this paper, we thus wish to test the hypothesis of dynamical coherence in the $E-L_z$ and the other above cited spaces, by analysing \textit{N}-body simulations of the accretion of satellite galaxies, and their globular cluster systems, onto a Milky Way-type galaxy, containing its own system of clusters. We focus on mergers with mass ratios of about 1:10, because Milky Way-type galaxies are expected to have experienced mergers of similar mass \citep[relative to the Milky Way at the time of the accretion, see for example][]{deason2016eating, renaud21,  khoperskov22b} and also because this is roughly the mass ratio estimated for the Gaia Sausage Enceladus accretion event \citep[see][but also \citet{kruijssen2020kraken} for a lower estimate]{helmi18}. It is worth mentioning that some studies considered even higher Gaia Sausage Enceladus masses (e.g. a 1:5 mass ratio in \cite{naidu20}). As detailed in the next sections, our analysis demonstrates that, for such mass ratios, accreted globular clusters do not show the claimed grouping in energy-angular momentum space nor in the other kinematic spaces, in agreement with the results of numerical simulations modelling the population of field stars \citep[see][]{jean2017kinematic,amarante22, khoperskov22a, khoperskov22b}. This finding has strong implications for the reconstruction of their kinematic-based galaxy membership, and questions the accretion history of our Galaxy, as traced so far in the literature \citep{massari2019origin, myeong19, forbes20, kruijssen2020kraken, malhan2022global}. 
The paper is organised as follows. In Sect.~\ref{method}, we present the numerical method, and the characteristics of the simulations. In Sect.~\ref{results}, we present the results of this study. We focus our  analysis, firstly, on the spatial distribution of accreted  and in-situ globular clusters (Sect.~\ref{spatial}), secondly on their distribution in the $E-L_z$ plane (Sect.~\ref{ELz}), and we generalise our results showing also how GCs redistribute in other kinematic spaces, such as the $L_{perp}-L_z$ and the $eccentricity - L_z$ planes, as well as the action space (see Sec.~\ref{kin_spaces}). We finally apply a Gaussian Mixture Model (GMM) to the outcomes to check whether such a procedure is able to retrieve the real accretion history of the simulated galaxies (Sec.~\ref{GMM}). Motivated by our findings, we reconsider the classification of Galactic globular clusters made on the basis of their distribution in the $E-L_z$ plane  \citep{massari2019origin}, showing that this classification cannot be supported by the age-metallicity relation described by these clusters (see Sect.~\ref{discussion}). Finally, we present our conclusions (Sect.~\ref{conclusions}).

\section{Numerical method}\label{method}

\subsection{Main simulations}

\begin{table}
\resizebox{\columnwidth}{!}{
\begin{tabular}{|l|c|c|c|c|c|}
\hline
& \textit{M} & \textit{a} & \textit{h} & \textit{N} & $\overline{m}$\\ \hline
MW: thin disc                   & 16.21     & 4.8        & 0.25       & $10^7$    &  $1.6 \times 10^{-6}$ \\
MW: intermediate disc           & 11.69      & 2.0        & 0.60       & $6 \times 10^6$  & $1.9 \times 10^{-6}$\\
MW: thick disc                  & 8.43       & 2.0        & 0.80       & $4 \times 10^6$ & $2.1 \times 10^{-6}$ \\
MW: GC system                   & 0.07       & 2.0        & 0.80       & 100        & $6.5 \times 10^{-4}$\\
MW: dark halo                   & 160        & -          & 20         & $5 \times 10^6$  & $3.2 \times 10^{-5}$\\ \hline
Satellite(s): thin disc         & 1.69       & 1.52       & 0.08       & $10^6$      & $1.6 \times 10^{-6}$\\
Satellite(s): intermediate disc & 1.17       & 0.63       & 0.19       & $6 \times 10^5$  & $1.9 \times 10^{-6}$\\
Satellite(s): thick disc        & 0.84       & 0.63       & 0.25       & $4 \times 10^5$  & $2.1 \times 10^{-6}$\\
Satellite(s): GC system         & 0.007      & 0.63       & 0.25       & 10        & $6.5 \times 10^{-4}$\\
Satellite(s): dark halo         & 16         & -          & 6.32       & $5 \times 10^5$  & $3.2 \times 10^{-5}$\\ \hline \hline
\end{tabular}}
\caption{Masses, characteristic scale lengths and heights, number and mean masses of particles for the different components of the Milky Way-type galaxy and the satellite(s). All masses are in units of $2.3\times10^9M_{\odot}$, distances in kpc.}
\label{tab:1}
\end{table}

\begin{table*}
\resizebox{\textwidth}{!}{
\begin{tabular}{|l | c c c c c c c | c c c c c c c |}
\hline
\small
Simulation ID    & & & & sat1 & & & &     & & & sat2 & & & \\
& $x_{sat}$ & $y_{sat}$ & $z_{sat}$ &  $v_{x,sat}$ & $v_{y,sat}$ & $v_{z,sat}$ & $\Phi_{orb}$ & $x_{sat}$ & $y_{sat}$ & $z_{sat}$ &  $v_{x,sat}$ & $v_{y,sat}$ & $v_{z,sat}$ & $\Phi_{orb}$ \\[0.1cm]  \hline
MWsat\_n1\_$\Phi$0 & 100.00 &        0.00 &      0.00 &    -2.06 &       0.42 &        0.00    & 0.   &--& --&--&--&--&--&--       \\
MWsat\_n1\_$\Phi$30 &  86.60 &       0.00 &     -50.00 &    -1.78 &       0.42 &      1.03 & 30.  &--& --&--&--&--&--&--  \\ 
MWsat\_n1\_$\Phi$60 & 50.00 &       0.00 &      -86.60 &      -1.03 &     0.42 &       1.78  & 60.   &--& --&--&--&--&--&--  \\
MWsat\_n1\_$\Phi$90 & 0.00 & 0.00 & -100.00 & 0.00 & 0.42 &       2.06  & 90.  &--& --&--&--&--&--&--   \\
MWsat\_n1\_$\Phi$120 &  -50.00 &       0.00 &      -86.60 &       1.03 &      0.42 &       1.78 & 120.  &--& --&--&--&--&--&--   \\
MWsat\_n1\_$\Phi$150 &  -86.60 &      0.00 &      -50.00 &       1.78 &     0.42 &       1.03. & 150.  &--& --&--&--&--&--&--   \\
MWsat\_n1\_$\Phi$180 &  -100.00 &       0.00 &     0.00 &   2.06 &     0.42 &      0.00 & 180.  &--& --&--&--&--&--&--   \\[0.1cm]  \hline 

MWsat\_n2\_$\Phi$0-180 & 100.00 &        0.00 &      0.00 &    -2.06 &       0.42 &        0.00    & 0. &  -100.00 &       0.00 &     0.00 &   2.06 &     0.42 &      0.00 & 180. \\
MWsat\_n2\_$\Phi$30-150 & 86.60 &       0.00 &     -50.00 &    -1.78 &       0.42 &      1.03 & 30. &  -86.60 &      0.00 &      -50.00 &       1.78 &     0.42 &       1.03. & 150.\\
MWsat\_n2\_$\Phi$60-120 & 50.00 &       0.00 &      -86.60 &      -1.03 &     0.42 &       1.78  & 60. &  -50.00 &       0.00 &      -86.60 &       1.03 &      0.42 &       1.78 & 120.\\
MWsat\_n2\_$\Phi$90-0 & 0.00 & 0.00 & -100.00 & 0.00 & 0.42 &       2.06  & 90. & 100.00 &        0.00 &      0.00 &    -2.06 &       0.42 &        0.00    & 0. \\
MWsat\_n2\_$\Phi$120-30 &  -50.00 &       0.00 &      -86.60 &       1.03 &      0.42 &       1.78 & 120. &  86.60 &       0.00 &     -50.00 &    -1.78 &       0.42 &      1.03 & 30.\\
MWsat\_n2\_$\Phi$150-60&  -86.60 &      0.00 &      -50.00 &       1.78 &     0.42 &       1.03. & 150. & 50.00 &       0.00 &      -86.60 &      -1.03 &     0.42 &       1.78  & 60. \\
MWsat\_n2\_$\Phi$180-90&  -100.00 &       0.00 &     0.00 &   2.06 &     0.42 &      0.00 & 180. &  0.00 & 0.00 & -100.00 & 0.00 & 0.42 &       2.06  & 90. \\ [0.1cm]  \hline \hline

\end{tabular}}
\caption{Initial positions and velocities of the barycentres of the simulated satellite galaxies. The initial inclination of the satellite orbital plane is also given. Distances are given in kpc, masses in units of $2.3\times10^9\,M_{\odot}$, velocities in units of $100\,km/s$ and $G = 1$. Energies are thus given in units of $10^4\,km^2/s^2$ and time is in unit of $10^7$ years.}
\label{tab:2}
\end{table*}

In this paper we analyse 14 dissipationless, high-resolution, \textit{N}-body simulations of a Milky Way-type galaxy accreting one or two satellites.  These simulations are similar to those presented in \citet{jean2017kinematic}. As comprehensively explained in that paper, in these simulations both the satellite(s) and the Milky Way galaxy can react to the interaction, experiencing kinematical heating, tidal effects and dynamical friction. Each satellite has a mass which is 1:10 of the mass of the Milky Way-like galaxy and this is motivated by the fact that mergers with these mass ratios are inevitable for Milky Way-type galaxies in $\Lambda$CDM simulations (\cite{read2008thin}; \cite{stewart2008merger}; \cite{de2008galaxy}; \cite{deason2016eating}).
Both the main galaxy and the satellite(s) are embedded in a dark matter halo and contain a thin, an intermediate and a thick stellar disc – mimicking the Galactic thin disc, the young thick disc and the old thick disc, respectively (see \cite{haywood2013age}; \cite{di2016disc}). A population of globular clusters initially redistributed in a disc is also taken into account and it is represented as point masses\footnote{Resolving both the simulated galaxies and the globular clusters as \textit{N}-body systems is currently not possible due to large range of spatial scales that should be modelled - from sub-parsec resolution to hundreds of kpc. In \citet{renaud2015flexible} a numerical method to describe the dynamical evolution of clusters in tidal fields has been carried out. We report to future works that employ the more generic method described in \citet{renaud11}, which is particularly adapted to study the dynamics of some of our clusters -- modelled as \textit{N}-body systems -- in the evolving tidal field generated by some of these accretion events.}.  The total number of particles used in these simulations is $N_{tot} = 27\,500\,110$, for the case of a single accretion, and $N_{tot} = 30\,000\,120$, when two accretions are modeled. The total number of stellar (thin, intermediate, thick disc) particles in the main galaxy is $20\,000\,000$, the number of globular cluster particles is 100 and the number of dark matter particles is $5\,000\,000$. The satellite galaxies are rescaled versions of the main galaxy, with masses and total number of particles 10 times smaller and sizes reduced by a factor $\sqrt{10}$. All these values are listed in Table~\ref{tab:1}. The discs are modeled with Miyamoto-Nagai density distributions (\cite{miyamoto1975three}) of the form:

\begin{equation*}
\rho_{*}(R,z) = \Bigg(\frac{h_{*}^2M_{*}}{4\pi}\Bigg)\frac{a_{*}R^2 + (a_{*} + 3\sqrt{z^2 + h_{*}^2})(a_{*} + \sqrt{z^2 + h_{*}^2})^2}{\Big[a_{*}^2 + \Big(a_{*} + \sqrt{z^2 + h_{*}^2}\Big)^2\Big]^{5/2} (z^2 + h_{*}^2)^{3/2}}
\end{equation*}

where masses ($M_{*}$), characteristic lengths ($a_{*})$ and heights ($h_{*}$) vary for the thin, intermediate and thick disc populations (see Table~\ref{tab:1}); the system of disc globular clusters has scale length and scale height equal to those of the thick disc and the dark matter halo is modeled as a Plummer sphere:
\begin{equation}
    \rho_{dm}(r) = \Bigg(\frac{3M_{dm}}{4\pi a_{dm}^3}\Bigg)\Bigg(1 + \frac{r^2}{a_{dm}^2}\Bigg)^{-5/2}
\end{equation}
where the characteristic mass ($M_{dm}$) and radius ($a_{dm}$) are listed in Table~\ref{tab:1}. The choice of using a core dark matter halo for both the Milky Way-type galaxy and the satellite(s) comes from a number of observational evidence that seem to be more consistent with a dark halo profile with a nearly flat density core (\cite{flores1994observational}; \cite{de2001mass}; \cite{marchesini2002halpha}; \cite{gentile2005dwarf}; \cite{de2008mass}). The final \textit{N}-body models of the main disc galaxy and of the satellite are then built by using the iterative method described in \cite{rodionov2009iterative}. Note that we do not include any significant spheroidal bulge in our model, since the contribution of a classical bulge to the total stellar mass of the Milky Way has been shown to be limited to few percent \citep[see, for example][]{shen10, kunder12, dimatteo15, gomez18}, nor any population of halo clusters before the interaction(s) (as previously stated, GCs are all initially confined in the disc). The inner regions of the Galaxy are indeed dominated by a stellar bar, whose impact on the kinematics and distribution of globular clusters  has recently started to be explored \citep[see for instance][]{perez2020}.

Once the \textit{N}-body systems which represent the main galaxy and the satellite are generated, we translate the barycentre of the satellite galaxy to the  $(x_{sat}, y_{sat}, z_{sat})$ positions and assign it velocities $(v_{x,sat}, v_{y,sat}, v_{z,sat})$ in order to have a parabolic orbit in the case of a 1x(1:10) interaction, with the satellite initially at a distance of 100~kpc from the Milky Way-type galaxy. For each  1x(1:10) interaction, we also vary the initial orientation of the satellite orbital plane ($\Phi_{orb}$), by rotating it about the y-axis, in such a way to span a range of possible orientations, from prograde orbits ($\Phi_{orb}=0^\circ, 30^\circ, 60^\circ$, with  $\Phi_{orb}=0^\circ$ being the planar, prograde orbit) to a polar orbit  ($\Phi_{orb}=90^\circ$) to retrograde orbits ($\Phi_{orb}=120^\circ, 150^\circ, 180^\circ$, with  $\Phi_{orb}=180^\circ$ being the planar, retrograde orbit). For the 2x(1:10) interactions, we use as initial orbital conditions for the two satellites a combination of those adopted for single 1x(1:10) mergers. All these initial values, for the 14 simulations (seven 1x(1:10) and seven 2x(1:10)) are reported in Table~\ref{tab:2}. In this table, each simulation has been given an identifier of the form "MWsat\_n$N$\_$\Phi \alpha_1(-\alpha_2)$", with $N$ being the number of satellites in the simulation and $\alpha_1$ ($\alpha_2$) the angles of the satellite(s) orbital plane(s) relative to the $x-y$ plane.  \\
As for the initial inclination of the satellite disc, each satellite has an internal angular momentum whose initial orientation has been assigned parallel to the $z$-axis of the host.\\
\begin{figure*}
\begin{centering}
\begin{multicols}{3}
     \includegraphics[width=.9\linewidth]{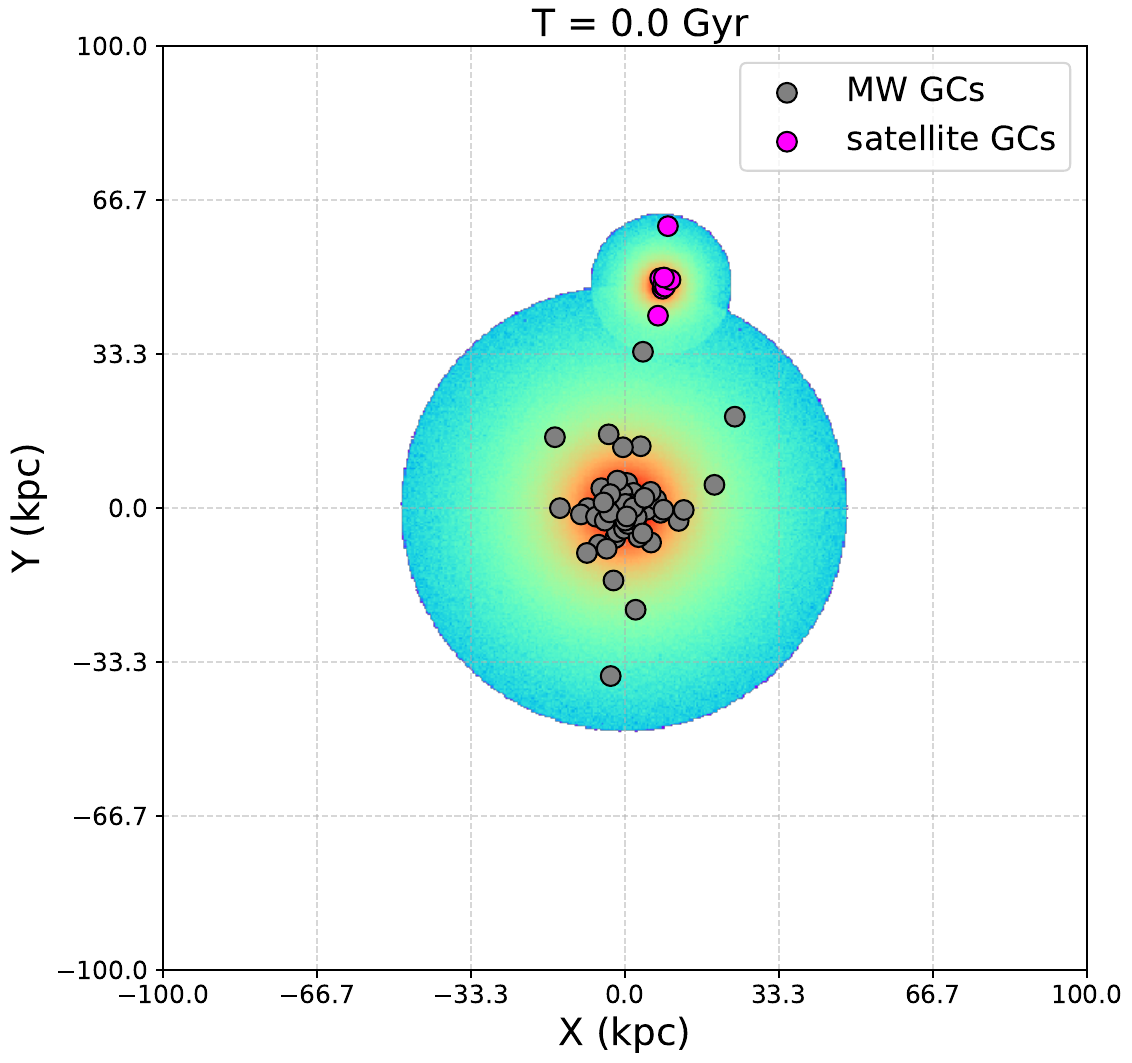}\par
         \includegraphics[width=.9\linewidth]{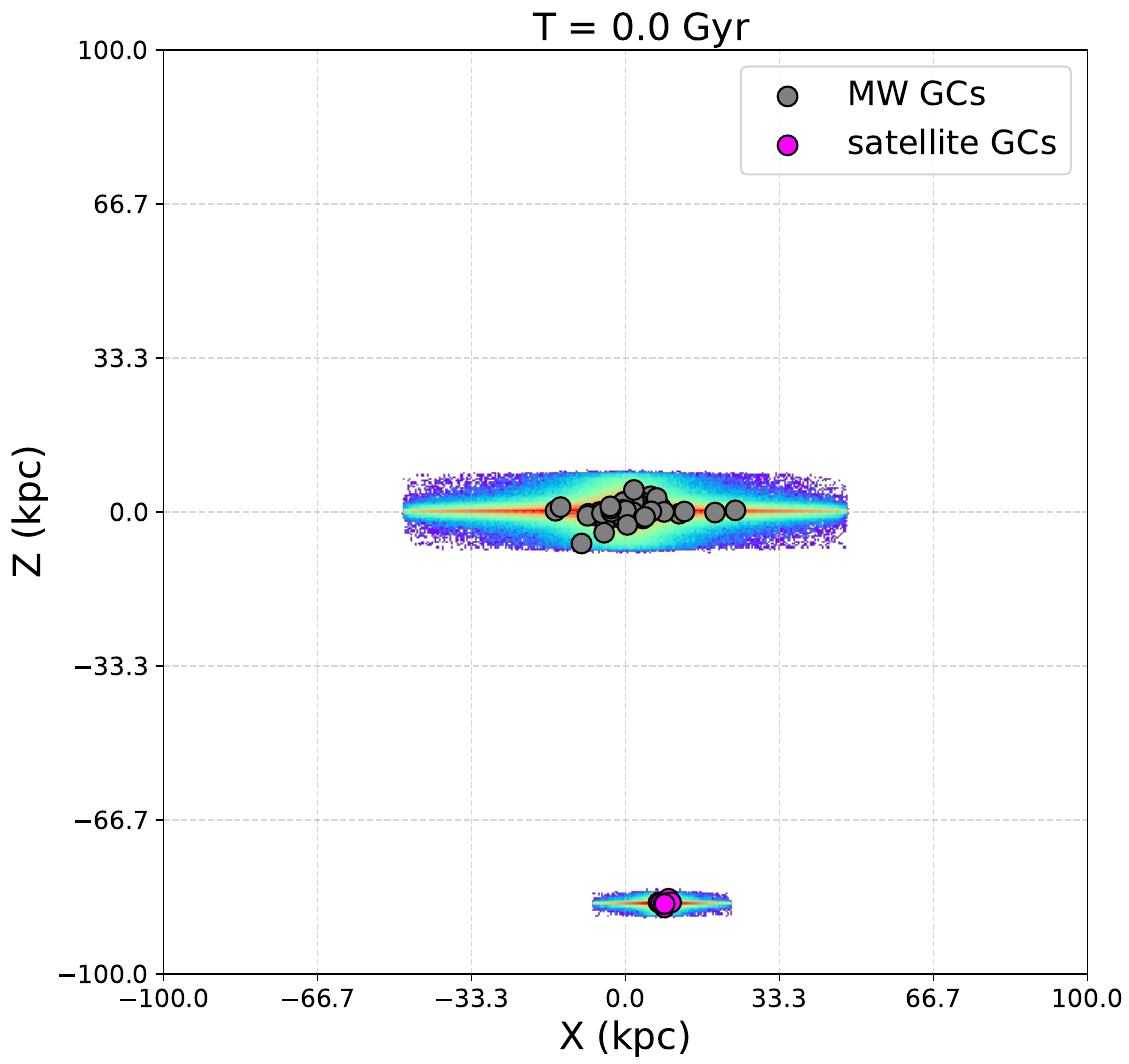} \par
             \includegraphics[width=.95\linewidth]{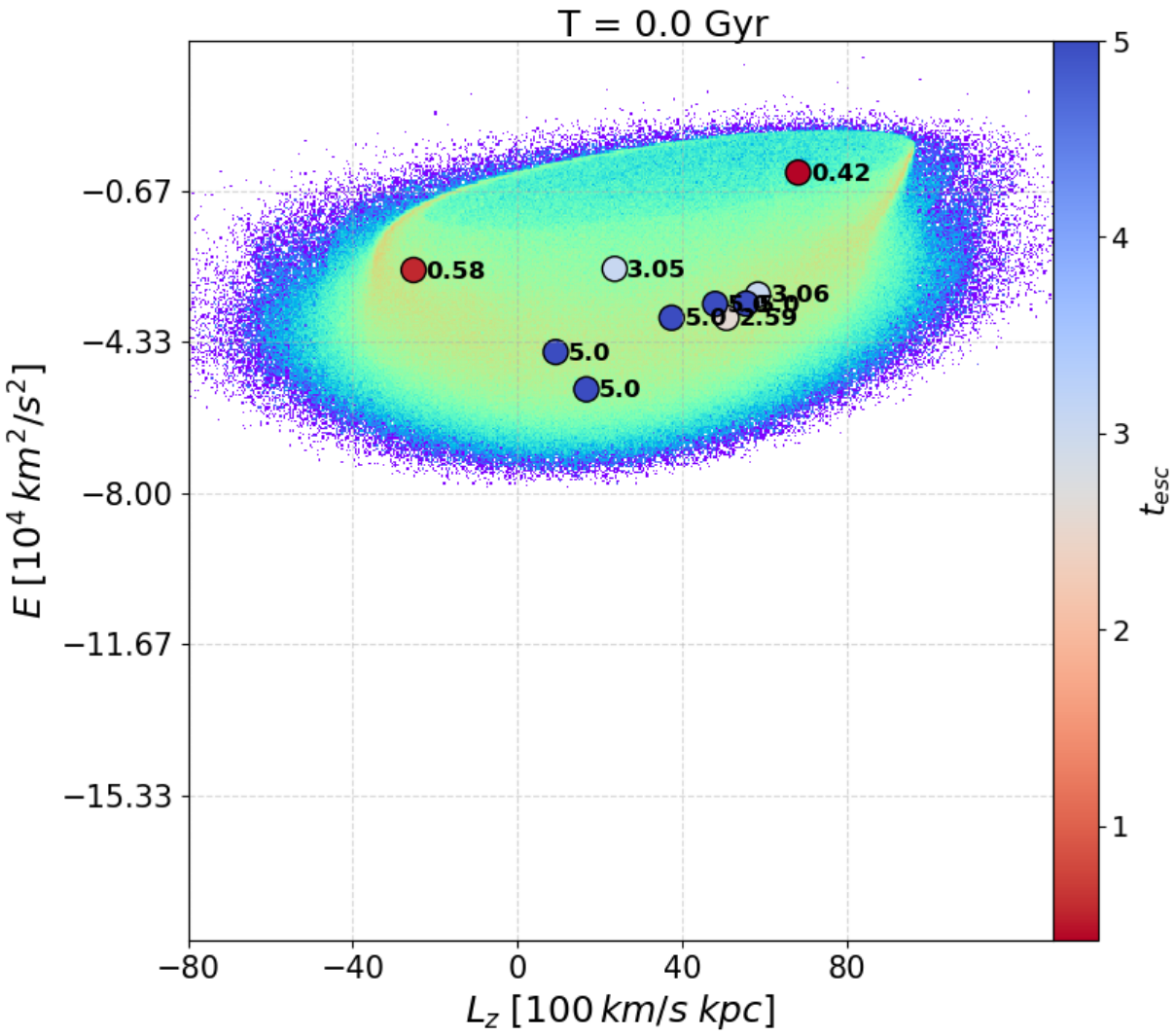} 
    \end{multicols}
    \begin{multicols}{3}
\includegraphics[width=.9\linewidth]{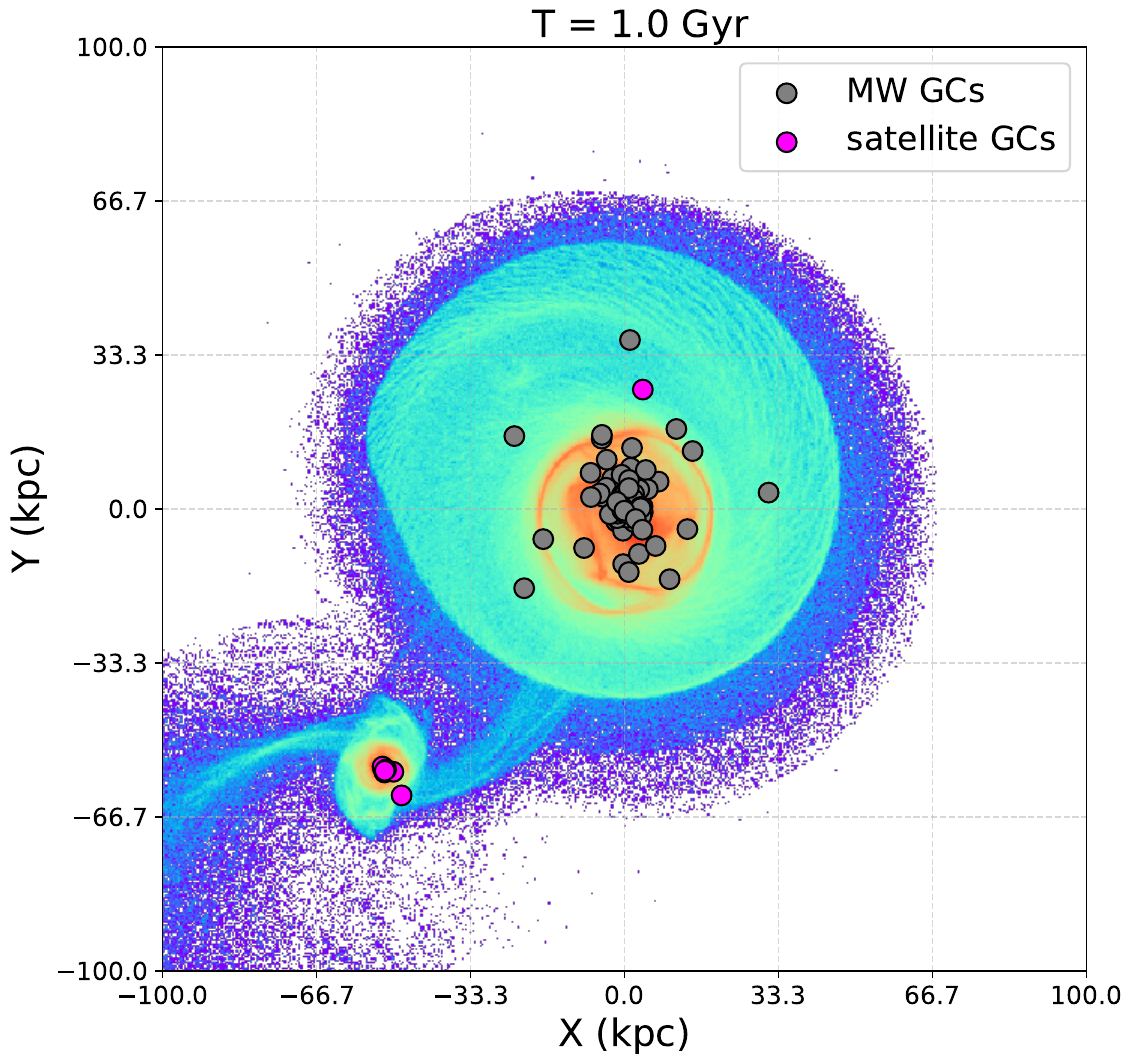}\par
\includegraphics[width=.9\linewidth]{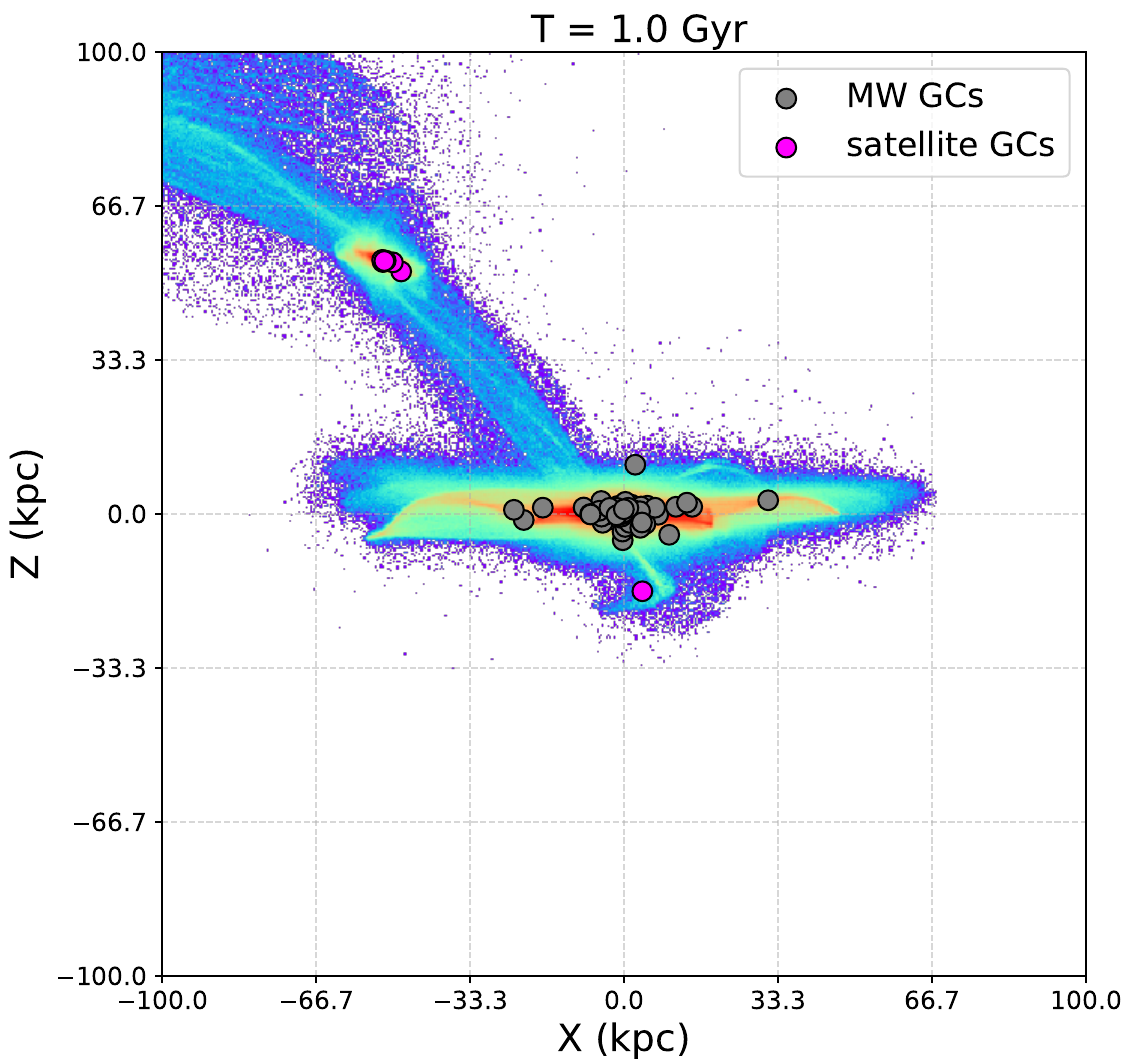}\par
\includegraphics[width=.95\linewidth]{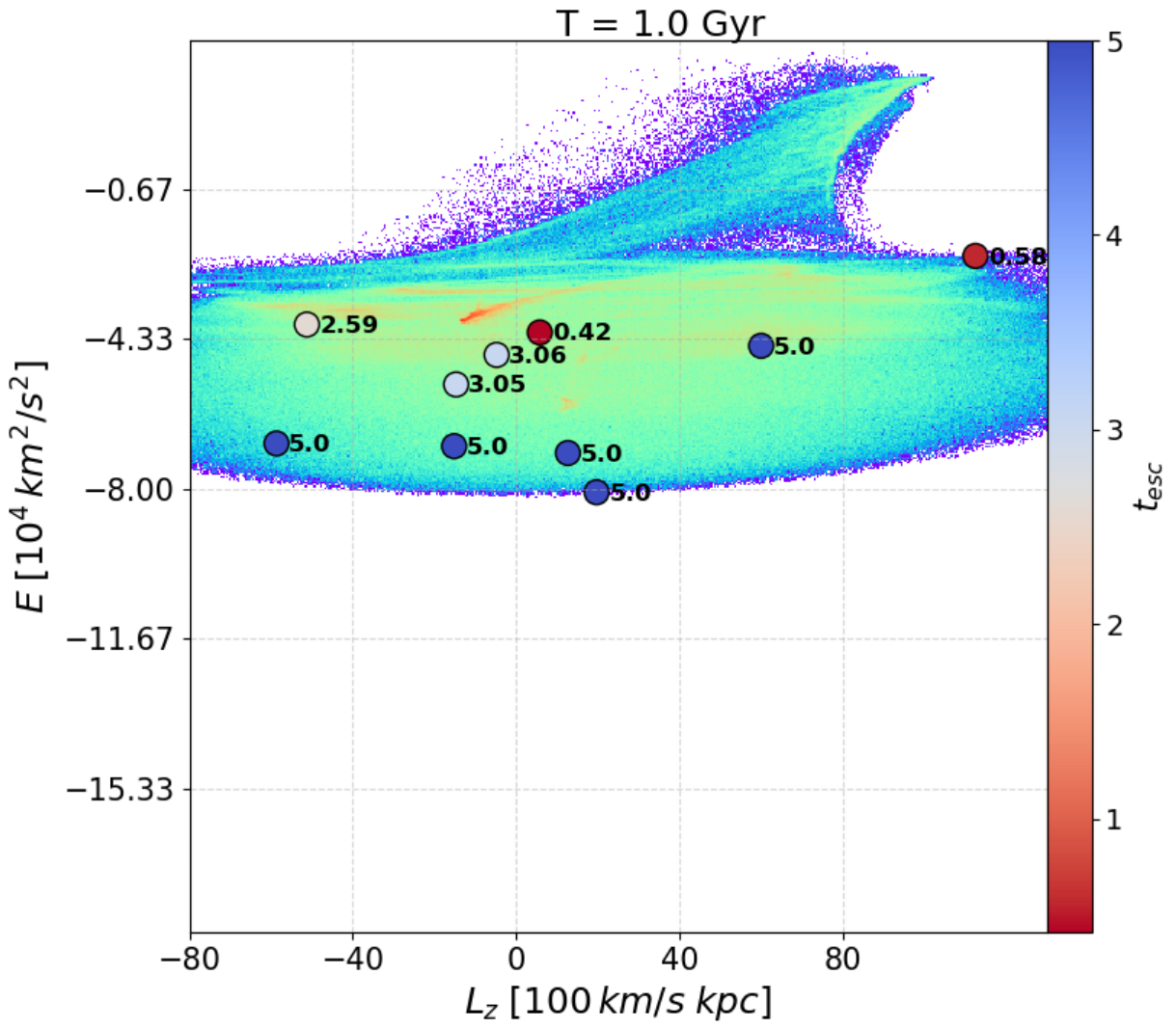}

    \end{multicols}
    \begin{multicols}{3}
\includegraphics[width=.9\linewidth]{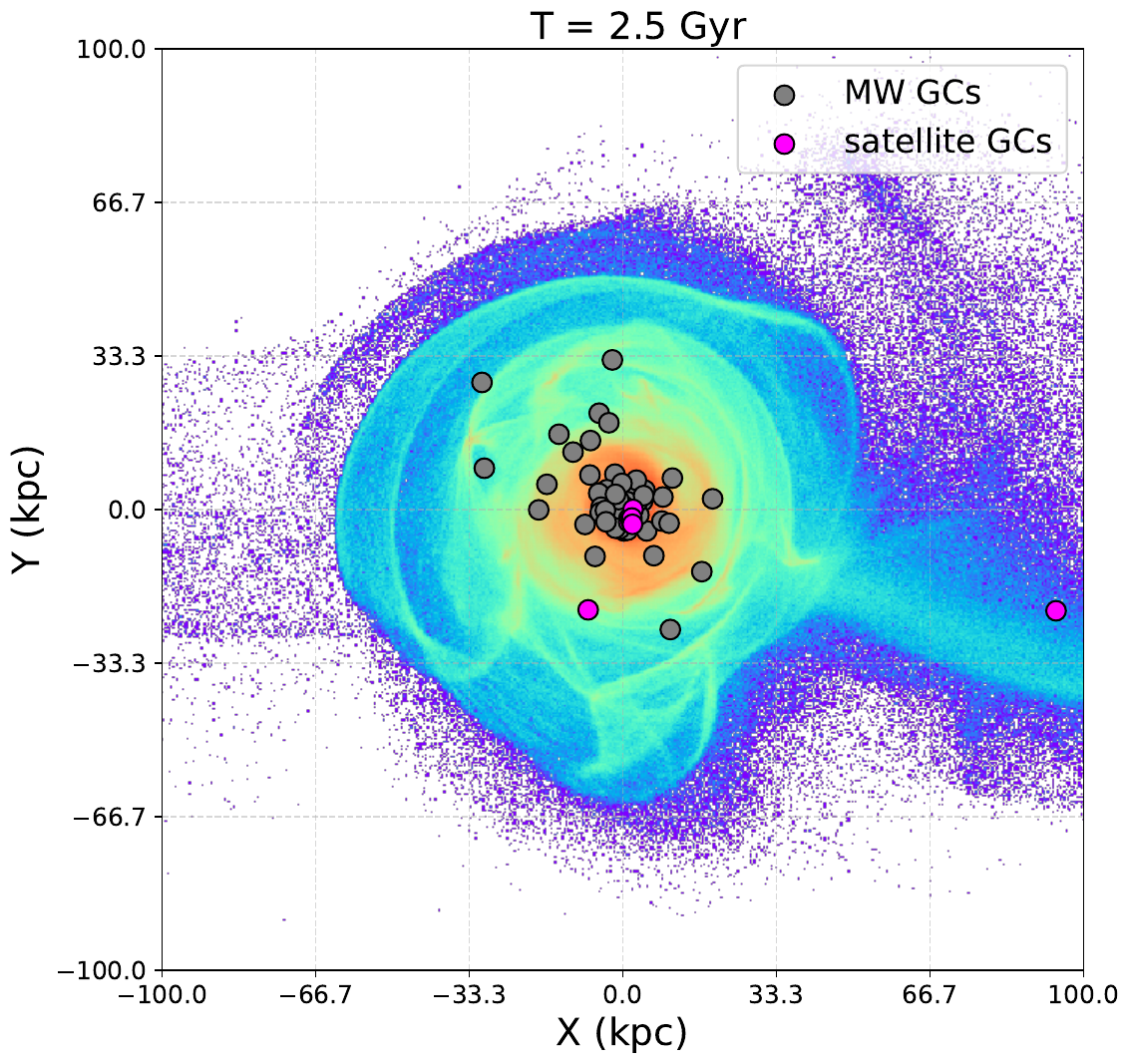}\par
\includegraphics[width=.9\linewidth]{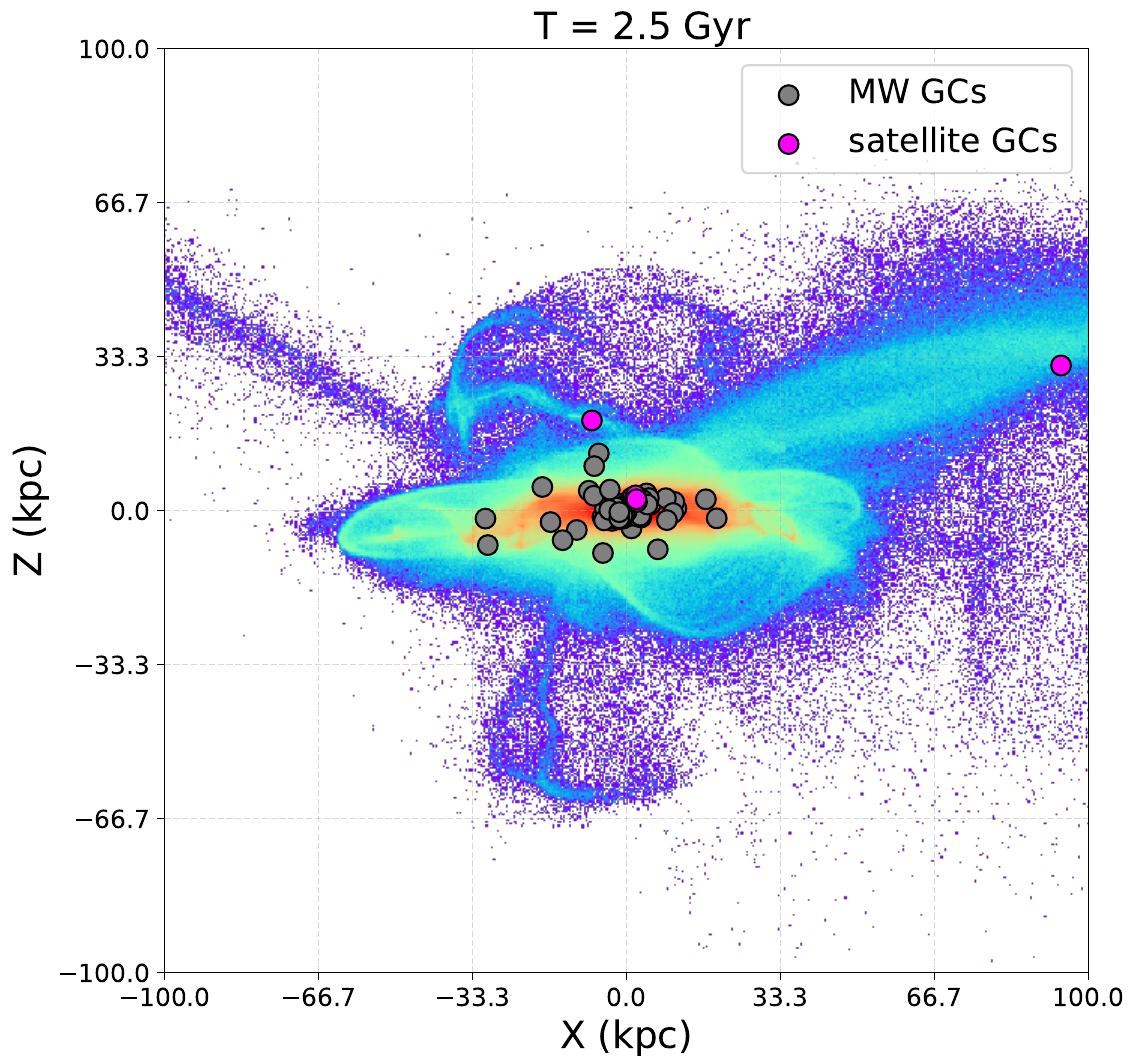}\par
\includegraphics[width=.95\linewidth]{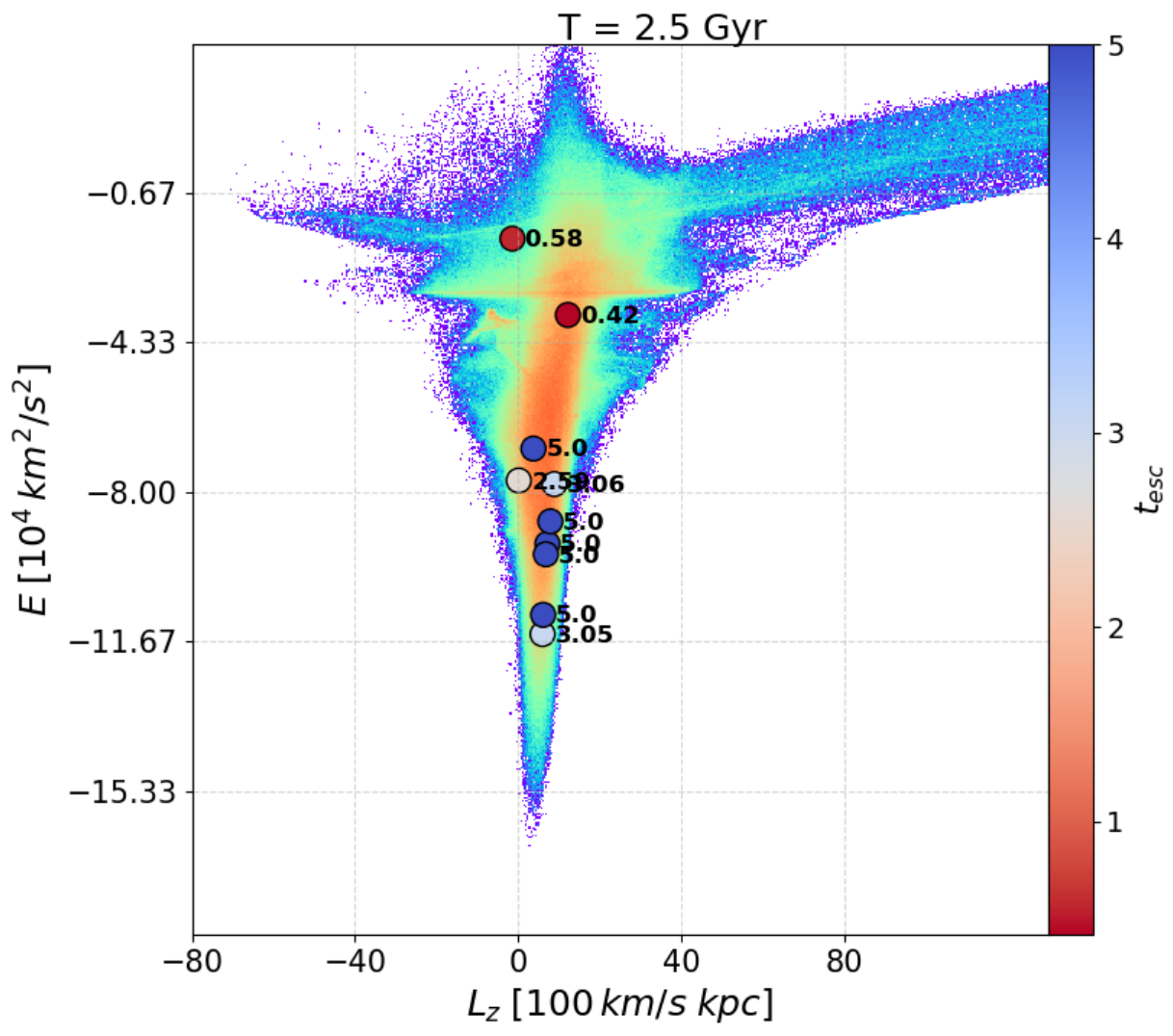}

    \end{multicols}
        \begin{multicols}{3}
\includegraphics[width=.9\linewidth]{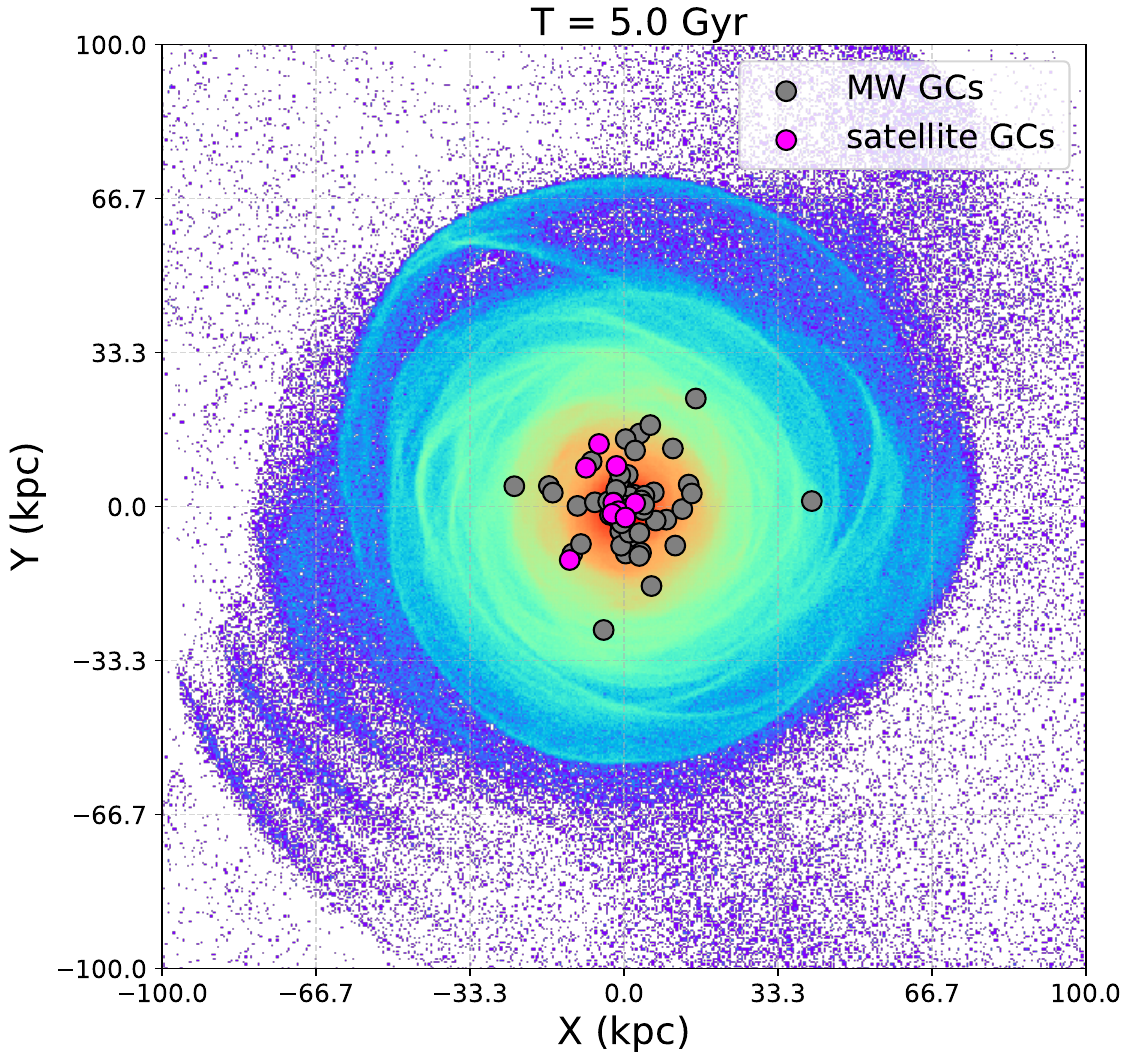}\par
\includegraphics[width=.9\linewidth]{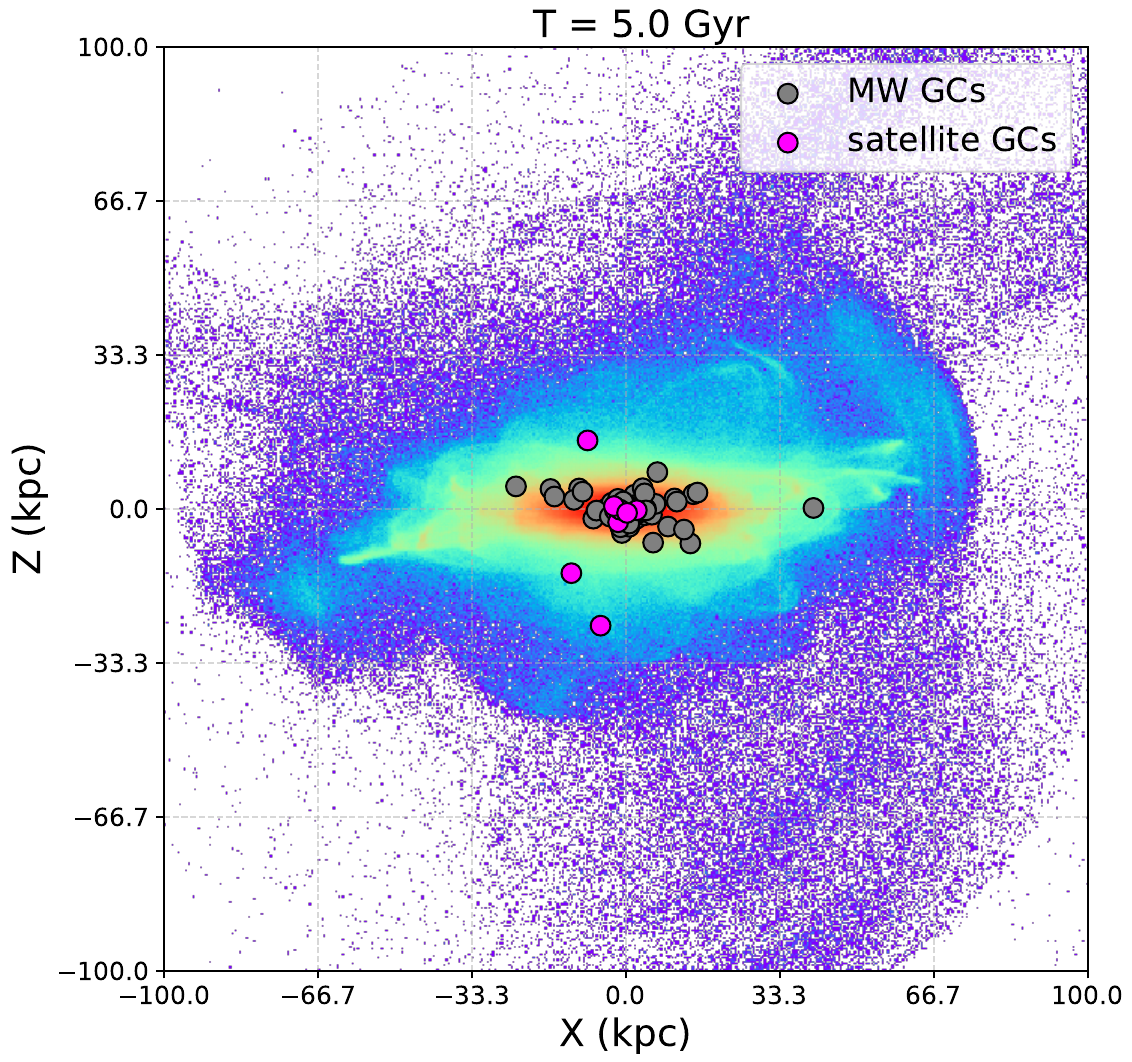}\par
\includegraphics[width=.95\linewidth]{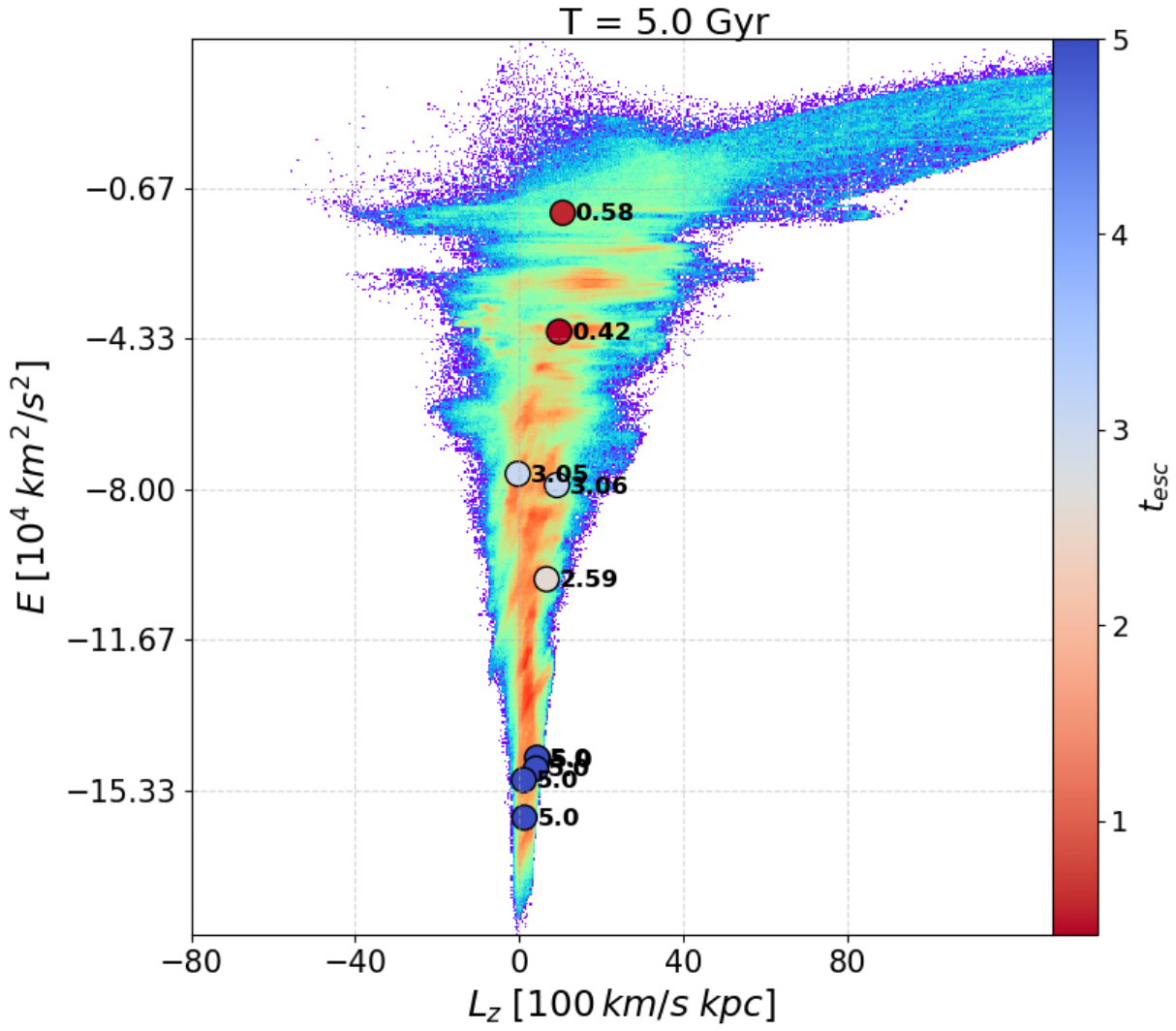} 
    \end{multicols}
\end{centering}
\vspace{-20pt}
\caption{\textit{Left, middle columns}: projections of the simulated globular clusters positions on the $xy$ and $xz$ planes, for different times (increasing from top to bottom) of the single-accretion simulation with $\Phi_{orb} = 60 \degree$ (MWsat\_n1\_$\Phi$60, see Tab.~\ref{tab:2} for the initial parameters). The in-situ clusters are represented by grey circles and the accreted clusters by magenta circles. In the background, the surface density of the totality of the stars of the simulation is also shown.
\textit{Right column}: Distributions of accreted GCs and field stars of the same satellite in the $E - L_{z}$ space, for different times (increasing from top to bottom) of the single-accretion simulation MWsat\_n1\_$\Phi$60.
Each globular cluster is colour-coded according to its escape time from the progenitor satellite which is also specified by the number on the top right and increases from 0.42 Gyr for GC1 to 5 Gyr for GC10 (see Sec~\ref{ELz} for the definition of $t_{esc}$).}
\label{fig:xy_maps}
\end{figure*}

\subsection{Additional simulations}

A couple of simulations with mass ratio 1:10 (namely the simulations with ID MWsat\_n1\_$\Phi$60 and MWsat\_n2\_$\Phi$30-150) have been rerun in a static Galactic mass distribution, artificially imposing that this latter is not influenced by the accretion of the satellite(s). This experiment allows us to study how the results of the present work would change if dynamical friction exerted by the Milky Way-like galaxy on the satellite and on the clusters was neglected (see Appendix~\ref{app_a}).\\
Finally, the 1:10 mass ratio simulations are complemented by a set of 7 simulations with mass ratio 1:100, whose initial conditions and  analysis are presented in Appendix ~\ref{app_b}, and which are aimed to show the distribution in kinematic spaces of clusters from low-mass galaxies.\\

All simulations have been run by making use of the TreeSPH code described in \cite{semelin2002formation}. Gravitational forces are calculated using a tolerance parameter\footnote{The tolerance parameter $\theta$, or opening angle, determines the precision of the force calculation in Tree-codes since it discriminates whether a group of particles sufficiently distant from another particle can be considered as one entity, or whether it has to be further subdivided into subgroups.}  $\theta = 0.7$ and include terms up to the quadrupole order in the multiple expansion. A Plummer potential is used to soften gravitational forces, with constant softening lengths for different species of particles. In all the simulations described here, we adopt $\epsilon = 50 $ pc. The equations of motion are integrated using a leapfrog algorithm with a fixed time step of $\Delta t = 2.5 \times 10^5 yr$.

In this work, we make use of the following set of units: distances are given in kpc, masses in units of $2.3\times10^9\,M_{\odot}$, velocities in units of $100\,km/s$ and $G = 1$. Energies are thus given in units of $10^4\,km^2/s^2$ and time is in unit of $10^7$ years. With this choice of units, the stellar mass of the Milky Way-type galaxy, at the beginning of the simulation, is $8.4\times10^{10}\,M_{\odot}$.

\section{Results}\label{results}


In this section, we present and discuss the effect of one or two 1:10 mass ratio accretions on the Milky Way galaxy, focusing our attention on the resulting globular cluster system. We will concentrate our analysis to understand:  how accreted and in-situ clusters distribute themselves spatially in the final galaxy; what information about the in-situ or accreted nature of clusters can be extracted from the so-called integrals-of-motion spaces; whether we can use the kinematic information contained in the Galactic globular cluster system today to trace the accretion history of the Galaxy.

We show and discuss, firstly, the spatial distribution of accreted and in-situ globular clusters (Sect.~\ref{spatial}), secondly their distribution in the $E-L_z$ plane (Sect.~\ref{ELz}), and finally we generalise our results showing also how GCs redistribute in other kinematic spaces and applying a Gaussian Mixture Model to the outcomes to check whether such an approach is able to provide robust information about the accretion history experienced by the remnant galaxy (Sec.~\ref{kin_spaces}).
\begin{figure}
\centering
\includegraphics[width=\linewidth]{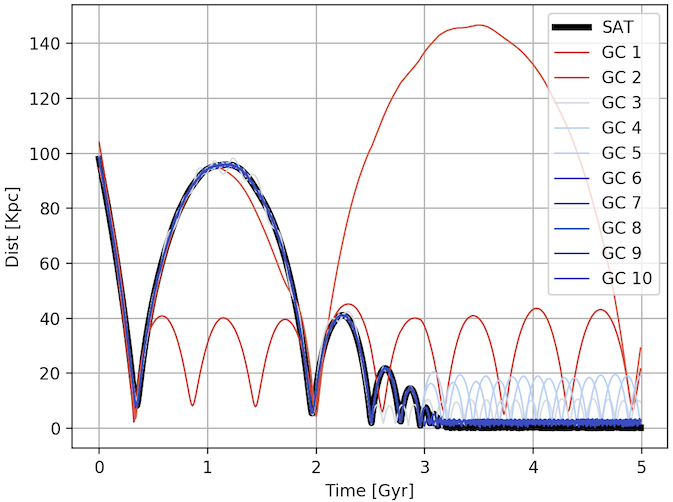} 
\includegraphics[width=\linewidth]{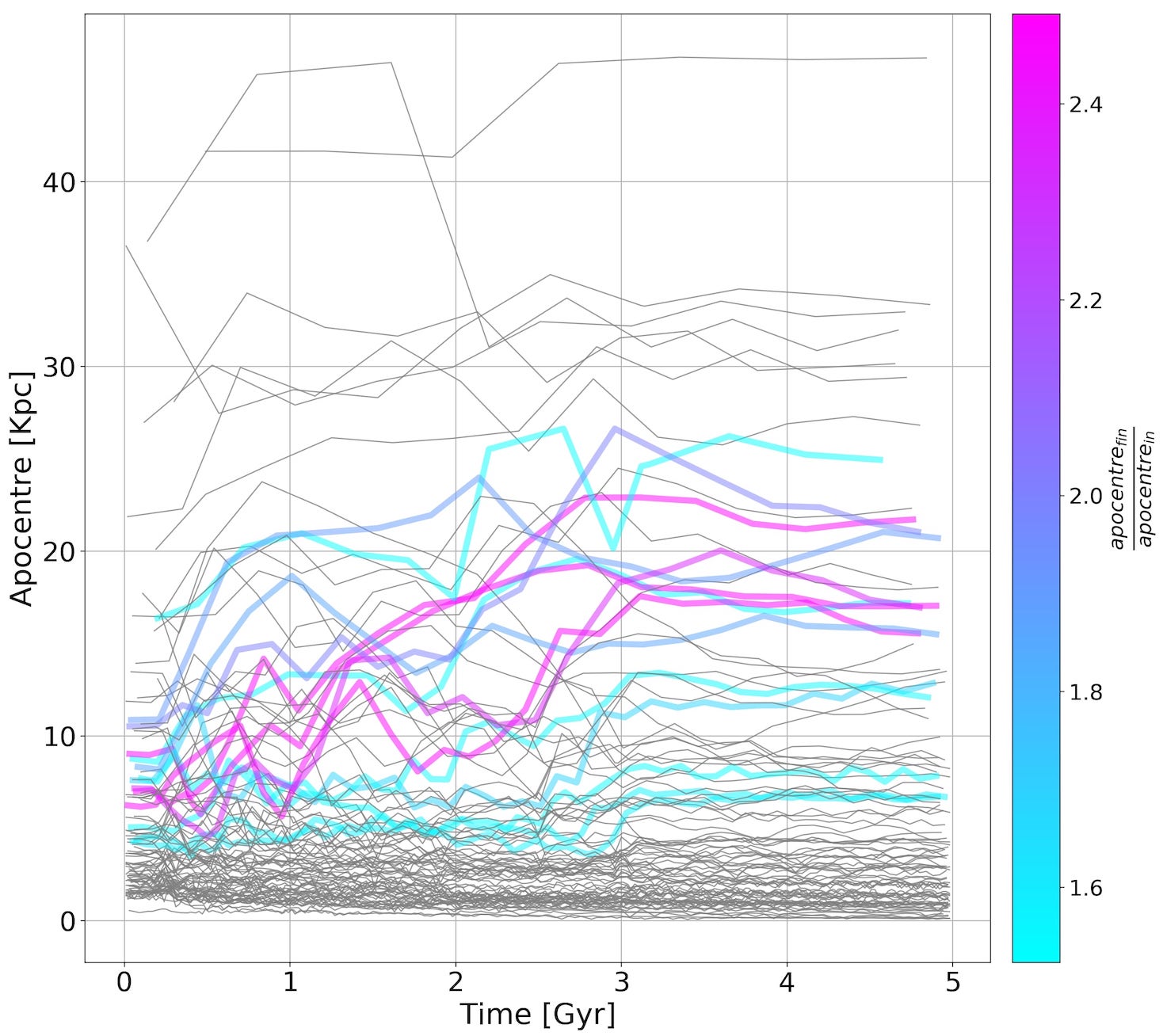}
\caption{\textit{Top panel}: Time evolution of the distances of the satellite (black line) and its globular clusters (coloured lines) from the Milky Way-type galaxy, for the simulation MWsat\_n1\_$\Phi$60. Each globular cluster is colour-coded according to its escape time from the progenitor satellite. \textit{Bottom panel}: Time evolution of the apocentres of the orbits of all in-situ clusters in the Milky Way-type galaxy, for the simulation MWsat\_n1\_$\Phi$60. Thick and coloured lines indicate clusters whose orbital apocentres, at the final time of the simulation, are at least 1.5 larger than their corresponding initial values; these are colour-coded according to the ratio of the final apocentre over the initial one.}
\label{fig:dist_evol}
\end{figure}
In the following of this analysis all quantities are evaluated in a reference frame whose origin is at the centre of the Milky Way-type galaxy with the spin of the Milky Way-type galaxy being oriented as the $z$-axis and positive.
The centre is evaluated, at each snapshot of the simulations, as the density centre, following the method described in \cite{casertano1985core}.
\begin{figure}
\centering
\includegraphics[width=\linewidth]{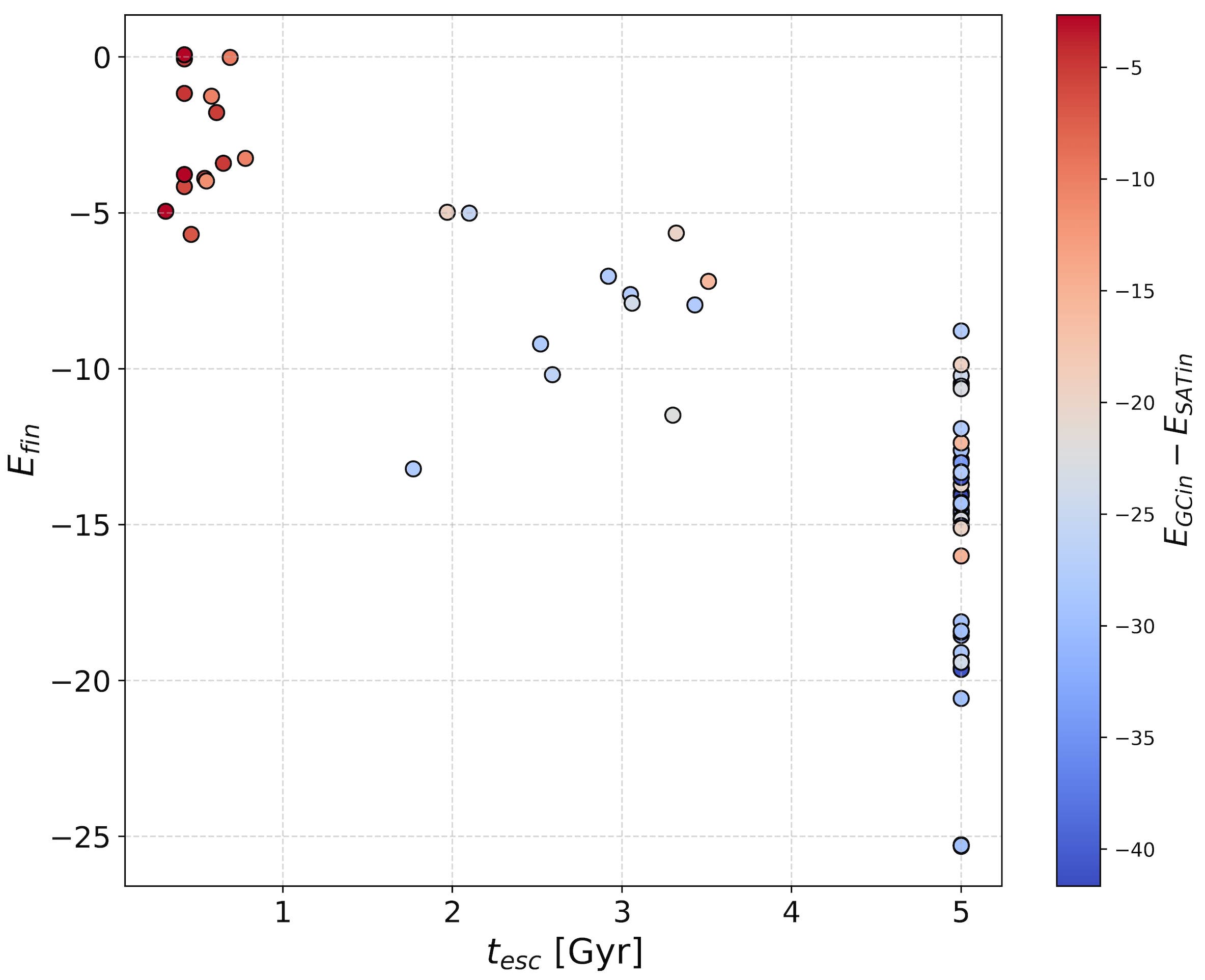}
\caption{Final energy of all the accreted clusters in the single accretion simulations as a function of their escape time from their satellites ($t_{esc}$); the color-coding indicates the clusters' initial energy in the reference frame of their progenitor satellite.}
\label{fig:e_tesc}
\end{figure}

\subsection{Spatial distribution of accreted and in-situ clusters}\label{spatial}

Fig.~\ref{fig:xy_maps} shows the globular clusters projections in the ($x, y$) and ($x, z$) planes, from one of the 7 single-accretion simulations ($\Phi_{orb} = 60 \degree$, i.e. simulation ID = MWsat\_n1\_$\Phi$60), at different times: the initial ($T = 0$~Gyr), two intermediate times ($T = 1$~Gyr, 2.5~Gyr), and at the final time (T$ = 5$~Gyr). In all the plots, the in-situ globular cluster system, i.e. originally in the Milky Way-like galaxy, is represented by grey circles, and the globular cluster system initially linked to the satellite by magenta circles. The distribution of field (in-situ and accreted) stars is also shown in the background.

At the beginning of the simulation, the Galactic and satellite globular cluster systems have both an axisymmetric disc distribution, as the corresponding field stellar particles.
This initial axisymmetric configuration of the clusters and field stars is rapidly disturbed by the first satellite passage (occurring at $T\simeq0.3$ Gyr, see Fig.~\ref{fig:dist_evol}): at time $T = 1.0$~Gyr, the satellite has disturbed the outer parts of the disc of the Milky Way-type galaxy, which has developed an extended spiral-like structure.
In the meantime, the outer parts of the satellite are tidally disrupted by the main galaxy and also one of the clusters originally associated with the satellite has escaped to join the outer parts of the massive galaxy. We can also appreciate that the vertical structure of the disc of the massive galaxy and the corresponding cluster system start to be kinematically ``heated'' at this time: seen from the edge, the disc has thickened, it shows tidal plumes in the outer parts, partly dragged by tidal pulls of the satellite, partly made of in-situ stars, disturbed by the interaction. This kinematic heating is the result of the redistribution of a fraction of the orbital energy of the satellite into internal energy of the stars in the merger remnant \citep[see, for example][]{quinn93, walker96, villalobos08, zolotov09, purcell10, dimatteo11, qu11, mccarthy12, cooper15, jean2017kinematic}. At $T = 2.5$~Gyr the merging process is toward the end and this is also visible in the plots: the stellar distribution of the satellite is no longer clearly separated from that of the main galaxy, and also most of its globular clusters are now part of the bulk of the Milky Way population. However, we can still appreciate a series of tidal plumes and two globular clusters originating from the satellite that have not yet been trapped by the Milky Way-type galaxy. At the final time of the simulation ($T = 5$~Gyr), we can see that the majority of the clusters (in-situ and accreted) are located in the inner densest regions of the final galaxy and if it was not for the different colors, they would not be distinguishable from each other. \\
The decay of part of the accreted clusters in the innermost regions of the remnant can be understood because these are the clusters that remain gravitationally bound to the satellite until the final phases of the merger. It is primarily the dynamical friction experienced by the satellite to which these clusters belong that make them decay towards the centre of the main galaxy. It is only when these clusters become gravitationally unbound to the satellite, that their motion from their progenitor galaxy decouples. This can be appreciated in Fig.~\ref{fig:dist_evol}, where we show -- for the case of the simulation MWsat\_n1\_$\Phi$60 -- the temporal evolution of the distance of the satellite to the main galaxy, together with the corresponding evolution of all the clusters initially bound to the satellite. We can see that soon after the first pericentre passage, the satellite loses one of its clusters, which is trapped on an orbit with apocentre at about 40~kpc, while the satellite itself moves to its first apocentre, at about a distance of 95~kpc from the main galaxy centre.  A second cluster is lost by the satellite at its next pericentre passage ($T = 2$~Gyr). The cluster lost at this time is ejected on a high energy orbit, which has an apocentre at more than 140~kpc from the centre of the Milky Way-type galaxy.   Over the entire duration of the merging process, this loss of the satellite  clusters from their progenitor leads to their redistribution over a large ranges of distances (pericentres and apocentres) from the merger remnant at the final time. While clusters associated to the satellite galaxy are lost at different times of the interaction, and hence contribute to populate both the outer and inner regions of the remnant galaxy, clusters initially in the Milky Way-type galaxy -- which, we remind the reader, are initially distributed in a disc-like configuration -- have their orbits perturbed by the interaction. In Fig.~\ref{fig:dist_evol} (bottom panel), we show indeed the evolution with time of the orbital apocentres - computed as the maxima of the 3D distance - of all the 100 clusters in the Milky Way-type galaxy. GCs whose orbital apocentres, at the end of the simulation, are at least 1.5 larger than their corresponding initial values are shown as thick and colour coded lines while the others as grey lines. We can see from this plot that 14\% of the clusters have their orbital apocentre increased of a factor 1.5 at least, and for some of them this heating coincides with the last phases of accretion of the satellite (around $T=2$~Gyr). The number of kinematically heated clusters increases to 27\% when we consider GCs that have their final orbital apocentre increased of at least a factor of 1.2 with respect to the initial one. 
\begin{figure*}
\begin{centering}
\begin{multicols}{3}
     \includegraphics[width=\linewidth]{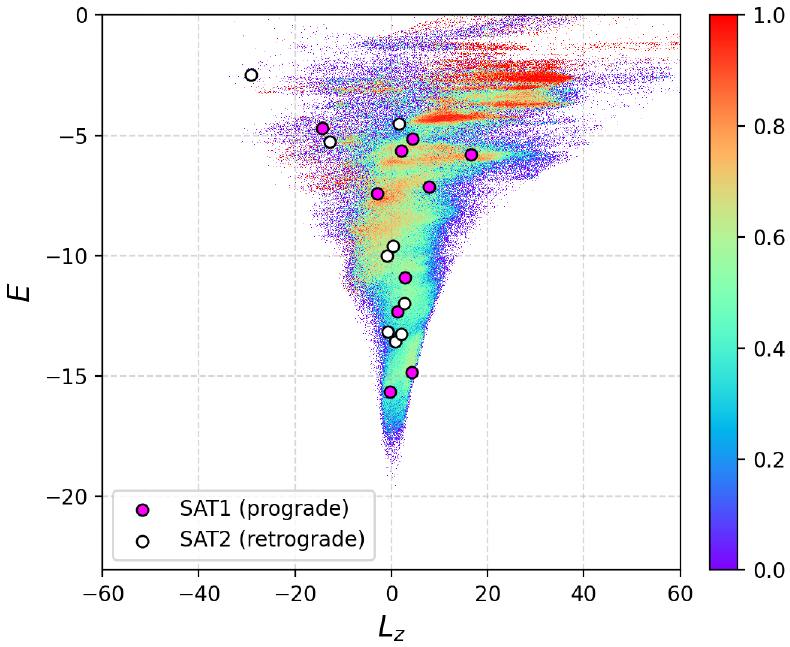} \par
         \includegraphics[width=\linewidth]{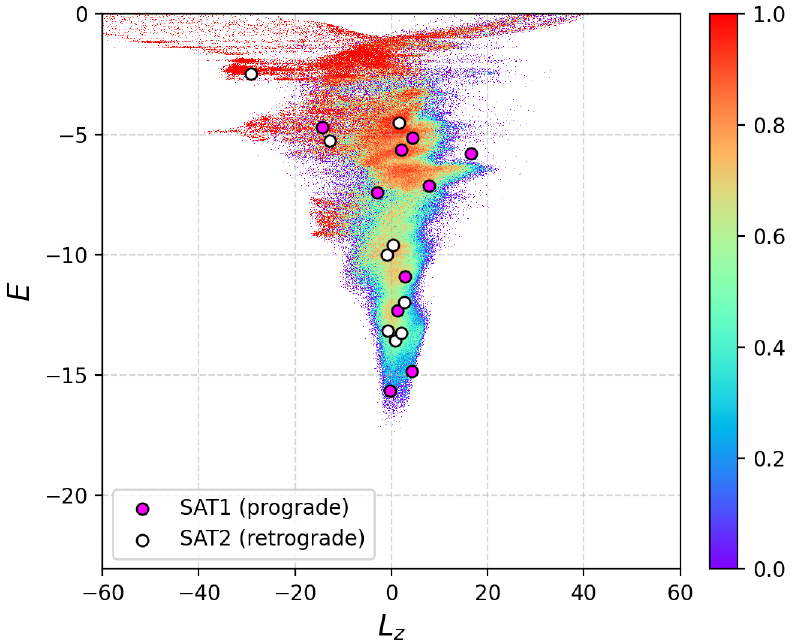} \par 
             \includegraphics[width=\linewidth]{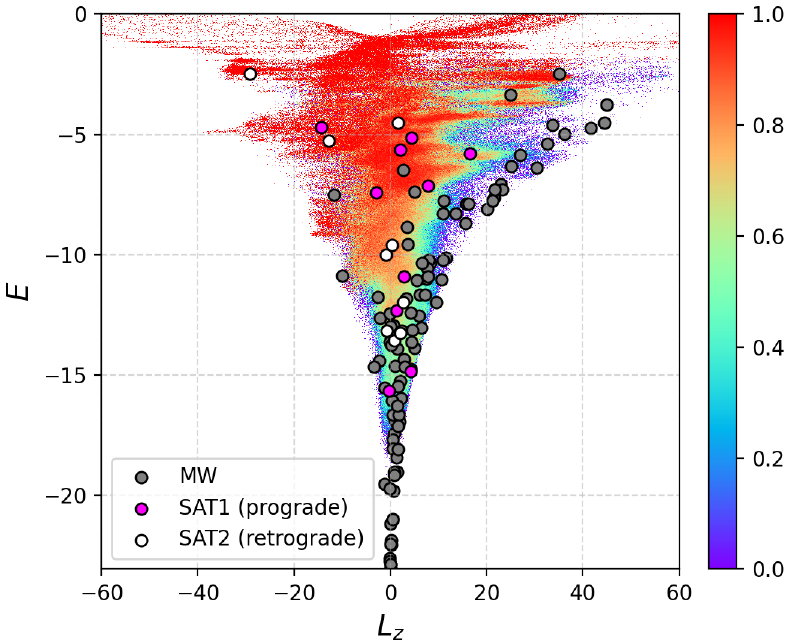} 
    \end{multicols}
    \caption{Accretion of two satellites on the Milky Way-type galaxy (MWsat\_n2\_$\Phi$30-150). \textit{Left, middle panel}: Distribution in the $E - L_z$ space of globular clusters originally belonging to the two satellites (magenta and white colors, respectively). The density maps show the fractional contribution of stars from satellite 1 (left panel) and satellite 2 (middle panel) relative to the totality of stars.
     \textit{Righ panel}: Distribution in the $E - L_z$ space of globular clusters originally belonging to the two satellites (magenta and white colors, respectively) and to the Milky Way-type galaxy (grey circles). The density map shows the fractional contribution of stars from the two satellites relative to the totality of stars.}
     \label{fig:gcs_stars}
    \end{centering}
    \end{figure*}

\subsection{Energy - angular momentum space}\label{ELz}
All the results presented in this Section concern the distribution of globular clusters in the energy - angular momentum space ($E - L_{z}$) in the case of a Milky Way-type galaxy accreting one or two satellites, since this space has been proposed as the natural one where to look for the signatures of past accretion events \citep{helmi2000mapping}.


\subsubsection{Accreted clusters}
Let us start with the case of a single accretion. 
In Fig.~\ref{fig:xy_maps} (right column), we show the distribution in $E - L_{z}$ space of the accreted clusters in the case of the $\Phi_{orb} = 60^\circ$ simulation (ID=MWsat\_n1\_$\Phi$60), at different times, T= 0, 1, 2.5, 5 Gyr (these times correspond to those for which the $x-y$ and $x-z$ maps are shown in the left and middle columns of the same figure). In these $E - L_{z}$ plots, we also report, for each cluster, the escape time from its progenitor satellite. This escape time has been estimated by means of a simple spatial criterion,  i.e. it is defined as the time when the distance between the GC and the centre of mass of the parent satellite is larger than 15 kpc. 
For comparison, the distribution in $E - L_{z}$ space of field stars of the satellite is also plotted in the background. 
At the beginning of the interaction, the distribution of satellite stars and globular clusters is clumped in the $E - L_{z}$ space and characterized by high energy. This is understandable because the satellite is, at the beginning of the simulation, a gravitationally bound system. At $T = 1$~Gyr the satellite is close to an apocentre passage (see Fig.~\ref{fig:dist_evol}), resulting in a broad extent in $L_{z}$\footnote{Before the end of the merging process, a fraction of satellite stars and globular clusters are still gravitationally bound to the system. For this reason, together with the motion of the centre of mass of the satellite relative to the Milky Way centre, one needs to take into account also the motion of satellite stars/GCs relative to the satellite centre. Since the angular momentum $L_z$ depends on the distance from the centre of the main galaxy, the effect of the peculiar velocities is particularly evident at large distances from the Milky Way centre (i.e. at the apocentre), and it is reduced when the satellite is at its pericentre.}.
Globally, as an effect of dynamical friction, the energy and the absolute value of the angular momentum decrease and the satellite penetrates deeper and deeper in the potential well of the main galaxy. This results in a distribution with a funnel-like shape as satellite's stars and GCs tend to be spread in energy but to converge in angular momentum. Table~\ref{tab:3}, that lists the mean and standard deviation of the initial and final $L_z$ and $E$ of the accreted GCs, quantifies well this trend. In fact, the final $L_z$ results on average closer to zero and less spread, while the energy is on average lower and more dispersed.  In general, clusters lost in late phases of the interaction have low orbital energies, that is they tend to be found in the potential well of the merger remnant, while clusters lost in the early phases of the satellite accretion tend to be positioned in the upper part of the $E - L_{z}$ plane, i.e. at high energies (we refer the reader to Appendix~\ref{app_a} to see how this behaviour changes if we consider a static MW potential where the satellite does not experiences dynamical friction). This tendency is clear in Fig.~\ref{fig:e_tesc}, where the final energy of all the accreted clusters in all the 1x(1:10) simulations is shown as a function of the escape time from their satellites ($t_{esc}$): clusters lost in the early phases of the interaction (low $t_{esc}$) tend to have higher energies than clusters lost at more advanced stages of the merging process. Here we can also identify three main groups: the first consisting of GCs with $t_{esc}<1$~Gyr, the second with $1.8$~Gyr$\,<t_{esc}<3.6$~Gyr and the last with $t_{esc}=5$~Gyr. The first two groups are associated with globular clusters lost at the first and subsequent pericentric passages, respectively. 
Globular clusters with $t_{esc}=5$ Gyr are those which did not escape from the satellite before the end of the merging process. From the color-coding of globular clusters in Fig.~\ref{fig:e_tesc}, it is possible also to notice that the clusters that escape the earliest from the progenitor satellite are those which are initially less bound to their progenitor. They are indeed the clusters with the highest values of $E_{int}$, where  $E_{int}$ is the sum of the kinetic energy of clusters in the satellite and of their potential energies, both estimated relative to the satellite centre. As expected, globular clusters that are more tightly bound to the satellite tend to escape later and have a lower final energy relative to the Milky Way-type galaxy reference system. However, from Fig.~\ref{fig:xy_maps} (bottom, right panel) we can see that this trend shows a number of exceptions: for instance, the cluster lost at $\simeq 2.59$ Gyr, at the end of the simulation has lower energy ($E\approx -9.5$) with respect to the clusters lost later at $\simeq 3.05$ Gyr ($E\approx -8.0$). The globular cluster lost at 0.58 Gyr also ends up at higher energy than the cluster lost at 0.42 Gyr. This happens because after leaving their parent satellite, the energy and angular momentum of globular clusters may change due to changes in the gravitational potential induced by the final phases of accretion of the same satellite.

\begin{figure}
\centering
\includegraphics[width=\linewidth]{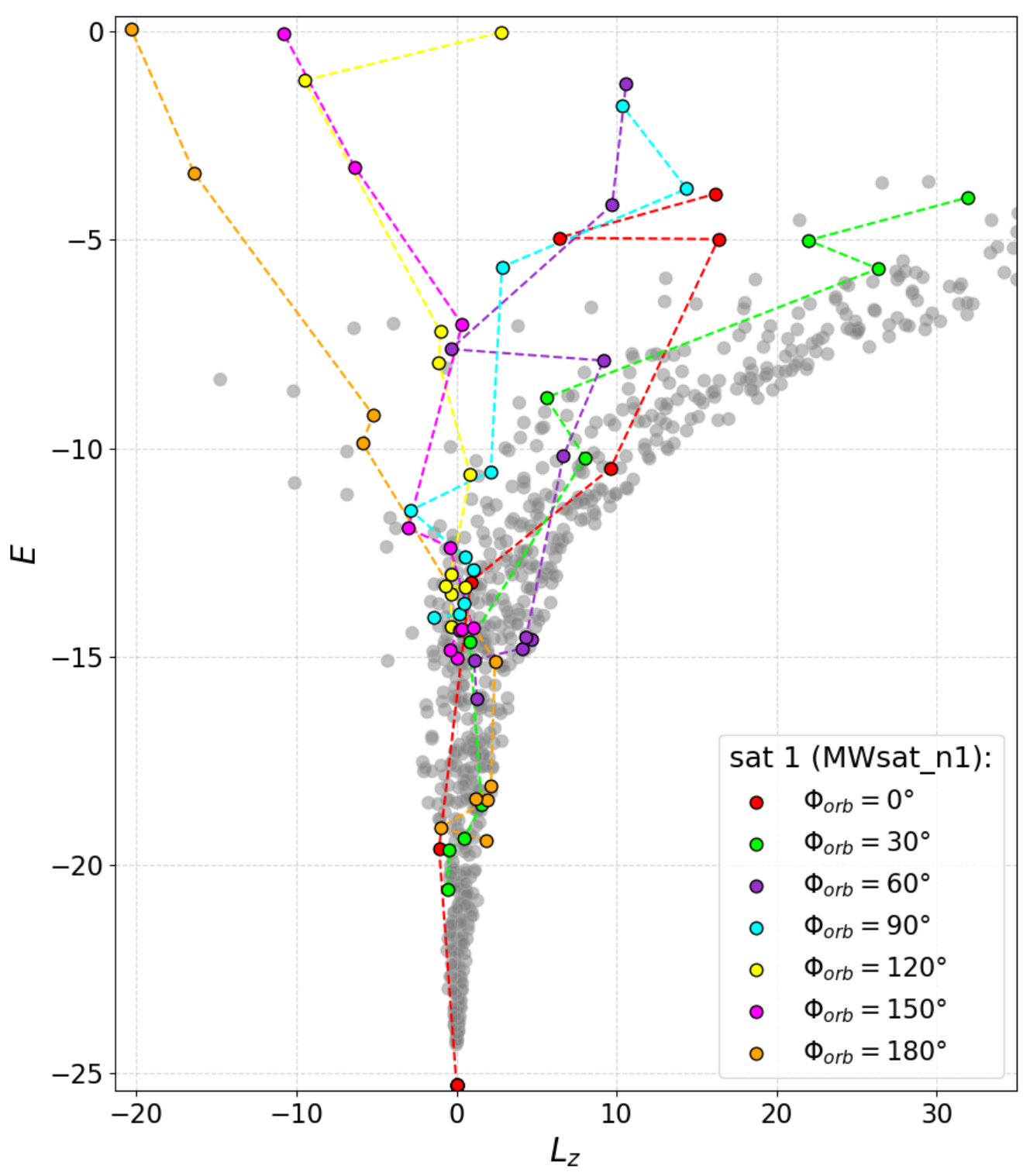}
\caption{$E - L_z$ distribution for the whole set of 1x(1:10) merger simulations at T = 5 Gyr. The distribution of the in-situ population of globular clusters has been stacked and is shown by the grey circles. The color-coding of accreted
GCs is different according to the initial inclination of the satellite orbital plane with respect to the Milky Way-like reference frame, $\Phi_{orb}$.}
\label{fig:MW1}
\end{figure}

\tiny
\begin{table*}
\centering
\resizebox{.8\textwidth}{!}{
\begin{tabular}{|l|cclc|cclc|}
\hline
\small
Simulation ID &
 \multicolumn{1}{l}{$\overline{L}_{z_{in}}$} &
  \multicolumn{1}{l}{$\sigma_{z_{in}}$} &
  $\overline{L}_{z_{fin}}$ &
  \multicolumn{1}{l|}{$\sigma_{z_{fin}}$} &
  \multicolumn{1}{l}{$\overline{E}_{in}$} &
  \multicolumn{1}{l}{$\sigma_{E_{in}}$} &
  $\overline{E}_{fin}$ &
  \multicolumn{1}{l|}{$\sigma_{E_{fin}}$} \\ \hline
MWsat\_n1\_$\Phi$0   & 66.17  & 52.51 & \multicolumn{1}{c}{4.84} & 6.57  & -3.35 & 1.75 & -15.83 & 8.87 \\ 
MWsat\_n1\_$\Phi$30  & 57.62  & 45.64 & \multicolumn{1}{c}{9.58} & 11.76 & -3.30 & 1.52 & -12.65 & 6.31 \\ 
MWsat\_n1\_$\Phi$60  & 34.27  & 26.91 & \multicolumn{1}{c}{5.13} & 3.65  & -3.28 & 1.32 & -10.61 & 4.94 \\ 
MWsat\_n1\_$\Phi$90  & 2.39   & 4.10  & \multicolumn{1}{c}{2.77} & 5.12  & -3.33 & 1.37 & -10.06 & 4.35 \\ 
MWsat\_n1\_$\Phi$120 & -29.49 & 25.22 & \multicolumn{1}{c}{-0.91} & 3.07  & -3.42 & 1.61 & -9.44  & 4.99 \\ 
MWsat\_n1\_$\Phi$150 & -52.83 & 43.94 & \multicolumn{1}{c}{-1.91} & 3.62  & -3.53 & 1.80 & -10.75 & 5.12 \\ 
MWsat\_n1\_$\Phi$180 & -61.41 & 50.81 & \multicolumn{1}{c}{-3.93} & 7.80  & -3.62 & 1.82 & -13.10 & 6.75 \\ \hline \hline
\end{tabular}}
\caption{Mean and standard deviation of the initial and final $L_z$ and $E$ for the populations of accreted GCs within the different single accretion simulations.}
\label{tab:3}
\end{table*}
\normalsize
When two satellites are accreted, the interpretation of the energy-angular momentum space becomes even more tricky \citep[see also][]{trelles2022concurrent}. Figure~\ref{fig:gcs_stars} (left, middle panel) shows the distribution in the $E - L_z$ space of globular clusters originally belonging to the two satellites (magenta and white colors, respectively) for one of the two accretion simulations (MWsat\_n2\_$\Phi30-150$). The background density maps show the fractional contribution of stars from satellite 1 (left panel) and satellite 2 (middle panel) relative to the totality of stars. As we can see from the figures, the globular cluster populations of the two satellites mix and overlap with each other. Furthermore, the cluster distribution does not necessarily coincide with the distribution of stars of the same satellite. For example, nearly half of the globular cluster population belonging to satellite 1 is redistributed in a region of the $E - L_z$ space dominated by stars of satellite 2 (see middle panel). The contribution of stars belonging to satellite 1 is still significant in this region, but not exclusive. Therefore, the stellar halo substructures and accreted GCs located in the same $E - L_z$ regions can not be directly associated with the same dwarf galaxy progenitor accreted in the past: regions of the $E-L_z$ dominated by stellar populations of a given progenitor can indeed contain a significant fraction of clusters (and stars) originating from a different progenitor.

\subsubsection{In-situ clusters}
The right panel of Fig.~\ref{fig:gcs_stars} shows the same distribution in the $E - L_z$ space of globular clusters originally belonging to the two satellites with the addition of GCs belonging to the Milky Way-type galaxy (grey circles). The density map illustrates the fractional contribution of stars from the two satellites relative to the totality of stars. 
This panel allows us to show an additional result: part of the in-situ cluster population overlaps with the accreted clusters in $E-L_z$ space. This in-situ population is made of disc clusters which have been perturbed enough by the interaction to be "pushed" into the halo (see also Fig.~\ref{fig:dist_evol}, bottom panel), following the same dynamical mechanism extensively discussed for in-situ field stars \citep[see, for example, ][]{zolotov09, purcell10, qu11, jean2017kinematic, khoperskov22a}. We emphasize that, by construction, our modelled Milky Way-type galaxy does not contain, before the interaction(s), any population of halo clusters (they are all initially confined in the disc). This implies that the overlap of in-situ GCs with satellites GCs could be even more significant, if part of the in-situ population had  halo-like kinematics already before the accretion(s). We further address this issue in Sect.~\ref{insitu}.
Interestingly, if we compare the distribution of globular clusters with the distribution of stars in the right panel of Fig.~\ref{fig:gcs_stars}, we can notice that a part of the in-situ GC population ends up in regions of $E - L_z$ space dominated by stars accreted from the two satellites. 
All this has as a consequence that, in the interpretation of the $E-L_z$ space, we cannot simply associate stars and clusters to the same origin (i.e.  the same progenitor galaxy) simply because they are found at similar values of $E$ and $L_z$: clusters from a satellite can be found in regions where the stellar density distribution is dominated by another satellite (middle panel, Fig.~\ref{fig:gcs_stars}), and clusters originally in the main galaxy can be found in regions of the $E-L_z$ space dominated by accreted stars (right panel Fig.~\ref{fig:gcs_stars}). \\

About the possibility to separate in-situ from accreted clusters, it is interesting to cite the numerical work by \citet{callingham2022chemo}, who constructed mock catalogues of accreted and in-situ clusters from the AURIGA simulations, reporting that ``the total in-situ population is, in general, well recovered with very high purity. Rarely does the methodology misidentify an accreted GC as an in-situ one in our mock tests, with a median purity of 98\%. ” This conclusion is clearly in contradiction with our results, but can be explained by the fact that in \citet{callingham2022chemo} work,  while their accreted GC populations are drawn from the AURIGA simulations, the in-situ GCs is added a posteriori (that is, it is not extracted self-consistently from the AURIGA simulations, as for the accreted population). In their work, the in-situ GCs  are constructed to be either in the bulge or in the disc, that is they do not allow the possibility that part of the in-situ GC population can have a halo-like kinematics (as it is the case, as we have shown, when part of a pre-existing disc GCs population is heated by a merger). The reason why they can separate so well in-situ disc GCs from accreted GCs in kinematic spaces (see for example the $E-L_z$ diagram shown in their Fig 3) is due to the way they construct the in-situ GC population, and it is not the result of a self-consistent evolution of the in-situ population itself during the merger(s). It would have been interesting, and more realistic, to draw also the in-situ GC population from their AURIGA simulations, by randomly extracting stars formed in the AURIGA MW-like progenitors. If this was the case, we argue that the separation between in-situ and accreted components would have been more challenging also in their models.

\subsubsection{Overlap in the $E - L_z$ space}
Figure~\ref{fig:MW1} shows the final globular cluster distribution in the $E - L_z$ space for the whole set of 1x(1:10) accretion simulations.
The distribution of the in-situ globular clusters from all the simulations at T = 5 Gyr has been stacked and is shown by grey circles. Accreted clusters are colour-coded circles according to the different $\Phi_{orb}$ and linked by a dashed line.
This figure summarise the results presented so far: regardless of the initial inclination of the satellite orbital plane, the overlap between accreted and in-situ globular clusters is clear and becomes critical in the most gravitationally bound regions. This may be partly caused by the narrowing of the $L_z$ range at high binding energies, but also may reflect the tendency of high-mass-ratio mergers to radialize their orbits \citep[e.g.][]{amorisco2017contributions, naidu20, vasiliev2022}. The figure also shows that, in the case of a single accretion, the angular momenta of clusters at high energies ($E \gtrsim -10$ in our units) correlate with the inclination of the orbital plane of the infalling satellite at early times: the higher the orbital inclination $\Phi_{orb}$ of the parent satellite,  the less prograde (lower $L_z$) the orbits of the clusters are. We will see in the following that this trend is less prominent in the case of two accretions.

\begin{figure*}
\begin{multicols}{2}
 \subfloat{\includegraphics[width=\linewidth]{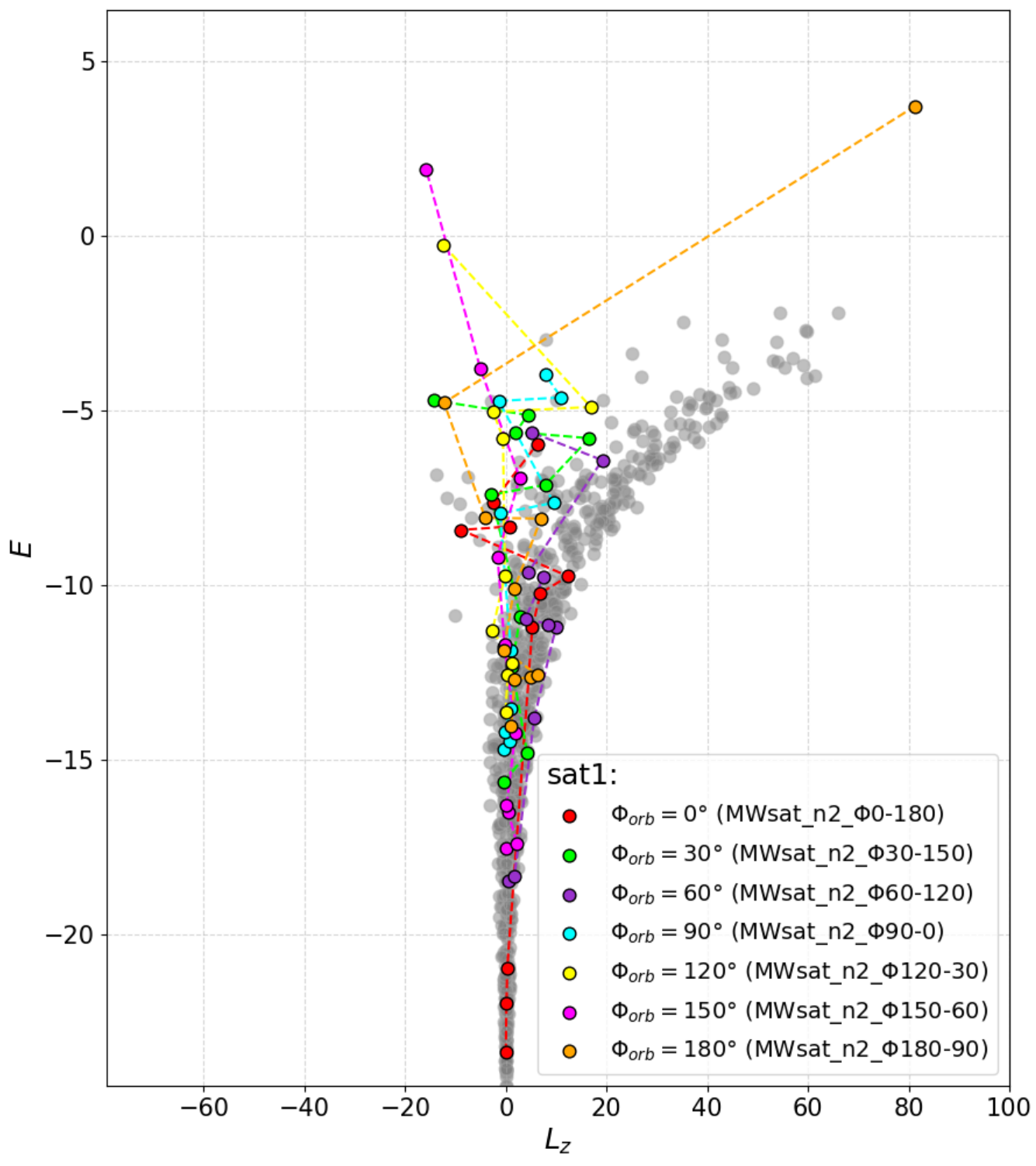}}
 \subfloat{\includegraphics[width=\linewidth]{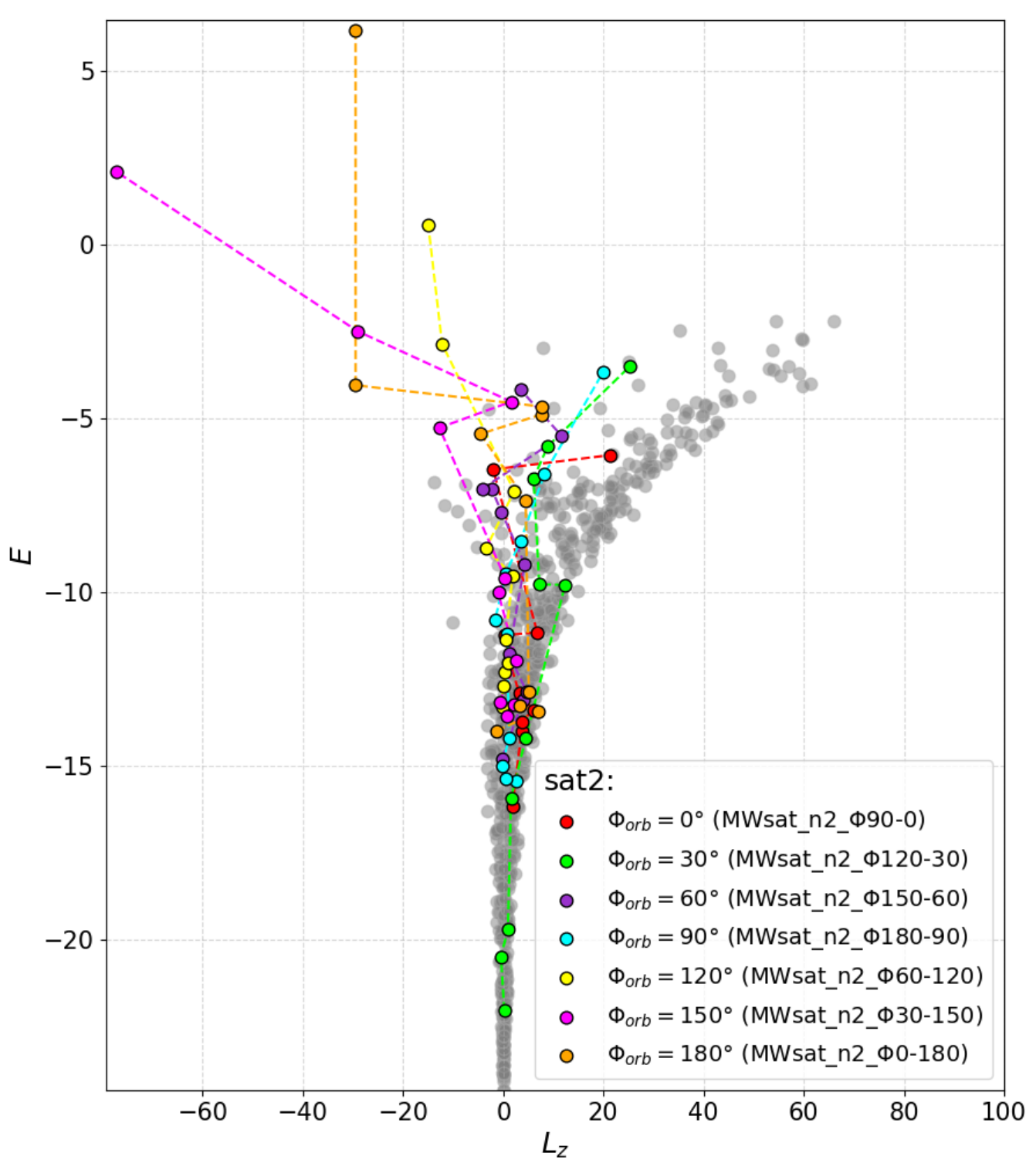}} 
\end{multicols}
\caption{$E - L_z$ distribution for the whole set of 2x(1:10) merger simulations at T = 5 Gyr. The distribution of the in-situ population of globular clusters has been stacked and is shown by the grey circles. The color-coding of satellites GCs (satellite 1: left panel, satellite 2: right panel) is different according to the initial inclination of the satellite orbital plane with respect to the Milky Way-like reference frame, $\Phi_{orb}$.\\
}
\label{fig:MW2}
\end{figure*}

Fig.~\ref{fig:MW2} shows the distribution in the energy-angular momentum plane of the whole set of globular clusters in the 2x(1:10) accretions simulations. The distribution of the in-situ population of globular clusters has been stacked and is shown by grey circles, while accreted GCs (belonging to satellite 1: left panel, to satellite 2: right panel) are shown as colour-coded circles according to the different $\Phi_{orb}$ and linked by a dashed line. 
As we can see in Fig.~\ref{fig:MW2}, in-situ and accreted globular clusters overlap almost everywhere in the $E - L_{z}$ space. Only for $E\gtrsim-6$ and/or $L_z\lesssim-3$ in principle we can distinguish the two groups. Even in that case, however, it is not possible to identify from which of the two satellites the clusters originate because, as mentioned above, these clusters appear mixed, that is clusters originating from different satellites can end up having similar energies and angular momenta. Moreover, the trend between $\Phi_{orb}$ and the final $L_z$ found for high-energy clusters in the case of a single accretion is clearly washed out in the case of two mergers. Looking, for example, at the case of satellites accreted with an initial orbital inclination of $\Phi_{orb}=180^\circ$ (as for the simulations MWsat\_n2\_$\Phi$0-180 and MWsat\_n2\_$\Phi$180-90), the final distribution of their clusters in the $E-L_z$ plane is different for the two 2x(1:10) simulations (compare the top-left and top-right panels of Fig.~\ref{fig:MW2}). This distribution appears also different from the one generated by clusters whose progenitor is on a $\Phi_{orb}=180^\circ$ inclination orbit, in the case of a single accretion (Fig.~\ref{fig:MW1}). This indicates that -- unless there are strong reasons to think that the Galaxy experienced only a main massive accretion (and not a few) -- we cannot infer from the current distribution of its accreted globular cluster population the inclination of the progenitor satellite. \\

From this analysis, we conclude that the only merger events that in principle could be distinguished in the $E - L_{z}$ plane are: (1) those that are still occurring and for which the population of GCs that the satellite is bringing with itself populates a rather delimited region at very high energies \citep[as it is the case for the Sagittarius dwarf galaxy and its globular cluster system, see][]{bellazzini2020globular}; (2) those that occurred in the past but which were characterised by a smaller mass ratio, typically 1:100 and lower \citep[as discussed in][]{jean2017kinematic}. In this case, stellar debris tend to be distributed in the high-energy regions of the $E-L_z$ plane \citep[see Sect.~4.4 in][and also \citet{pfeffer20}]{jean2017kinematic}, and a similar behaviour is found also for the associated globular cluster population. We refer the reader to Appendix \ref{app_b} for the distribution in the $E - L_{z}$ space obtained when considering the Milky Way accreting four 1:100 mass ratio satellites.  
\\[6pt]


\subsection{Other kinematic spaces: $L_{perp} - L_z$, $eccentricity - L_z$, and action space}\label{kin_spaces}
\begin{figure*}
\begin{centering}
\begin{multicols}{2}
     \includegraphics[width=.95\linewidth]{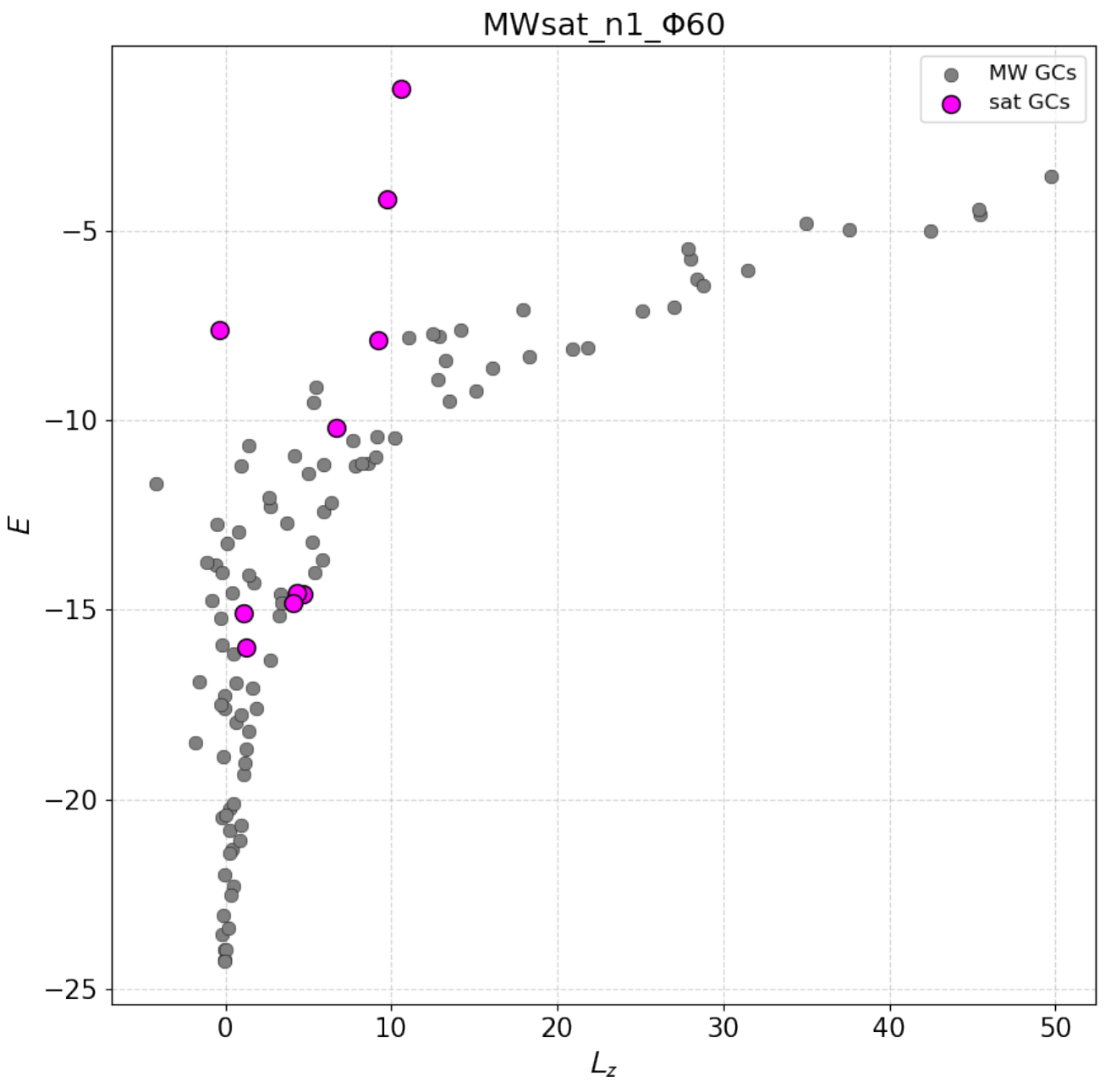}\par
         \includegraphics[width=.95\linewidth]{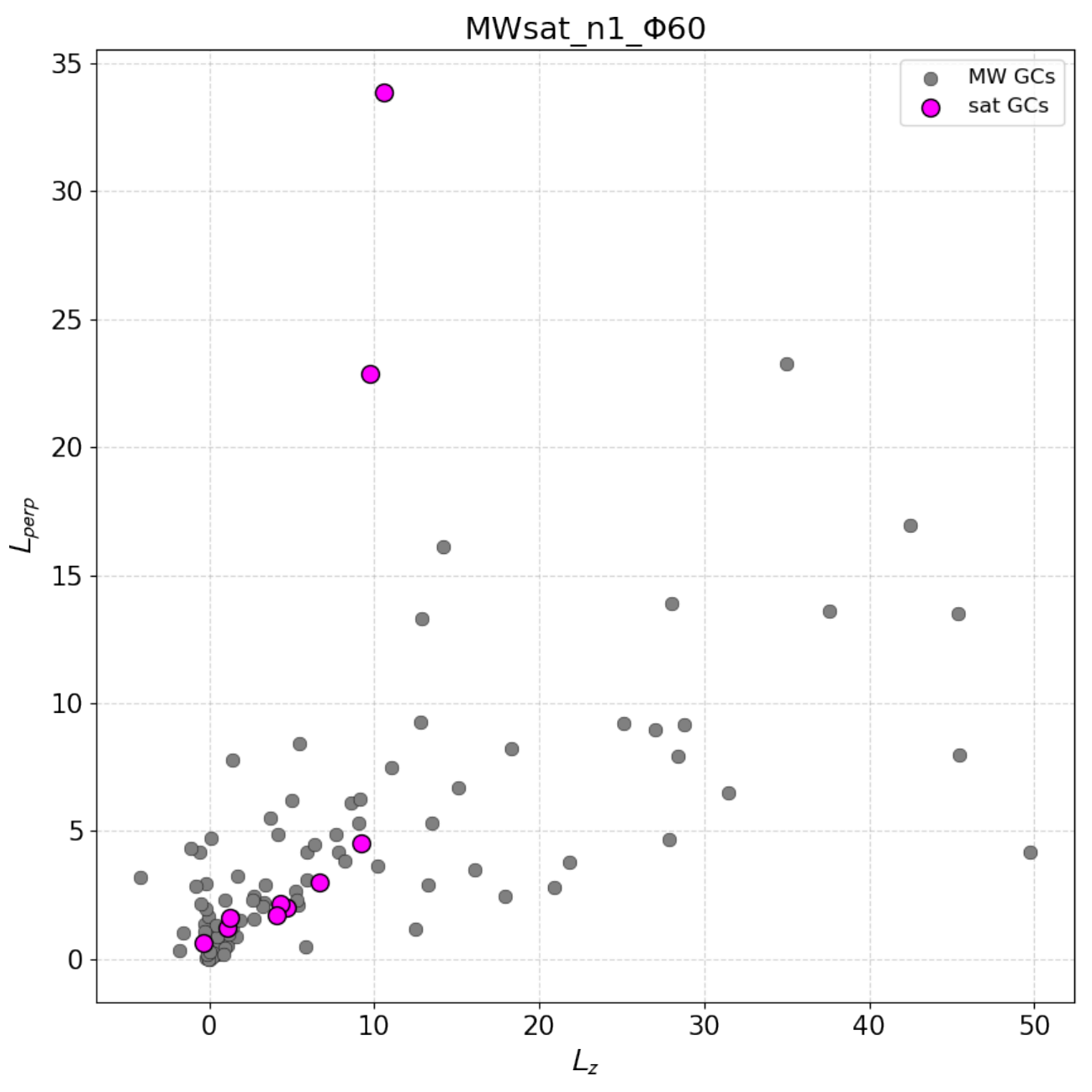} 
    \end{multicols}
    \vspace{-20pt}
    \begin{multicols}{2}
\includegraphics[width=.95\linewidth]{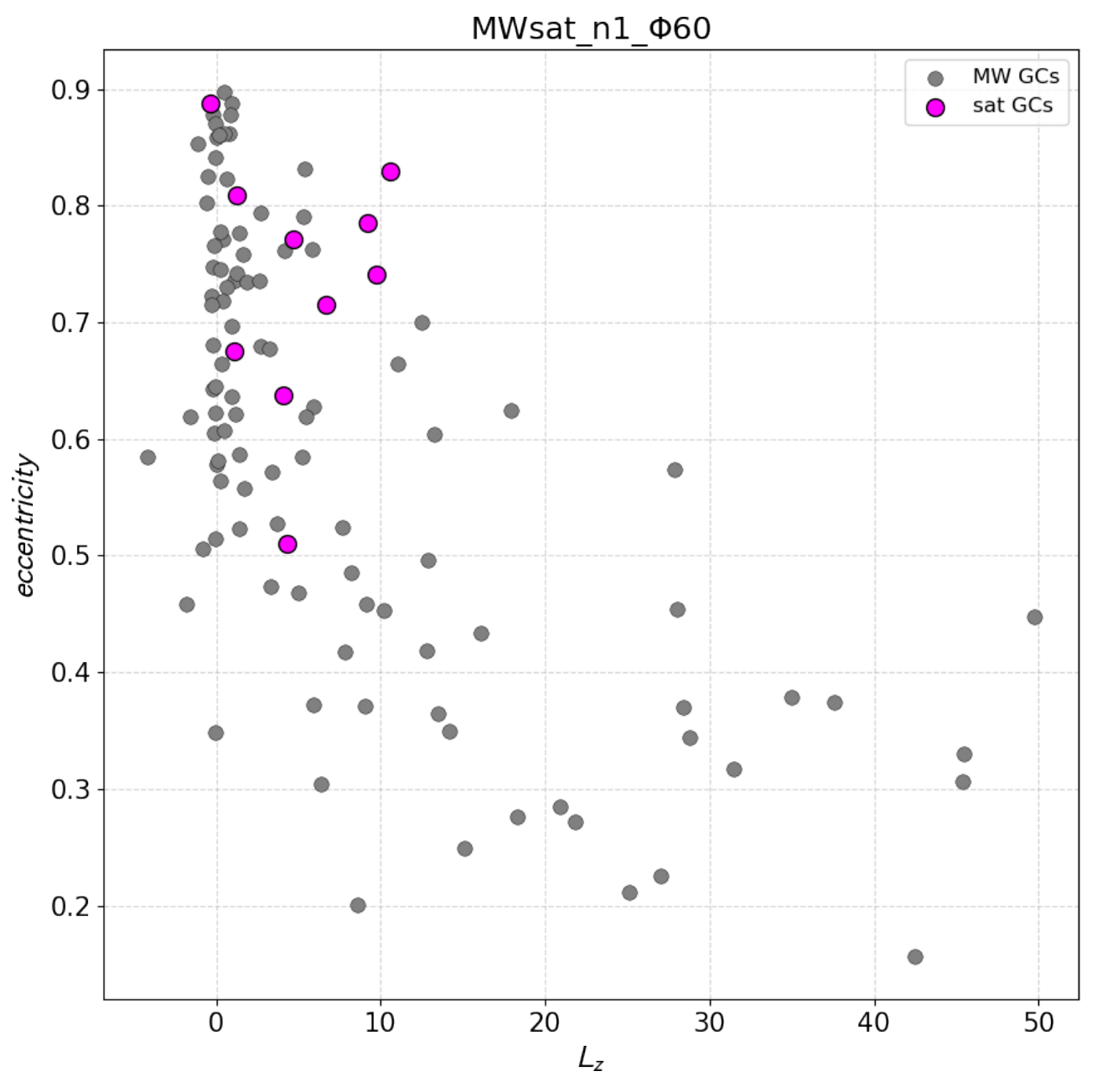}\par
\includegraphics[width=\linewidth]{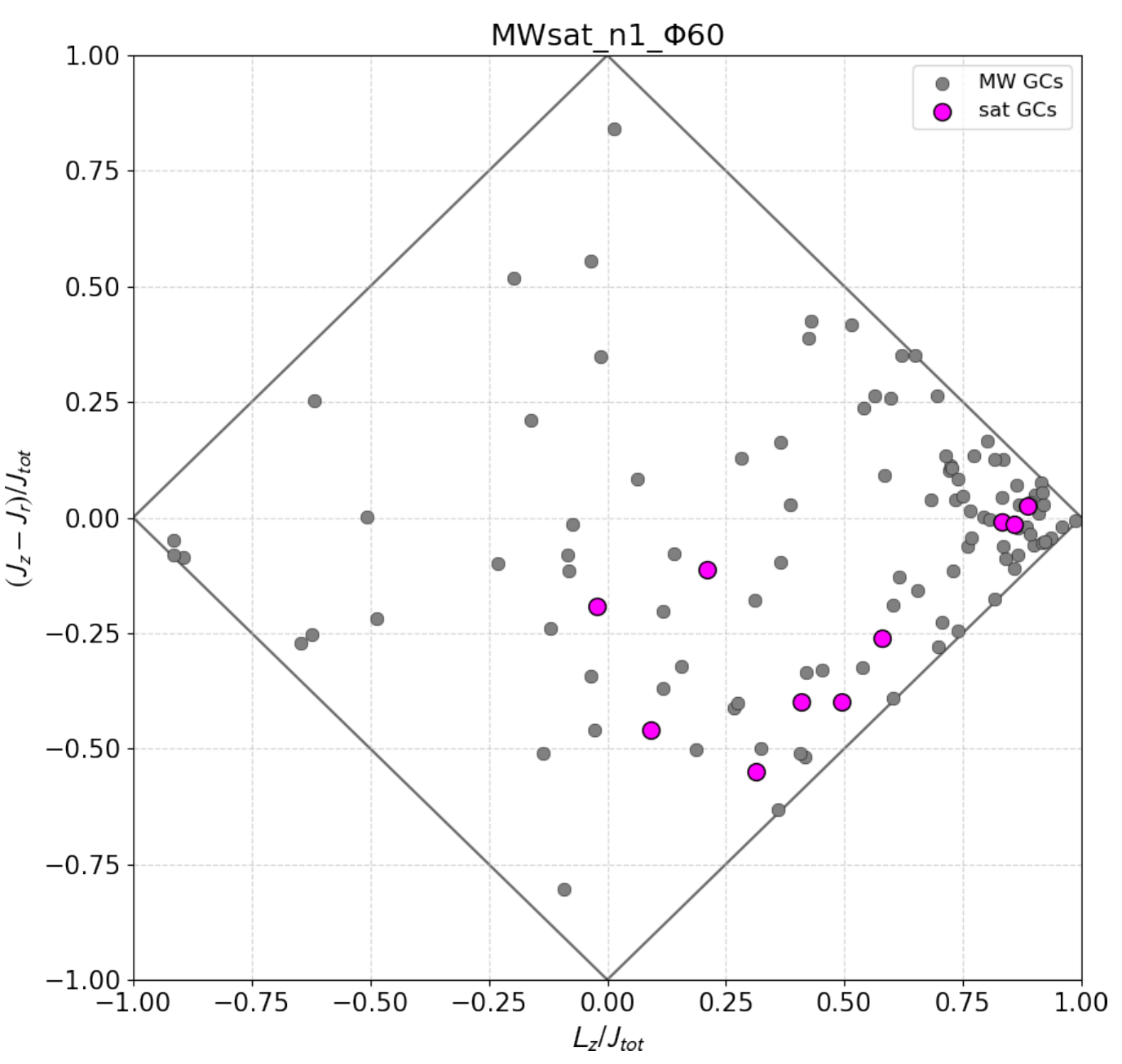}
    \end{multicols}
\end{centering}
\caption{Final globular clusters distribution in kinematic spaces namely $E - L_z$, $L_{perp} - L_z$, $eccentricity - L_z$ and action space for the simulation MWsat\_n1\_$\Phi$60. The in-situ population of globular clusters is represented by grey circles while the accreted one by magenta circles.}
\label{fig:kin_1sat}
\end{figure*}

Let us now examine how the populations of globular clusters in our simulations redistribute in other kinematic spaces. In addition to the $E - L_z$ space, we have analysed the $L_{perp} - L_z$ space where $L_{perp}$ is the projection of the total angular momentum onto the Galactic plane and is defined as: $L_{perp} = \sqrt{L_x^2 + L_y^2}$ ($L_x$, $L_y$ being respectively the $x$ and $y$ component of the angular momentum space in a reference frame with the Galactic disc in the $x-y$ plane).
Eccentricity ($e$), defined as $e = \frac{R_{apo} - R_{peri}}{R_{apo} + R_{peri}}$ ($R_{apo}$ and $R_{peri}$ being respectively the apocentre and the pericentre of the orbit) is another important parameter in describing an orbit. Here we use the eccentricity by combining it with $L_z$ as suggested in \citet{lane2022kinematic} \citep[see also][]{cordoni2021exploring}. The last kinematic space we analysed is the action space where the horizontal axis is the (normalized) azimuthal action ($J_{\phi}/J_{tot} \equiv L_z/J_{tot}$, where $J_{tot}=\sqrt{J_R^2 + J_z^2 + J_{\phi}^2}$), while the vertical axis is the (normalized) difference between the vertical and radial actions ($(J_z - J_R )/J_{tot}$). The right and left points of the space ($|L_z| = J_{tot}$, see bottom right panel in Fig.~\ref{fig:kin_1sat}) are in-plane prograde and in-plane retrograde orbits respectively. The top point ($J_z = J_{tot}$) is a circular polar orbit, and the bottom point ($J_R = J_{tot}$) is a radial orbit. The bottom-right and bottom-left edges are prograde and retrograde in-plane orbits. The top-right and top-left edges are prograde and retrograde circular orbits. 

\begin{figure*}
\begin{centering}
\begin{multicols}{2}
     \includegraphics[width=.95\linewidth]{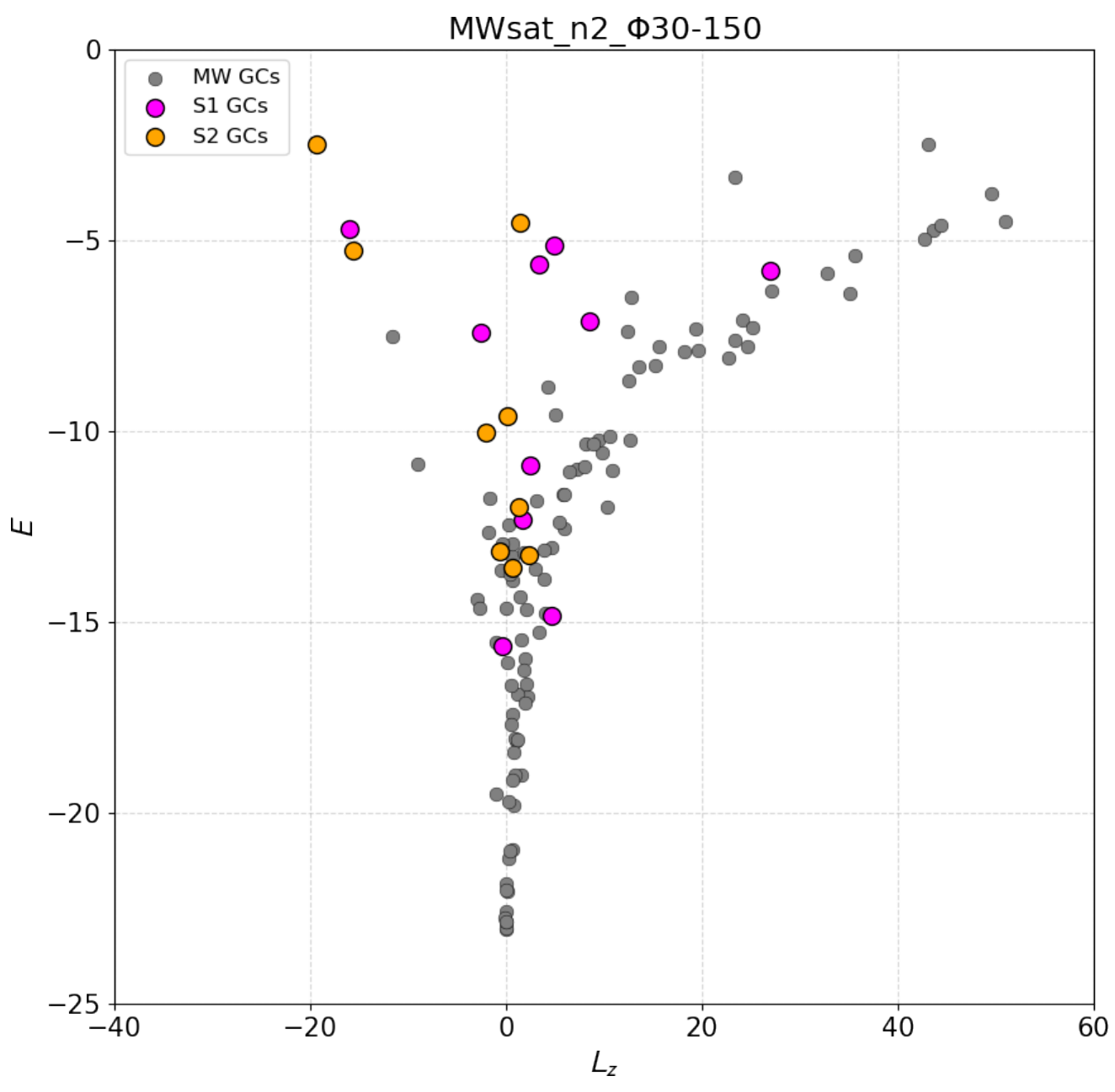}\par
         \includegraphics[width=.95\linewidth]{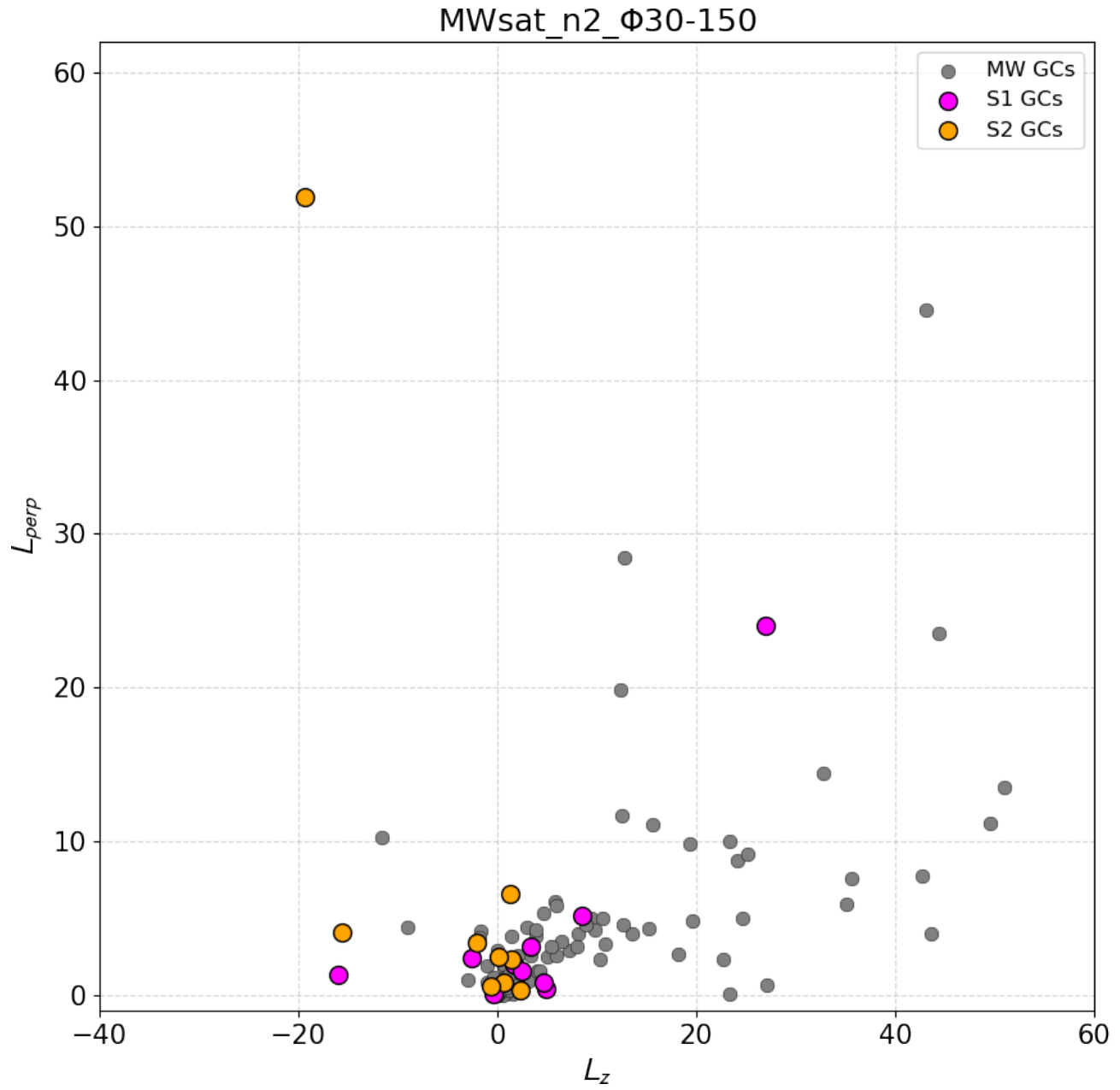} 
    \end{multicols}
    \vspace{-20pt}
    \begin{multicols}{2}
\includegraphics[width=.95\linewidth]{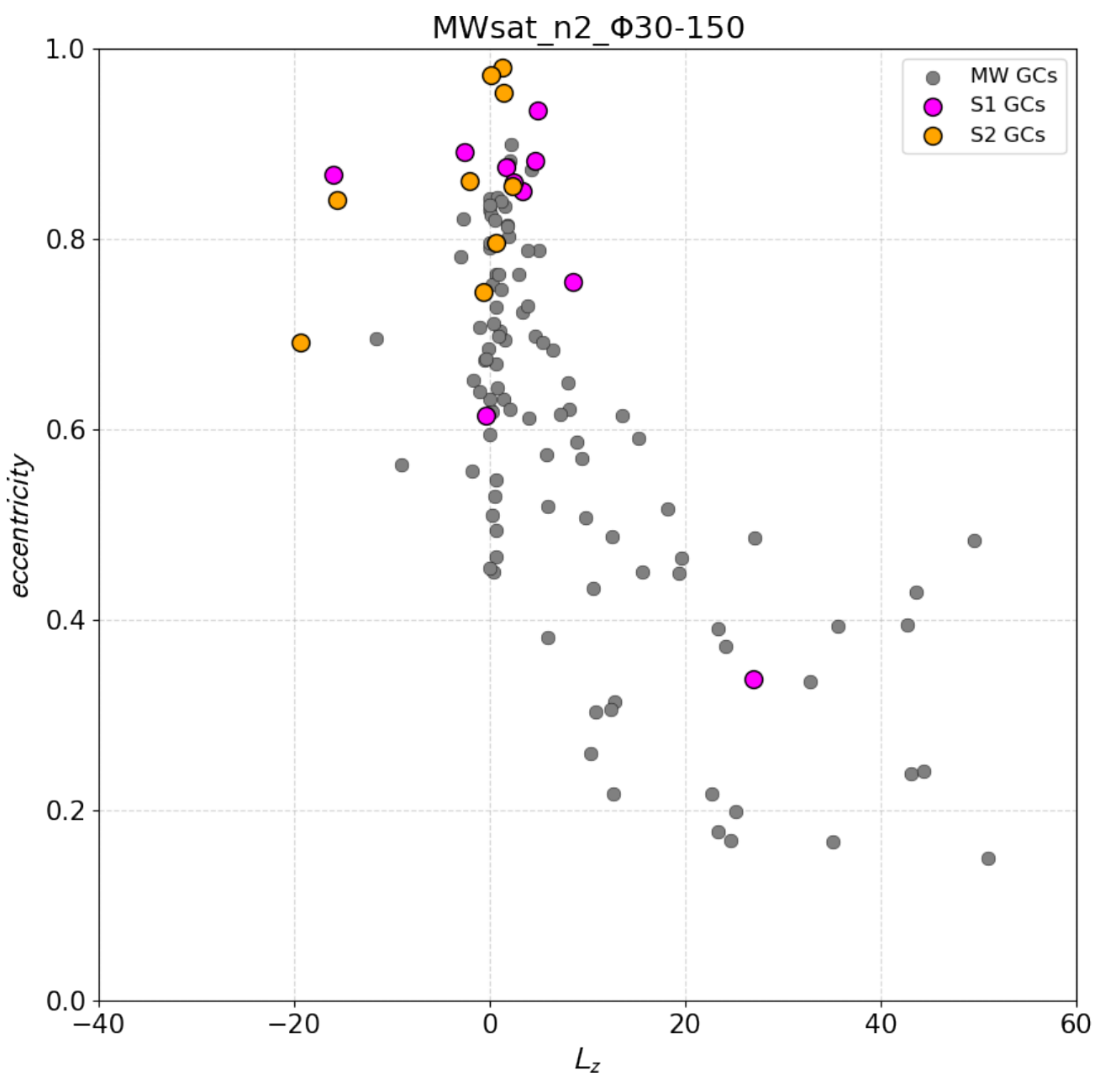}\par
\includegraphics[width=\linewidth]{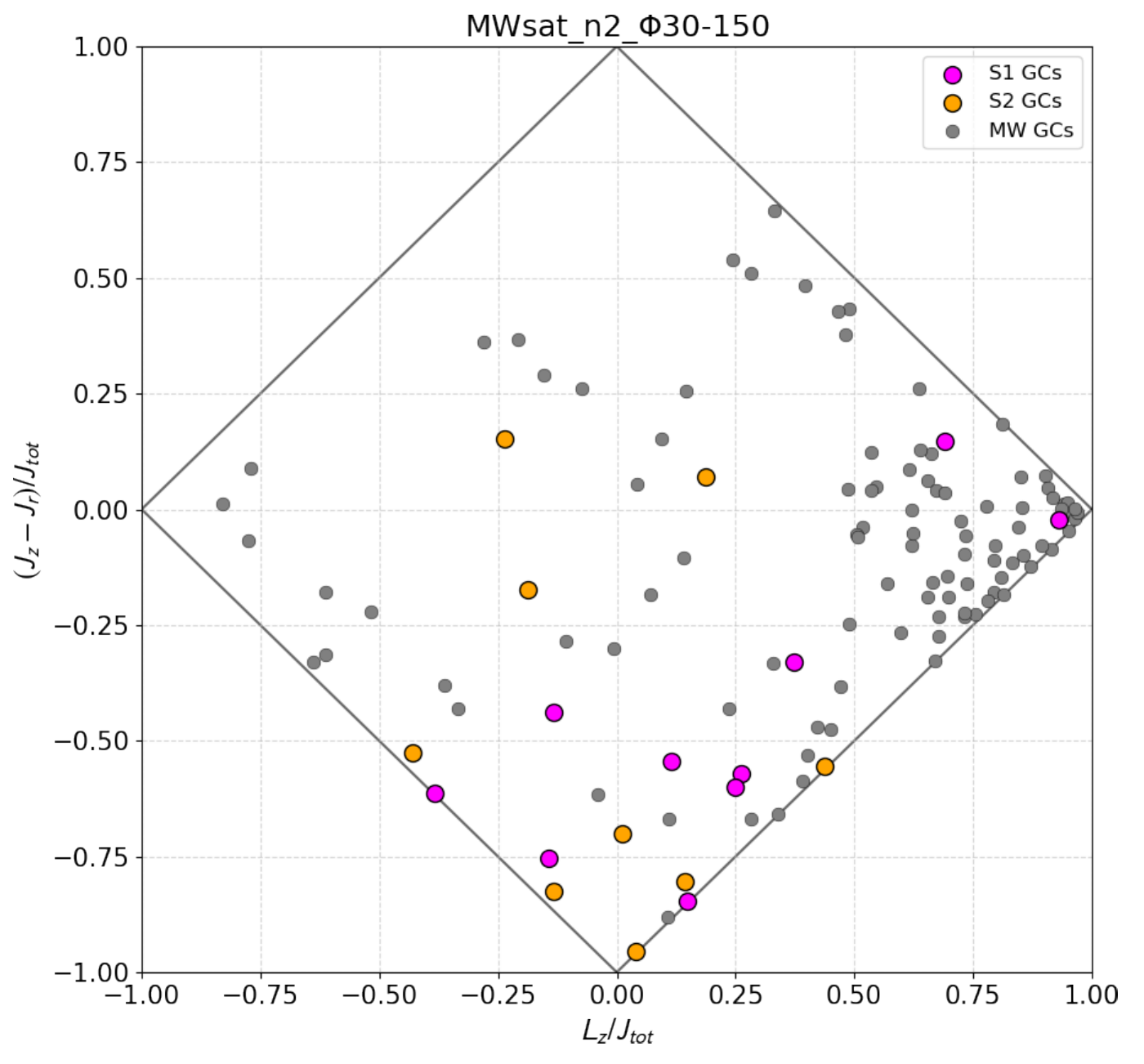}
    \end{multicols}
\end{centering}
\caption{Final globular clusters distribution in kinematic spaces namely, $E - L_z$, $L_{perp} - L_z$, $eccentricity - L_z$ and action space for the simulation MWsat\_n2\_$\Phi$30-150. The in-situ population of globular clusters is represented by grey circles while the accreted one by magenta and orange circles respectively for satellite 1 and satellite 2.}
\label{fig:kin_2sat}
\end{figure*}

Figures \ref{fig:kin_1sat}, \ref{fig:kin_2sat} show the final globular clusters distribution in the four kinematic spaces described above i.e. the $E - L_z$, $L_{perp} - L_z$, $eccentricity - L_z$ and action space for the two previously examined simulations with one or two accretions respectively (MWsat\_n1\_$\Phi$60 and MWsat\_n2\_$\Phi$30-150). The population of globular clusters originally belonging to the MW-type galaxy is represented by grey circles while the accreted one by magenta (and orange) circles for satellite 1 (and satellite 2). Actions are computed using the St$\rm\ddot{a}$ckel fudge approach \citep{binney2012actions}, as implemented in the AGAMA code \citep{vasiliev2019agama}.
The addition of $L_{perp}$ (see top right panel in Fig.~\ref{fig:kin_1sat}) to the previous analysis of the $E - L_z$ space only confirms the conclusions drawn in the previous paragraph: at the end of the simulation, overall the superposition between the in-situ and accreted population is considerable everywhere. Only GCs lost by the satellite at the first pericentric passage ($t_{esc}\simeq 0.5$) which are therefore at high energies ($E \gtrsim -5$) are in principle distinguishable from the rest.  Furthermore, the accretion event generates a population of in-situ halo clusters (consisting of disk GCs heated by the interaction), whose distribution extends towards high $L_{perp}$. From the $eccentricity-L_z$ and action spaces (see bottom panels in Fig.~\ref{fig:kin_1sat}), it is not possible to make any difference between accreted or in-situ globular clusters. In the action space we can notice a higher density in the right-hand corner related to the more prograde orbits where in-situ and accreted GCs definitely overlap. Overall clusters from the satellite are preferentially located towards the bottom-right edge, i.e. towards prograde in-plane orbits, but are not found in distinct groups from in-situ clusters. The strength of the action space should lie in the fact that distinct types of orbits occupy different loci in the space \citep[see][]{lane2022kinematic}, but this is not sufficient to reconstruct the origin of the globular clusters (i.e. accreted or in-situ), since during the merger process they are scattered and mixed over most of the space.

As we have already mentioned, when we consider the merging with two satellites (see Fig.~\ref{fig:kin_2sat}), the interpretation of kinematic spaces obviously does not improve. We can confirm that overall GCs that end at higher energy in the $E - L_z$ plane, are accreted. In this specific simulation (MWsat\_n2\_$\Phi$30-150) they also stand out as the most retrograde (see top and bottom left panels of Fig.~\ref{fig:kin_2sat}). Despite this, it is still hardly feasible to distinguish clusters coming from the first or second satellite and it is evident that also in-situ clusters originally in the disc can be found at these energy levels. If the points were not coloured, it would not be possible to assign a membership to the different GCs, on the basis of their location in the $E - L_z$ space. Other kinematic spaces added in the analysis seem not to add any relevant information for this purpose. In the $L_{perp} - L_z$ and $eccentricity - L_z$ spaces, in fact, the only clusters that are detached from the rest of the mixed population of accreted and in-situ clusters are those with a more retrograde orbit comprising 2 MW GCs, two clusters originating from satellite 2 and one GC from satellite 1. In the action space, again, the clusters are scattered and mixed over most of the plane; the accreted clusters are predominantly in the region characterised by radial in-plane orbits, but we cannot identify groups with the same galactic membership: the small clumps detectable by eye are all composed of GCs of different origins. In this respect, it is worth recalling the work by  \citet{lane2022kinematic} which extensively investigates the best kinematic spaces to separate radially anisotropic from isotropic halo populations, concluding that, to this aim, the “action diamond” space is superior to other spaces used in the literature. The problem is that clusters of different nature (accreted or in-situ, and/or belonging to different satellite progenitors) can end up having similar orbital anisotropy (see, for example Figs \ref{fig:kin_1sat} and \ref{fig:kin_2sat}), thus a kinematic space can be very good at separating stars (or clusters) with specific orbital properties, but without being able, however, to assess their origin. This is the limitation of the analysis presented in the literature so far: to associate a specific region of a kinematic space, and hence specific orbital characteristics, with a specific accretion origin.
\begin{figure}
\begin{centering}
\includegraphics[width=.82\linewidth]{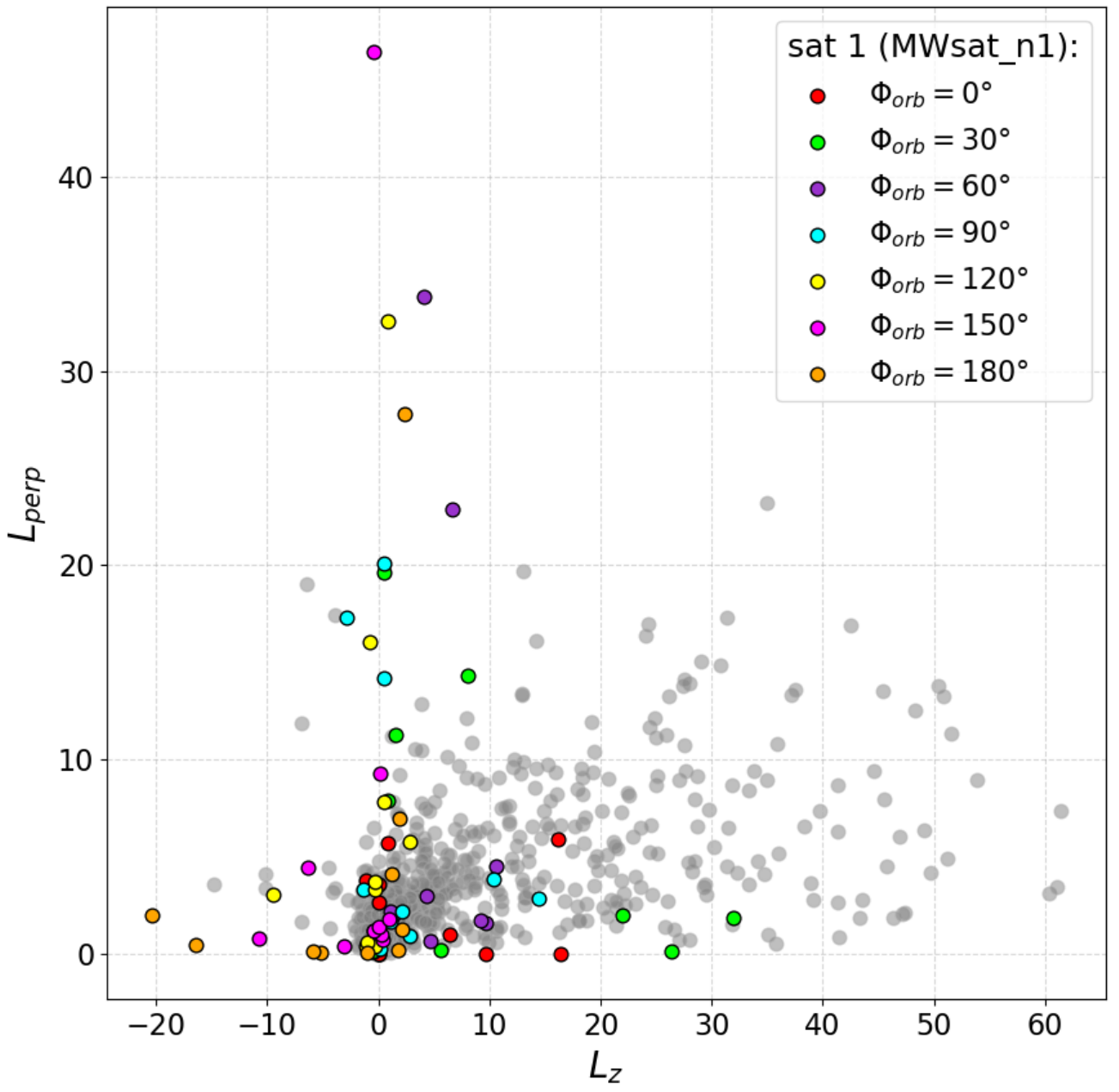} \par
\includegraphics[width=.82\linewidth]{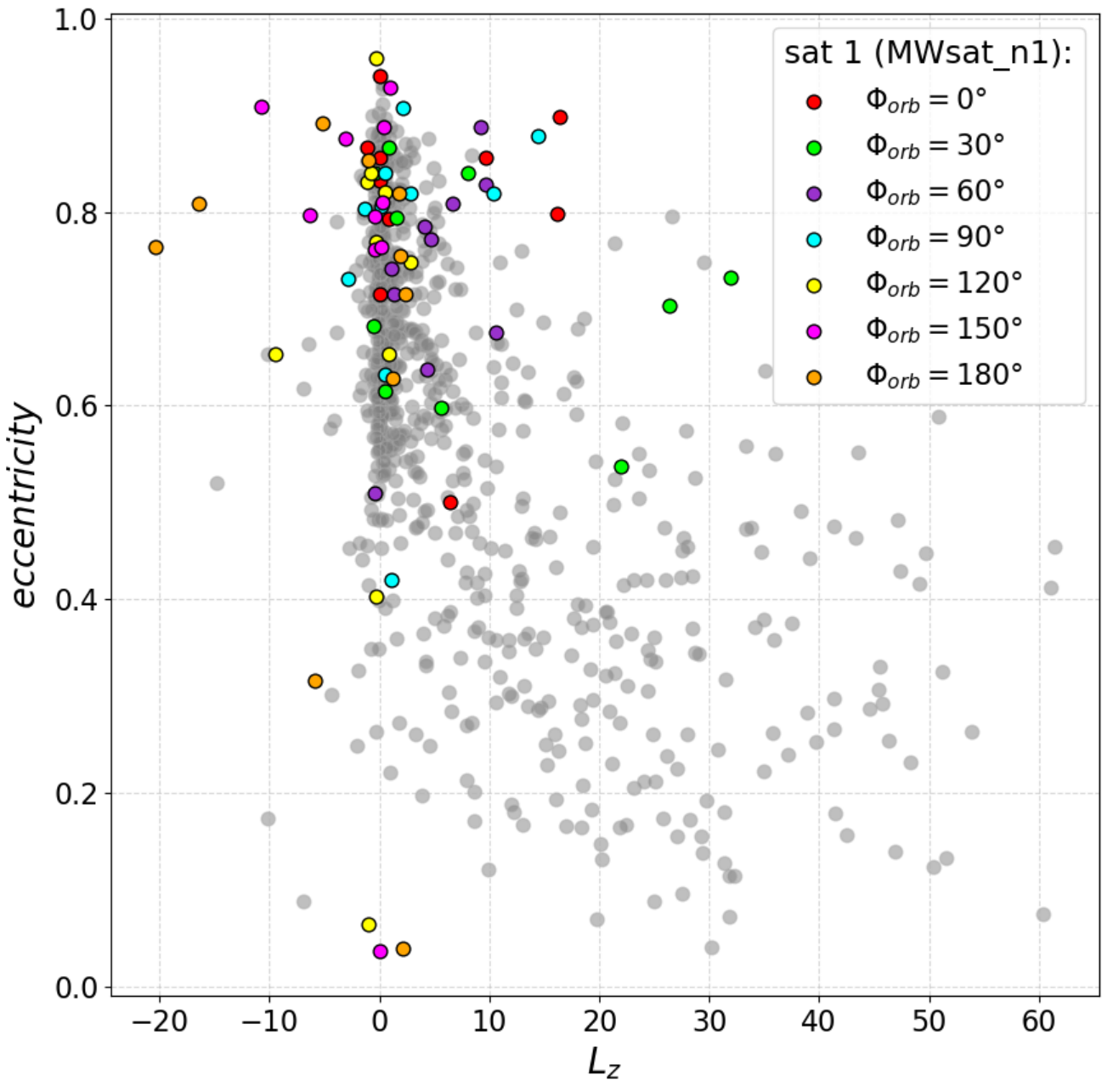} \par 
\includegraphics[width=.85\linewidth]{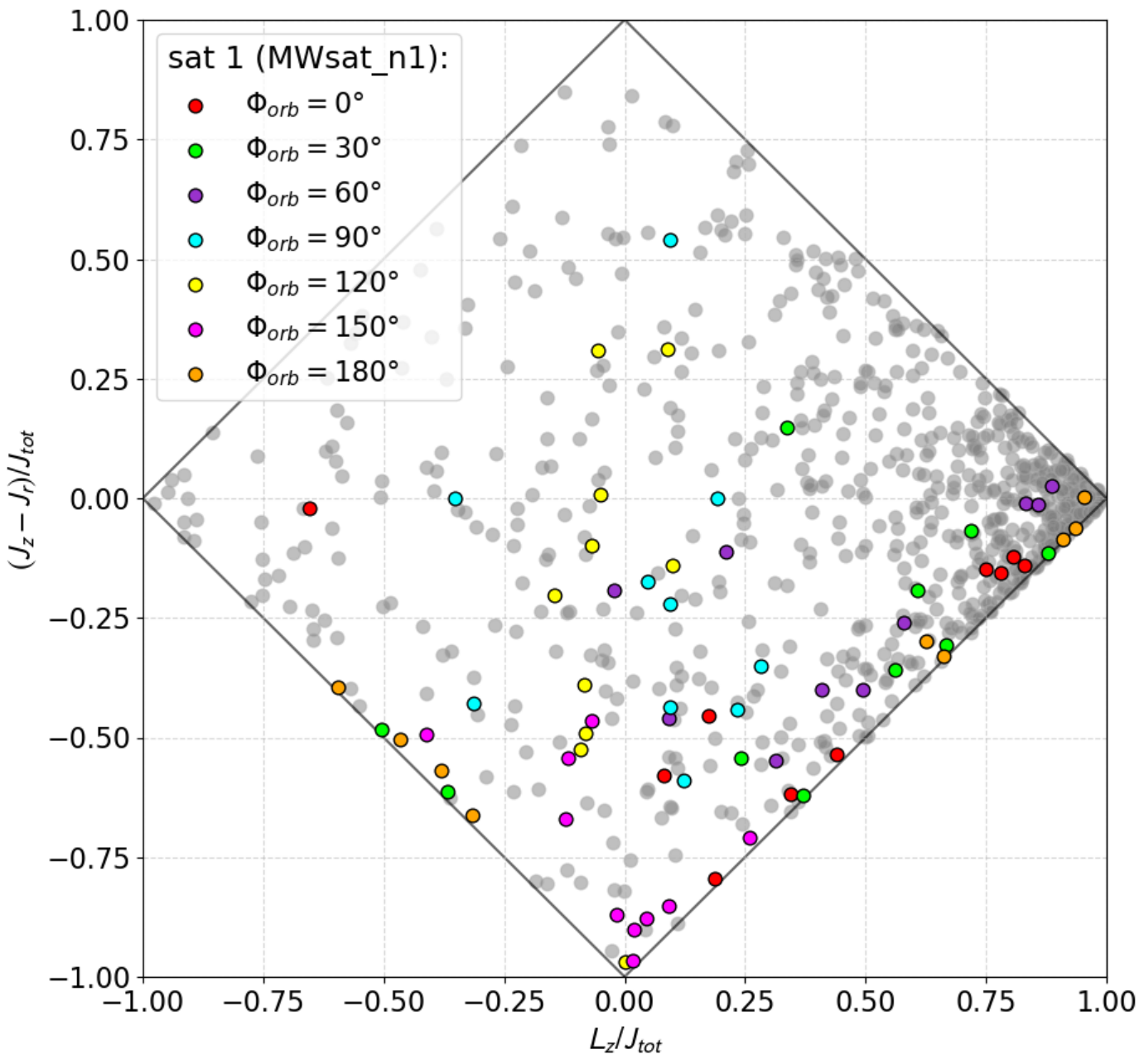} 

\caption{Distribution in $L_{perp} - L_z$, $eccentricity - L_z$ and action space for the whole set of 1x(1:10) merger simulations at T = 5 Gyr. The distribution of the in-situ population of globular clusters has been stacked and is shown by the grey circles. The color-coding of satellites GCs is different according to the initial inclination of the satellite orbital plane with respect to the Milky Way-like reference frame, $\Phi_{orb}$.}
\label{fig:kin_stack}
\end{centering}
\end{figure}

\begin{figure*}
\begin{centering}
\begin{multicols}{2}
     \includegraphics[width=.9\linewidth]{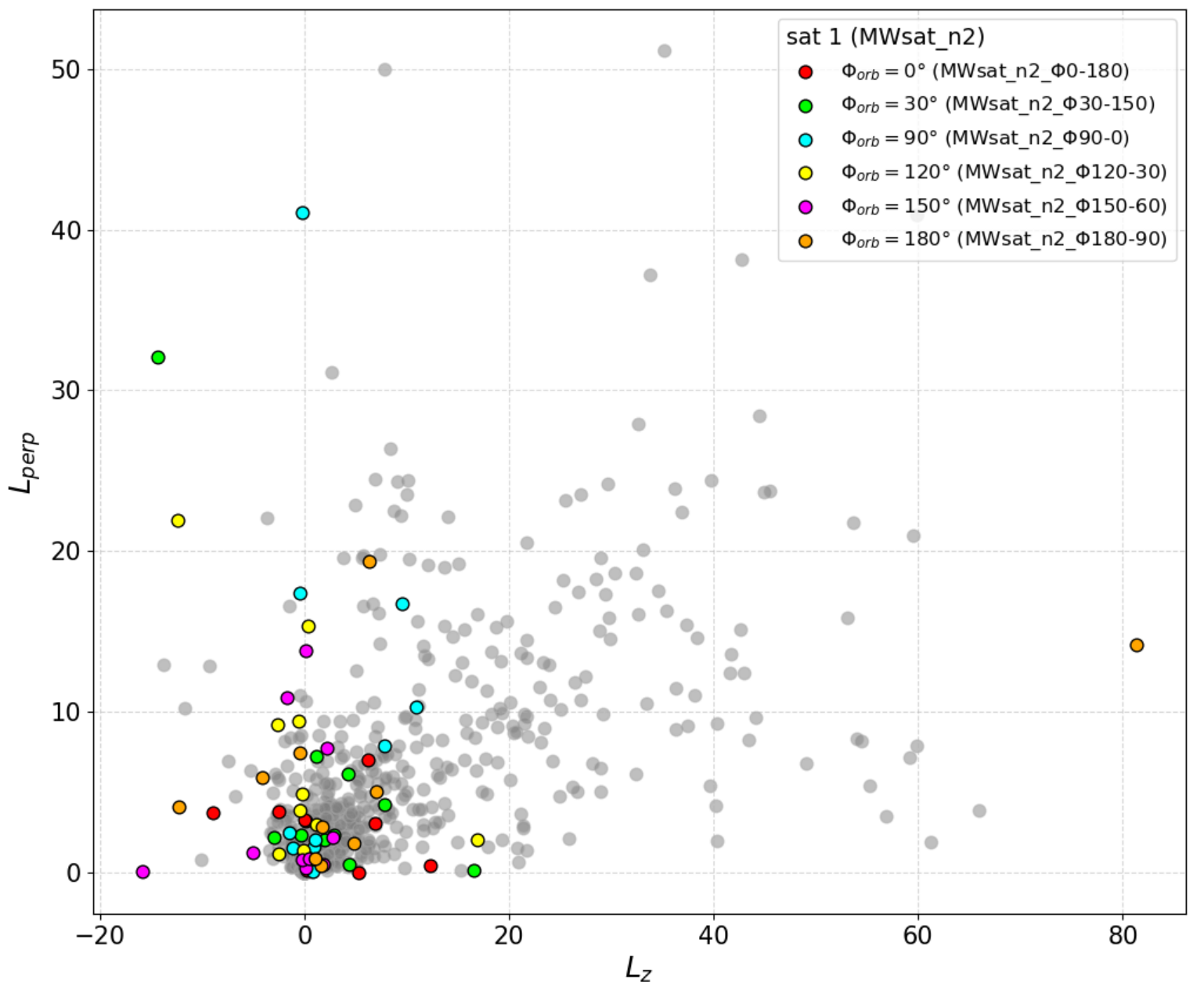} \par
    \includegraphics[width=.9\linewidth]{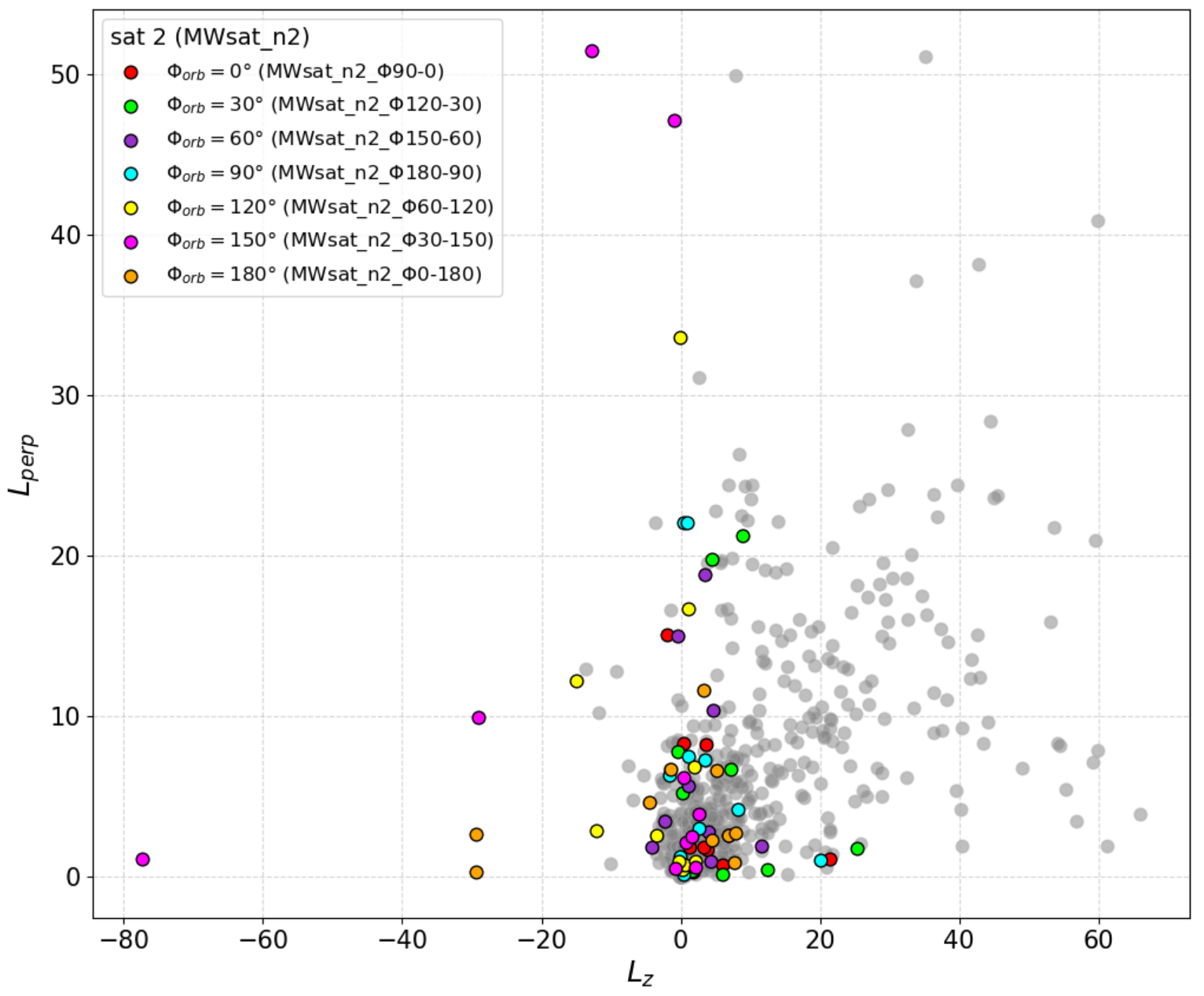}  

    \end{multicols}
    \begin{multicols}{2}
\includegraphics[width=.9\linewidth]{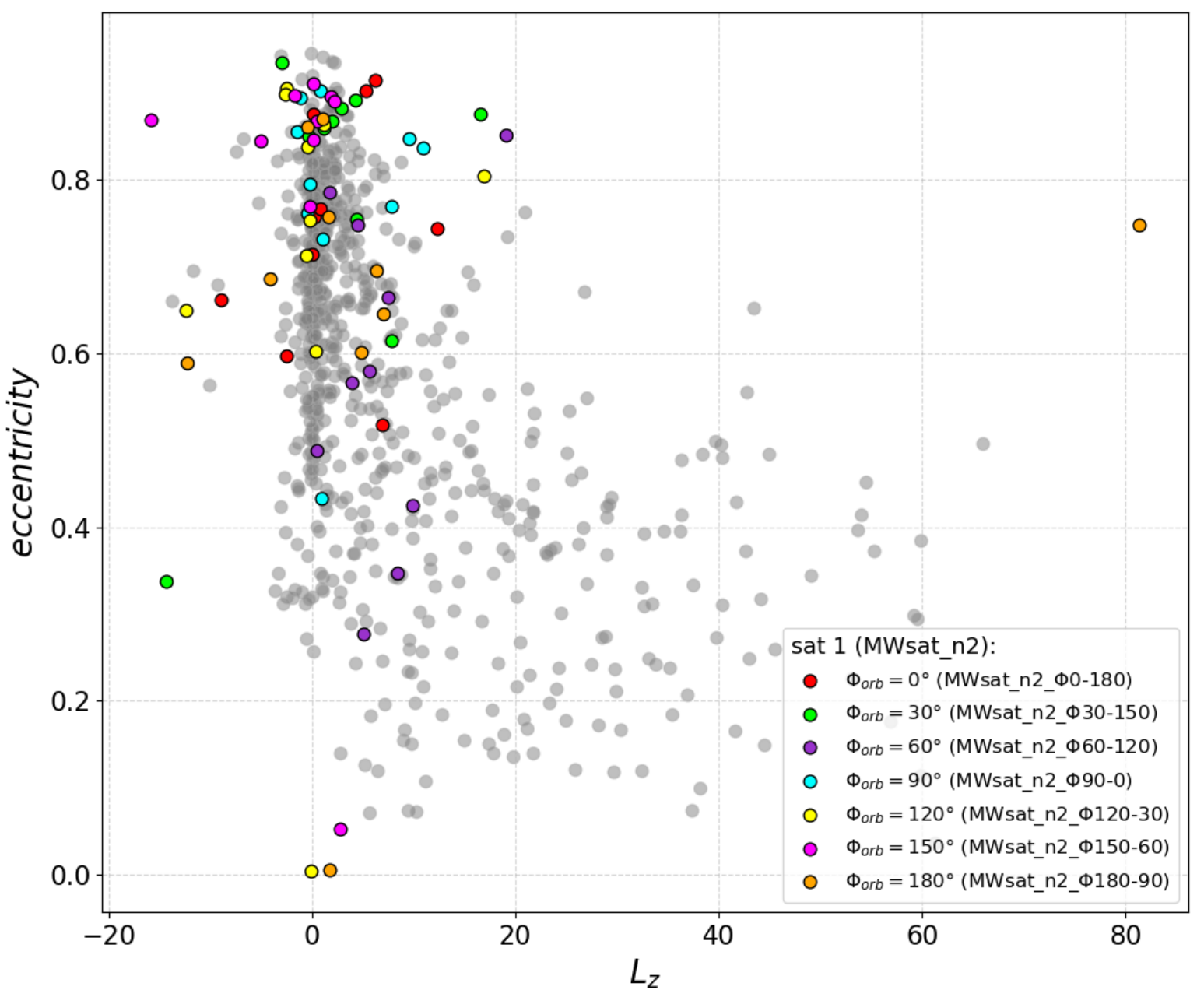} \par
\includegraphics[width=.9\linewidth]{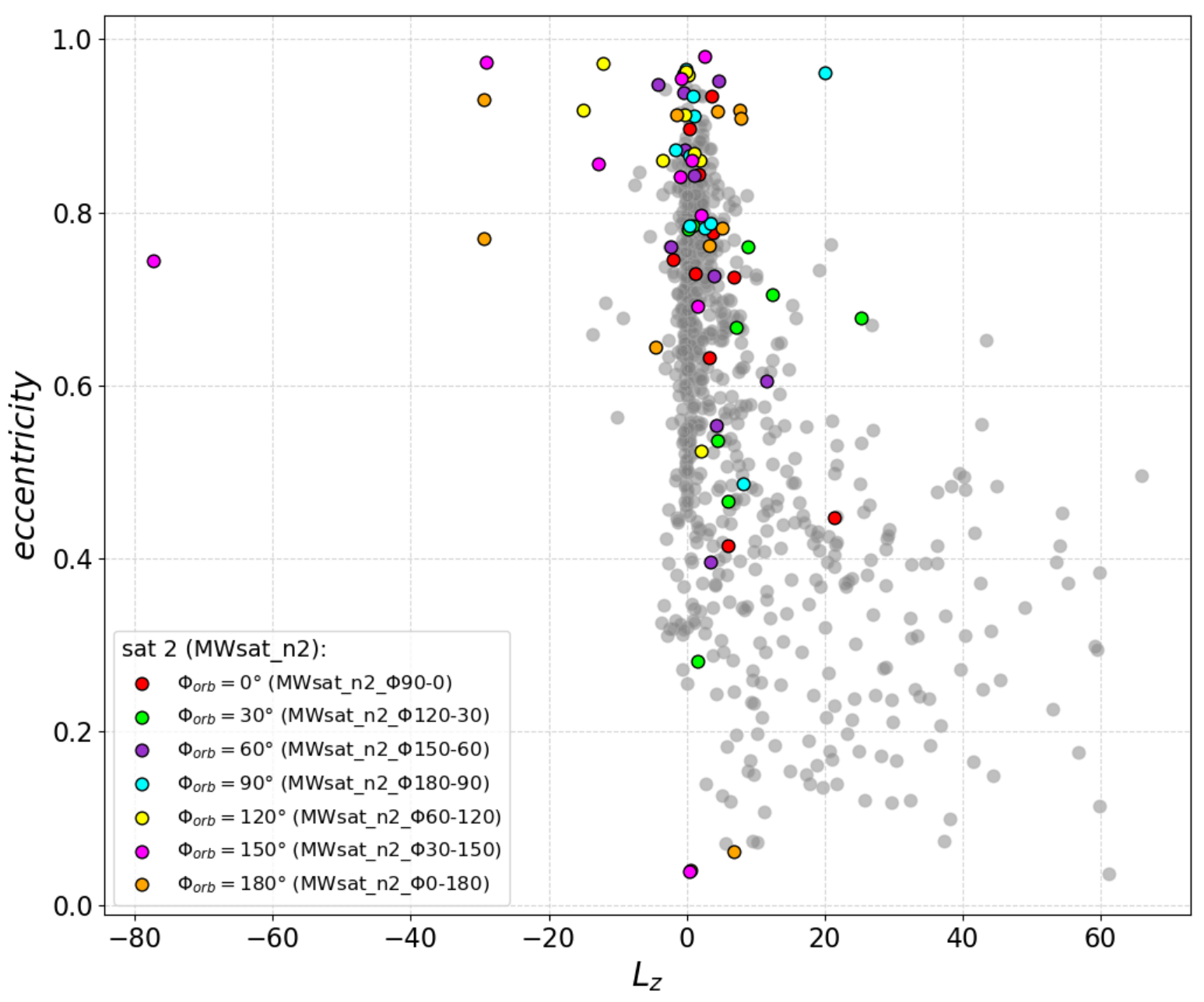} \par 
    \end{multicols}
    \begin{multicols}{2}
    \includegraphics[width=.97\linewidth]{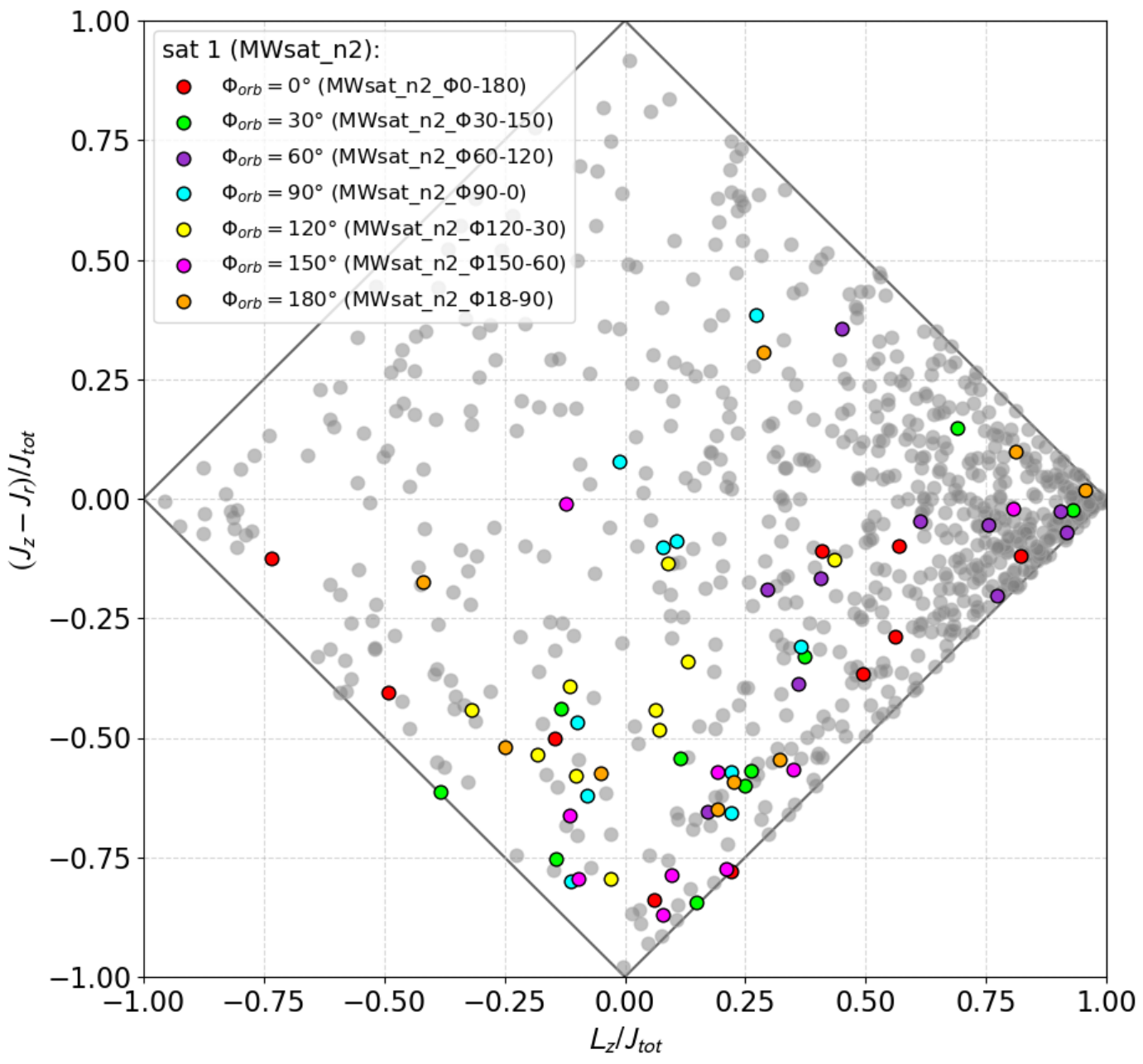} \par
    \includegraphics[width=.97\linewidth]{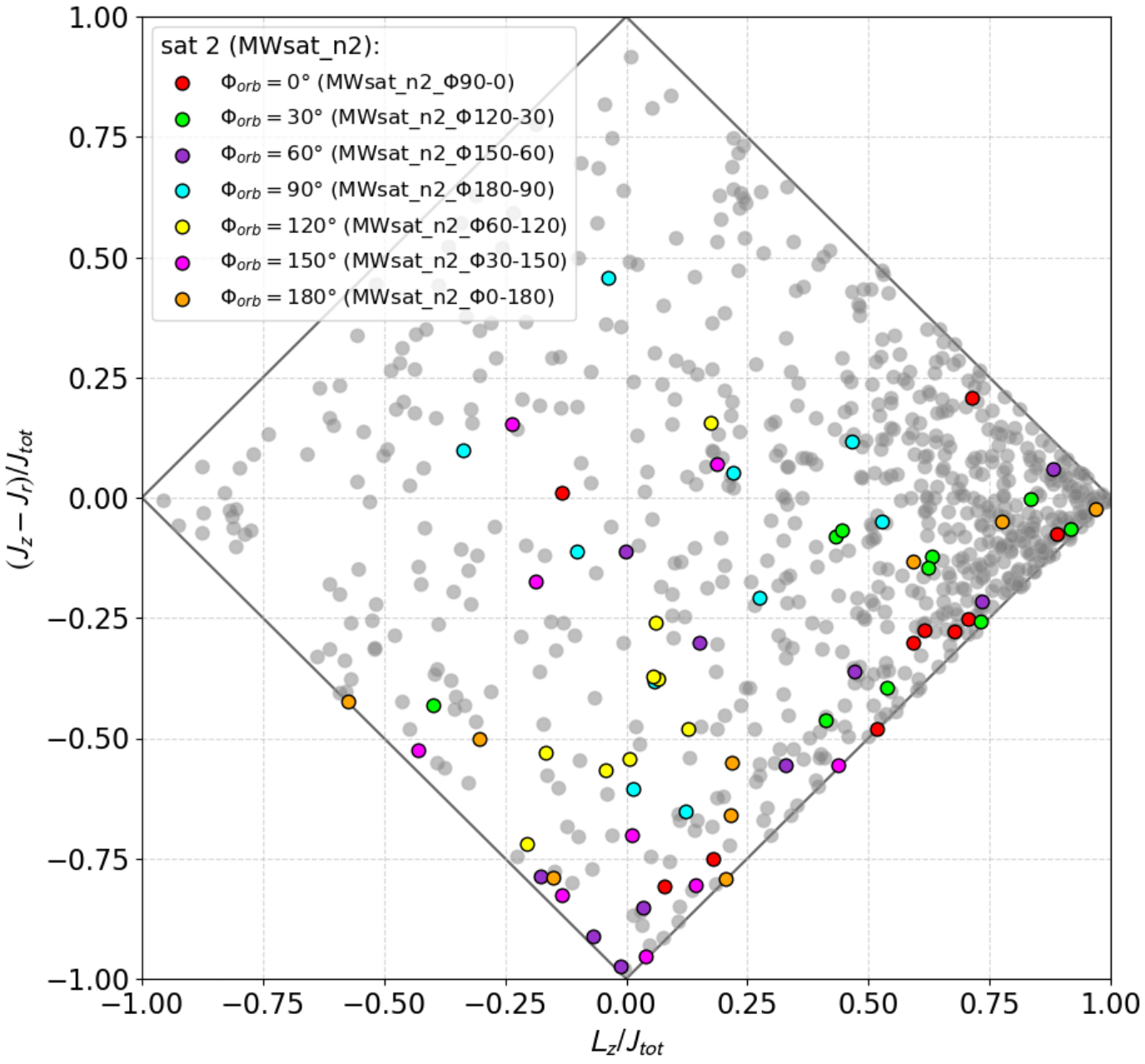} 
        \end{multicols}
    
    \caption{Distribution in $L_{perp} - L_z$, $eccentricity - L_z$ and action space for the whole set of 2x(1:10) merger simulations at T = 5 Gyr. The distribution of the in-situ population of globular clusters has been stacked and is shown by the grey circles. The color-coding of satellites GCs (satellite 1: left column, satellite 2: right column) is different according to the initial inclination of the satellite orbital plane with respect to the Milky Way-like reference frame, $\Phi_{orb}$.}
     \label{fig:kin_stack_2sat}
    \end{centering}
    \end{figure*}
    
In Fig.~\ref{fig:kin_stack} and \ref{fig:kin_stack_2sat} we have stacked together the GCs distributions in $L_{perp} - L_z$, $eccentricity - L_z$ and $action - space$ for the whole set of 1x(1:10) and 2x(1:10) simulations respectively as in Fig.~\ref{fig:MW1} and \ref{fig:MW2}. All in-situ globular clusters are shown with gray circles while accreted GCs are colour-coded according to the initial inclination of the satellite of membership with respect to the Milky Way-like reference frame, $\Phi_{orb}$.
These figures allow us to generalise the conclusions just drawn for the two examples of single and double accretion, the most important of which concerns the fact that the kinematic spaces considered in addition to the $E - L_z$ space do not add any useful insight for discriminating between globular clusters originating from satellites accreted over time by the MW or formed in the MW itself. Regarding the $L_{perp} - L_z$ plane, we note that in the case of a single accretion, if we select the region with $L_{perp} \geq 20$ and $L_z \lesssim 30$, we will certainly find satellite clusters, an argument that no longer holds if we analyse simulations with double accretion: here indeed we find several in-situ clusters heated so much kinematically that they end up having $L_{perp} > 20$. Looking at the $eccentricity-L_z$ plane, the only noteworthy point is the fact that with both one and two accreted satellites, at fixed $|L_z|>0$, we find accreted GCs with more eccentric orbits than those in-situ, having $e \geq 0.8$ while for $|L_z|\simeq0$ accreted and in-situ clusters share the same more extended range of eccentricities. As far as action space is concerned, we can say that in general accreted clusters tend to have in-plane prograde or radial orbits. The distribution of in-situ globular clusters is denser in the right corner corresponding to prograde in-plane orbits and this makes sense since, by hypothesis, the clusters formed in the MW-type galaxy initially have disc orbits. Despite this, we also find in-situ GCs at the opposite extreme - i.e. with strongly retrograde orbits - and indeed, the most retrograde clusters are actually in-situ in all cases. If we had also considered in-situ GCs with halo-like orbits, this feature would have been even more evident (see Sect.~\ref{insitu}).




    


\subsection{Does a clustering analysis allow to recover the history of accretion?}\label{GMM}
\begin{figure*}
\begin{centering}
\begin{multicols}{2}
     \includegraphics[width=.9\linewidth]{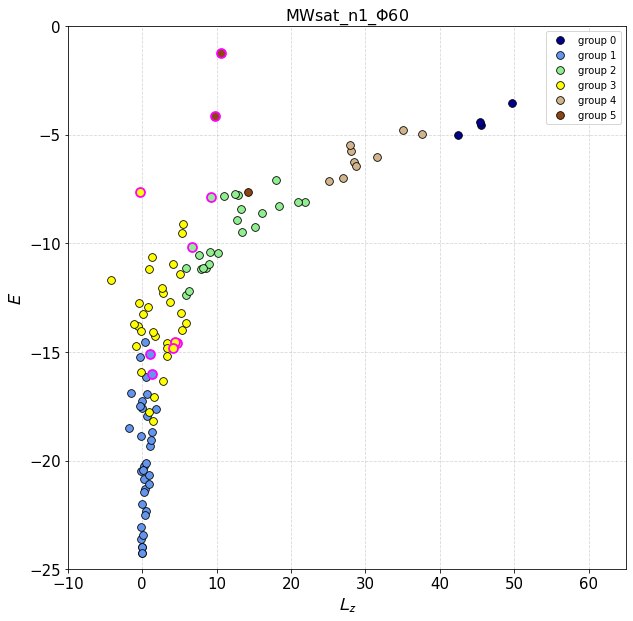}\par
         \includegraphics[width=.9\linewidth]{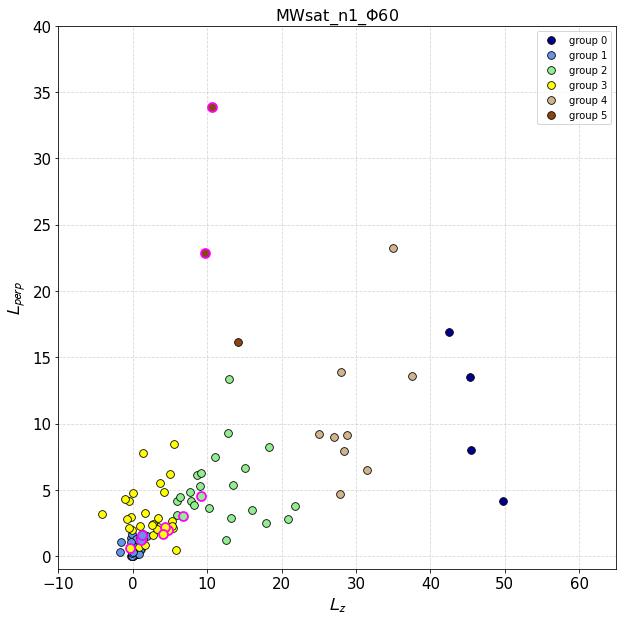} 
    \end{multicols}
    \vspace{-20pt}
    \begin{multicols}{2}
\includegraphics[width=.9\linewidth]{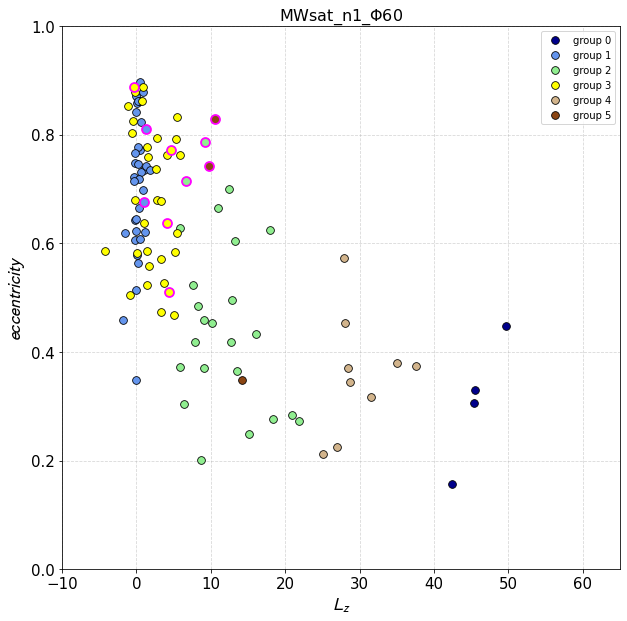}\par
\includegraphics[width=.95\linewidth]{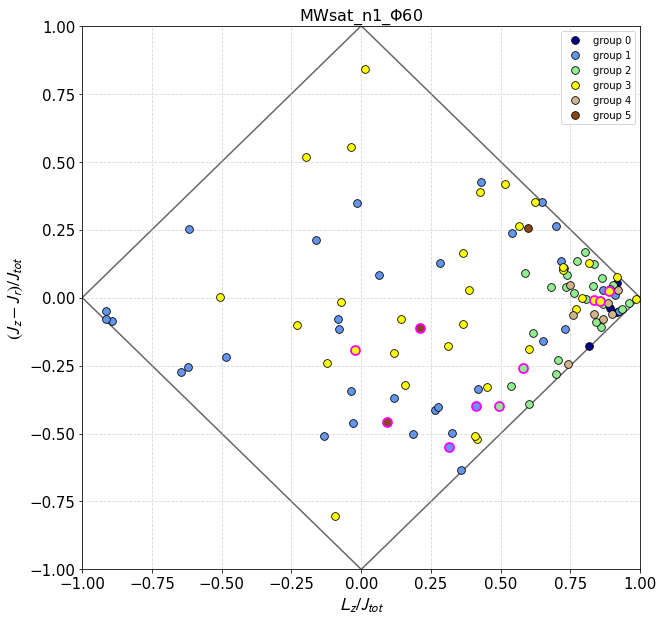}

    \end{multicols}
\includegraphics[width=.5\linewidth]{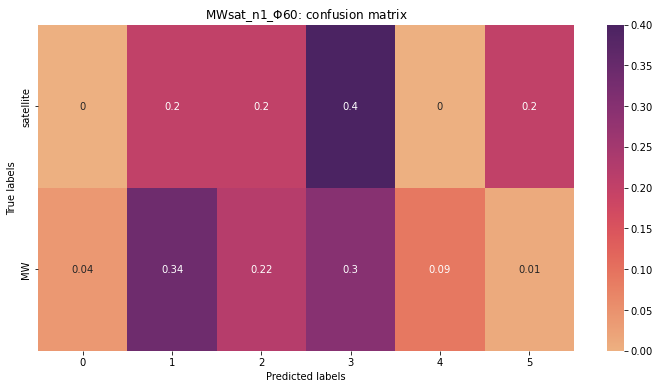}

\end{centering}

\caption{\emph{Top, middle row}: Final globular clusters distribution in kinematic spaces ($E - L_z$, $L_{perp} - L_z$, $eccentricity - L_z$ and action space) for the simulation MWsat\_n1\_$\Phi$60. The colour-coding of GCs is related to the different components retrieved applying the minimum BIC criterion in the Gaussian Mixture Model. Truly accreted globular clusters (i.e. with the true label given by our simulation) are bordered by magenta circles.
\emph{Bottom panel}: Confusion matrix obtained by the GMM with values normalised to the total number of GCs in each true class.}
\label{fig:gmm_1sat}
\end{figure*}

\begin{figure*}
\begin{centering}
\begin{multicols}{2}
     \includegraphics[width=.9\linewidth]{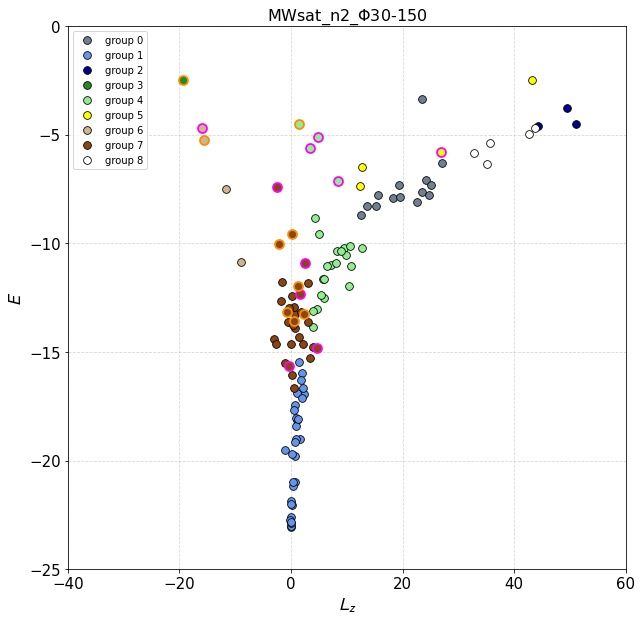}\par
         \includegraphics[width=.9\linewidth]{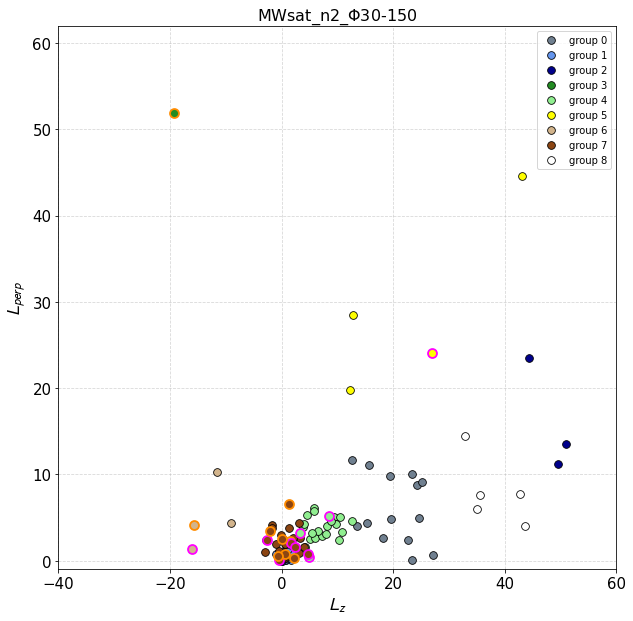} 
    \end{multicols}
    \vspace{-20pt}
    \begin{multicols}{2}
\includegraphics[width=.9\linewidth]{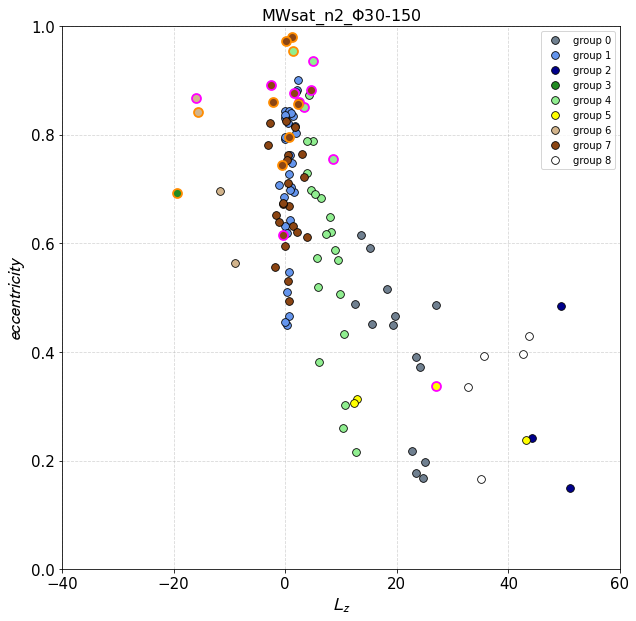}\par
\includegraphics[width=.95\linewidth]{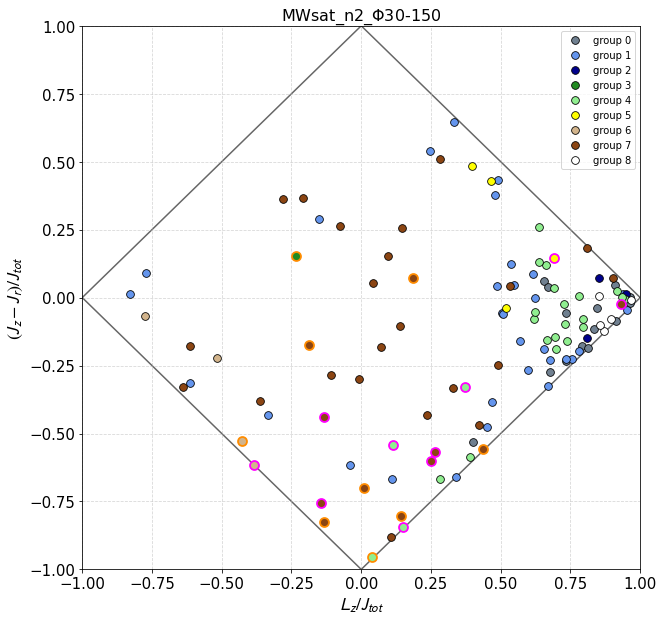}

    \end{multicols}
\includegraphics[width=.5\linewidth]{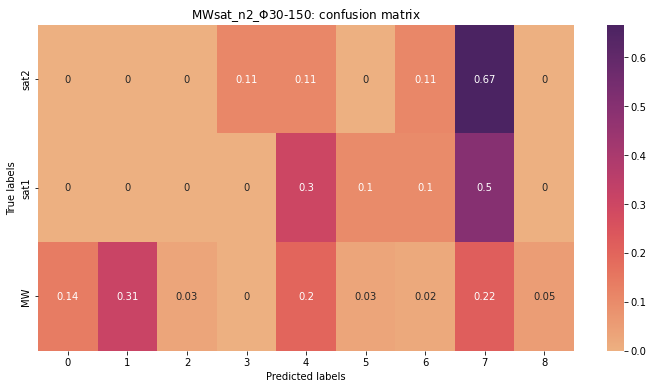}

\end{centering}

\caption{\emph{Top, middle row}: Final globular clusters distribution in kinematic spaces ($E - L_z$, $L_{perp} - L_z$, $eccentricity - L_z$ and $action - space$) for the simulation MWsat\_n2\_$\Phi$30-150. The colour-coding of GCs is related to the different components retrieved applying the minimum BIC criterion in the Gaussian Mixture Model. Truly accreted globular clusters (i.e. with the true label given by our simulation) are bordered by magenta and orange circles respectively for satellite 1 and satellite 2.
\emph{Bottom panel}: Confusion matrix obtained by the GMM with values normalised to the total number of GCs in each true class.}
\label{fig:gmm_2sat}
\end{figure*}

To make our analysis more quantitative, we exploited Gaussian Mixture Models both as a clustering algorithm but mainly for its fundamental purpose, which is to model the overall distribution of the input data. The purpose of this analysis was to investigate whether the algorithm can grasp any properties of the final distributions of GCs in kinematic spaces not visible by eye and in a more quantitative manner. Lately, several works have started to use this type of tool both to distinguish between accreted and in-situ stars/globular clusters and to retrieve the number of accretions undergone by the Milky Way \citep[see for instance][]{Donlon2022:2211.12576v1, callingham2022chemo}. We then proceeded as follows. For each simulation we considered a six-dimensional space consisting of the globular clusters' kinematic quantities previously examined, namely: the $z$ component of the angular momentum $L_z$, the total orbital energy $E$, the projection of the angular momentum onto the galactic plane $L_{perp}$, the eccentricity $e$, the normalized azimuthal action  $L_z/J_{tot}$, and the difference between the vertical and radial actions  $(J_z - J_R)/J_{tot}$. We tried to fit this with a two/three-component GMM (depending on the number of accreted satellites considered in addition to the in-situ component) viewed as a clustering model, but the results were not particularly useful as the real memberships were not retrieved. Therefore, we fitted several models with increasing number of components and determined the optimal number of components for our dataset by minimising the Bayesian information criterion (BIC, \citet{schwarz1978estimating}). Figures~\ref{fig:gmm_1sat} and \ref{fig:gmm_2sat} show the same GCs distribution in the four kinematic spaces respectively for the simulations MWsat\_n1\_$\Phi$60 and  MWsat\_n2\_$\Phi$30-150 as in Figs.~\ref{fig:kin_1sat},\ref{fig:kin_2sat} but with clusters colour-coded according to the groups retrieved by the minimum BIC criterion in the GMM. Truly accreted GCs, i.e. with the true label given by our simulation, are bordered by magenta and orange circles respectively for satellite 1 and satellite 2 labels. The bottom panels of Figs.~\ref{fig:gmm_1sat} and \ref{fig:gmm_2sat} show the confusion matrix where each row represents the true labels given by the simulations (MW, satellite or sat1 and sat2) while each column represents the predicted group identified by the model. Values are normalised to the total number of GCs in each true class. Interestingly, for both types of simulation, there is a predicted class that includes both the majority of accreted and in-situ clusters. This class (group 3 in Fig.~\ref{fig:kin_1sat} and group 7 in Fig.~\ref{fig:kin_2sat}) covers the region of the $E-L_z$ space where we have the largest overlap between GCs from different galactic progenitors, namely the region with $-17 \lesssim E \lesssim -12$. Overall, the groups that do not contain accreted clusters are those that trace highly prograde disc orbits in $E-L_z$ space, while the rest of the groups into which the MW is divided also contain a high fraction of satellite(s) clusters. The GMM, as it is designed, practically cuts the different kinematic spaces (with the exception of the action space) into separate slices that overlap only slightly. This behaviour therefore tends to support the interpretation suggested so far based on grouping in kinematic spaces but fails in grasping the underlying physical process.
We acknowledge the good prospects in using this type of tool to describe, in a quantitative manner, the distribution of stars and clusters in kinematic spaces, but also suggest caution in interpreting the results, as the components retrieved by the algorithm are  not directly related to the accretion events experienced by our Galaxy: the number of groups does not trace the number of satellites accreted over time, and groups are in general made of a mixture of in-situ and accreted populations.

\subsection{Adding an in-situ halo population}\label{insitu}
As we have stressed several times in previous sections, by construction, our modelled Milky Way-type galaxy does not contain any population of halo clusters before the interaction(s), since in-situ GCs are all initially confined in a Miyamoto-Nagai disc (see Sec.~\ref{method}). In literature, the presence of an in-situ halo is still being debated. For instance the only in-situ population that \citet{haywood18, di2019milky} find in great proportions in Gaia DR2 and APOGEE data, is the thick disc i.e. the early disc of the Galaxy heated to hot kinematics. On the other hand, \cite{belokurov2022} support a period of chaotic pre-disc evolution when stars are born in dense clumps scattered on all kinds of orbits and thus populating a hot halo. If the latter scenario were the case (both for stars and globular clusters), this would imply that the overlap between in-situ and satellites GCs in kinematic spaces could be even more significant. To test this claim with our simulations, we randomly selected a group of 20 particles from the dark matter halo of the MW-type galaxy, initially modelled with a Plummer distribution, and assumed these to be globular clusters formed in-situ, originally with halo kinematics. This was feasible since, as shown in Table~\ref{tab:1}, the mass of dark matter particles in our simulations is $\simeq7.4\times10^4\,M_{\odot}$ and so consistent with the mass range of Milky Way GCs \citep[see, for example, ][]{baumgardt19}. 

\begin{figure*}
\begin{centering}
\begin{multicols}{2}
     \includegraphics[width=.95\linewidth]{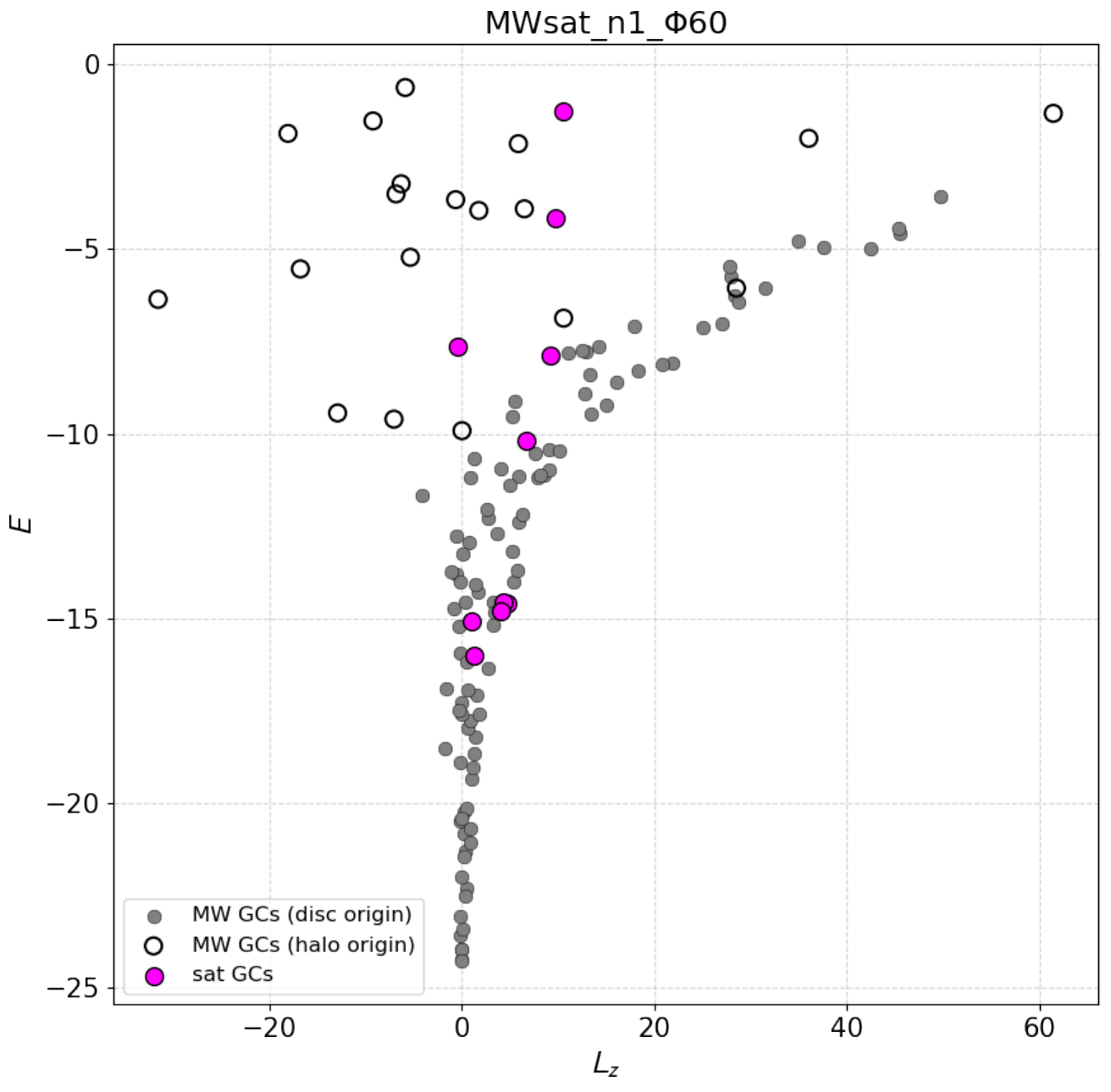}\par
         \includegraphics[width=.95\linewidth]{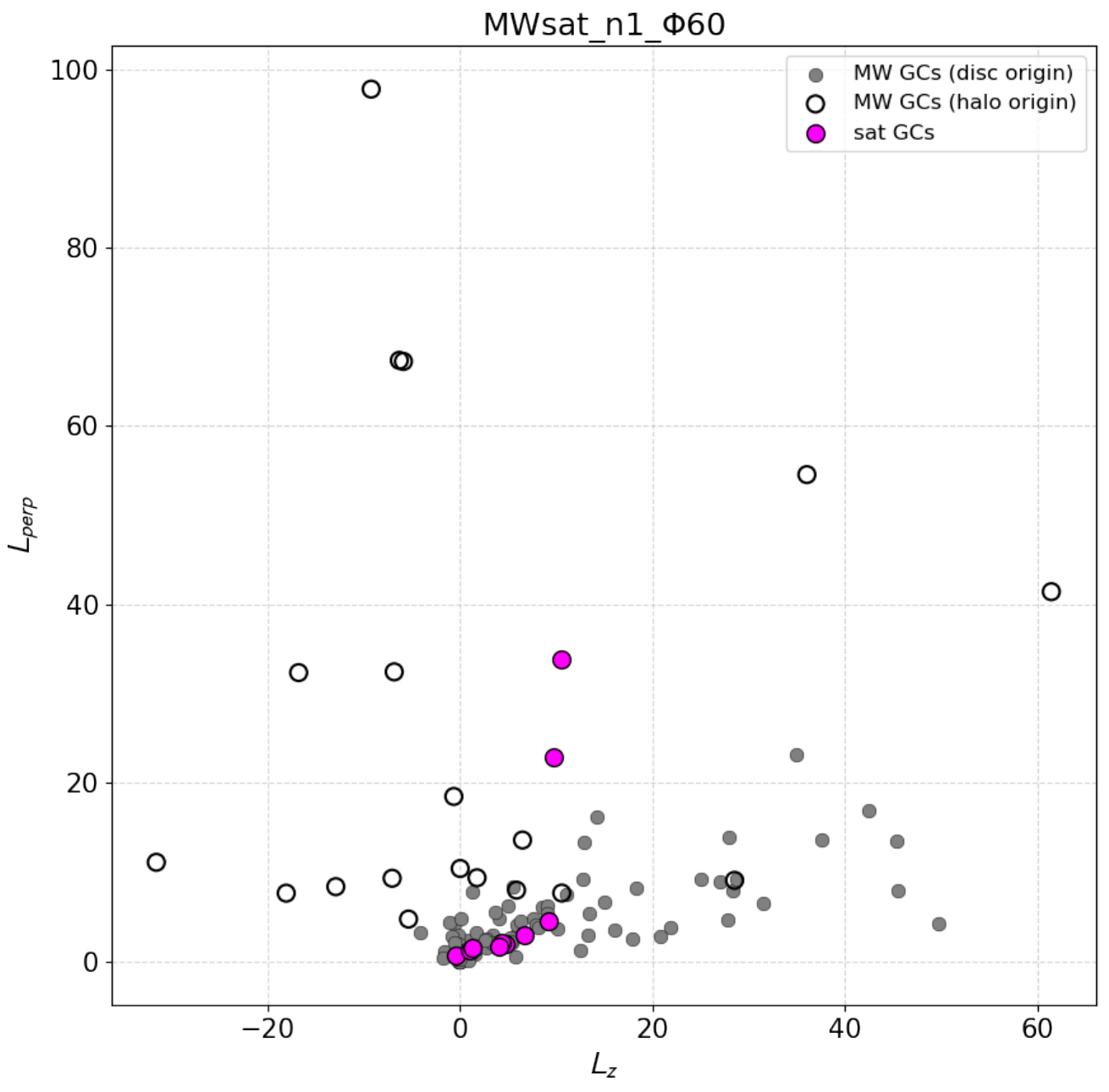} 
    \end{multicols}
    \vspace{-20pt}
    \begin{multicols}{2}
\includegraphics[width=.95\linewidth]{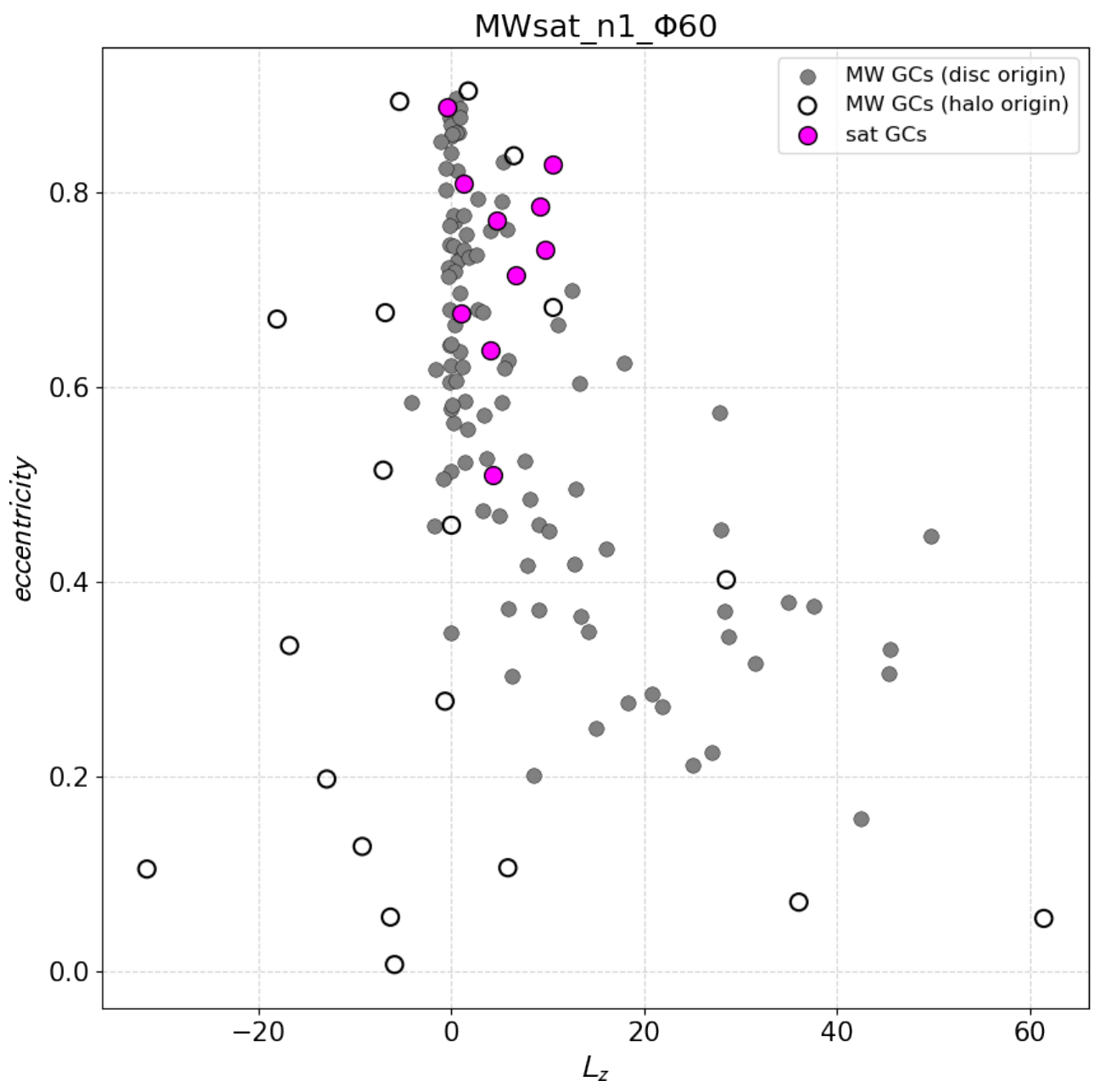}\par
\includegraphics[width=\linewidth]{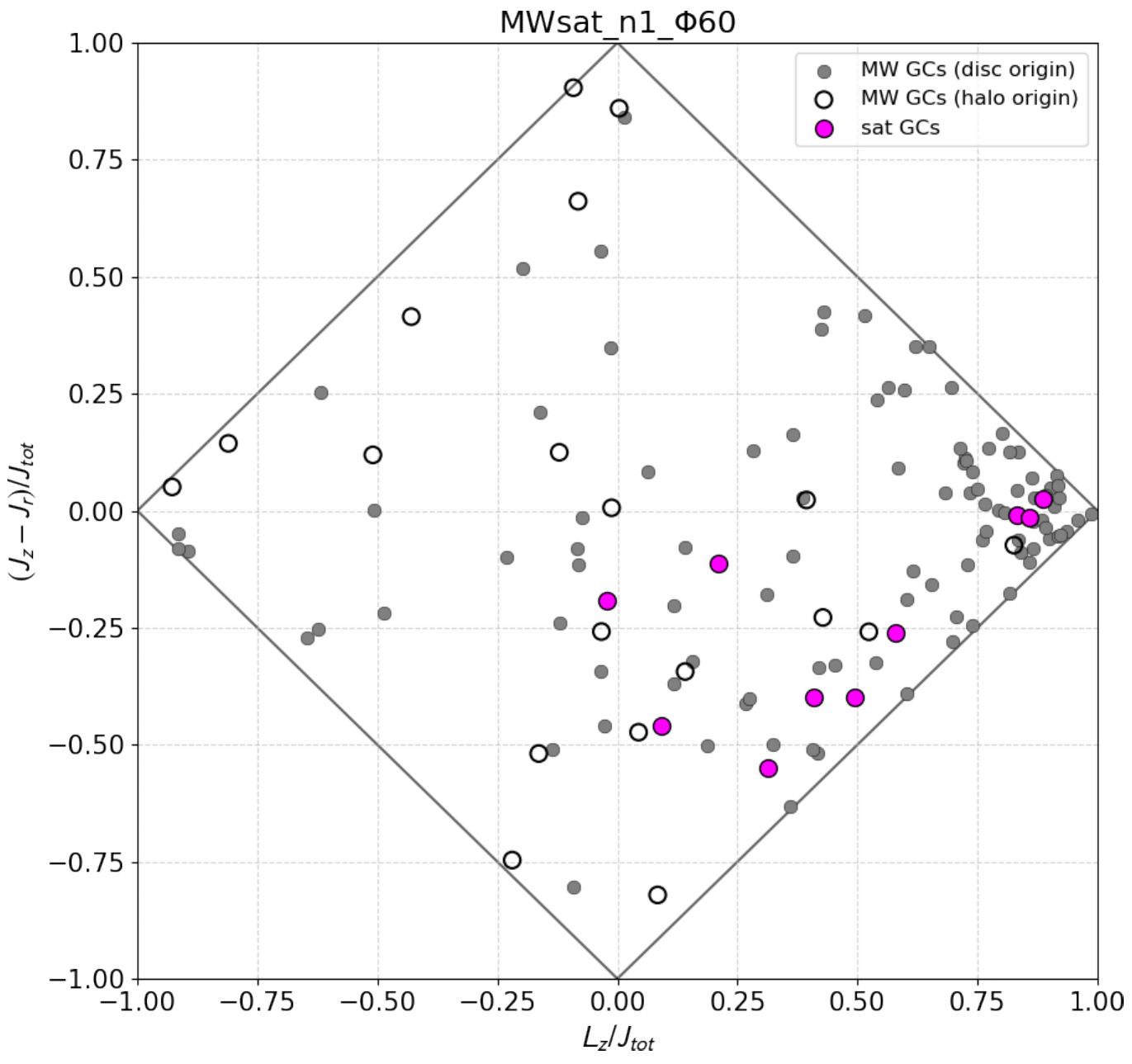}
    \end{multicols}
\end{centering}
\caption{Same as Fig.~\ref{fig:kin_1sat} with the addition of in-situ GCs with halo kinematics at the beginning of the simulation, represented as empty circles.}
\label{fig:halo_insitu}
\end{figure*}

Figure \ref{fig:halo_insitu} shows the final globular clusters distribution in the four kinematic spaces analysed i.e. the $E - L_z$, $L_{perp} - L_z$, $eccentricity - L_z$ and action space for the simulation MWsat\_n1\_$\Phi$60 - as Figure ~\ref{fig:kin_1sat} - with the addition of the mock halo in-situ GCs as empty circles. As expected, the population of in-situ GCs that initially has a halo kinematics, maintains the same type of kinematics: in $E - L_z$ space they end up in a rather high energy region since the dynamical friction on the individual clusters is not strong enough to allow them to lose energy and we also find many GCs with negative $L_z$. As a result, the different kinematic spaces - in particular the $E - L_z$ space - become even more indecipherable and at this point we are no longer sure whether even high-energy or highly retrograde clusters are accreted. In this scenario therefore, with only kinematic information, not even GCs coming from less massive satellites - as for instance with a 1:100 mass ratio (see App.~\ref{app_b}) - would be distinguishable from in-situ GCs. 

\section{Discussion: a new look at Galactic globular clusters}\label{discussion}

The results presented in the previous section show that we do not expect globular clusters accreted with their own parent satellite to clump in specific regions of the $E-L_z$ plane nor in the other analysed kinematic spaces, unless there are reasons to assume that the Galaxy has not experienced any significant (i.e. mass ratio $\sim 1:10$) merger -- assumption which would contradict the current estimates of the mass ratio of the Gaia Sausage-Enceladus merger \citep[see, for example,][]{helmi18}. Moreover, if the Galaxy has experienced more than one significant merger during its evolution \citep[see, for example,][]{kruijssen2020kraken}, clusters associated to these mergers can overlap and mix in $E-L_z$ space, particularly in the most gravitationally bound regions of the diagram (i.e. low energies). A suite of mergers would also muddle the traces left by previous accretion events in this space. Finally, a region of the $E-L_z$ space dominated by stars associated to a progenitor satellite is not necessarily dominated by clusters associated to the same accretion event. \\
 However, the current reconstruction of the merger tree of the Galaxy is exactly based on these assumptions: (1) dynamical coherence of globular clusters in the kinematic spaces, (2) negligible overlap of clusters originating in different satellite galaxies in all the kinematic spaces, (3) no kinematic heating of the in-situ population, (4)  straightforward correspondance between field stars and clusters in the $E-L_z$ space, meaning that the regions of this space where field stars associated to Sequoia, Gaia Sausage-Enceladus, Helmi stream -- and other possible accretions -- have been identified  are also the regions used as  boundaries to assign clusters to each of these progenitors. \\
 
\begin{figure*}
\centering
\includegraphics[clip=true,width=0.48\linewidth]{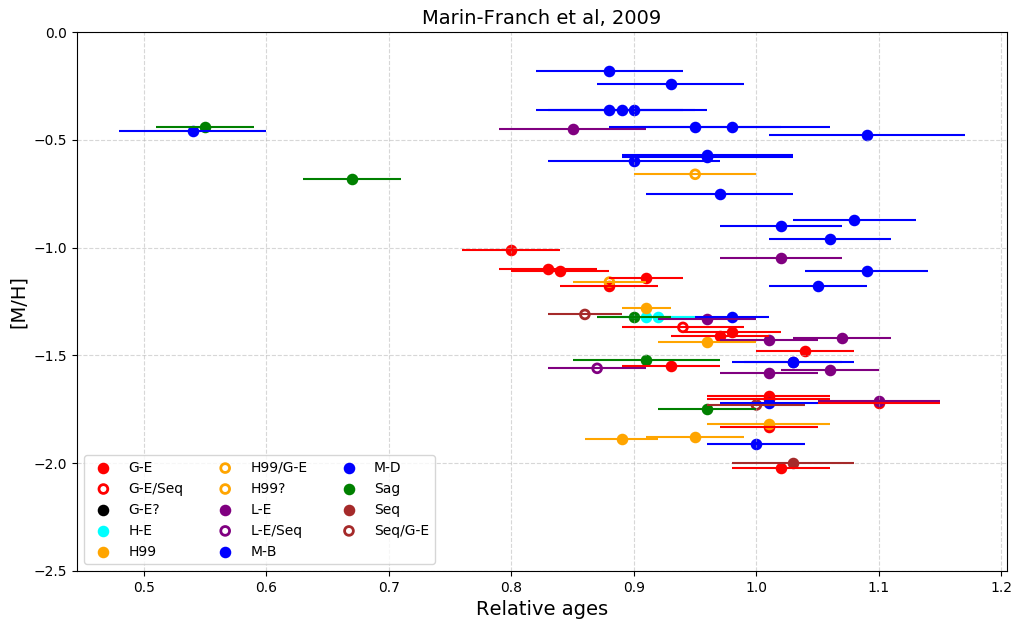}
\includegraphics[clip=true,width=0.48\linewidth]{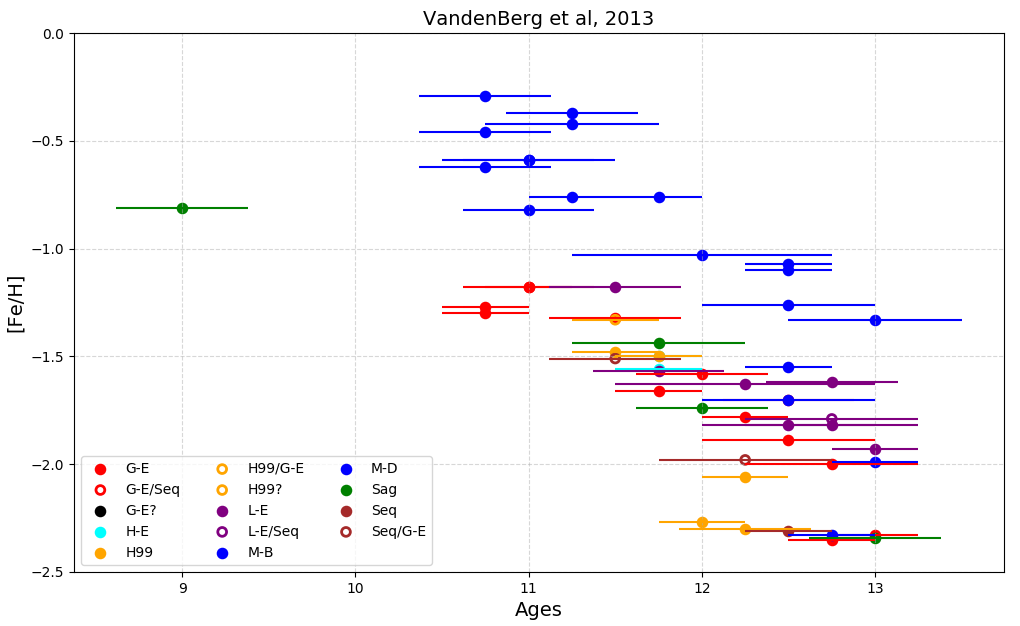}\\
\includegraphics[clip=true,width=0.48\linewidth]{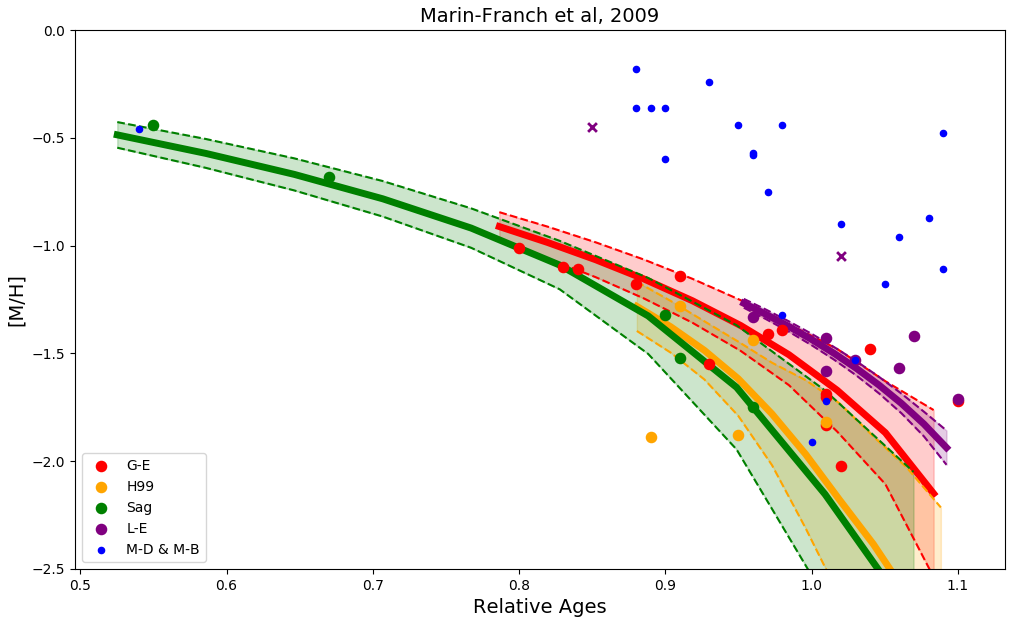}
\includegraphics[clip=true,width=0.48\linewidth]{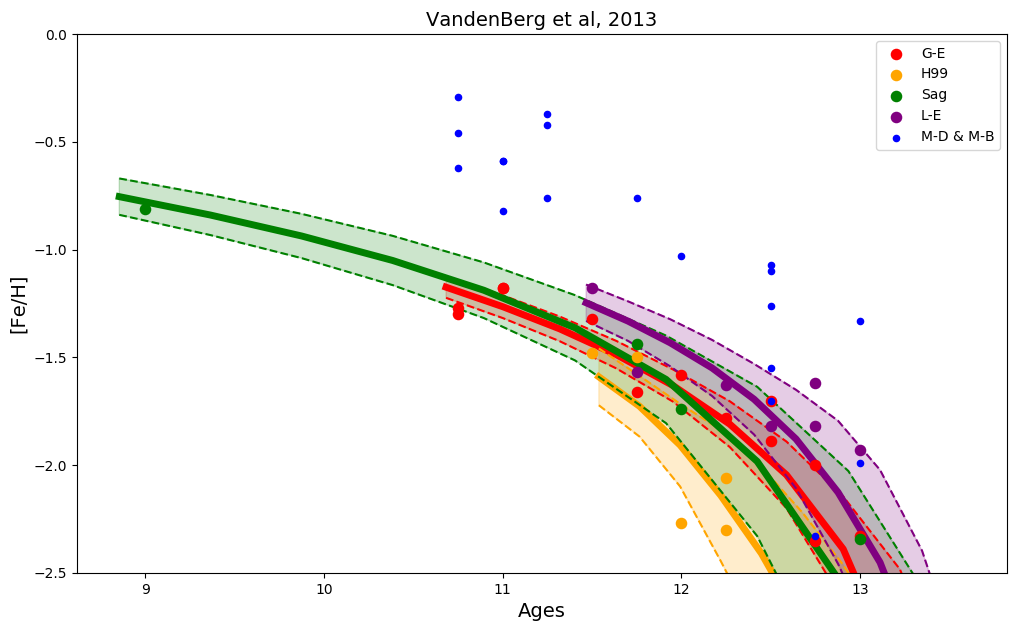}\\
\caption{\emph{Top panels:} Age-metallicity relations of Galactic globular clusters from the literature: (\emph{Left panel}): \citet{marin09}; (\emph{Right panel}): \citet{vandenberg13}. Different colours and symbols in each plot indicate the different galaxy progenitors of these clusters, following the classification given by \citet{massari2019origin}: Gaia Sausage Enceladus (G-E), Sequoia (Seq), Helmi stream (H99), Sagittarius (Sag), Low-Energy (L-E), High-Energy clusters (H-E) are shown by solid circles, while tentative associations are shown by empty circles. Clusters classified as in-situ by \citet{massari2019origin} (M-D and M-B groups) are shown with blue dots.  Errors on ages are also reported. \emph{Bottom panels:} Fits to the age-metallicity relations of accreted globular clusters in the G-E, H99, Sag and L-E groups. For each group, the mean fit to the data, as a function of metallicity, is shown, together with the corresponding standard deviation. Metallicities and ages as provided by   \citet{marin09} are used for the fits shown in the left panel, while in the right panel we make use of ages and metallicities as reported by \citet{vandenberg13}.  The in-situ clusters are also shown for comparison (blue points). The two clusters identified by a purple cross in the bottom-left panel are NGC~6441 and NGC~6121, which have been excluded from the fit, as motivated in the text.  Error bars are not reported in these two panels, to avoid having overly complex figures, but they have been taken into account in the fitting procedure (see text).}\label{AMR}
\end{figure*}

Our results show that there  is no physical reason to proceed in this way.  If we want to look for the remnants of these accretion events, we cannot require that the associated clusters satisfy any dynamical coherence in kinematic spaces. In this respect, let us recall the results by \citet{kruijssen2019formation}, who identified traces of an ancient accretion event in the Galaxy, called Kraken, studying the system of Galactic globular clusters. More specifically,  \citet{kruijssen2019formation} based their conclusions on the study of the age-metallicity relation of these clusters which was compared to that of a suite of cosmological zoom-in simulations of Milky Way-mass galaxies from the E-MOSAICS project.
This association of the possible clusters associated to Kraken was then questioned by  \citet{massari2019origin}, who noted  that most of the clusters reported in \citet{kruijssen2019formation} as possible members of Kraken were not dynamically coherent\footnote{See however the caution expressed by \citet{kruijssen2019formation} themselves on this association.}, since associated to different regions of the $E-L_z$ diagram. They thus reconsidered the initial suggestion of Kraken-like globular clusters made by \citet{kruijssen2019formation}, proposing a new one where Kraken-like clusters are a group of dynamically-coherent clusters found in the low-energy part of the $E-L_z$ plane, and concluded that ".. taking into account the dynamical properties is fundamental to establishing the origin of the different GCs of our Galaxy."  Note that it is on the basis of this new classification that  \citet{kruijssen2020kraken} subsequently based the reconstruction of the merger tree of the Milky Way \citep[see also][]{forbes20}.
Our work urges to reconsider this overall approach, since it shows that the remnants of massive accretion events are not expected to show any global clustering in this space, as already shown to be the case also for field stars \citep{jean2017kinematic, amarante22, khoperskov22b}: it is not because of a lack of dynamical coherence in the $E-L_z$ space that a subset of Galactic globular clusters cannot be associated to the same progenitor. \\
If we return to  the classification of the Galactic globular clusters made by \citet{massari2019origin} on the basis of their  kinematics (energies, and angular momenta), we can take a new look at the age-metallicity relation(s) (AMRs) of clusters which in their study are associated to different progenitors. It is important to rediscuss these age-metallicity relations briefly here, because they have been used in the literature to further justify kinematically-based classifications. For example, \citet{massari2019origin} concluded that quite remarkably, “the dynamical identification of associations of GC results in AMRs that are all well-defined and depict different shapes or amplitudes”. However, in their work, no actual fit to the data was done (as acknowledged by the authors themselves). It is worth thus reconsidering whether the kinematic classification by \citet{massari2019origin} effectively leads to groups whose AMRs have different shapes and amplitudes, especially considering that all age estimates come with associated errors, that each sample has a finite and limited number of clusters for which ages are available (typically less than 10) and that different ages and metallicity estimates exist in literature.  Do kinematically-based classifications effectively lead to select clusters with different enrichment histories? And how dependent are the retrieved enrichment histories on the chosen set of ages and metallicities? These are fundamental questions, since the apparent difference between AMRs of groups of clusters identified on the basis of their kinematics has been used as an additional probe of the robustness of the classification itself.   
To rediscuss this issue, we make use of ages and metallicities from literature data, as reported by \citet{marin09} and \citet{vandenberg13}. We recall the reader some main differences and similarities among these studies. \citet{marin09} measured relative ages of a sample of 64 Galactic globular clusters, observed in the framework of the HST/ACS Survey of Galactic globular clusters. The corresponding ages, and errors, are reported in Table~4 of their paper, for a set of different theoretical isochrones,  and metallicities for two abundance scales. We adopt in the following the ages corresponding to the theoretical isochrones of \citet{dotter07} using the \citet{zinn1984globular} abundance scale.  
\citet{vandenberg13} analyse 55 clusters -- many of which are also in the \citet{marin09} -- whose ages and metallicities are reported in Table~2 of their paper. 
The resulting age-metallicity relations are shown in Fig.~\ref{AMR}, where colours indicate the different galaxy progenitors to which \citet{massari2019origin} associate the clusters: Gaia-Sausage Enceladus, Sagittarius, Helmi Streams, Sequoia, together with two additional groups, the Low-Energy clusters, and High-Energy clusters (see Table~\ref{TableAGEMET}). 
Despite the differences in the shape and extension of the age-metallicity relations resulting from these two datasets, we notice that:
\begin{enumerate}
\item some of the clusters classified as accreted clusters, either associated to the Low-Energy group or to the Helmi stream -- namely E3, NGC~6441, and 
NGC~6121 -- have ages and metallicities compatible with being in-situ disc clusters heated to halo-kinematics \citep[see also][for similar conclusions]{forbes20, horta20}. They are indeed on the old, metal-rich branch of the age-metallicity relation (see for example the left panel of Fig.~\ref{AMR}), where most of the disc-bulge clusters lies, but have hotter kinematics than that expected from disc-bulge systems (and this is the reason why they have not been classified by \citet{massari2019origin} as in-situ clusters)\footnote{Note that these clusters are either absent from the analysis of the age-metallicity relation done in \citet{massari2019origin} (see their Appendix A.2 for a discussion about the metal-rich clusters analysed in their study) or, as it is the case for NGC~6121, only the age by \citet{vandenberg13} has been taken into account in their study. This age brings this cluster half-way between the young and old branches (see right panel of Fig.~\ref{AMR}), while the estimates by \citet{marin09} suggest an older age for it.};
\item the young branch at [M/H]$\gtrsim -1.4$ (see left panel) or [Fe/H]$\gtrsim -1.4 $ (see right panel) is made of clusters which have been associated to different progenitors, but which have age and metallicities that are indistinguishable one from another. If we look at the left panel of Fig.~\ref{AMR}, indeed,  in between the group of clusters associated to Gaia Sausage-Enceladus (such as NGC~5286, NGC~6205, NGC~7089, NGC~2808, NGC~288, NGC~362, NGC~1261, NGC~1851, red colours in Fig.~\ref{AMR}, and possibly NGC~5139, red empty circle in the same figure) we find a cluster identified as a disc cluster (NGC~6752, blue point), two clusters which have been associated to the Helmi stream (NGC~5272, NGC~6981, orange points, and possibly NGC~5904, orange empty circle), two High-Energy clusters (NGC~6584, and NGC~6934, cyan points), one cluster associated to Sagittarius (NGC~6715, green point) and one possible to Sequoia (NGC~3201, brown empty circle). At higher [M/H]$> -0.8$, the Pal~1 cluster, which is classified as a disc cluster, lies in between two clusters associated to the Sagittarius galaxy, Palomar~12 and Terzan~7. The only three clusters in this region that seem to slightly separate from the bulk of the distribution are NGC~4147, Arp~2 and NGC~6535 (relative ages lower than 1.0 and $-1.6 < \rm{[M/H]} \le -1.5$), but their are still compatible within $2\sigma$ with the other clusters;
\item at [M/H]$< -1.4$, no distinction can be made, on the basis of the age-metallicity relation, among the clusters that have been classified as Sequoia, Gaia Sausage Enceladus, Bulge/Disc or Sagittarius clusters.
\end{enumerate}
To further probe that the accreted groups in the \citet{massari2019origin} classification - which, we remind the reader, constitutes the backbone of many other classifications that have been published afterwards in the literature \citep[see also][]{forbes20, pfeffer20, callingham2022chemo} - do not depict specific AMRs in terms of shapes or amplitude, in the bottom panels of Fig.~\ref{AMR} we report a fit to the AMR of each of the groups  \citet{massari2019origin} identified. Note that we restricted this analysis only to the groups for which the association was considered robust (i.e. Sag, H99, G-E, L-E, but not groups as H99/G-E, H99?, G-E?, etc) excluding Sequoia (Seq) and the high-energy group (H-E) since for them the number of clusters for which ages are provided by \citet{marin09} or \citet{vandenberg13} is less than 3, making any fit meaningless. In the case of the AMRs from \citet{marin09} data, we have also excluded two clusters which were assigned by \citet{massari2019origin} to the L-E group (NGC~6121 and NGC~6441) since -- as discussed previously --  their ages, metallicities and chemical abundances are more compatible with them being in-situ clusters. In performing the fit, we have first bootstrapped the data and for each bootstrapped sample we have drawn ages from gaussian distributions with means and dispersions as those reported in \citet{marin09} (bottom-left panel) and \citet{vandenberg13} (bottom-right panel). In repeating this procedure a hundred times, the limited statistics, and the uncertainties on ages,  have been both taken into account in the analysis. The functional form of the curve fitted to the data is consistent with a leaky-box evolution of the systems, and it takes the form:
\begin{equation}
\rm [Fe/H] = p \times ln \left( \frac{t_f-t}{t_f} \right)
\end{equation}
$p$ being the effective yield, $t$ the look-back time and $t_f$ the time in the past where the chemical enrichment started.

\begin{figure}
\centering
\includegraphics[clip=true,width=0.8\linewidth]{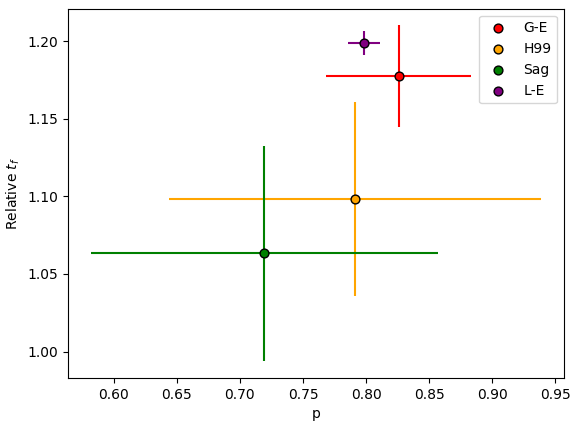}
\includegraphics[clip=true,width=0.8\linewidth]{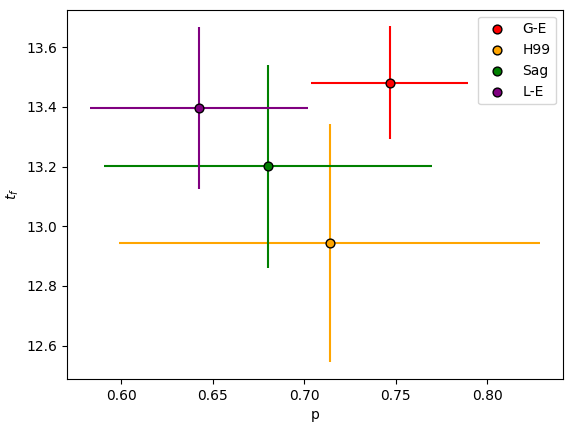}
\caption{Time, $t_f$, when the chemical enrichment started, as a function of the effective yield $p$ for the four different groups reported in Fig.~\ref{AMR}. \emph{Top panel}: $p$ and $t_f$ resulting from fitting the AMR of  \citet{marin09}; \emph{Bottom panel:} $p$ and $t_f$ resulting from fitting the AMR of  \citet{vandenberg13}. }\label{ptf}
\end{figure}

The result of this analysis are shown in the bottom panels of Fig.~\ref{AMR} and in Fig.~\ref{ptf}. If we use the AMRs by \citet{marin09}, we find that the chemical evolutions of the Sag and H99 groups have similar yields (respectively $p=0.72 \pm 0.14$ and $p=0.79\pm0.15$) and formation times (respectively $t_f=1.06\pm0.07$ and $t_f=1.10\pm0.06$). The L-E and G-E groups have slightly higher $p$ and $t_f$ (respectively $p=0.80 \pm 0.01$ and $p=0.83\pm 0.06$, $t_f=1.20\pm 0.01$ and $t_f=1.18 \pm 0.03$), however the evolution of G-E is still compatible -- within $1 \sigma$ -- with those found for Sag and H99 groups. If we use the AMRs by \citet{vandenberg13}, the retrieved chemical enrichment histories are all indistinguishable within $1\sigma$ (H99: $p=0.71\pm0.12, t_f=12.94\pm 0.40$, Sag: $p=0.68\pm0.09, t_f=13.20\pm0.34$, L-E: $p=0.64\pm 0.06, t_f=13.40\pm 0.27$,  G-E: $p=0.75\pm0.04, t_f=13.48\pm0.19$). Thus, the conclusion that kinematically-based classifications lead to groups of clusters with different chemical enrichment histories is not supported by a robust analysis of the data, and this for two independent datasets of ages and metallicities. \\
	

To summarise, we have demonstrated that the assumption of ``dynamical coherence'' for the interpretation of globular clusters in kinematic spaces is not supported by physical arguments (unless a very specific merger history for our Galaxy is assumed), and indeed we do not find in the observational data any confirmation that merger histories based on this assumption identify clusters with specific age-metallicity relations, and hence star formation histories of their progenitor systems. There is room for a new interpretation of AMR and/or chemical abundance spaces. Dynamical coherence should not be the "a priori" assumption to analyse globular cluster data, as done essentially by all studies published in the literature so far \citep[see, for example,][]{forbes20, callingham2022chemo} with the exception of \citet{kruijssen2019formation, horta20}. Even if uncertainties on ages are still significant, and even if possible overlaps among different evolutionary sequences can be challenging\footnote{See for example the overlap between the in-situ and accreted branches in the AMRs or abundance planes at low metallicities.}, it is by finding features in the age-metallicity relations or in abundance-spaces -- which are conserved quantities through time -- that we can hope to solve the question of the accretion history of the Galaxy.  

\section{Conclusions}\label{conclusions}

In this paper, we have analysed dissipationless $N$-body simulations of a Milky Way-type galaxy accreting one or two satellites with mass ratio 1:10, as well as some 1:100. Each galactic system has a population of disc globular clusters represented by point masses.
We have analysed this set of simulations to investigate the possibility to make use of kinematics information to find accreted globular clusters in the Galaxy, remnants of past accretion events and which have lost their spatial coherence. In particular, we have examined the energy - angular momentum ($E - L_{z}$) space, and shown that:
\begin{itemize}
\item[(1)] clusters originating from the same progenitor generally do not group together in $E - L_{z}$ space and thus GCs populating a large range of energy and/or angular momentum could have a common origin;
\item[(2)] if several satellites are accreted, their globular cluster populations can overlap, in particular in the region  $E - L_{z}$ plane where the  most gravitationally bound clusters are found (i.e. low energies);
\item[(3)] the clusters distribution does not necessarily match that of the field stars of the same progenitor galaxy: in several cases we find regions of $E-L_z$ space populated by clusters originating from one satellite, but dominated by field stars from another satellite. 
This implies that the correspondence between field stars and associated clusters in the $E-L_z$ diagram is not necessarily trivial: if a certain region of this space is dominated by the stellar debris of a satellite, this does not imply that clusters found in the same region all originated from the same satellite;   
\item[(4)] the in-situ population of globular clusters (that in our models is initially confined into a disc) can be heated up by the accretion, and a fraction of it acquires halo-like kinematics, thus becoming indistinguishable -- on the basis of the kinematics alone -- from the accreted populations.
\end{itemize}

We find that accreted clusters are confined in a tight range of energies only when they originate in low mass progenitors (mass ratio of 1:100). For such mass ratios, the distribution of angular momenta is still very extended, and these clusters tend to be all in the high energy part of the $E-L_z$ diagram, since dynamical friction is not able to bring them to the inner regions of the Milky Way-type galaxy before their parent satellite is  severely affected by the gravitational tides exerted by the main galaxy. Because clusters associated to low-mass progenitors are all found at high-energies in our simulations, a significant overlap is found also for these systems. Note that, besides the effects studied in this paper i.e. dynamical friction, perturbations induced by the other ongoing accretions \citep[see also][]{garrow2020}, other processes can contribute to the non-conservation of energy and angular momenta of globular clusters, as for example the mass growth of the Galaxy with time, which is expected to be significant especially in the first $4-5$~Gyr of its evolution \citep{snaith14}. Interestingly, simulations run in a cosmological context \citep{khoperskov22b, khoperskov22a} have recently confirmed the results of tailored $N$-body simulations \citep{jean2017kinematic, amarante22} about the large spread of accreted stars in $E-L_z$ plane. For these reasons, we are confident that these results are robust and would find confirmation also in a cosmological framework.  

We have also exploited other kinematic spaces suggested in the literature to reconstruct the accretion history of our Galaxy, namely $L_{perp} - L_z$, $eccentricity - L_z$ and action space, and this additional analysis only confirms the problems encountered in disentangling GCs in the $E - L_z$ space: clusters with different origins appear scattered and mixed together also in those spaces. By means of Gaussian Mixture Models, we have demonstrated that the overlap of clusters is not only a projection effect on specific planes but it is found also when the whole set of kinematic properties (i.e. $E, L_z, L_{perp}$, eccentricity, radial and vertical actions) is taken into account. Consequently, applying algorithms such as  Gaussian Mixture Models - a method lately used to identify groups of Galactic GCs in kinematic spaces - is conceptually wrong if the results of such methods are used to infer the merger tree of the Galaxy, and the interpretation of the results of these models tends to support the classification based on dynamical coherence.
These findings, together, question the history of accretions experienced by the
Galaxy, as it has been reconstructed so far by analysing its globular cluster system \citep[see][] {massari2019origin,kruijssen2020kraken,malhan2022global}. Indeed, this reconstruction usually assumes a dynamical coherence for clusters originating by the same accreted galaxy -- assumption theoretically motivated by the \citet{helmi2000mapping} work, which however was based on a number of over-simplifications, as extensively discussed in \citet{jean2017kinematic}. When more realistic simulations are analysed, indeed, the dynamical coherence of the accreted stars and globular clusters is not guaranteed anymore. When we test one of this merger histories \citep[specifically the first one, proposed by][]{massari2019origin} we find no confirmation that the association that has been  made so far in the literature between clusters and their progenitor galaxies is confirmed by the analysis of the age-metallicity relation of Galactic globular clusters.\\ 
\\

To understand the origin of the globular clusters population in the Milky Way, we  need to exploit the information from other dimensions like detailed chemical abundances and ages coming from the spectroscopic surveys such as the APOGEE survey and the soon operational WEAVE@WHT and MOONS@VLT surveys. The kinematic information could potentially be still useful, for example, in retrieving the paths traced in kinematic planes by GCs having the same origin, putting constraints on the orbital history of the progenitor satellites. However, this can be done only at the point when the distinction between accreted and in-situ globular clusters, and also the association with the different satellites, have already been made.

\onecolumn
\tiny
\begin{longtable}{llllllll}
\caption{\label{TableAGEMET}Metallicities, ages and errors on ages (when available) for all Galactic globular clusters studied in \citet{massari2019origin}. Ages and metallicities are taken from (1) \citet{marin09}, (2) \citet{vandenberg13}. Note that in the case of (1) ages are relative ones, while for (2) they are expressed in units of Gyr. For each cluster, its progenitor galaxy, as reported by \citet{massari2019origin}, is also given. }\\
\hline\hline
\multicolumn{1}{l}{GC name}& 
\multicolumn{1}{l}{Progenitor} &
 \multicolumn{1}{l}{[M/H]$^{(1)}$} &
  \multicolumn{1}{l}{Age$^{(1)}$} &
  \multicolumn{1}{l}{Age err$^{(1)}$} &

  \multicolumn{1}{l}{[Fe/H]$^{(2)}$} &
  \multicolumn{1}{l}{Age$^{(2)}$} &
  \multicolumn{1}{l}{Age err$^{(2)}$} \\
\hline
\endfirsthead
\caption{continued.}\\
\hline\hline
\multicolumn{1}{l}{GC name}& 
\multicolumn{1}{l}{Progenitor} &
 \multicolumn{1}{l}{[M/H]$^{(1)}$} &
  \multicolumn{1}{l}{Age$^{(1)}$} &
  \multicolumn{1}{l}{Age err$^{(1)}$} &

  \multicolumn{1}{l}{[Fe/H]$^{(2)}$} &
  \multicolumn{1}{l}{Age$^{(2)}$} &
  \multicolumn{1}{l}{Age err$^{(2)}$} \\
  
  \hline
\endhead
\hline
\endfoot

   NGC104 &        M-D &     -0.57 &    0.96 &        0.07 &  -0.76 &    11.75 &         0.25  \\
   NGC288 &        G-E &     -1.18 &    0.88 &        0.04 &-1.32 &    11.50 &         0.38  \\
   NGC362 &        G-E &     -1.11 &    0.84 &        0.04 & -1.30 &    10.75 &         0.25 \\
 Whiting1 &        Sag &       -- &     -- &         -- & -- &      -- &          --  \\
  NGC1261 &        G-E &     -1.10 &    0.83 &        0.04 & -1.27 &    10.75 &         0.25  \\
     Pal1 &        M-D &     -0.46 &    0.54 &        0.06 &  -- &      -- &          --  \\
      AM1 &        H-E &       -- &     -- &         -- & -- &      -- &          --  \\
  Eridanus &        H-E &       -- &     -- &         -- &  -- &      -- &          --  \\
     Pal2 &       G-E? &       -- &     -- &         -- &   -- &      -- &          --  \\
  NGC1851 &        G-E &     -1.01 &    0.80 &        0.04 &  -1.18 &    11.00 &         0.25  \\
  NGC1904 &        G-E &       -- &     -- &         -- & -- &      -- &          --  \\
  NGC2298 &        G-E &     -1.69 &    1.01 &        0.05 &-- &      -- &          -- \\
  NGC2419 &        Sag &       -- &     -- &         -- & -- &      -- &          --  \\
      Ko2 &        XXX &       -- &     -- &         -- &         -- &      -- &          --  \\
     Pyxis &        H-E &       -- &     -- &         -- &         -- &      -- &          -- \\
  NGC2808 &        G-E &     -1.14 &    0.91 &        0.03 & -1.18 &    11.00 &         0.38 \\
       E3 &       H99? &     -0.66 &    0.95 &        0.05 &  -- &      -- &          -- \\
     Pal3 &        H-E &       -- &     -- &         -- & -- &      -- &          --  \\
  NGC3201 &    Seq/G-E &     -1.31 &    0.86 &        0.03 & -1.51 &    11.50 &         0.38  \\
     Pal4 &        H-E &       -- &     -- &         -- &  -- &      -- &          --  \\
      Ko1 &        XXX &       -- &     -- &         -- &  -- &      -- &          -- \\
  NGC4147 &        G-E &     -1.55 &    0.93 &        0.04 &  -1.78 &    12.25 &         0.25 \\
  NGC4372 &        M-D &       -- &     -- &         -- & -- &      -- &          --  \\
   Rup106 &       H99? &       -- &     -- &         -- & -- &      -- &          -- \\
  NGC4590 &        H99 &     -1.89 &    0.89 &        0.03 & -2.27 &    12.00 &         0.25 \\
  NGC4833 &        G-E &     -1.70 &    1.01 &        0.05 & -1.89 &    12.50 &         0.50  \\
  NGC5024 &        H99 &     -1.82 &    1.01 &        0.05 & -2.06 &    12.25 &         0.25  \\
  NGC5053 &        H99 &     -1.88 &    0.95 &        0.04 & -2.30 &    12.25 &         0.38 \\
  NGC5139 &    G-E/Seq &     -1.37 &    0.94 &        0.05 & -- &      -- &          -- \\
  NGC5272 &        H99 &     -1.44 &    0.96 &        0.04 & -1.50 &    11.75 &         0.25  \\
  NGC5286 &        G-E &     -1.48 &    1.04 &        0.04 &  -1.70 &    12.50 &         0.38  \\
      AM4 &        XXX &       -- &     -- &         -- &  -- &      -- &          -- \\
  NGC5466 &        Seq &     -2.00 &    1.03 &        0.05 &  -2.31 &    12.50 &         0.25  \\
  NGC5634 &    H99/G-E &       -- &     -- &         -- &  -- &      -- &          --  \\
  NGC5694 &        H-E &       -- &     -- &         -- &  -- &      -- &          -- \\
   IC4499 &        Seq &       -- &     -- &         -- &  -- &      -- &          --  \\
  NGC5824 &        Sag &       -- &     -- &         -- &   -- &      -- &          -- \\
     Pal5 &       H99? &       -- &     -- &         -- &  -- &      -- &          --\\
  NGC5897 &        G-E &       -- &     -- &         -- &   -- &      -- &          -- \\
  NGC5904 &    H99/G-E &     -1.16 &    0.88 &        0.03 &   -1.33 &    11.50 &         0.25  \\
  NGC5927 &        M-D &     -0.18 &    0.88 &        0.06 &   -0.29 &    10.75 &         0.38  \\
  NGC5946 &        L-E &       -- &     -- &         -- &  -- &      -- &          --  \\
    BH176 &        M-D &       -- &     -- &         -- &          -- &      -- &          --\\
  NGC5986 &        L-E &     -1.43 &    1.01 &        0.04 &  -1.63 &    12.25 &         0.75  \\
   Lynga7 &        M-D &     -0.48 &    1.09 &        0.08 &  -- &      -- &          --  \\
    Pal14 &        H-E &       -- &     -- &         -- &  -- &      -- &          --  \\
  NGC6093 &        L-E &     -1.53 &    1.03 &        0.04 &    -- &      -- &          -- \\
  NGC6121 &        L-E &     -1.05 &    1.02 &        0.05 &  -1.18 &    11.50 &         0.38 \\
  NGC6101 &    Seq/G-E &     -1.73 &    1.00 &        0.04 &   -1.98 &    12.25 &         0.50  \\
  NGC6144 &        L-E &     -1.71 &    1.10 &        0.05 &   -1.82 &    12.75 &         0.50 \\
  NGC6139 &        L-E &       -- &     -- &         -- &          -- &      -- &          -- \\
  Terzan3 &        M-D &       -- &     -- &         -- &          -- &      -- &          --  \\
  NGC6171 &        M-B &     -0.87 &    1.08 &        0.05 &   -1.03 &    12.00 &         0.75  \\
 ESO452-11 &        M-B &       -- &     -- &         -- &  -- &      -- &          --  \\
  NGC6205 &        G-E &     -1.41 &    0.97 &        0.04 &  -1.58 &    12.00 &         0.38  \\
  NGC6229 &        G-E &       -- &     -- &         -- &          -- &      -- &          --\\
  NGC6218 &        M-D &     -1.18 &    1.05 &        0.04 &   -1.33 &    13.00 &         0.50 \\
  FSR1735 &        L-E &       -- &     -- &         -- &          -- &      -- &          -- \\
  NGC6235 &        G-E &       -- &     -- &         -- &     -- &      -- &          -- \\
  NGC6254 &        L-E &     -1.33 &    0.96 &        0.04 &    -1.57 &    11.75 &         0.38  \\
  NGC6256 &        L-E &       -- &     -- &         -- &          -- &      -- &          --  \\
    Pal15 &       G-E? &       -- &     -- &         -- &          -- &      -- &          --\\
  NGC6266 &        M-B &       -- &     -- &         -- &     -- &      -- &          --\\
  NGC6273 &        L-E &       -- &     -- &         -- &      -- &      -- &          --  \\
  NGC6284 &        G-E &       -- &     -- &         -- &   -- &      -- &          --\\
  NGC6287 &        L-E &       -- &     -- &         -- &    -- &      -- &          -- \\
  NGC6293 &        M-B &       -- &     -- &         -- &           -- &      -- &          --  \\
  NGC6304 &        M-B &     -0.24 &    0.93 &        0.06 &  -0.37 &    11.25 &         0.38  \\
  NGC6316 &        M-B &       -- &     -- &         -- &           -- &      -- &          --  \\
  NGC6341 &        G-E &     -2.02 &    1.02 &        0.04 &  -2.35 &    12.75 &         0.25  \\
  NGC6325 &        M-B &       -- &     -- &         -- &           -- &      -- &          --  \\
  NGC6333 &        L-E &       -- &     -- &         -- &            -- &      -- &          --  \\
  NGC6342 &        M-B &       -- &     -- &         -- &       -- &      -- &          -- \\
  NGC6356 &        M-D &       -- &     -- &         -- &           -- &      -- &          --  \\
  NGC6355 &        M-B &       -- &     -- &         -- &           -- &      -- &          -- \\
  NGC6352 &        M-D &     -0.36 &    0.90 &   12.67 &       -0.62 &    10.75 &         0.38  \\
   IC1257 &        G-E &       -- &     -- &         -- &          -- &      -- &          --  \\
  Terzan2 &        M-B &       -- &     -- &         -- &           -- &      -- &          -- \\
  NGC6366 &        M-D &     -0.44 &    0.95 &        0.07 &   -0.59 &    11.00 &         0.50  \\
  Terzan4 &        M-B &       -- &     -- &         -- &           -- &      -- &          --  \\
      HP1 &        M-B &       -- &     -- &         -- &           -- &      -- &          -- \\
  NGC6362 &        M-D &     -0.96 &    1.06 &        0.05 &   -1.07 &    12.50 &         0.25  \\
  Liller1 &        XXX &       -- &     -- &         -- &          -- &      -- &          --  \\
  NGC6380 &        M-B &       -- &     -- &         -- &          -- &      -- &          -- \\
  Terzan1 &        M-B &       -- &     -- &         -- &          -- &      -- &          --  \\
     Ton2 &        L-E &       -- &     -- &         -- &           -- &      -- &          --\\
  NGC6388 &        M-B &     -0.60 &    0.90 &        0.07 &   -- &      -- &          --\\
  NGC6402 &        L-E &       -- &     -- &         -- &          -- &      -- &          --\\
  NGC6401 &        L-E &       -- &     -- &         -- &          -- &      -- &          --\\
  NGC6397 &        M-D &     -1.72 &    1.01 &        0.04 &  -1.99 &    13.00 &         0.25  \\
     Pal6 &        L-E &       -- &     -- &         -- &           -- &      -- &          --  \\
  NGC6426 &        H-E &       -- &     -- &         -- &  -- &      -- &          --  \\
   Djorg1 &        G-E &       -- &     -- &         -- &          -- &      -- &          --  \\
  Terzan5 &        M-B &       -- &     -- &         -- &           -- &      -- &          --  \\
  NGC6440 &        M-B &       -- &     -- &         -- &           -- &      -- &          --  \\
  NGC6441 &        L-E &     -0.45 &    0.85 &        0.06 &   -- &      -- &          --  \\
  Terzan6 &        M-B &       -- &     -- &         -- &          -- &      -- &          --  \\
  NGC6453 &        L-E &       -- &     -- &         -- &    -- &      -- &          --\\
     UKS1 &        XXX &       -- &     -- &         -- &          -- &      -- &          -- \\
  NGC6496 &        M-D &     -0.36 &    0.88 &        0.06 & -0.46 &    10.75 &         0.38  \\
  Terzan9 &        M-B &       -- &     -- &         -- &         -- &      -- &          -- \\
   Djorg2 &        M-B &       -- &     -- &         -- &          -- &      -- &          --  \\
  NGC6517 &        L-E &       -- &     -- &         -- &           -- &      -- &          --  \\
 Terzan10 &        G-E &       -- &     -- &         -- &          -- &      -- &          --  \\
  NGC6522 &        M-B &       -- &     -- &         -- &          -- &      -- &          --  \\
  NGC6535 &    L-E/Seq &     -1.56 &    0.87 &        0.04 &       -1.51 &  12.75 &         0.50  \\
  NGC6528 &        M-B &       -- &     -- &         -- &          -- &      -- &          -- \\
  NGC6539 &        M-B &       -- &     -- &         -- &           -- &      -- &          -- \\
  NGC6540 &        M-B &       -- &     -- &         -- &          -- &      -- &          --  \\
  NGC6544 &        L-E &       -- &     -- &         --  &         -- &      -- &          -- \\
  NGC6541 &        L-E &     -1.57 &    1.06 &        0.04 &   -1.82 &    12.50 &         0.50  \\
2MASSGC01 &        XXX &       -- &     -- &         -- &    -- &      -- &          --\\
 ESO280-06 &        G-E &       -- &     -- &         -- &    -- &      -- &          -- \\
  NGC6553 &        M-B &       -- &     -- &         -- &    -- &      -- &          -- \\
2MASSGC02 &        XXX &       -- &     -- &         -- &    -- &      -- &          --  \\
  NGC6558 &        M-B &       -- &     -- &         -- &     -- &      -- &          --  \\
   IC1276 &        M-D &       -- &     -- &         -- &    -- &      -- &          --  \\
 Terzan12 &        M-D &       -- &     -- &         -- &   -- &      -- &          -- \\
  NGC6569 &        M-B &       -- &     -- &         -- &     -- &      -- &          -- \\
    BH261 &        M-B &       -- &     -- &         -- &   -- &      -- &          -- \\
GLIMPSE02 &        XXX &       -- &     -- &         -- &     -- &      -- &          -- \\
  NGC6584 &        H-E &     -1.32 &    0.92 &        0.03 &    -1.50 &    11.75 &         0.25  \\
  NGC6624 &        M-B &     -0.36 &    0.89 &        0.06 &  -0.42 &    11.25 &         0.50  \\
  NGC6626 &        M-B &       -- &     -- &         -- &          -- &      -- &          --  \\
  NGC6638 &        M-B &       -- &     -- &         -- &           -- &      -- &          --  \\
  NGC6637 &        M-B &     -0.58 &    0.96 &        0.07 &    -0.59 &    11.00 &         0.38  \\
  NGC6642 &        M-B &       -- &     -- &         -- &           -- &      -- &          --  \\
  NGC6652 &        M-B &     -0.75 &    0.97 &        0.06 &    -0.76 &    11.25 &         0.25  \\
  NGC6656 &        M-D &     -1.53 &    1.03 &        0.05 &  -1.70 &    12.50 &         0.50 \\
     Pal8 &        M-D &       -- &     -- &         -- &           -- &      -- &          --\\
  NGC6681 &        L-E &     -1.42 &    1.07 &        0.04 & -1.62 &    12.75 &         0.38 \\
GLIMPSE01 &        XXX &       -- &     -- &         -- &    -- &      -- &          -- \\
  NGC6712 &        L-E &       -- &     -- &         -- &   -- &      -- &          --  \\
  NGC6715 &        Sag &     -1.32 &    0.90 &        0.03 &   -1.44 &    11.75 &         0.50  \\
  NGC6717 &        M-B &     -1.11 &    1.09 &        0.05 &   -1.26 &    12.50 &         0.50  \\
  NGC6723 &        M-B &     -0.90 &    1.02 &        0.05 &  -1.10 &    12.50 &         0.25 \\
  NGC6749 &        M-D &       -- &     -- &         -- &     -- &      -- &          -- \\
  NGC6752 &        M-D &     -1.32 &    0.98 &        0.03 &   -1.55 &    12.50 &         0.25  \\
  NGC6760 &        M-D &       -- &     -- &         -- &          -- &      -- &          --  \\
  NGC6779 &        G-E &     -1.72 &    1.10 &   13.70 &       -2.00 &    12.75 &         0.50  \\
  Terzan7 &        Sag &     -0.44 &    0.55 &        0.04 &  -- &      -- &          --  \\
    Pal10 &        M-D &       -- &     -- &         -- &    -- &      -- &          --  \\
     Arp2 &        Sag &     -1.52 &    0.91 &        0.06 &  -1.74 &    12.00 &         0.38 \\
  NGC6809 &        L-E &     -1.58 &    1.01 &        0.04 &  -1.93 &    13.00 &         0.25 \\
  Terzan8 &        Sag &     -1.75 &    0.96 &        0.04 &  -2.34 &    13.00 &         0.38  \\
    Pal11 &        M-D &       -- &     -- &         -- &          -- &      -- &          --  \\
  NGC6838 &        M-D &     -0.44 &    0.98 &        0.08 &   -0.82 &    11.00 &         0.38  \\
  NGC6864 &        G-E &       -- &     -- &         -- &  -- &      -- &          -- \\
  NGC6934 &        H-E &     -1.32 &    0.91 &        0.04 &  -1.56 &    11.75 &         0.25 \\
  NGC6981 &        H99 &     -1.28 &    0.91 &        0.02 &   -1.48 &    11.50 &         0.25  \\
  NGC7006 &        Seq &       -- &     -- &         -- &          -- &      -- &          -- \\
  NGC7078 &        M-D &     -1.91 &    1.00 &        0.04 &  -2.33 &    12.75 &         0.25  \\
  NGC7089 &        G-E &     -1.39 &    0.98 &        0.04 &  -1.66 &    11.75 &         0.25  \\
  NGC7099 &        G-E &     -1.83 &    1.01 &        0.04 &   -2.33 &    13.00 &         0.25  \\
    Pal12 &        Sag &     -0.68 &    0.67 &        0.04 &  -0.81 &     9.00 &         0.38  \\
    Pal13 &        Seq &       -- &     -- &         -- &          -- &      -- &          --  \\
  NGC7492 &        G-E &       -- &     -- &         -- &    -- &      -- &          --\\
    Crater &        H-E &       -- &     -- &         -- &          -- &      -- &          -- \\
  FSR1716 &        M-D &       -- &     -- &         -- &          -- &      -- &          -- \\
  FSR1758 &        Seq &       -- &     -- &         -- &           -- &      -- &          -- \\
\end{longtable}
\normalsize
\twocolumn

\begin{acknowledgements}
We are grateful to the referee for their report, which much improved the presentation of the results. We also wish to thank J. Pfeffer for his comments to a first version of this manuscript. GP and PDM thank P. Boldrini and D. Valls-Gabaud, for their comments on this work.
This work has made use of the computational resources obtained through the DARI grant A0120410154.
AMB acknowledges funding from the European Union\textquotesingle s Horizon 2020 research and innovation programme under the Marie Sk\l{}odowska-Curie grant agreement No 895174. FR acknowledges support from the Knut and Alice Wallenberg Foundation. 
\end{acknowledgements}

%
%
\bibliographystyle{aa}
\bibliography{bibliography}

\begin{thebibliography}{120}
\expandafter\ifx\csname natexlab\endcsname\relax\def\natexlab#1{#1}\fi

\bibitem[{{Amarante} {et~al.}(2022){Amarante}, {Debattista}, {Beraldo e Silva},
  {Laporte}, \& {Deg}}]{amarante22}
{Amarante}, J. A.~S., {Debattista}, V.~P., {Beraldo e Silva}, L., {Laporte}, C.
  F.~P., \& {Deg}, N. 2022, \apj, 937, 12

\bibitem[{Amorisco(2017)}]{amorisco2017contributions}
Amorisco, N. 2017, Monthly Notices of the Royal Astronomical Society, 464, 2882

\bibitem[{{Baumgardt} \& {Hilker}(2018)}]{baumgardt18}
{Baumgardt}, H. \& {Hilker}, M. 2018, \mnras, 478, 1520

\bibitem[{{Baumgardt} {et~al.}(2019){Baumgardt}, {Hilker}, {Sollima}, \&
  {Bellini}}]{baumgardt19}
{Baumgardt}, H., {Hilker}, M., {Sollima}, A., \& {Bellini}, A. 2019, \mnras,
  482, 5138

\bibitem[{{Baumgardt} \& {Vasiliev}(2021)}]{baumgardt21}
{Baumgardt}, H. \& {Vasiliev}, E. 2021, \mnras, 505, 5957

\bibitem[{Bellazzini {et~al.}(2020)Bellazzini, Ibata, Malhan, Martin, Famaey,
  \& Thomas}]{bellazzini2020globular}
Bellazzini, M., Ibata, R., Malhan, K., {et~al.} 2020, Astronomy \&
  Astrophysics, 636, A107

\bibitem[{Belokurov {et~al.}(2018)Belokurov, Erkal, Evans, Koposov, \&
  Deason}]{belokurov2018co}
Belokurov, V., Erkal, D., Evans, N., Koposov, S., \& Deason, A. 2018, Monthly
  Notices of the Royal Astronomical Society, 478, 611

\bibitem[{{Belokurov} \& {Kravtsov}(2022)}]{belokurov2022}
{Belokurov}, V. \& {Kravtsov}, A. 2022, \mnras, 514, 689

\bibitem[{Binney(2012)}]{binney2012actions}
Binney, J. 2012, Monthly Notices of the Royal Astronomical Society, 426, 1324

\bibitem[{Binney \& Spergel(1982)}]{binney1982spectral}
Binney, J. \& Spergel, D. 1982, The Astrophysical Journal, 252, 308

\bibitem[{Callingham {et~al.}(2022)Callingham, Cautun, Deason, Frenk, Grand, \&
  Marinacci}]{callingham2022chemo}
Callingham, T.~M., Cautun, M., Deason, A.~J., {et~al.} 2022, Monthly Notices of
  the Royal Astronomical Society, 513, 4107

\bibitem[{Carretta {et~al.}(2010)Carretta, Bragaglia, Gratton, Recio-Blanco,
  Lucatello, D'Orazi, \& Cassisi}]{carretta2010properties}
Carretta, E., Bragaglia, A., Gratton, R.~G., {et~al.} 2010, Astronomy \&
  Astrophysics, 516, A55

\bibitem[{Casertano \& Hut(1985)}]{casertano1985core}
Casertano, S. \& Hut, P. 1985, The Astrophysical Journal, 298, 80

\bibitem[{{Chen} \& {Gnedin}(2022)}]{chen_gnedin2022}
{Chen}, Y. \& {Gnedin}, O.~Y. 2022, \mnras, 514, 4736

\bibitem[{Cirasuolo {et~al.}(2014)Cirasuolo, Afonso, Carollo, Flores, Maiolino,
  Oliva, Paltani, Vanzi, Evans, Abreu, {et~al.}}]{cirasuolo2014moons}
Cirasuolo, M., Afonso, J., Carollo, M., {et~al.} 2014, in Ground-based and
  airborne instrumentation for astronomy V, Vol. 9147, International Society
  for Optics and Photonics, 91470N

\bibitem[{{Cirasuolo} {et~al.}(2020){Cirasuolo}, {Fairley}, {Rees}, {Gonzalez},
  {Taylor}, {Maiolino}, {Afonso}, {Evans}, {Flores}, {Lilly}, {Oliva},
  {Paltani}, {Vanzi}, {Abreu}, {Accardo}, {Adams}, {{\'A}lvarez M{\'e}ndez},
  {Amans}, {Amarantidis}, {Atek}, {Atkinson}, {Banerji}, {Barrett},
  {Barrientos}, {Bauer}, {Beard}, {B{\'e}chet}, {Belfiore}, {Bellazzini},
  {Benoist}, {Best}, {Biazzo}, {Black}, {Boettger}, {Bonifacio}, {Bowler},
  {Bragaglia}, {Brierley}, {Brinchmann}, {Brinkmann}, {Buat}, {Buitrago},
  {Burgarella}, {Burningham}, {Buscher}, {Cabral}, {Caffau}, {Cardoso},
  {Carnall}, {Carollo}, {Castillo}, {Castignani}, {Catelan}, {Cicone},
  {Cimatti}, {Cioni}, {Clementini}, {Cochrane}, {Coelho}, {Colling}, {Contini},
  {Contreras}, {Conzelmann}, {Cresci}, {Cropper}, {Cucciati}, {Cullen},
  {Cumani}, {Curti}, {Da Silva}, {Daddi}, {Dalessandro}, {Dalessio}, {Dauvin},
  {Davidson}, {de Laverny}, {Delplancke-Str{\"o}bele}, {De Lucia}, {Del
  Vecchio}, {Dessauges-Zavadsky}, {Di Matteo}, {Dole}, {Drass}, {Dunlop},
  {D{\"u}nner}, {Eales}, {Ellis}, {Enriques}, {Fasola}, {Ferguson}, {Ferruzzi},
  {Fisher}, {Flores}, {Fontana}, {Forchi}, {Francois}, {Franzetti}, {Gargiulo},
  {Garilli}, {Gaudemard}, {Gieles}, {Gilmore}, {Ginolfi}, {Gomes}, {Guinouard},
  {Gutierrez}, {Haigron}, {Hammer}, {Hammersley}, {Haniff}, {Harrison},
  {Haywood}, {Hill}, {Hubin}, {Humphrey}, {Ibata}, {Infante}, {Ives}, {Ivison},
  {Iwert}, {Jablonka}, {Jakob}, {Jarvis}, {King}, {Kneib}, {Laporte},
  {Lawrence}, {Lee}, {Li Causi}, {Lorenzoni}, {Lucatello}, {Luco}, {Macleod},
  {Magliocchetti}, {Magrini}, {Mainieri}, {Maire}, {Mannucci}, {Martin},
  {Matute}, {Maurogordato}, {McGee}, {Mcleod}, {McLure}, {McMahon}, {Melse},
  {Messias}, {Mucciarelli}, {Nisini}, {Nix}, {Norberg}, {Oesch}, {Oliveira},
  {Origlia}, {Padilla}, {Palsa}, {Pancino}, {Papaderos}, {Pappalardo}, {Parry},
  {Pasquini}, {Peacock}, {Pedichini}, {Pello}, {Peng}, {Pentericci}, {Pfuhl},
  {Piazzesi}, {Popovic}, {Pozzetti}, {Puech}, {Puzia}, {Raichoor}, {Randich},
  {Recio-Blanco}, {Reis}, {Reix}, {Renzini}, {Rodrigues}, {Rojas},
  {Rojas-Arriagada}, {Rota}, {Royer}, {Sacco}, {Sanchez-Janssen}, {Sanna},
  {Santos}, {Sarzi}, {Schaerer}, {Schiavon}, {Schnell}, {Schultheis},
  {Scodeggio}, {Serjeant}, {Shen}, {Simmonds}, {Smoker}, {Sobral}, {Sordet},
  {Sp{\'e}rone}, {Strachan}, {Sun}, {Swinbank}, {Tait}, {Tereno}, {Tojeiro},
  {Torres}, {Tosi}, {Tozzi}, {Tresiter}, {Valenti}, {Valenzuela Navarro},
  {Vanzella}, {Vergani}, {Verhamme}, {Vernet}, {Vignali}, {Vinther}, {Von
  Dran}, {Waring}, {Watson}, {Wild}, {Willesme}, {Woodward}, {Wuyts}, {Yang},
  {Zamorani}, {Zoccali}, {Bluck}, \& {Trussler}}]{cirasuolo20}
{Cirasuolo}, M., {Fairley}, A., {Rees}, P., {et~al.} 2020, The Messenger, 180,
  10

\bibitem[{Conroy {et~al.}(2021)Conroy, Naidu, Garavito-Camargo, Besla,
  Zaritsky, Bonaca, \& Johnson}]{conroy2021all}
Conroy, C., Naidu, R.~P., Garavito-Camargo, N., {et~al.} 2021, Nature, 592, 534

\bibitem[{{Cooper} {et~al.}(2015){Cooper}, {Parry}, {Lowing}, {Cole}, \&
  {Frenk}}]{cooper15}
{Cooper}, A.~P., {Parry}, O.~H., {Lowing}, B., {Cole}, S., \& {Frenk}, C. 2015,
  \mnras, 454, 3185

\bibitem[{Cordoni {et~al.}(2021)Cordoni, Da~Costa, Yong, Mackey, Marino, Monty,
  Nordlander, Norris, Asplund, Bessell, {et~al.}}]{cordoni2021exploring}
Cordoni, G., Da~Costa, G., Yong, D., {et~al.} 2021, Monthly Notices of the
  Royal Astronomical Society, 503, 2539

\bibitem[{Dalton {et~al.}(2012)Dalton, Trager, Abrams, Carter, Bonifacio,
  Aguerri, MacIntosh, Evans, Lewis, Navarro, {et~al.}}]{dalton2012weave}
Dalton, G., Trager, S.~C., Abrams, D.~C., {et~al.} 2012, in Ground-based and
  Airborne Instrumentation for Astronomy IV, Vol. 8446, SPIE, 220--231

\bibitem[{De~Blok {et~al.}(2001)De~Blok, McGaugh, Bosma, \& Rubin}]{de2001mass}
De~Blok, W., McGaugh, S.~S., Bosma, A., \& Rubin, V.~C. 2001, The Astrophysical
  Journal, 552, L23

\bibitem[{de~Jong {et~al.}(2012)de~Jong, Chiappini, \& Schnurr}]{de20124most}
de~Jong, R.~S., Chiappini, C., \& Schnurr, O. 2012, in EPJ Web of Conferences,
  Vol.~19, EDP Sciences, 09004

\bibitem[{De~Lucia \& Helmi(2008)}]{de2008galaxy}
De~Lucia, G. \& Helmi, A. 2008, Monthly Notices of the Royal Astronomical
  Society, 391, 14

\bibitem[{De~Naray {et~al.}(2008)De~Naray, McGaugh, \& De~Blok}]{de2008mass}
De~Naray, R.~K., McGaugh, S.~S., \& De~Blok, W. 2008, The Astrophysical
  Journal, 676, 920

\bibitem[{{Deason} {et~al.}(2011){Deason}, {Belokurov}, \& {Evans}}]{deason11}
{Deason}, A.~J., {Belokurov}, V., \& {Evans}, N.~W. 2011, \mnras, 416, 2903

\bibitem[{Deason {et~al.}(2019)Deason, Belokurov, \& Sanders}]{deason2019total}
Deason, A.~J., Belokurov, V., \& Sanders, J.~L. 2019, Monthly Notices of the
  Royal Astronomical Society, 490, 3426

\bibitem[{Deason {et~al.}(2016)Deason, Mao, \& Wechsler}]{deason2016eating}
Deason, A.~J., Mao, Y.-Y., \& Wechsler, R.~H. 2016, The Astrophysical Journal,
  821, 5

\bibitem[{Di~Matteo(2016)}]{di2016disc}
Di~Matteo, P. 2016, Publications of the Astronomical Society of Australia, 33

\bibitem[{{Di Matteo} {et~al.}(2015){Di Matteo}, {G{\'o}mez}, {Haywood},
  {Combes}, {Lehnert}, {Ness}, {Snaith}, {Katz}, \& {Semelin}}]{dimatteo15}
{Di Matteo}, P., {G{\'o}mez}, A., {Haywood}, M., {et~al.} 2015, \aap, 577, A1

\bibitem[{Di~Matteo {et~al.}(2019)Di~Matteo, Haywood, Lehnert, Katz,
  Khoperskov, Snaith, G{\'o}mez, \& Robichon}]{di2019milky}
Di~Matteo, P., Haywood, M., Lehnert, M., {et~al.} 2019, Astronomy \&
  Astrophysics, 632, A4

\bibitem[{{Di Matteo} {et~al.}(2011){Di Matteo}, {Lehnert}, {Qu}, \& {van
  Driel}}]{dimatteo11}
{Di Matteo}, P., {Lehnert}, M.~D., {Qu}, Y., \& {van Driel}, W. 2011, \aap,
  525, L3

\bibitem[{Donlon \& Newberg(2022)}]{Donlon2022:2211.12576v1}
Donlon, T. \& Newberg, H.~J. 2022 [\eprint{Arxiv:2211.12576v1}]

\bibitem[{{Dotter} {et~al.}(2007){Dotter}, {Chaboyer}, {Jevremovi{\'c}},
  {Baron}, {Ferguson}, {Sarajedini}, \& {Anderson}}]{dotter07}
{Dotter}, A., {Chaboyer}, B., {Jevremovi{\'c}}, D., {et~al.} 2007, \aj, 134,
  376

\bibitem[{{Dotter} {et~al.}(2010){Dotter}, {Sarajedini}, {Anderson},
  {Aparicio}, {Bedin}, {Chaboyer}, {Majewski}, {Mar{\'\i}n-Franch}, {Milone},
  {Paust}, {Piotto}, {Reid}, {Rosenberg}, \& {Siegel}}]{dotter10}
{Dotter}, A., {Sarajedini}, A., {Anderson}, J., {et~al.} 2010, \apj, 708, 698

\bibitem[{Erkal {et~al.}(2021)Erkal, Deason, Belokurov, Xue, Koposov, Bird,
  Liu, Simion, Yang, Zhang, {et~al.}}]{erkal2021detection}
Erkal, D., Deason, A.~J., Belokurov, V., {et~al.} 2021, Monthly Notices of the
  Royal Astronomical Society, 506, 2677

\bibitem[{Flores \& Primack(1994)}]{flores1994observational}
Flores, R.~A. \& Primack, J.~R. 1994, arXiv preprint astro-ph/9402004

\bibitem[{{Forbes}(2020)}]{forbes20}
{Forbes}, D.~A. 2020, \mnras, 493, 847

\bibitem[{{Forbes} \& {Bridges}(2010)}]{forbes10}
{Forbes}, D.~A. \& {Bridges}, T. 2010, \mnras, 404, 1203

\bibitem[{{Gaia Collaboration}(2022)}]{gaiaDR3cat}
{Gaia Collaboration}. 2022, VizieR Online Data Catalog, I/355

\bibitem[{{Gaia Collaboration} {et~al.}(2018{\natexlab{a}}){Gaia
  Collaboration}, {Brown}, {Vallenari}, {Prusti}, {de Bruijne}, {Babusiaux},
  {Bailer-Jones}, {Biermann}, {Evans}, {Eyer}, {Jansen}, {Jordi}, {Klioner},
  {Lammers}, {Lindegren}, {Luri}, {Mignard}, {Panem}, {Pourbaix}, {Randich},
  {Sartoretti}, {Siddiqui}, {Soubiran}, {van Leeuwen}, {Walton}, {Arenou},
  {Bastian}, {Cropper}, {Drimmel}, {Katz}, {Lattanzi}, {Bakker}, {Cacciari},
  {Casta{\~n}eda}, {Chaoul}, {Cheek}, {De Angeli}, {Fabricius}, {Guerra},
  {Holl}, {Masana}, {Messineo}, {Mowlavi}, {Nienartowicz}, {Panuzzo},
  {Portell}, {Riello}, {Seabroke}, {Tanga}, {Th{\'e}venin}, {Gracia-Abril},
  {Comoretto}, {Garcia-Reinaldos}, {Teyssier}, {Altmann}, {Andrae}, {Audard},
  {Bellas-Velidis}, {Benson}, {Berthier}, {Blomme}, {Burgess}, {Busso},
  {Carry}, {Cellino}, {Clementini}, {Clotet}, {Creevey}, {Davidson}, {De
  Ridder}, {Delchambre}, {Dell'Oro}, {Ducourant},
  {Fern{\'a}ndez-Hern{\'a}ndez}, {Fouesneau}, {Fr{\'e}mat}, {Galluccio},
  {Garc{\'\i}a-Torres}, {Gonz{\'a}lez-N{\'u}{\~n}ez}, {Gonz{\'a}lez-Vidal},
  {Gosset}, {Guy}, {Halbwachs}, {Hambly}, {Harrison}, {Hern{\'a}ndez},
  {Hestroffer}, {Hodgkin}, {Hutton}, {Jasniewicz}, {Jean-Antoine-Piccolo},
  {Jordan}, {Korn}, {Krone-Martins}, {Lanzafame}, {Lebzelter}, {L{\"o}ffler},
  {Manteiga}, {Marrese}, {Mart{\'\i}n-Fleitas}, {Moitinho}, {Mora}, {Muinonen},
  {Osinde}, {Pancino}, {Pauwels}, {Petit}, {Recio-Blanco}, {Richards},
  {Rimoldini}, {Robin}, {Sarro}, {Siopis}, {Smith}, {Sozzetti}, {S{\"u}veges},
  {Torra}, {van Reeven}, {Abbas}, {Abreu Aramburu}, {Accart}, {Aerts},
  {Altavilla}, {{\'A}lvarez}, {Alvarez}, {Alves}, {Anderson}, {Andrei},
  {Anglada Varela}, {Antiche}, {Antoja}, {Arcay}, {Astraatmadja}, {Bach},
  {Baker}, {Balaguer-N{\'u}{\~n}ez}, {Balm}, {Barache}, {Barata}, {Barbato},
  {Barblan}, {Barklem}, {Barrado}, {Barros}, {Barstow}, {Bartholom{\'e}
  Mu{\~n}oz}, {Bassilana}, {Becciani}, {Bellazzini}, {Berihuete}, {Bertone},
  {Bianchi}, {Bienaym{\'e}}, {Blanco-Cuaresma}, {Boch}, {Boeche}, {Bombrun},
  {Borrachero}, {Bossini}, {Bouquillon}, {Bourda}, {Bragaglia}, {Bramante},
  {Breddels}, {Bressan}, {Brouillet}, {Br{\"u}semeister}, {Brugaletta},
  {Bucciarelli}, {Burlacu}, {Busonero}, {Butkevich}, {Buzzi}, {Caffau},
  {Cancelliere}, {Cannizzaro}, {Cantat-Gaudin}, {Carballo}, {Carlucci},
  {Carrasco}, {Casamiquela}, {Castellani}, {Castro-Ginard}, {Charlot},
  {Chemin}, {Chiavassa}, {Cocozza}, {Costigan}, {Cowell}, {Crifo}, {Crosta},
  {Crowley}, {Cuypers}, {Dafonte}, {Damerdji}, {Dapergolas}, {David}, {David},
  {de Laverny}, {De Luise}, {De March}, {de Martino}, {de Souza}, {de Torres},
  {Debosscher}, {del Pozo}, {Delbo}, {Delgado}, {Delgado}, {Di Matteo},
  {Diakite}, {Diener}, {Distefano}, {Dolding}, {Drazinos}, {Dur{\'a}n},
  {Edvardsson}, {Enke}, {Eriksson}, {Esquej}, {Eynard Bontemps}, {Fabre},
  {Fabrizio}, {Faigler}, {Falc{\~a}o}, {Farr{\`a}s Casas}, {Federici},
  {Fedorets}, {Fernique}, {Figueras}, {Filippi}, {Findeisen}, {Fonti},
  {Fraile}, {Fraser}, {Fr{\'e}zouls}, {Gai}, {Galleti}, {Garabato},
  {Garc{\'\i}a-Sedano}, {Garofalo}, {Garralda}, {Gavel}, {Gavras}, {Gerssen},
  {Geyer}, {Giacobbe}, {Gilmore}, {Girona}, {Giuffrida}, {Glass}, {Gomes},
  {Granvik}, {Gueguen}, {Guerrier}, {Guiraud}, {Guti{\'e}rrez-S{\'a}nchez},
  {Haigron}, {Hatzidimitriou}, {Hauser}, {Haywood}, {Heiter}, {Helmi}, {Heu},
  {Hilger}, {Hobbs}, {Hofmann}, {Holland}, {Huckle}, {Hypki}, {Icardi},
  {Jan{\ss}en}, {Jevardat de Fombelle}, {Jonker}, {Juh{\'a}sz}, {Julbe},
  {Karampelas}, {Kewley}, {Klar}, {Kochoska}, {Kohley}, {Kolenberg},
  {Kontizas}, {Kontizas}, {Koposov}, {Kordopatis}, {Kostrzewa-Rutkowska},
  {Koubsky}, {Lambert}, {Lanza}, {Lasne}, {Lavigne}, {Le Fustec}, {Le
  Poncin-Lafitte}, {Lebreton}, {Leccia}, {Leclerc}, {Lecoeur-Taibi},
  {Lenhardt}, {Leroux}, {Liao}, {Licata}, {Lindstr{\o}m}, {Lister}, {Livanou},
  {Lobel}, {L{\'o}pez}, {Managau}, {Mann}, {Mantelet}, {Marchal}, {Marchant},
  {Marconi}, {Marinoni}, {Marschalk{\'o}}, {Marshall}, {Martino}, {Marton},
  {Mary}, {Massari}, {Matijevi{\v{c}}}, {Mazeh}, {McMillan}, {Messina},
  {Michalik}, {Millar}, {Molina}, {Molinaro}, {Moln{\'a}r}, {Montegriffo},
  {Mor}, {Morbidelli}, {Morel}, {Morris}, {Mulone}, {Muraveva}, {Musella},
  {Nelemans}, {Nicastro}, {Noval}, {O'Mullane}, {Ord{\'e}novic},
  {Ord{\'o}{\~n}ez-Blanco}, {Osborne}, {Pagani}, {Pagano}, {Pailler},
  {Palacin}, {Palaversa}, {Panahi}, {Pawlak}, {Piersimoni}, {Pineau}, {Plachy},
  {Plum}, {Poggio}, {Poujoulet}, {Pr{\v{s}}a}, {Pulone}, {Racero}, {Ragaini},
  {Rambaux}, {Ramos-Lerate}, {Regibo}, {Reyl{\'e}}, {Riclet}, {Ripepi}, {Riva},
  {Rivard}, {Rixon}, {Roegiers}, {Roelens}, {Romero-G{\'o}mez}, {Rowell},
  {Royer}, {Ruiz-Dern}, {Sadowski}, {Sagrist{\`a} Sell{\'e}s}, {Sahlmann},
  {Salgado}, {Salguero}, {Sanna}, {Santana-Ros}, {Sarasso}, {Savietto},
  {Schultheis}, {Sciacca}, {Segol}, {Segovia}, {S{\'e}gransan}, {Shih},
  {Siltala}, {Silva}, {Smart}, {Smith}, {Solano}, {Solitro}, {Sordo}, {Soria
  Nieto}, {Souchay}, {Spagna}, {Spoto}, {Stampa}, {Steele},
  {Steidelm{\"u}ller}, {Stephenson}, {Stoev}, {Suess}, {Surdej}, {Szabados},
  {Szegedi-Elek}, {Tapiador}, {Taris}, {Tauran}, {Taylor}, {Teixeira},
  {Terrett}, {Teyssandier}, {Thuillot}, {Titarenko}, {Torra Clotet}, {Turon},
  {Ulla}, {Utrilla}, {Uzzi}, {Vaillant}, {Valentini}, {Valette}, {van Elteren},
  {Van Hemelryck}, {van Leeuwen}, {Vaschetto}, {Vecchiato}, {Veljanoski},
  {Viala}, {Vicente}, {Vogt}, {von Essen}, {Voss}, {Votruba}, {Voutsinas},
  {Walmsley}, {Weiler}, {Wertz}, {Wevers}, {Wyrzykowski}, {Yoldas},
  {{\v{Z}}erjal}, {Ziaeepour}, {Zorec}, {Zschocke}, {Zucker}, {Zurbach}, \&
  {Zwitter}}]{brown18gaia}
{Gaia Collaboration}, {Brown}, A.~G.~A., {Vallenari}, A., {et~al.}
  2018{\natexlab{a}}, \aap, 616, A1

\bibitem[{{Gaia Collaboration} {et~al.}(2018{\natexlab{b}}){Gaia
  Collaboration}, {Helmi}, {van Leeuwen}, {McMillan}, {Massari}, {Antoja},
  {Robin}, {Lindegren}, {Bastian}, {Arenou}, {Babusiaux}, {Biermann},
  {Breddels}, {Hobbs}, {Jordi}, {Pancino}, {Reyl{\'e}}, {Veljanoski}, {Brown},
  {Vallenari}, {Prusti}, {de Bruijne}, {Bailer-Jones}, {Evans}, {Eyer},
  {Jansen}, {Klioner}, {Lammers}, {Luri}, {Mignard}, {Panem}, {Pourbaix},
  {Randich}, {Sartoretti}, {Siddiqui}, {Soubiran}, {Walton}, {Cropper},
  {Drimmel}, {Katz}, {Lattanzi}, {Bakker}, {Cacciari}, {Casta{\~n}eda},
  {Chaoul}, {Cheek}, {De Angeli}, {Fabricius}, {Guerra}, {Holl}, {Masana},
  {Messineo}, {Mowlavi}, {Nienartowicz}, {Panuzzo}, {Portell}, {Riello},
  {Seabroke}, {Tanga}, {Th{\'e}venin}, {Gracia-Abril}, {Comoretto},
  {Garcia-Reinaldos}, {Teyssier}, {Altmann}, {Andrae}, {Audard},
  {Bellas-Velidis}, {Benson}, {Berthier}, {Blomme}, {Burgess}, {Busso},
  {Carry}, {Cellino}, {Clementini}, {Clotet}, {Creevey}, {Davidson}, {De
  Ridder}, {Delchambre}, {Dell'Oro}, {Ducourant},
  {Fern{\'a}ndez-Hern{\'a}ndez}, {Fouesneau}, {Fr{\'e}mat}, {Galluccio},
  {Garc{\'\i}a-Torres}, {Gonz{\'a}lez-N{\'u}{\~n}ez}, {Gonz{\'a}lez-Vidal},
  {Gosset}, {Guy}, {Halbwachs}, {Hambly}, {Harrison}, {Hern{\'a}ndez},
  {Hestroffer}, {Hodgkin}, {Hutton}, {Jasniewicz}, {Jean-Antoine-Piccolo},
  {Jordan}, {Korn}, {Krone-Martins}, {Lanzafame}, {Lebzelter}, {L{\"o}ffler},
  {Manteiga}, {Marrese}, {Mart{\'\i}n-Fleitas}, {Moitinho}, {Mora}, {Muinonen},
  {Osinde}, {Pauwels}, {Petit}, {Recio-Blanco}, {Richards}, {Rimoldini},
  {Sarro}, {Siopis}, {Smith}, {Sozzetti}, {S{\"u}veges}, {Torra}, {van Reeven},
  {Abbas}, {Abreu Aramburu}, {Accart}, {Aerts}, {Altavilla}, {{\'A}lvarez},
  {Alvarez}, {Alves}, {Anderson}, {Andrei}, {Anglada Varela}, {Antiche},
  {Arcay}, {Astraatmadja}, {Bach}, {Baker}, {Balaguer-N{\'u}{\~n}ez}, {Balm},
  {Barache}, {Barata}, {Barbato}, {Barblan}, {Barklem}, {Barrado}, {Barros},
  {Barstow}, {Bartholom{\'e} Mu{\~n}oz}, {Bassilana}, {Becciani}, {Bellazzini},
  {Berihuete}, {Bertone}, {Bianchi}, {Bienaym{\'e}}, {Blanco-Cuaresma}, {Boch},
  {Boeche}, {Bombrun}, {Borrachero}, {Bossini}, {Bouquillon}, {Bourda},
  {Bragaglia}, {Bramante}, {Bressan}, {Brouillet}, {Br{\"u}semeister},
  {Brugaletta}, {Bucciarelli}, {Burlacu}, {Busonero}, {Butkevich}, {Buzzi},
  {Caffau}, {Cancelliere}, {Cannizzaro}, {Cantat-Gaudin}, {Carballo},
  {Carlucci}, {Carrasco}, {Casamiquela}, {Castellani}, {Castro-Ginard},
  {Charlot}, {Chemin}, {Chiavassa}, {Cocozza}, {Costigan}, {Cowell}, {Crifo},
  {Crosta}, {Crowley}, {Cuypers}, {Dafonte}, {Damerdji}, {Dapergolas}, {David},
  {David}, {de Laverny}, {De Luise}, {De March}, {de Martino}, {de Souza}, {de
  Torres}, {Debosscher}, {del Pozo}, {Delbo}, {Delgado}, {Delgado}, {Di
  Matteo}, {Diakite}, {Diener}, {Distefano}, {Dolding}, {Drazinos},
  {Dur{\'a}n}, {Edvardsson}, {Enke}, {Eriksson}, {Esquej}, {Eynard Bontemps},
  {Fabre}, {Fabrizio}, {Faigler}, {Falc{\~a}o}, {Farr{\`a}s Casas}, {Federici},
  {Fedorets}, {Fernique}, {Figueras}, {Filippi}, {Findeisen}, {Fonti},
  {Fraile}, {Fraser}, {Fr{\'e}zouls}, {Gai}, {Galleti}, {Garabato},
  {Garc{\'\i}a-Sedano}, {Garofalo}, {Garralda}, {Gavel}, {Gavras}, {Gerssen},
  {Geyer}, {Giacobbe}, {Gilmore}, {Girona}, {Giuffrida}, {Glass}, {Gomes},
  {Granvik}, {Gueguen}, {Guerrier}, {Guiraud}, {Guti{\'e}rrez-S{\'a}nchez},
  {Hofmann}, {Holland}, {Huckle}, {Hypki}, {Icardi}, {Jan{\ss}en}, {Jevardat de
  Fombelle}, {Jonker}, {Juh{\'a}sz}, {Julbe}, {Karampelas}, {Kewley}, {Klar},
  {Kochoska}, {Kohley}, {Kolenberg}, {Kontizas}, {Kontizas}, {Koposov},
  {Kordopatis}, {Kostrzewa-Rutkowska}, {Koubsky}, {Lambert}, {Lanza}, {Lasne},
  {Lavigne}, {Le Fustec}, {Le Poncin-Lafitte}, {Lebreton}, {Leccia}, {Leclerc},
  {Lecoeur-Taibi}, {Lenhardt}, {Leroux}, {Liao}, {Licata}, {Lindstr{\o}m},
  {Lister}, {Livanou}, {Lobel}, {L{\'o}pez}, {Managau}, {Mann}, {Mantelet},
  {Marchal}, {Marchant}, {Marconi}, {Marinoni}, {Marschalk{\'o}}, {Marshall},
  {Martino}, {Marton}, {Mary}, {Matijevi{\v{c}}}, {Mazeh}, {Messina},
  {Michalik}, {Millar}, {Molina}, {Molinaro}, {Moln{\'a}r}, {Montegriffo},
  {Mor}, {Morbidelli}, {Morel}, {Morris}, {Mulone}, {Muraveva}, {Musella},
  {Nelemans}, {Nicastro}, {Noval}, {O'Mullane}, {Ord{\'e}novic},
  {Ord{\'o}{\~n}ez-Blanco}, {Osborne}, {Pagani}, {Pagano}, {Pailler},
  {Palacin}, {Palaversa}, {Panahi}, {Pawlak}, {Piersimoni}, {Pineau}, {Plachy},
  {Plum}, {Poggio}, {Poujoulet}, {Pr{\v{s}}a}, {Pulone}, {Racero}, {Ragaini},
  {Rambaux}, {Ramos-Lerate}, {Regibo}, {Riclet}, {Ripepi}, {Riva}, {Rivard},
  {Rixon}, {Roegiers}, {Roelens}, {Romero-G{\'o}mez}, {Rowell}, {Royer},
  {Ruiz-Dern}, {Sadowski}, {Sagrist{\`a} Sell{\'e}s}, {Sahlmann}, {Salgado},
  {Salguero}, {Sanna}, {Santana-Ros}, {Sarasso}, {Savietto}, {Schultheis},
  {Sciacca}, {Segol}, {Segovia}, {S{\'e}gransan}, {Shih}, {Siltala}, {Silva},
  {Smart}, {Smith}, {Solano}, {Solitro}, {Sordo}, {Soria Nieto}, {Souchay},
  {Spagna}, {Spoto}, {Stampa}, {Steele}, {Steidelm{\"u}ller}, {Stephenson},
  {Stoev}, {Suess}, {Surdej}, {Szabados}, {Szegedi-Elek}, {Tapiador}, {Taris},
  {Tauran}, {Taylor}, {Teixeira}, {Terrett}, {Teyssandier}, {Thuillot},
  {Titarenko}, {Torra Clotet}, {Turon}, {Ulla}, {Utrilla}, {Uzzi}, {Vaillant},
  {Valentini}, {Valette}, {van Elteren}, {Van Hemelryck}, {van Leeuwen},
  {Vaschetto}, {Vecchiato}, {Viala}, {Vicente}, {Vogt}, {von Essen}, {Voss},
  {Votruba}, {Voutsinas}, {Walmsley}, {Weiler}, {Wertz}, {Wevems},
  {Wyrzykowski}, {Yoldas}, {{\v{Z}}erjal}, {Ziaeepour}, {Zorec}, {Zschocke},
  {Zucker}, {Zurbach}, \& {Zwitter}}]{helmi2018gaia}
{Gaia Collaboration}, {Helmi}, A., {van Leeuwen}, F., {et~al.}
  2018{\natexlab{b}}, \aap, 616, A12

\bibitem[{{Garrow} {et~al.}(2020){Garrow}, {Webb}, \& {Bovy}}]{garrow2020}
{Garrow}, T., {Webb}, J.~J., \& {Bovy}, J. 2020, \mnras, 499, 804

\bibitem[{Gentile {et~al.}(2005)Gentile, Burkert, Salucci, Klein, \&
  Walter}]{gentile2005dwarf}
Gentile, G., Burkert, A., Salucci, P., Klein, U., \& Walter, F. 2005, The
  Astrophysical Journal, 634, L145

\bibitem[{{G{\'o}mez} {et~al.}(2018){G{\'o}mez}, {Di Matteo}, {Schultheis},
  {Fragkoudi}, {Haywood}, \& {Combes}}]{gomez18}
{G{\'o}mez}, A., {Di Matteo}, P., {Schultheis}, M., {et~al.} 2018, \aap, 615,
  A100

\bibitem[{Grand {et~al.}(2019)Grand, Deason, White, Simpson, G{\'o}mez,
  Marinacci, \& Pakmor}]{grand2019effects}
Grand, R.~J., Deason, A.~J., White, S.~D., {et~al.} 2019, Monthly Notices of
  the Royal Astronomical Society: Letters, 487, L72

\bibitem[{Gratton {et~al.}(2019)Gratton, Bragaglia, Carretta, D’Orazi,
  Lucatello, \& Sollima}]{gratton2019globular}
Gratton, R., Bragaglia, A., Carretta, E., {et~al.} 2019, The Astronomy and
  Astrophysics Review, 27, 1

\bibitem[{{Hammer} {et~al.}(2023){Hammer}, {Li}, {Mamon}, {Pawlowski},
  {Bonifacio}, {Jiao}, {Wang}, {Wang}, \& {Yang}}]{hammer2023}
{Hammer}, F., {Li}, H., {Mamon}, G.~A., {et~al.} 2023, \mnras, 519, 5059

\bibitem[{{Harris}(1996)}]{harris96}
{Harris}, W.~E. 1996, \aj, 112, 1487

\bibitem[{Harris {et~al.}(2013)Harris, Harris, \& Alessi}]{harris2013catalog}
Harris, W.~E., Harris, G.~L., \& Alessi, M. 2013, The Astrophysical Journal,
  772, 82

\bibitem[{Harris \& Racine(1979)}]{harris1979globular}
Harris, W.~E. \& Racine, R. 1979, Annual Review of Astronomy and Astrophysics,
  17, 241

\bibitem[{Hartwick(1987)}]{hartwick1987structure}
Hartwick, F. 1987, in the Galaxy (Springer), 281--290

\bibitem[{Haywood {et~al.}(2013)Haywood, Di~Matteo, Lehnert, Katz, \&
  G{\'o}mez}]{haywood2013age}
Haywood, M., Di~Matteo, P., Lehnert, M.~D., Katz, D., \& G{\'o}mez, A. 2013,
  Astronomy \& Astrophysics, 560, A109

\bibitem[{{Haywood} {et~al.}(2018){Haywood}, {Di Matteo}, {Lehnert}, {Snaith},
  {Khoperskov}, \& {G{\'o}mez}}]{haywood18}
{Haywood}, M., {Di Matteo}, P., {Lehnert}, M.~D., {et~al.} 2018, \apj, 863, 113

\bibitem[{{Helmi} {et~al.}(2018){Helmi}, {Babusiaux}, {Koppelman}, {Massari},
  {Veljanoski}, \& {Brown}}]{helmi18}
{Helmi}, A., {Babusiaux}, C., {Koppelman}, H.~H., {et~al.} 2018, \nat, 563, 85

\bibitem[{{Helmi} \& {de Zeeuw}(2000)}]{helmi2000mapping}
{Helmi}, A. \& {de Zeeuw}, P.~T. 2000, \mnras, 319, 657

\bibitem[{{Helmi} {et~al.}(1999){Helmi}, {White}, {de Zeeuw}, \&
  {Zhao}}]{helmi99}
{Helmi}, A., {White}, S. D.~M., {de Zeeuw}, P.~T., \& {Zhao}, H. 1999, \nat,
  402, 53

\bibitem[{{Horta} {et~al.}(2020){Horta}, {Schiavon}, {Mackereth}, {Beers},
  {Fern{\'a}ndez-Trincado}, {Frinchaboy}, {Garc{\'\i}a-Hern{\'a}ndez},
  {Geisler}, {Hasselquist}, {J{\"o}nsson}, {Lane}, {Majewski},
  {M{\'e}sz{\'a}ros}, {Bidin}, {Nataf}, {Roman-Lopes}, {Nitschelm},
  {Vargas-Gonz{\'a}lez}, \& {Zasowski}}]{horta20}
{Horta}, D., {Schiavon}, R.~P., {Mackereth}, J.~T., {et~al.} 2020, \mnras, 493,
  3363

\bibitem[{Ibata {et~al.}(1994)Ibata, Gilmore, \& Irwin}]{ibata1994dwarf}
Ibata, R., Gilmore, G., \& Irwin, M. 1994, Nature, 370, 194

\bibitem[{Jean-Baptiste {et~al.}(2017)Jean-Baptiste, Di~Matteo, Haywood,
  G{\'o}mez, Montuori, Combes, \& Semelin}]{jean2017kinematic}
Jean-Baptiste, I., Di~Matteo, P., Haywood, M., {et~al.} 2017, Astronomy \&
  Astrophysics, 604, A106

\bibitem[{Kepley {et~al.}(2007)Kepley, Morrison, Helmi, Kinman, Van~Duyne,
  Martin, Harding, Norris, \& Freeman}]{kepley2007halo}
Kepley, A.~A., Morrison, H.~L., Helmi, A., {et~al.} 2007, The Astronomical
  Journal, 134, 1579

\bibitem[{{Khoperskov} {et~al.}(2022{\natexlab{a}}){Khoperskov}, {Minchev},
  {Libeskind}, {Haywood}, {Di Matteo}, {Belokurov}, {Steinmetz}, {Gomez},
  {Grand}, {Knebe}, {Sorce}, {Sparre}, {Tempel}, \&
  {Vogelsberger}}]{khoperskov22b}
{Khoperskov}, S., {Minchev}, I., {Libeskind}, N., {et~al.} 2022{\natexlab{a}},
  arXiv e-prints, arXiv:2206.04521

\bibitem[{{Khoperskov} {et~al.}(2022{\natexlab{b}}){Khoperskov}, {Minchev},
  {Libeskind}, {Haywood}, {Di Matteo}, {Belokurov}, {Steinmetz}, {Gomez},
  {Grand}, {Knebe}, {Sorce}, {Sparre}, {Tempel}, \&
  {Vogelsberger}}]{khoperskov22a}
{Khoperskov}, S., {Minchev}, I., {Libeskind}, N., {et~al.} 2022{\natexlab{b}},
  arXiv e-prints, arXiv:2206.04522

\bibitem[{Koppelman {et~al.}(2020)Koppelman, Bos, \&
  Helmi}]{koppelman2020messy}
Koppelman, H.~H., Bos, R.~O., \& Helmi, A. 2020, Astronomy \& Astrophysics,
  642, L18

\bibitem[{{Koppelman} {et~al.}(2019){Koppelman}, {Helmi}, {Massari},
  {Roelenga}, \& {Bastian}}]{koppelman2019}
{Koppelman}, H.~H., {Helmi}, A., {Massari}, D., {Roelenga}, S., \& {Bastian},
  U. 2019, \aap, 625, A5

\bibitem[{Kruijssen {et~al.}(2020)Kruijssen, Pfeffer, Chevance, Bonaca,
  Trujillo-Gomez, Bastian, Reina-Campos, Crain, \&
  Hughes}]{kruijssen2020kraken}
Kruijssen, J.~D., Pfeffer, J.~L., Chevance, M., {et~al.} 2020, Monthly Notices
  of the Royal Astronomical Society, 498, 2472

\bibitem[{Kruijssen {et~al.}(2019)Kruijssen, Pfeffer, Reina-Campos, Crain, \&
  Bastian}]{kruijssen2019formation}
Kruijssen, J.~D., Pfeffer, J.~L., Reina-Campos, M., Crain, R.~A., \& Bastian,
  N. 2019, Monthly Notices of the Royal Astronomical Society, 486, 3180

\bibitem[{{Kunder} {et~al.}(2012){Kunder}, {Koch}, {Rich}, {de Propris},
  {Howard}, {Stubbs}, {Johnson}, {Shen}, {Wang}, {Robin}, {Kormendy}, {Soto},
  {Frinchaboy}, {Reitzel}, {Zhao}, \& {Origlia}}]{kunder12}
{Kunder}, A., {Koch}, A., {Rich}, R.~M., {et~al.} 2012, \aj, 143, 57

\bibitem[{Lane {et~al.}(2022)Lane, Bovy, \& Mackereth}]{lane2022kinematic}
Lane, J.~M., Bovy, J., \& Mackereth, J.~T. 2022, Monthly Notices of the Royal
  Astronomical Society, 510, 5119

\bibitem[{Leaman {et~al.}(2013)Leaman, VandenBerg, \&
  Mendel}]{leaman2013bifurcated}
Leaman, R., VandenBerg, D.~A., \& Mendel, J.~T. 2013, Monthly Notices of the
  Royal Astronomical Society, 436, 122

\bibitem[{{Mackereth} {et~al.}(2019){Mackereth}, {Schiavon}, {Pfeffer},
  {Hayes}, {Bovy}, {Anguiano}, {Allende Prieto}, {Hasselquist}, {Holtzman},
  {Johnson}, {Majewski}, {O'Connell}, {Shetrone}, {Tissera}, \&
  {Fern{\'a}ndez-Trincado}}]{mackereth2019}
{Mackereth}, J.~T., {Schiavon}, R.~P., {Pfeffer}, J., {et~al.} 2019, \mnras,
  482, 3426

\bibitem[{{Majewski} {et~al.}(2003){Majewski}, {Skrutskie}, {Weinberg}, \&
  {Ostheimer}}]{majewski03}
{Majewski}, S.~R., {Skrutskie}, M.~F., {Weinberg}, M.~D., \& {Ostheimer}, J.~C.
  2003, \apj, 599, 1082

\bibitem[{{Malhan}(2022)}]{malhan22pontus}
{Malhan}, K. 2022, \apjl, 930, L9

\bibitem[{Malhan {et~al.}(2022)Malhan, Ibata, Sharma, Famaey, Bellazzini,
  Carlberg, D’souza, Yuan, Martin, \& Thomas}]{malhan2022global}
Malhan, K., Ibata, R.~A., Sharma, S., {et~al.} 2022, The Astrophysical Journal,
  926, 107

\bibitem[{Marchesini {et~al.}(2002)Marchesini, D’Onghia, Chincarini, Firmani,
  Conconi, Molinari, \& Zacchei}]{marchesini2002halpha}
Marchesini, D., D’Onghia, E., Chincarini, G., {et~al.} 2002, The
  Astrophysical Journal, 575, 801

\bibitem[{{Mar{\'\i}n-Franch} {et~al.}(2009){Mar{\'\i}n-Franch}, {Aparicio},
  {Piotto}, {Rosenberg}, {Chaboyer}, {Sarajedini}, {Siegel}, {Anderson},
  {Bedin}, {Dotter}, {Hempel}, {King}, {Majewski}, {Milone}, {Paust}, \&
  {Reid}}]{marin09}
{Mar{\'\i}n-Franch}, A., {Aparicio}, A., {Piotto}, G., {et~al.} 2009, \apj,
  694, 1498

\bibitem[{Massari {et~al.}(2019)Massari, Koppelman, \&
  Helmi}]{massari2019origin}
Massari, D., Koppelman, H.~H., \& Helmi, A. 2019, Astronomy \& Astrophysics,
  630, L4

\bibitem[{{McCarthy} {et~al.}(2012){McCarthy}, {Font}, {Crain}, {Deason},
  {Schaye}, \& {Theuns}}]{mccarthy12}
{McCarthy}, I.~G., {Font}, A.~S., {Crain}, R.~A., {et~al.} 2012, \mnras, 420,
  2245

\bibitem[{{Meylan} \& {Heggie}(1997)}]{meylan97}
{Meylan}, G. \& {Heggie}, D.~C. 1997, \aapr, 8, 1

\bibitem[{Miyamoto \& Nagai(1975)}]{miyamoto1975three}
Miyamoto, M. \& Nagai, R. 1975, Publications of the Astronomical Society of
  Japan, 27, 533

\bibitem[{Myeong {et~al.}(2018)Myeong, Evans, Belokurov, Sanders, \&
  Koposov}]{myeong2018milky}
Myeong, G., Evans, N., Belokurov, V., Sanders, J., \& Koposov, S. 2018, The
  Astrophysical Journal Letters, 856, L26

\bibitem[{{Myeong} {et~al.}(2019){Myeong}, {Vasiliev}, {Iorio}, {Evans}, \&
  {Belokurov}}]{myeong19}
{Myeong}, G.~C., {Vasiliev}, E., {Iorio}, G., {Evans}, N.~W., \& {Belokurov},
  V. 2019, \mnras, 488, 1235

\bibitem[{{Naidu} {et~al.}(2020){Naidu}, {Conroy}, {Bonaca}, {Johnson}, {Ting},
  {Caldwell}, {Zaritsky}, \& {Cargile}}]{naidu20}
{Naidu}, R.~P., {Conroy}, C., {Bonaca}, A., {et~al.} 2020, \apj, 901, 48

\bibitem[{Ness \& Lang(2016)}]{ness2016x}
Ness, M. \& Lang, D. 2016, The Astronomical Journal, 152, 14

\bibitem[{{Newberg} {et~al.}(2002){Newberg}, {Yanny}, {Rockosi}, {Grebel},
  {Rix}, {Brinkmann}, {Csabai}, {Hennessy}, {Hindsley}, {Ibata}, {Ivezi{\'c}},
  {Lamb}, {Nash}, {Odenkirchen}, {Rave}, {Schneider}, {Smith}, {Stolte}, \&
  {York}}]{newberg02}
{Newberg}, H.~J., {Yanny}, B., {Rockosi}, C., {et~al.} 2002, \apj, 569, 245

\bibitem[{Newberg {et~al.}(2009)Newberg, Yanny, \& Willett}]{newberg09}
Newberg, H.~J., Yanny, B., \& Willett, B.~A. 2009, The Astrophysical Journal,
  700, L61

\bibitem[{{Nissen} \& {Schuster}(2010)}]{nissen10}
{Nissen}, P.~E. \& {Schuster}, W.~J. 2010, \aap, 511, L10

\bibitem[{Panithanpaisal {et~al.}(2021)Panithanpaisal, Sanderson, Wetzel,
  Cunningham, Bailin, \& Faucher-Gigu{\`e}re}]{panithanpaisal2021galaxy}
Panithanpaisal, N., Sanderson, R.~E., Wetzel, A., {et~al.} 2021, The
  Astrophysical Journal, 920, 10

\bibitem[{Penarrubia {et~al.}(2009)Penarrubia, Walker, \&
  Gilmore}]{penarrubia2009tidal}
Penarrubia, J., Walker, M.~G., \& Gilmore, G. 2009, Monthly Notices of the
  Royal Astronomical Society, 399, 1275

\bibitem[{{P{\'e}rez-Villegas} {et~al.}(2020){P{\'e}rez-Villegas}, {Barbuy},
  {Kerber}, {Ortolani}, {Souza}, \& {Bica}}]{perez2020}
{P{\'e}rez-Villegas}, A., {Barbuy}, B., {Kerber}, L.~O., {et~al.} 2020, \mnras,
  491, 3251

\bibitem[{{Pfeffer} {et~al.}(2020){Pfeffer}, {Trujillo-Gomez}, {Kruijssen},
  {Crain}, {Hughes}, {Reina-Campos}, \& {Bastian}}]{pfeffer20}
{Pfeffer}, J.~L., {Trujillo-Gomez}, S., {Kruijssen}, J.~M.~D., {et~al.} 2020,
  \mnras, 499, 4863

\bibitem[{Price-Whelan {et~al.}(2016)Price-Whelan, Johnston, Valluri, Pearson,
  K{\"u}pper, \& Hogg}]{price2016chaotic}
Price-Whelan, A.~M., Johnston, K.~V., Valluri, M., {et~al.} 2016, Monthly
  Notices of the Royal Astronomical Society, 455, 1079

\bibitem[{Prieto {et~al.}(2008)Prieto, Majewski, Schiavon, Cunha, Frinchaboy,
  Holtzman, Johnston, Shetrone, Skrutskie, Smith, {et~al.}}]{prieto2008apogee}
Prieto, C.~A., Majewski, S., Schiavon, R., {et~al.} 2008, Astronomische
  Nachrichten: Astronomical Notes, 329, 1018

\bibitem[{{Purcell} {et~al.}(2010){Purcell}, {Bullock}, \&
  {Kazantzidis}}]{purcell10}
{Purcell}, C.~W., {Bullock}, J.~S., \& {Kazantzidis}, S. 2010, \mnras, 404,
  1711

\bibitem[{{Qu} {et~al.}(2011){Qu}, {Di Matteo}, {Lehnert}, \& {van
  Driel}}]{qu11}
{Qu}, Y., {Di Matteo}, P., {Lehnert}, M.~D., \& {van Driel}, W. 2011, \aap,
  530, A10

\bibitem[{{Quinn} {et~al.}(1993){Quinn}, {Hernquist}, \& {Fullagar}}]{quinn93}
{Quinn}, P.~J., {Hernquist}, L., \& {Fullagar}, D.~P. 1993, \apj, 403, 74

\bibitem[{Read {et~al.}(2008)Read, Lake, Agertz, \& Debattista}]{read2008thin}
Read, J., Lake, G., Agertz, O., \& Debattista, V.~P. 2008, Monthly Notices of
  the Royal Astronomical Society, 389, 1041

\bibitem[{Renaud {et~al.}(2017)Renaud, Agertz, \& Gieles}]{renaud2016origin}
Renaud, F., Agertz, O., \& Gieles, M. 2017, Monthly Notices of the Royal
  Astronomical Society, 465, 3622

\bibitem[{{Renaud} {et~al.}(2021){Renaud}, {Agertz}, {Read}, {Ryde},
  {Andersson}, {Bensby}, {Rey}, \& {Feuillet}}]{renaud21}
{Renaud}, F., {Agertz}, O., {Read}, J.~I., {et~al.} 2021, \mnras, 503, 5846

\bibitem[{Renaud \& Gieles(2015)}]{renaud2015flexible}
Renaud, F. \& Gieles, M. 2015, Monthly Notices of the Royal Astronomical
  Society, 448, 3416

\bibitem[{{Renaud} {et~al.}(2011){Renaud}, {Gieles}, \& {Boily}}]{renaud11}
{Renaud}, F., {Gieles}, M., \& {Boily}, C.~M. 2011, \mnras, 418, 759

\bibitem[{Rodionov {et~al.}(2009)Rodionov, Athanassoula, \&
  Sotnikova}]{rodionov2009iterative}
Rodionov, S., Athanassoula, E., \& Sotnikova, N.~Y. 2009, Monthly Notices of
  the Royal Astronomical Society, 392, 904

\bibitem[{Schwarz(1978)}]{schwarz1978estimating}
Schwarz, G. 1978, The annals of statistics, 461

\bibitem[{Semelin \& Combes(2002)}]{semelin2002formation}
Semelin, B. \& Combes, F. 2002, Astronomy \& Astrophysics, 388, 826

\bibitem[{{Shen} {et~al.}(2010){Shen}, {Rich}, {Kormendy}, {Howard}, {De
  Propris}, \& {Kunder}}]{shen10}
{Shen}, J., {Rich}, R.~M., {Kormendy}, J., {et~al.} 2010, \apjl, 720, L72

\bibitem[{Simpson {et~al.}(2019)Simpson, Gargiulo, G{\'o}mez, Grand, Maffione,
  Cooper, Deason, Frenk, Helly, Marinacci, {et~al.}}]{simpson2019simulating}
Simpson, C.~M., Gargiulo, I., G{\'o}mez, F.~A., {et~al.} 2019, Monthly Notices
  of the Royal Astronomical Society: Letters, 490, L32

\bibitem[{{Snaith} {et~al.}(2014){Snaith}, {Haywood}, {Di Matteo}, {Lehnert},
  {Combes}, {Katz}, \& {G{\'o}mez}}]{snaith14}
{Snaith}, O.~N., {Haywood}, M., {Di Matteo}, P., {et~al.} 2014, \apjl, 781, L31

\bibitem[{Stewart {et~al.}(2008)Stewart, Bullock, Wechsler, Maller, \&
  Zentner}]{stewart2008merger}
Stewart, K.~R., Bullock, J.~S., Wechsler, R.~H., Maller, A.~H., \& Zentner,
  A.~R. 2008, The Astrophysical Journal, 683, 597

\bibitem[{Trelles {et~al.}(2022)Trelles, Valenzuela, Roca-F{\'a}brega, \&
  Vel{\'a}zquez}]{trelles2022concurrent}
Trelles, A., Valenzuela, O., Roca-F{\'a}brega, S., \& Vel{\'a}zquez, H. 2022,
  Astronomy \& Astrophysics, 668, A20

\bibitem[{{VandenBerg} {et~al.}(2013){VandenBerg}, {Brogaard}, {Leaman}, \&
  {Casagrande}}]{vandenberg13}
{VandenBerg}, D.~A., {Brogaard}, K., {Leaman}, R., \& {Casagrande}, L. 2013,
  \apj, 775, 134

\bibitem[{Vasiliev(2019{\natexlab{a}})}]{vasiliev2019agama}
Vasiliev, E. 2019{\natexlab{a}}, Monthly Notices of the Royal Astronomical
  Society, 482, 1525

\bibitem[{Vasiliev(2019{\natexlab{b}})}]{vasiliev2019proper}
Vasiliev, E. 2019{\natexlab{b}}, Monthly Notices of the Royal Astronomical
  Society, 484, 2832

\bibitem[{{Vasiliev} {et~al.}(2022){Vasiliev}, {Belokurov}, \&
  {Evans}}]{vasiliev2022}
{Vasiliev}, E., {Belokurov}, V., \& {Evans}, N.~W. 2022, \apj, 926, 203

\bibitem[{{Villalobos} \& {Helmi}(2008)}]{villalobos08}
{Villalobos}, {\'A}. \& {Helmi}, A. 2008, \mnras, 391, 1806

\bibitem[{Vogelsberger {et~al.}(2008)Vogelsberger, White, Helmi, \&
  Springel}]{vogelsberger2008fine}
Vogelsberger, M., White, S.~D., Helmi, A., \& Springel, V. 2008, Monthly
  Notices of the Royal Astronomical Society, 385, 236

\bibitem[{{Walker} {et~al.}(1996){Walker}, {Mihos}, \& {Hernquist}}]{walker96}
{Walker}, I.~R., {Mihos}, J.~C., \& {Hernquist}, L. 1996, \apj, 460, 121

\bibitem[{Wegg \& Gerhard(2013)}]{wegg2013milky}
Wegg, C. \& Gerhard, O. 2013, The Messenger, 154, 54

\bibitem[{White \& Rees(1978)}]{white1978core}
White, S.~D. \& Rees, M.~J. 1978, Monthly Notices of the Royal Astronomical
  Society, 183, 341

\bibitem[{{Yuan} {et~al.}(2020){Yuan}, {Chang}, {Beers}, \& {Huang}}]{yuan20}
{Yuan}, Z., {Chang}, J., {Beers}, T.~C., \& {Huang}, Y. 2020, \apjl, 898, L37

\bibitem[{Zinn \& West(1984)}]{zinn1984globular}
Zinn, R. \& West, M.~J. 1984, The Astrophysical Journal Supplement Series, 55,
  45

\bibitem[{{Zolotov} {et~al.}(2009){Zolotov}, {Willman}, {Brooks}, {Governato},
  {Brook}, {Hogg}, {Quinn}, \& {Stinson}}]{zolotov09}
{Zolotov}, A., {Willman}, B., {Brooks}, A.~M., {et~al.} 2009, \apj, 702, 1058

\end{thebibliography}

\begin{appendices}
\section{Static Milky Way potential}
\label{app_a}

To illustrate how the distribution in the $E-L_z$ plane changes when the dynamical friction experienced by the satellite during the interaction with the Milky Way is not taken into account, we show here the results obtained by considering a static potential for the MW. To this aim, we have run the same simulations with one or two accreted satellites this time keeping the positions of the MW particles fixed at the initial values. 

\begin{figure*}
\begin{centering}
\begin{multicols}{3}
     \includegraphics[width=.9\linewidth]{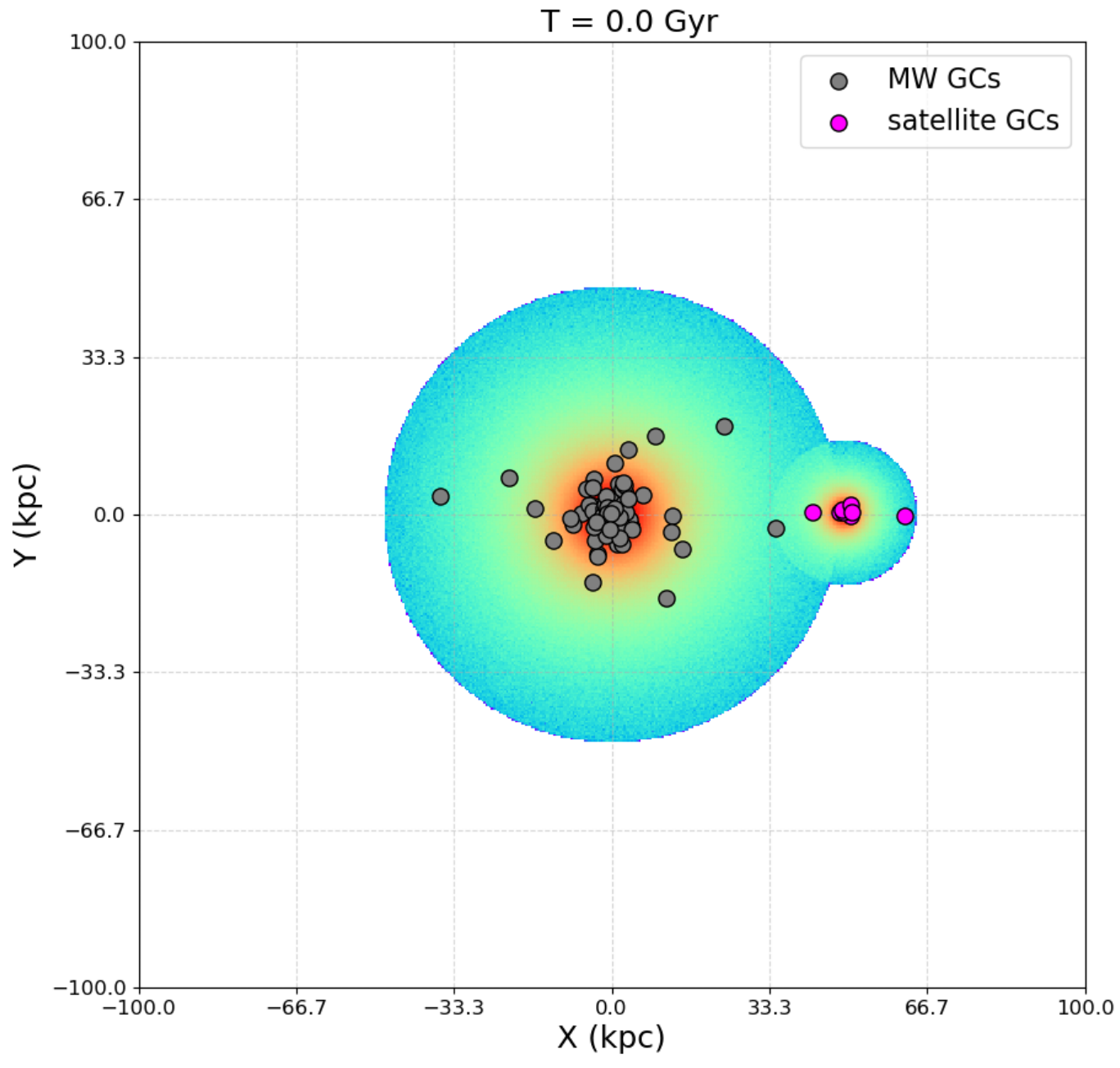}\par
         \includegraphics[width=.9\linewidth]{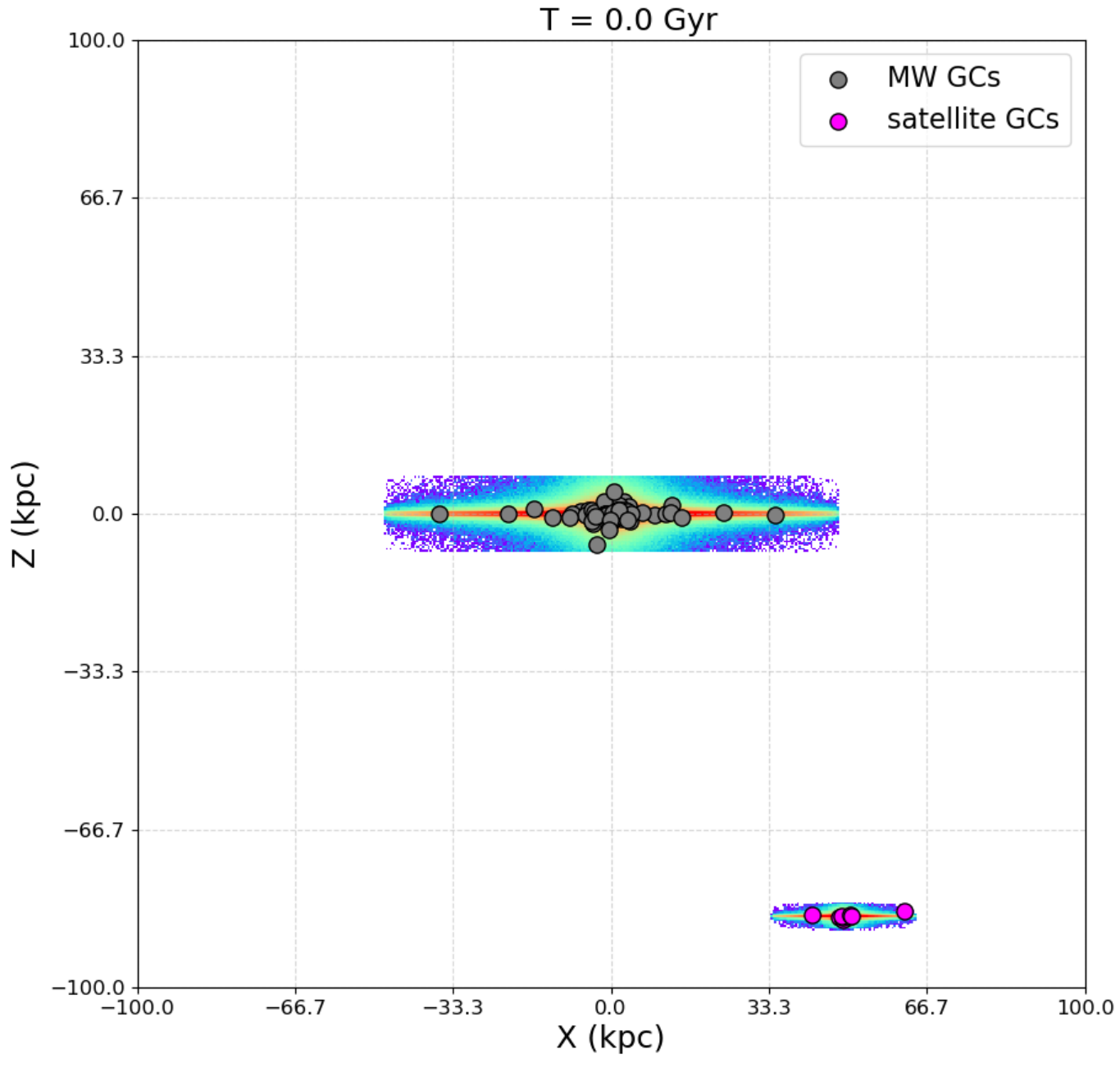} \par
             \includegraphics[width=.95\linewidth]{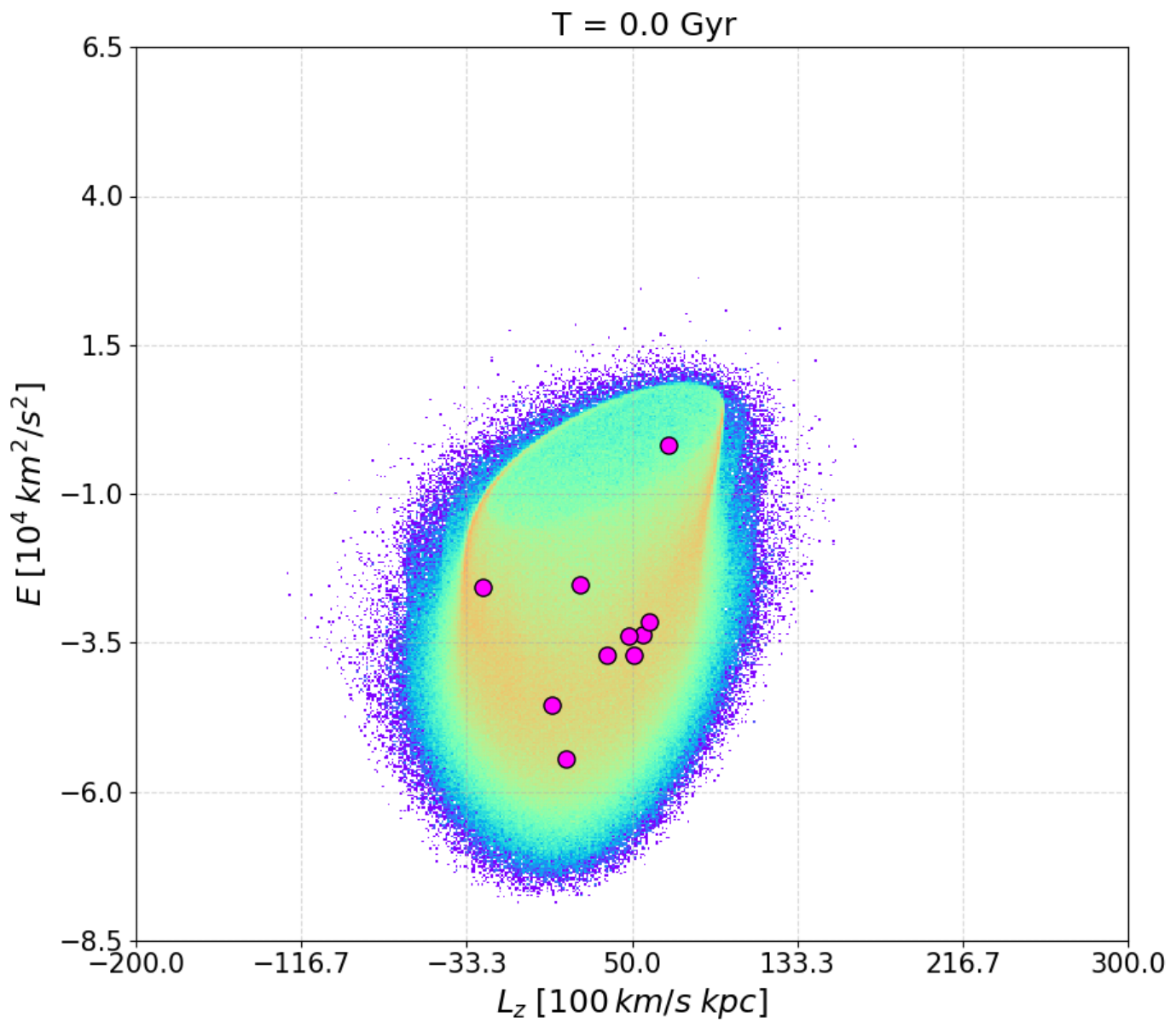} 
    \end{multicols}
    \begin{multicols}{3}
\includegraphics[width=.9\linewidth]{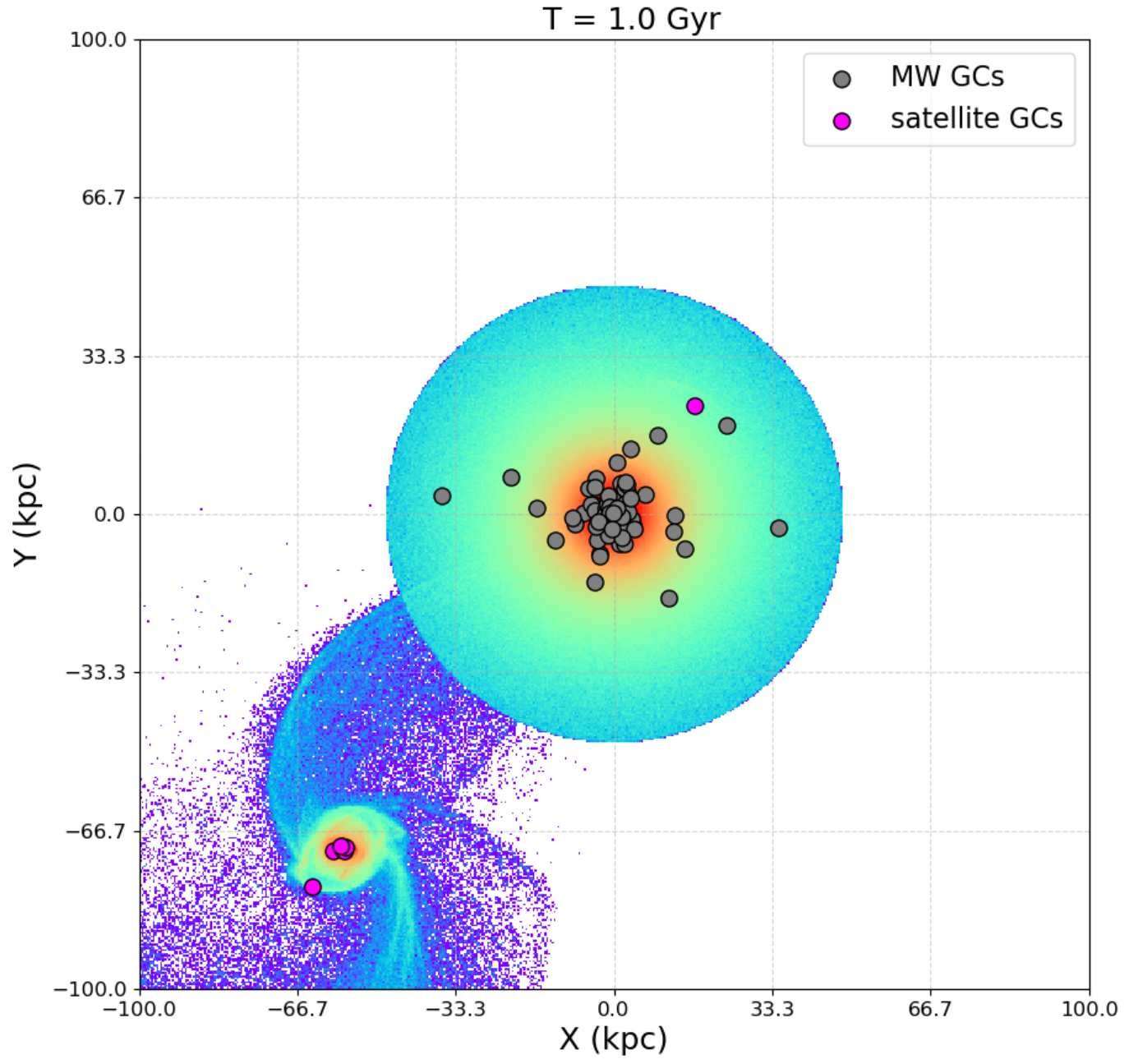}\par
\includegraphics[width=.9\linewidth]{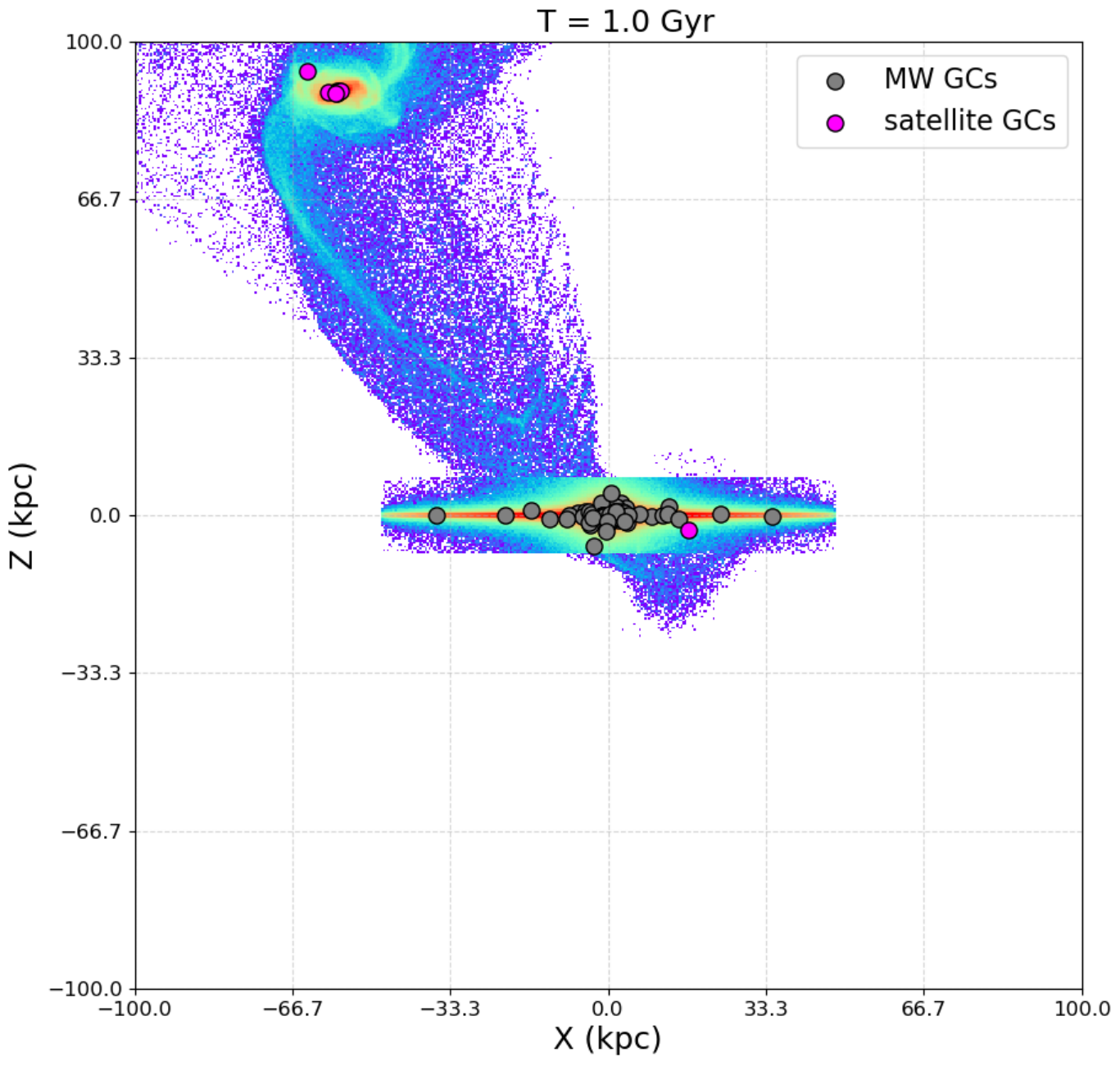}\par
\includegraphics[width=.95\linewidth]{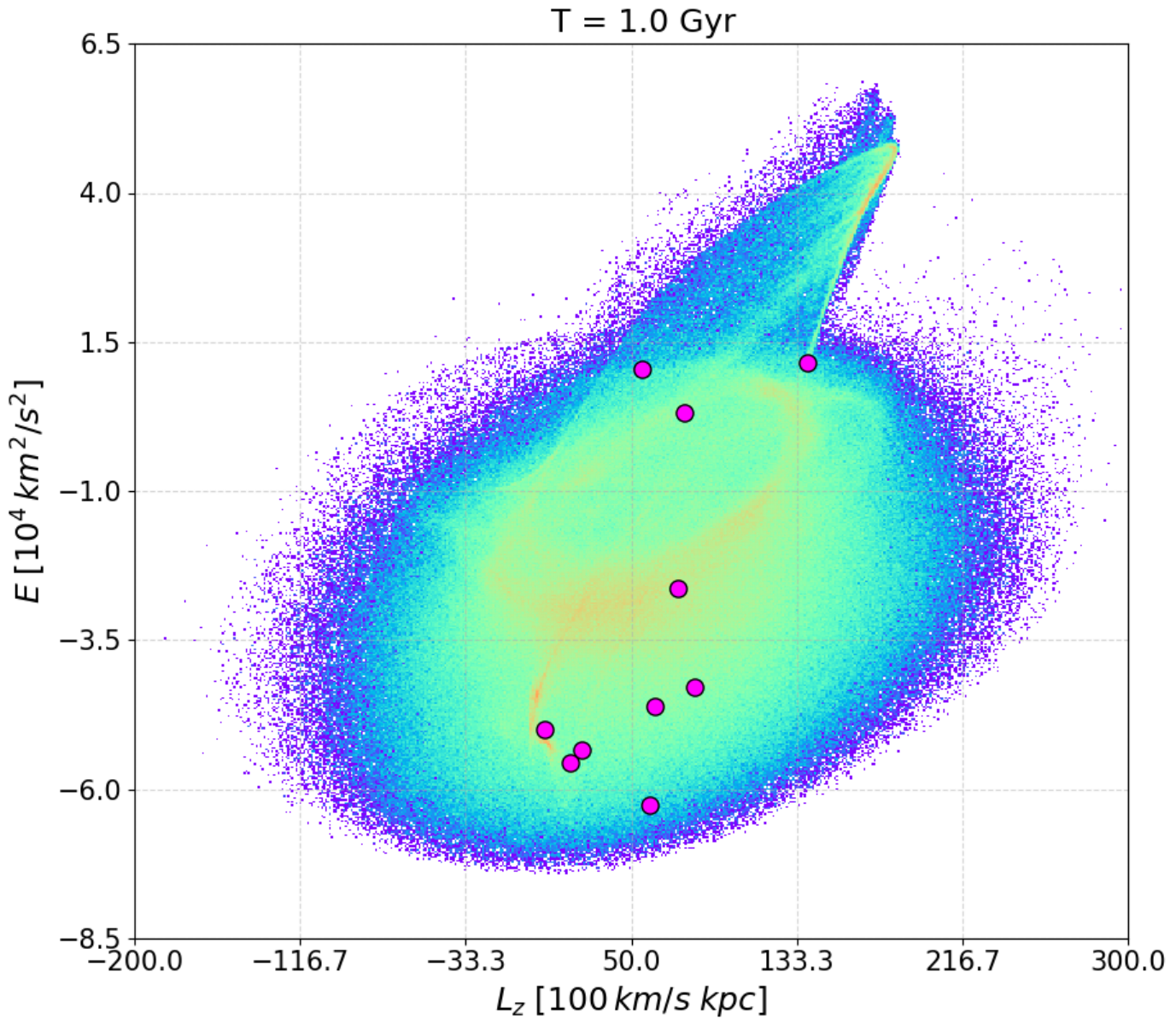}

    \end{multicols}
    \begin{multicols}{3}
\includegraphics[width=.9\linewidth]{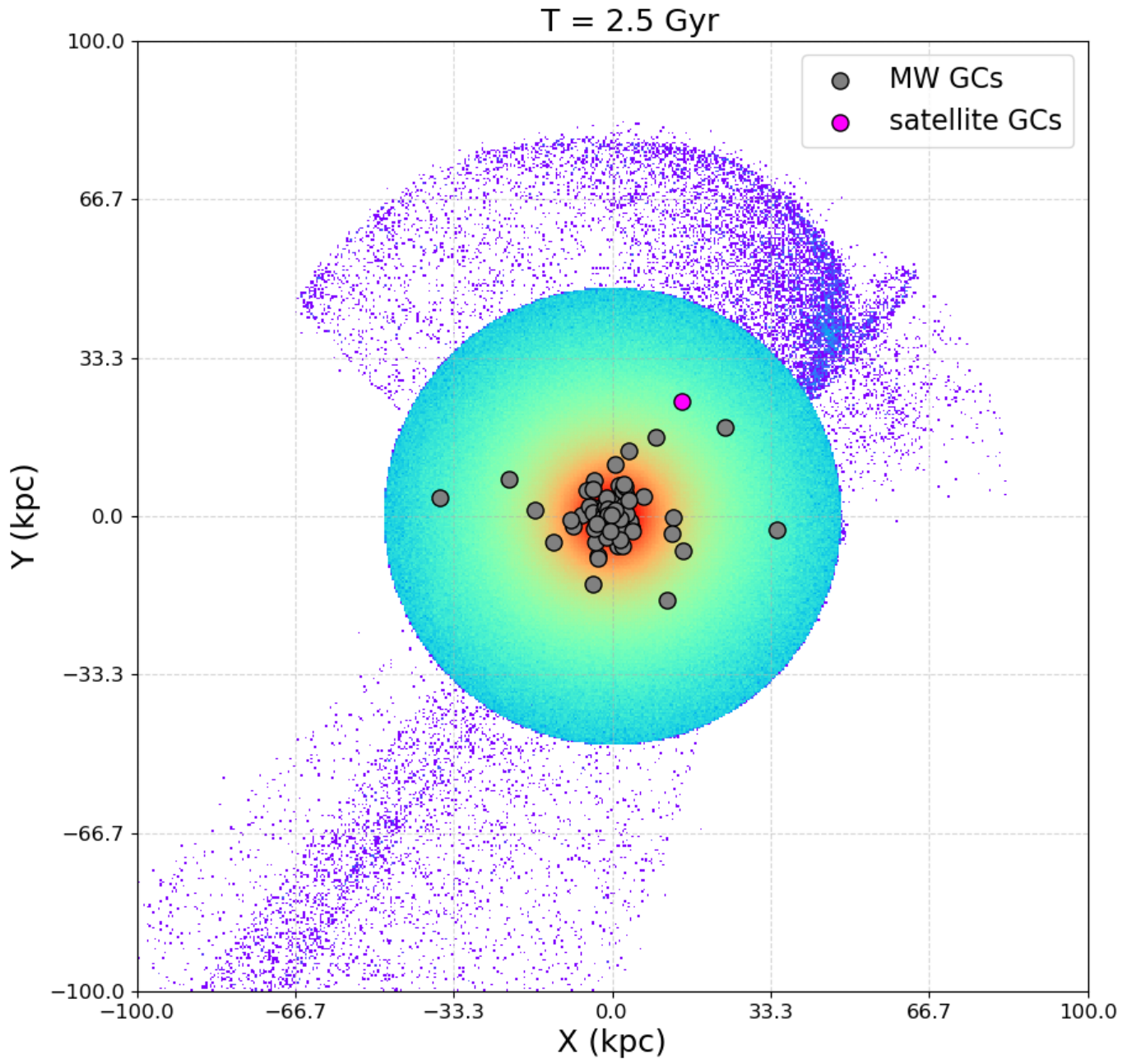}\par
\includegraphics[width=.9\linewidth]{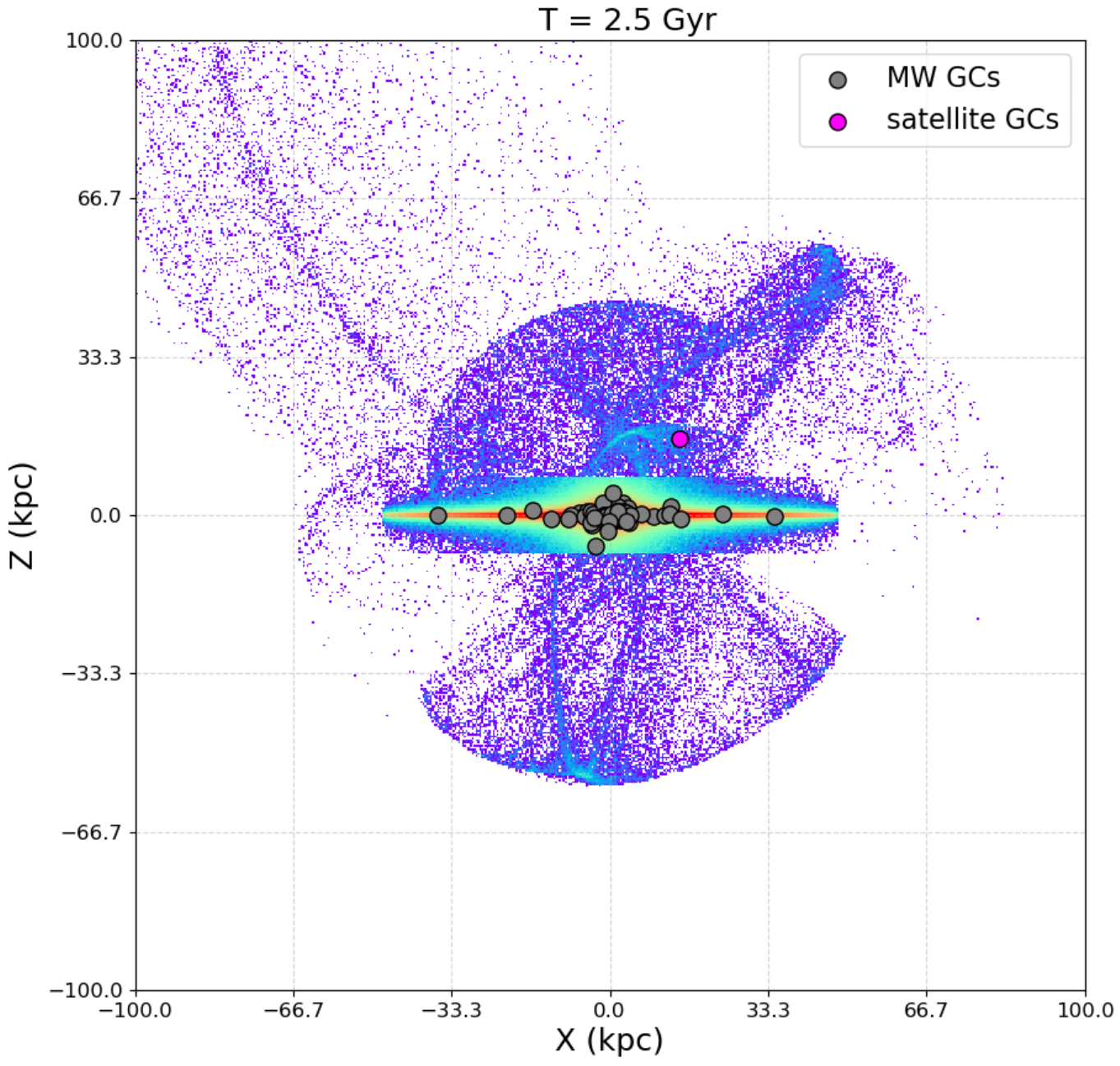}\par
\includegraphics[width=.95\linewidth]{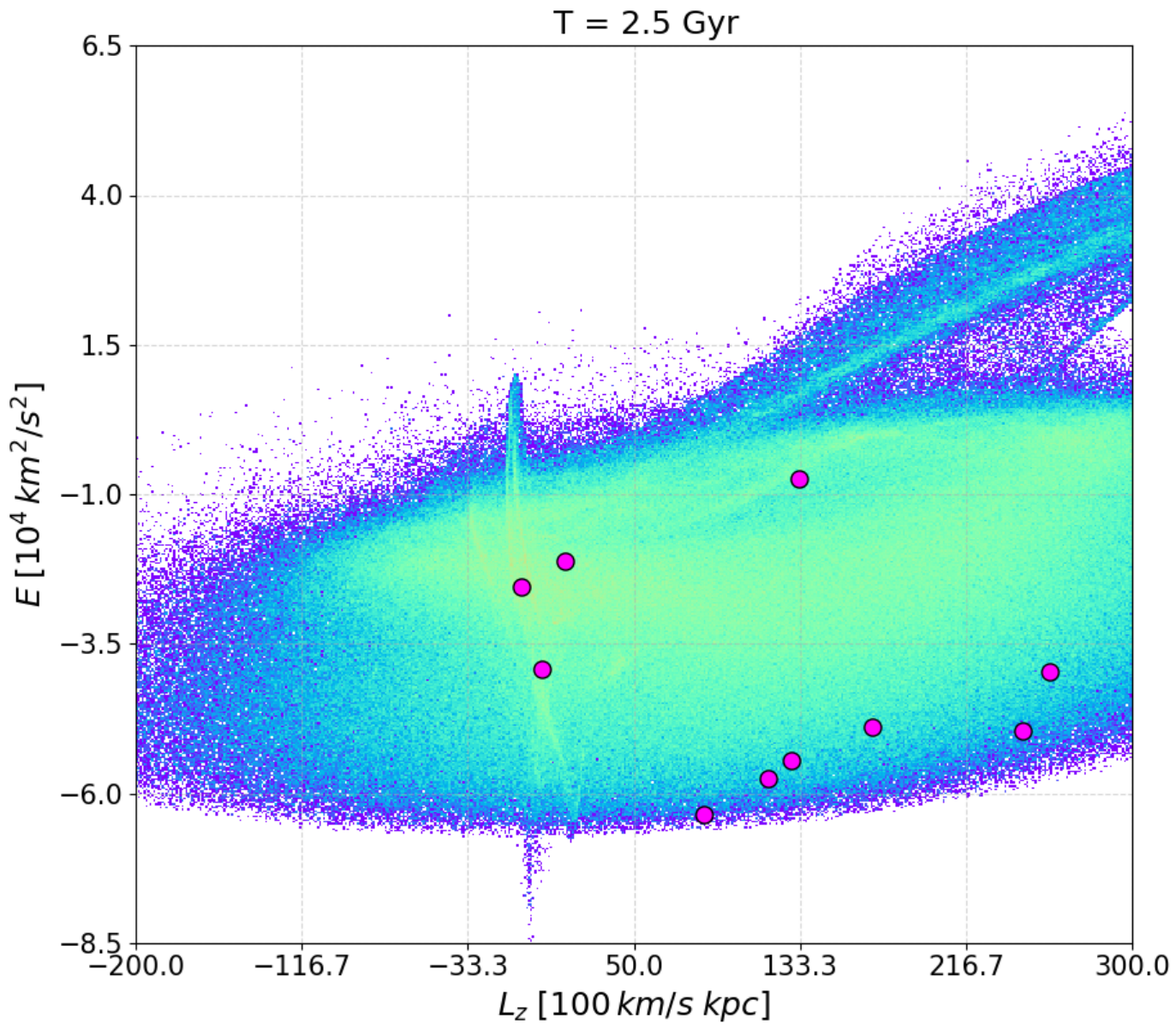}

    \end{multicols}
        \begin{multicols}{3}
\includegraphics[width=.9\linewidth]{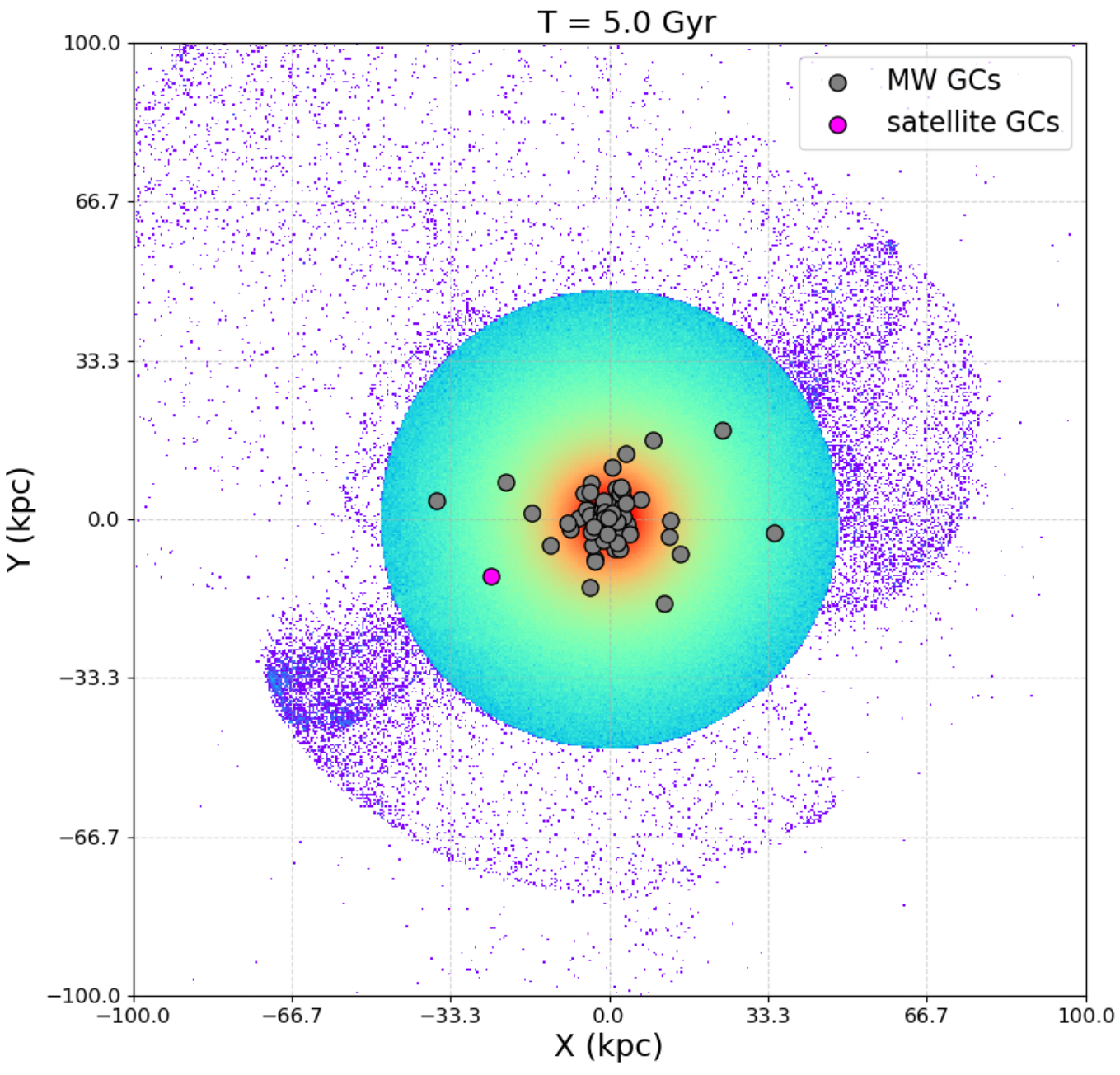}\par
\includegraphics[width=.9\linewidth]{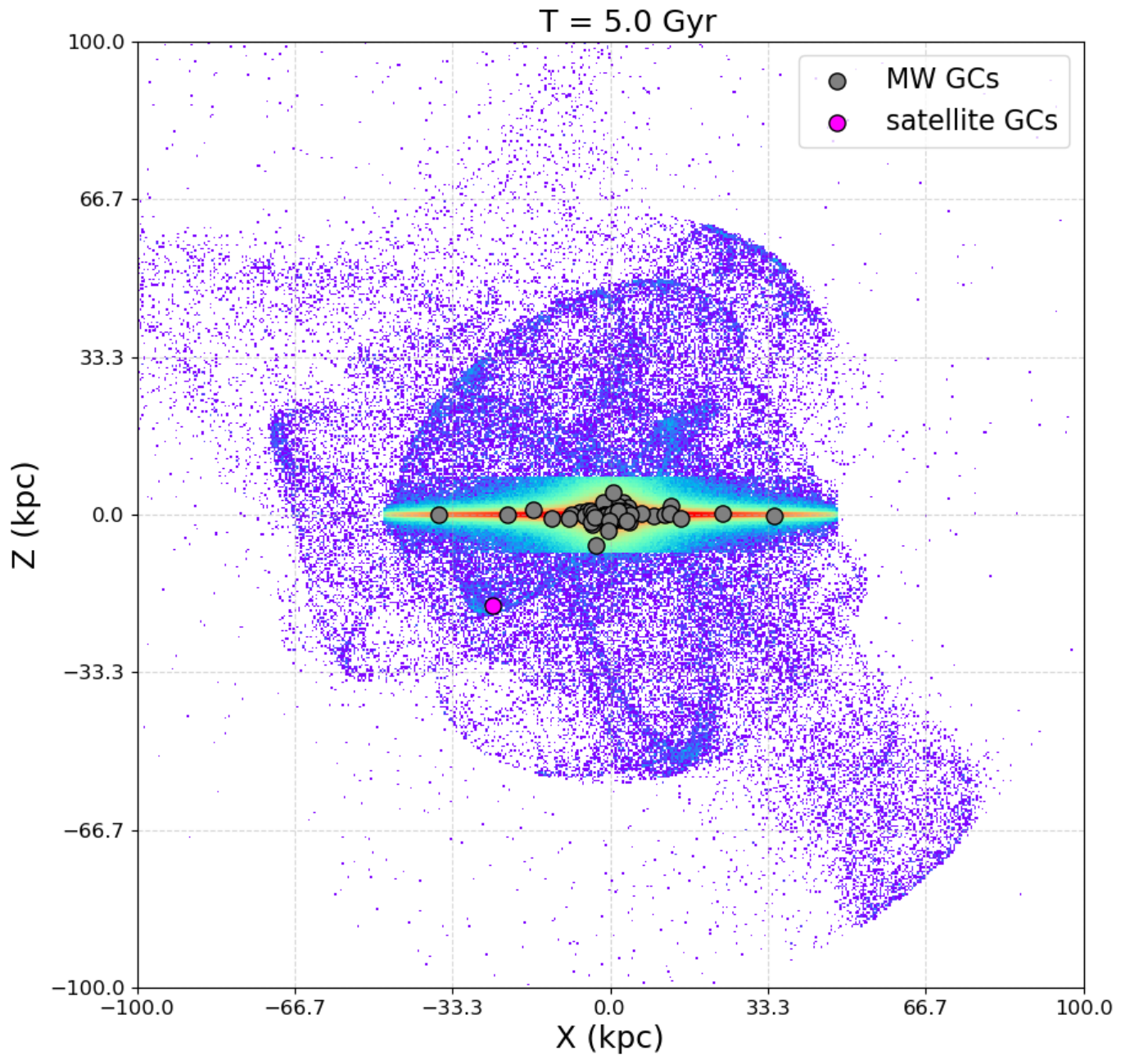}\par
\includegraphics[width=.95\linewidth]{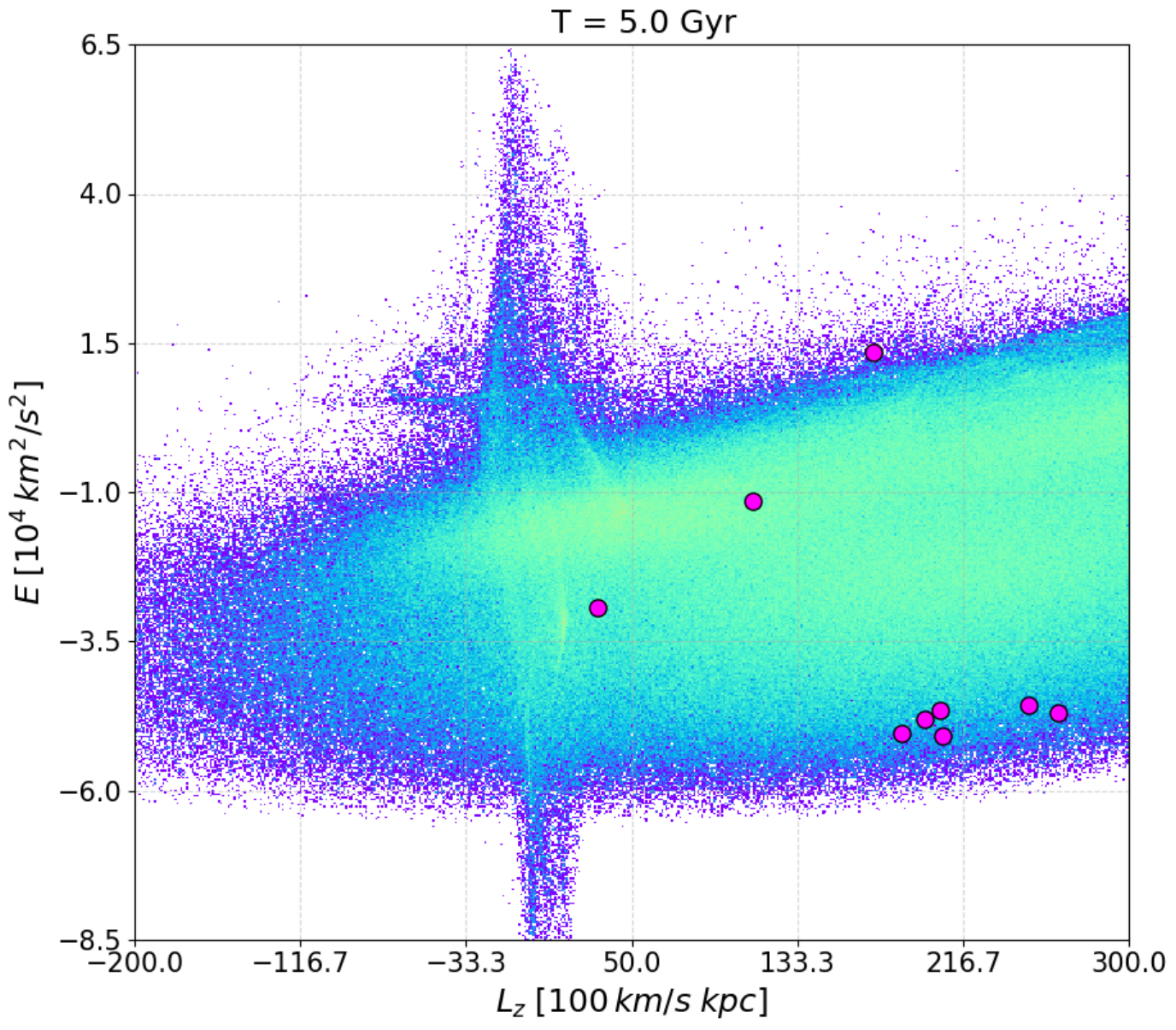} 
    \end{multicols}
\end{centering}
\caption{\textit{Left, middle columns}: projections of the simulated globular clusters positions on the $xy$ and $xz$ planes, for different times (increasing from top to bottom, same as Fig.~\ref{fig:xy_maps}) of the single-accretion simulation with $\Phi_{orb} = 60 \degree$ (MWsat\_n1\_$\Phi$60, see Tab.~\ref{tab:2} for the initial parameters). The in-situ clusters are represented by grey circles and the accreted clusters by magenta circles. In the background, the surface density of the totality of the stars of the simulation is also shown.
\textit{Right column}: Distributions of accreted globular clusters and field stars of the same satellite in the $E - L_{z}$ space, for different times (increasing from top to bottom) of the single-accretion simulation with $\Phi_{orb} = 60 \degree$.}
\label{fig:xy_maps_fixed}
\end{figure*}

\begin{figure}
\centering
\includegraphics[width=\linewidth]{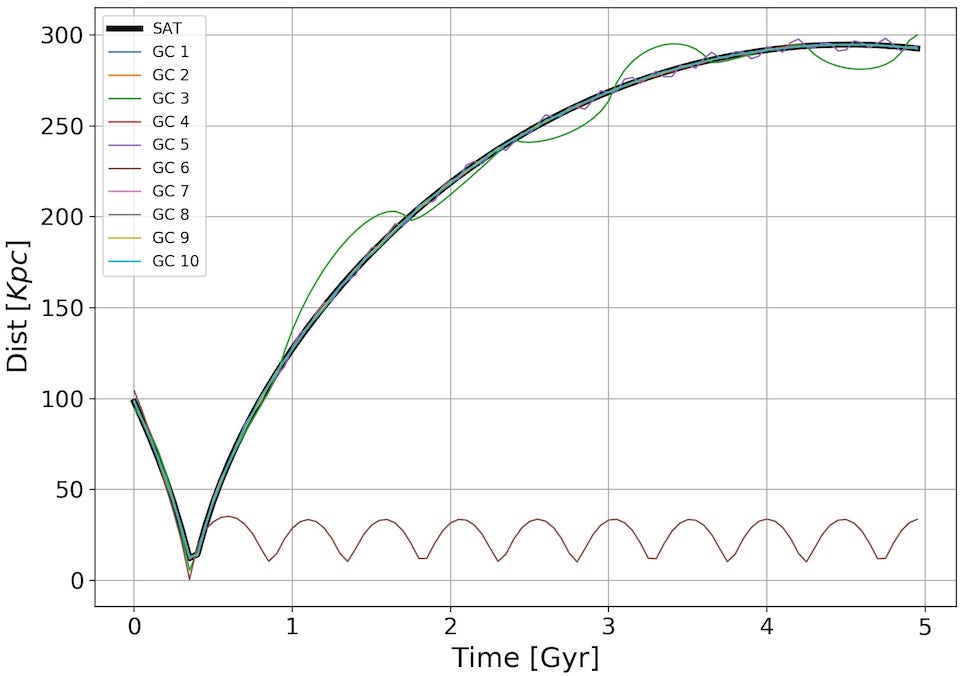}
\includegraphics[width=\linewidth]{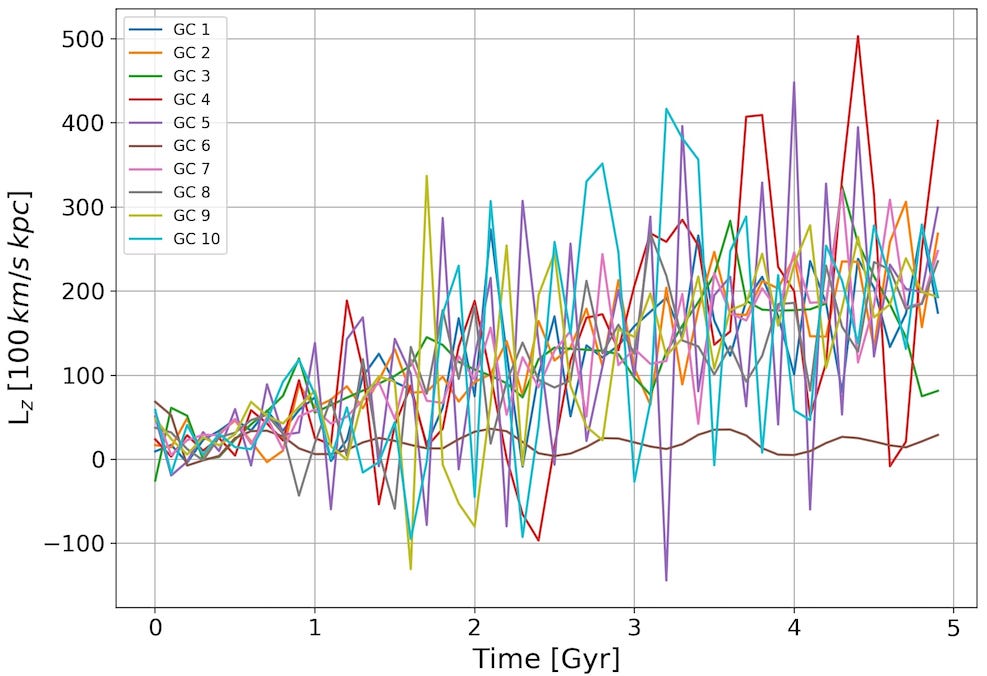}
\includegraphics[width=\linewidth]{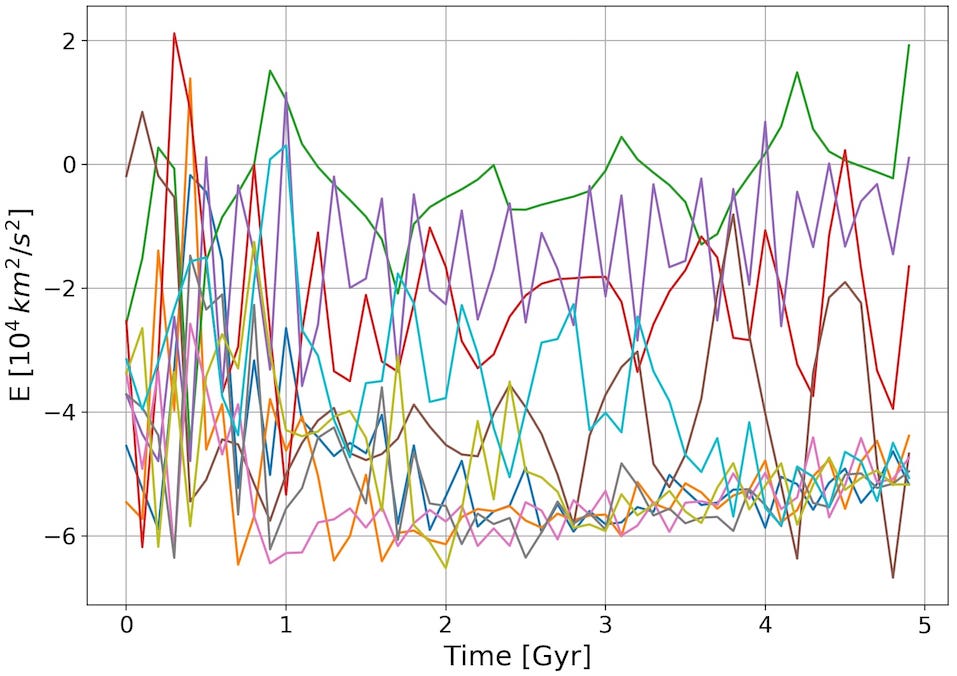}
\caption{\textit{Top panel:} time evolution of the distances of the satellite (black line) and its globular clusters (coloured lines) from the Milky Way-type galaxy, for the simulation with fixed MW potential MWsat\_n1\_$\Phi$60. \textit{Middle panel:} time evolution of the angular momenta of the satellite globular clusters.
\textit{Bottom panel:} time evolution of the orbital energies of the satellite globular clusters.}
\label{fig:d_vs_time}
\end{figure}

Figure~\ref{fig:xy_maps_fixed} is the analogue of Fig.~\ref{fig:xy_maps} thus showing the globular clusters projections in the $(x,y)$ and $(x,z)$ planes (left and central columns), for the single-accretion simulation with $\Phi_{orb} = 60\degree$ (i.e. simulation ID = MWsat\_n1\_$\Phi$60), at different times: the initial (T = 0 Gyr), two intermediate times (T = 1 Gyr, 2.5 Gyr), and at the final time (T= 5 Gyr). In all the plots, the in-situ globular cluster system is represented by grey circles, and the globular cluster system initially linked to the satellite by magenta circles. The distribution of field (in-situ and accreted) stars is also shown in the background.  Fig.~\ref{fig:d_vs_time} shows the temporal evolution of the distance of the satellite to the main galaxy, together with the corresponding evolution of all the clusters initially bound to the satellite.
Due to its quasi-parabolic orbit, in this scenario we can see the satellite approaching the fixed MW-type galaxy at T$\simeq$0.4 Gyr and then moving away again. At this approach, the satellite loses a globular cluster that is trapped by the Milky way on an orbit with apocentre at about 35 kpc and, as a result of its consequent departure, it retains the rest of its globular clusters.
This can also be seen in Fig.~\ref{fig:xy_maps_fixed} (left, middle columns) where indeed we notice that at T = 1 Gyr the satellite is receding (at T = 2.5, 5 Gyr it is no longer visible within the 100 kpc) after having deposited its globular cluster and part of its stars in the Milky Way, which at the end of the simulation (T = 5 Gyr) constitute the halo of the MW. 

The corresponding evolution of the distributions in the $E-L_z$ space (right column of Fig.~\ref{fig:xy_maps_fixed}) of globular clusters (magenta circles) and field stars (background density map) of the accreted satellite are very different from the full \textit{N}-body case (see Fig.~\ref{fig:xy_maps}, right column). At the beginning of the simulation the satellite, being still a bound system, appears to be clumped. With the passing of time, however, since in this case the positions of the MW particles are fixed and energy redistribution is not possible, i.e. there is no dynamical friction, the orbital energy of the satellite does not decrease as the satellite does not penetrate deeper in the potential well of the main galaxy. As a consequence, at the end of the simulation, we do not see a distribution of satellite stars and GCs in the $E-L_z$ space with a funnel-like shape elongated towards low energies as in the full \textit{N}-body scenario. The energy fluctuates instead of decreasing, as we can also see in Fig.~\ref{fig:d_vs_time} (bottom panel) showing the temporal evolution of the orbital energy of the 10 satellite globular clusters. In fact the mean energy of the GCs remains equal to -3.3 as the initial one. Angular momentum is not conserved since the mass distribution of the system is not axisymmetric and changes over time. Moreover, as we have seen, the satellite does not merge with the Milky Way and this implies a drift in time of the $L_z$ for the stars and clusters that remain attached to it. This trend is clearly visible in Fig.~\ref{fig:d_vs_time} (middle panel) where, except for the one GC trapped by the MW for which $L_z$ fluctuates, for all the satellite's clusters the $L_z$ increases over time as it moves away. The mean $L_z$ in fact goes from being equal to 34.5 at the beginning of the simulation to being equal to 203.1 at the end, and the spread in $L_z$ also rises (as the standard deviation increases from 27.0 to 92.7).

\begin{figure}
\centering
\includegraphics[width=\columnwidth]{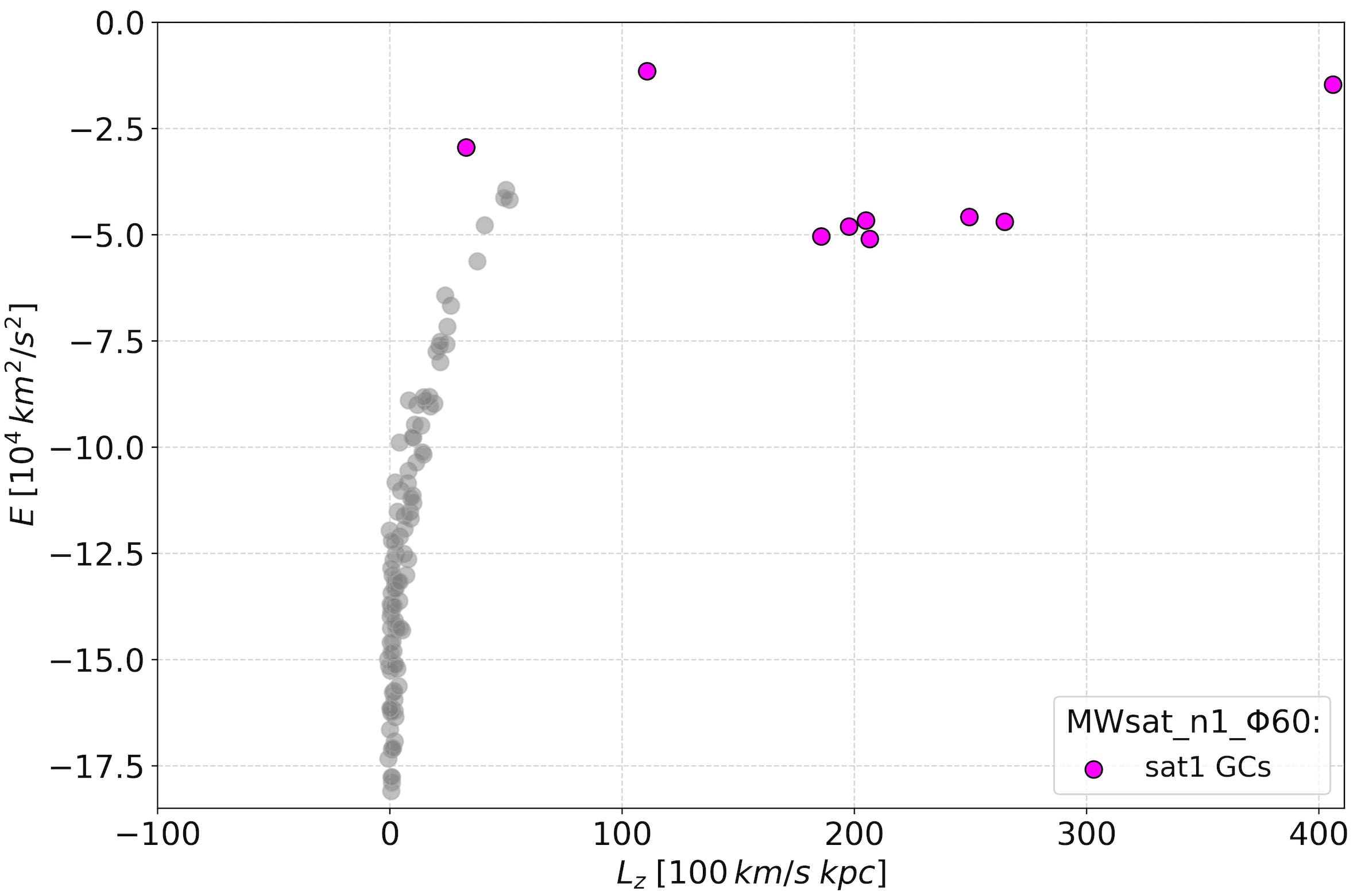}\
\includegraphics[width=\columnwidth]{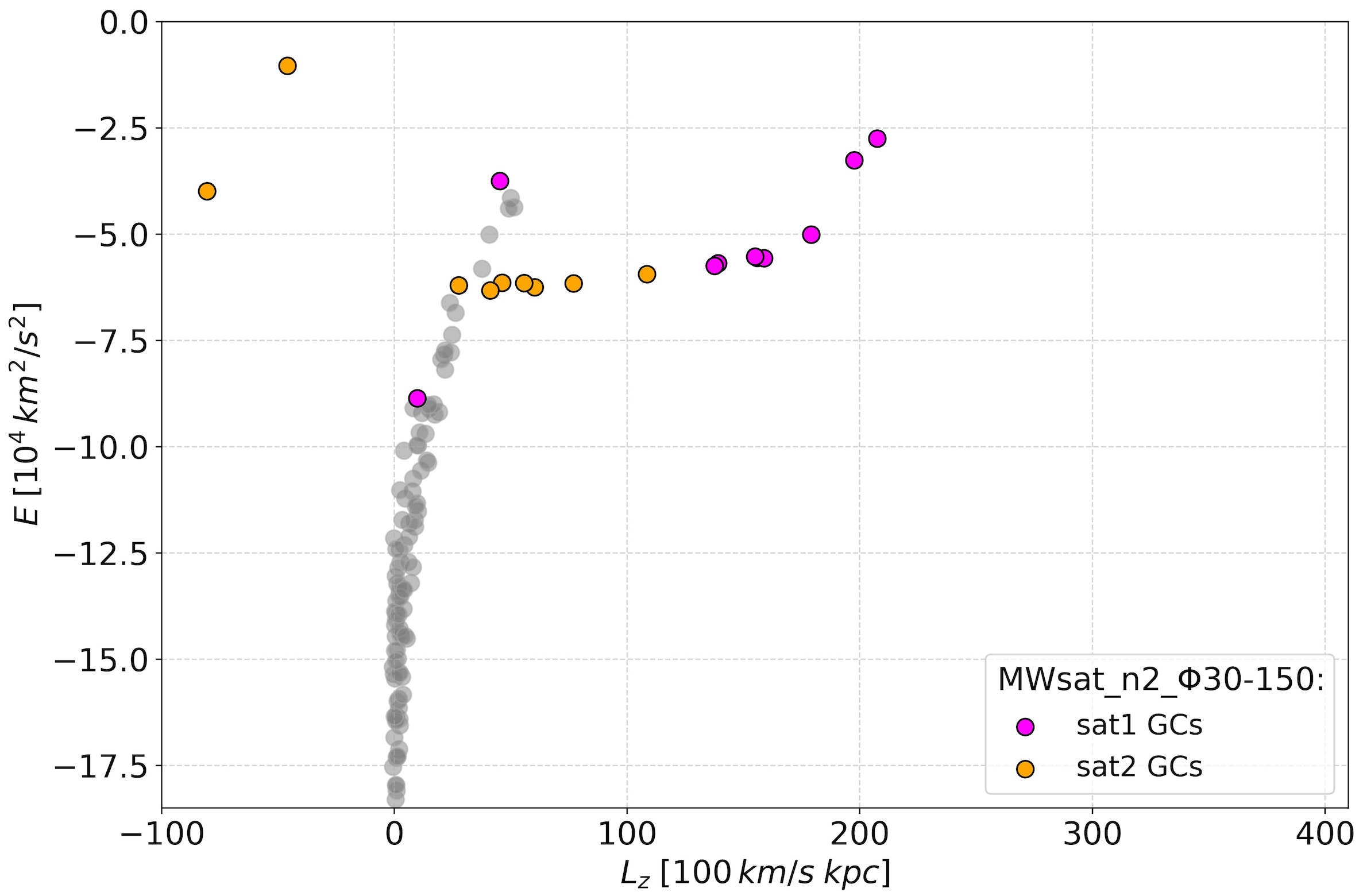} 
\caption{Distribution of globular clusters in the $E - L_{z}$ space at the end of the single (MWsat\_n1\_$\Phi$60, top panel) and double (MWsat\_n2\_$\Phi$30-150, bottom panel) accretions simulations. Milky Way's globular clusters are identified as grey circles while satellite(s) GCs are represented as magenta and orange circles.}
\label{fig:e_lz_fixed}
\end{figure}

As a simulation with two accretions, we re-run the one with ID = MWsat\_n2\_$\Phi$30 - 150 with a fixed MW potential.
Figure \ref{fig:e_lz_fixed} shows the final distribution in the $E-L_z$ space of the MW (grey circles) and satellite(s) GCs (magenta and orange circles) for the single and double accretion simulations with fixed galactic potential. These plots summarise the arguments just presented as the satellites' GCs are generally positioned at high energies and drifted towards large angular momenta. We do not find well-defined groups of clusters at different $E-L_z$ levels resembling the regions populated by different galactic progenitors in Fig.~4 in \citet{helmi2000mapping}.   
Furthermore, in this scenario, MW's clusters end up in a typical disc-like distribution since they are not heated up to halo-type kinematics by the redistribution of the energy, and this prevents us to account for a more realistic overlap between accreted and in-situ GCs.

\section{4x(1:100) mergers}
\label{app_b}
In this section we show how globular clusters distribute in the $E - L_{z}$ space in the full \textit{N}-body case when the MW-type galaxy accretes four satellites with relative mass ratios of 1:100 with respect to the Milky Way. In these simulations, the Milky Way-type galaxy has the same properties (in terms of masses and sizes of its components, and of number of particles adopted) as those of the main simulations described in Sect.~\ref{method}. As for the satellites, each of them is made of $25\,000$ particles and contains a population of 5 globular clusters. We have explored 7 different inclinations for the initial orbital plane of the satellite relative to the Milky Way disc. Here we present the results of two such simulations following the accretion over a period of 5 Gyrs that differ, as before, in the initial inclination of the various satellites $\Phi_{orb}$ and so are referred as MWsat\_n4\_$\Phi$150-60-0-30 and MWsat\_n4\_$\Phi$180-90-30-120. 

\begin{figure*}
\begin{centering}
\begin{multicols}{2}
     \includegraphics[width=.67\linewidth]{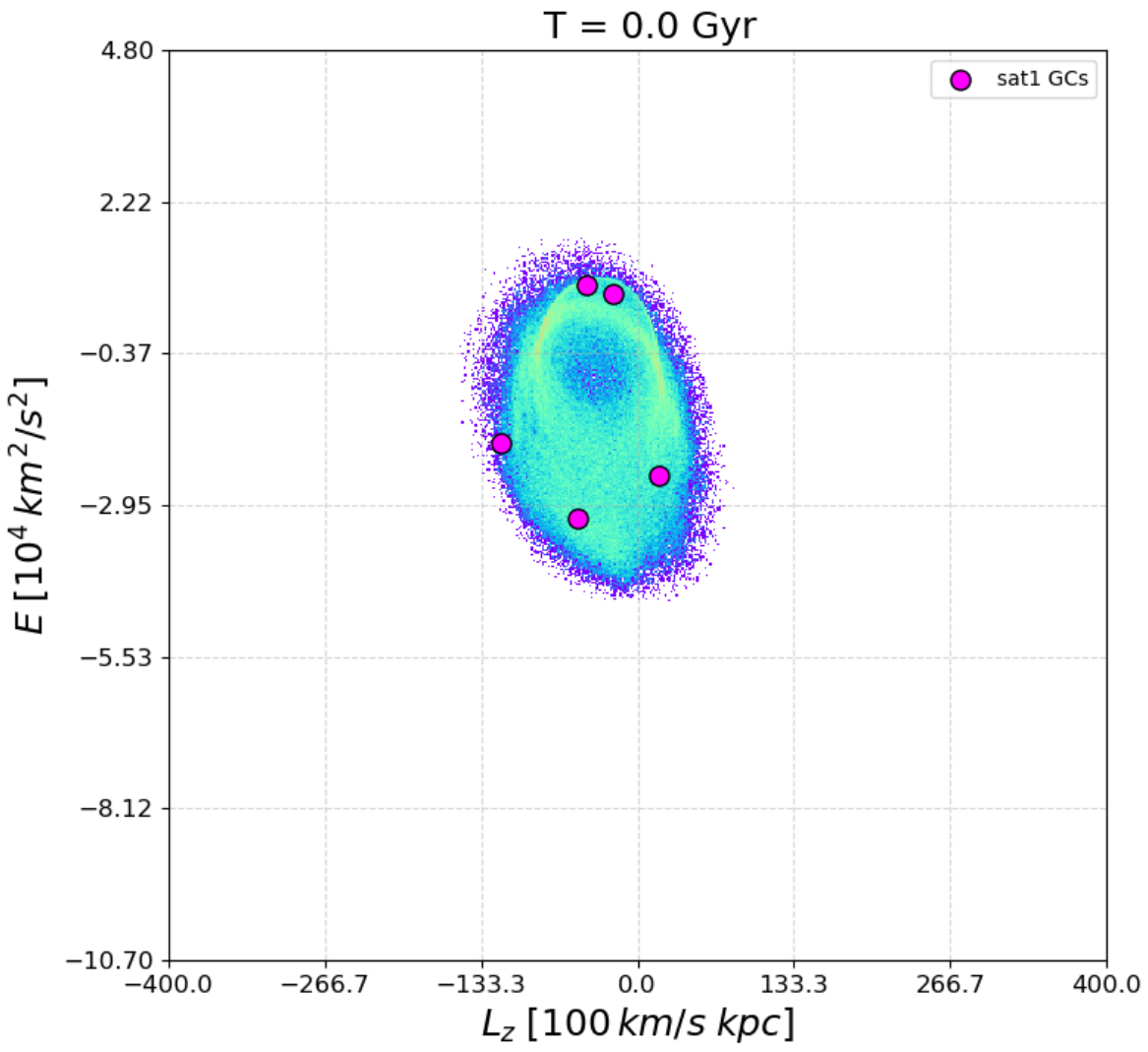}\par
         \includegraphics[width=.67\linewidth]{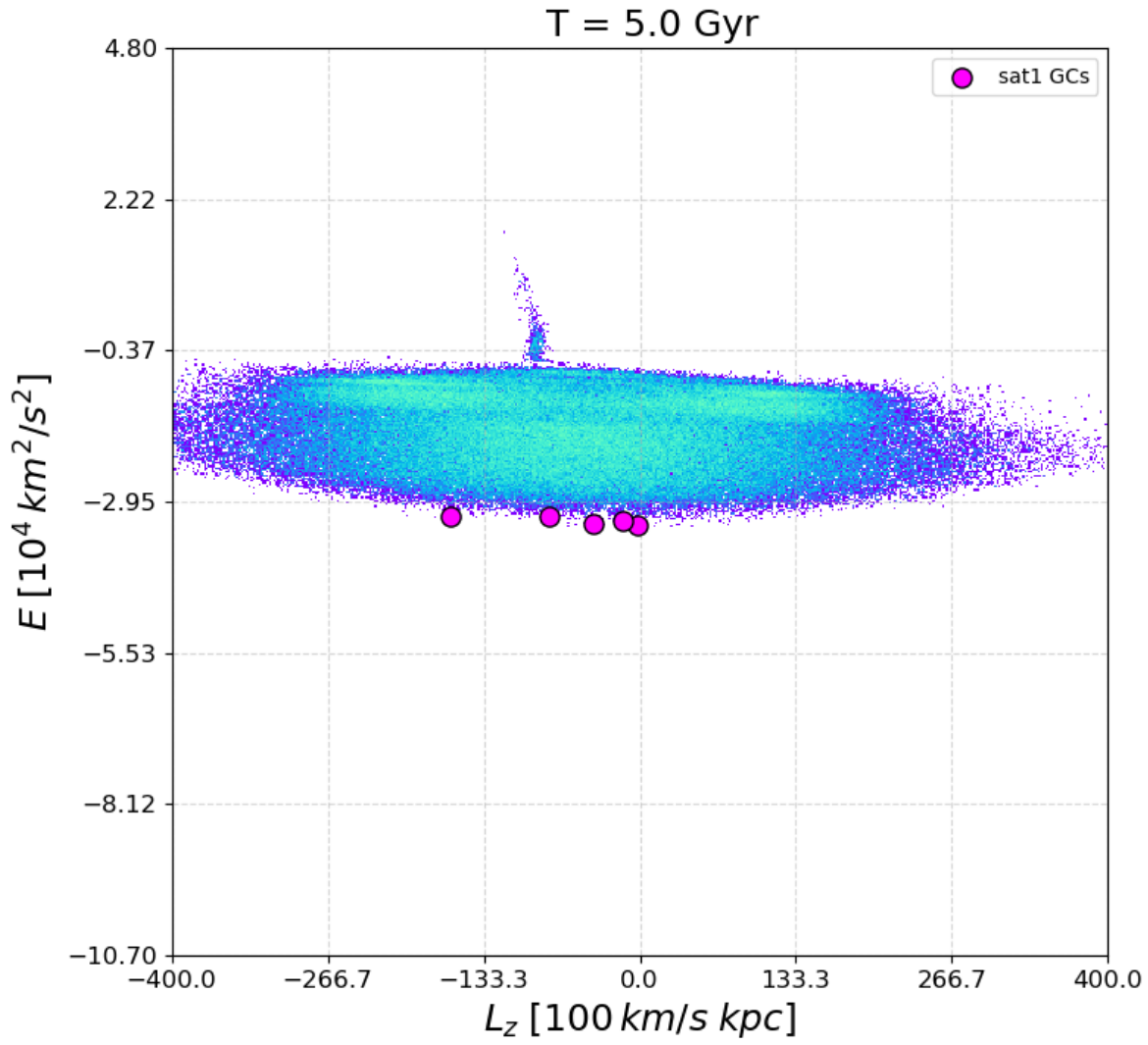} 
    \end{multicols}
        \vspace{-20pt}
    \begin{multicols}{2}
\includegraphics[width=.67\linewidth]{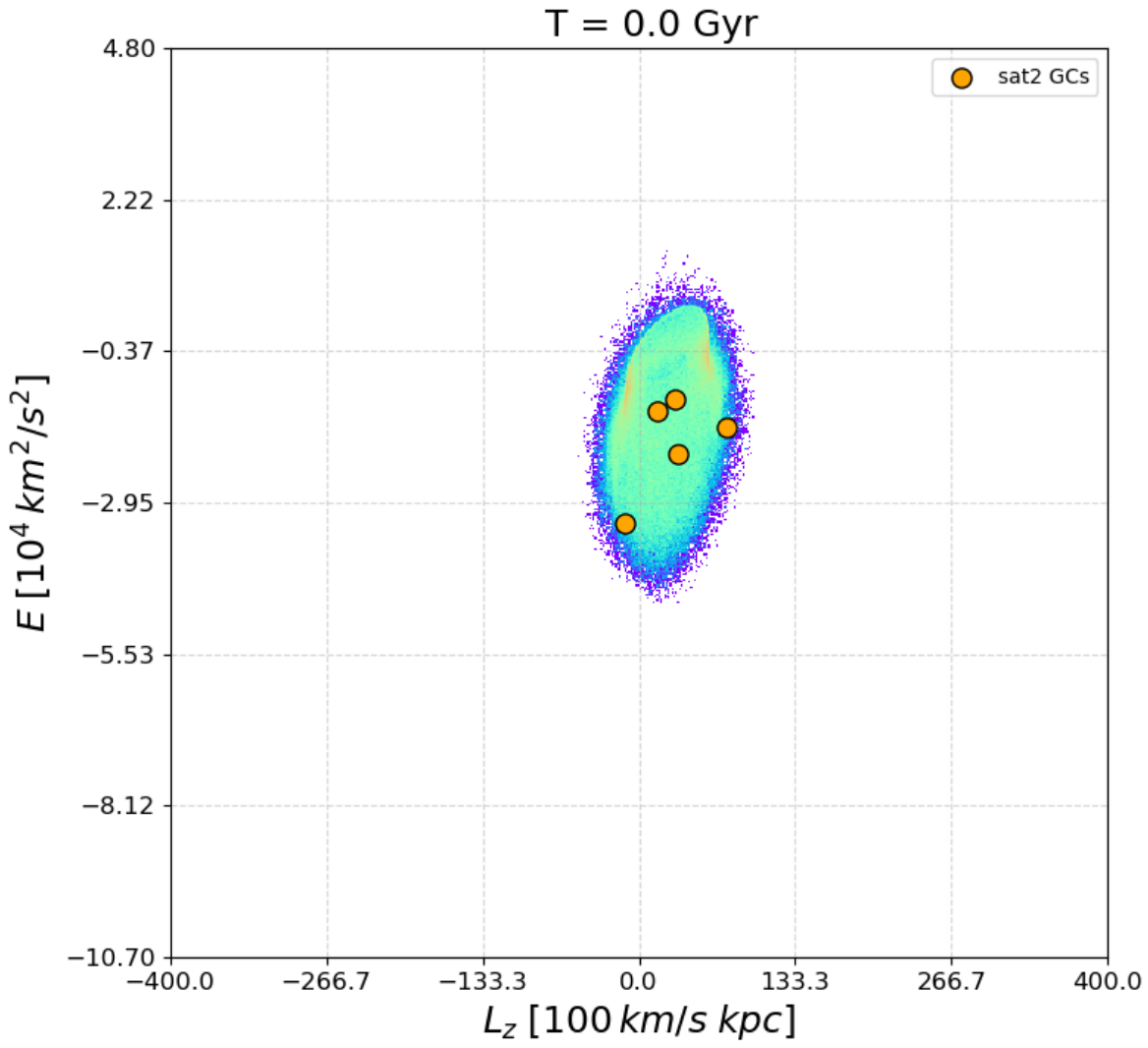}\par
\includegraphics[width=.67\linewidth]{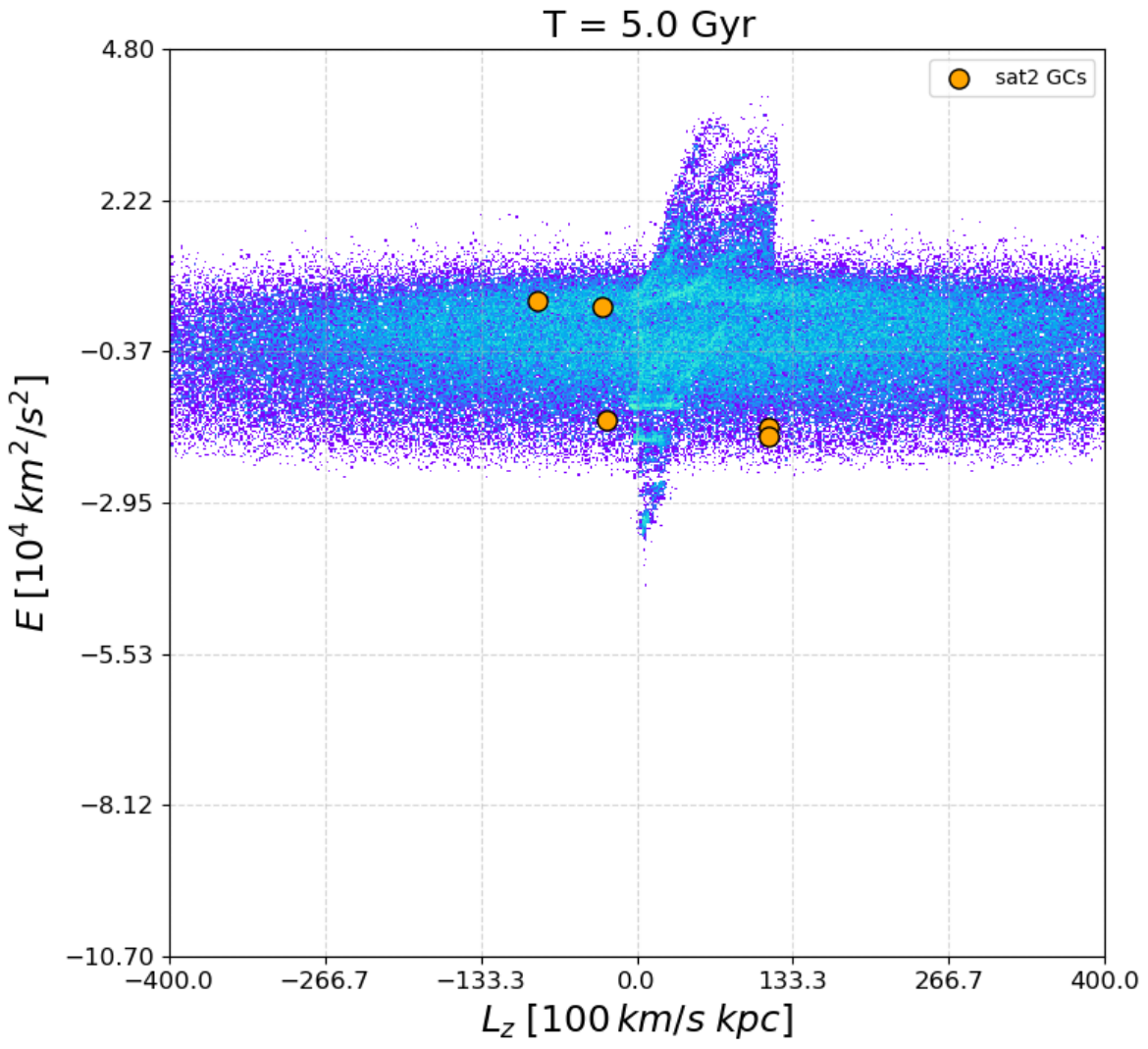}
    \end{multicols}
    \vspace{-20pt}
    \begin{multicols}{2}
\includegraphics[width=.67\linewidth]{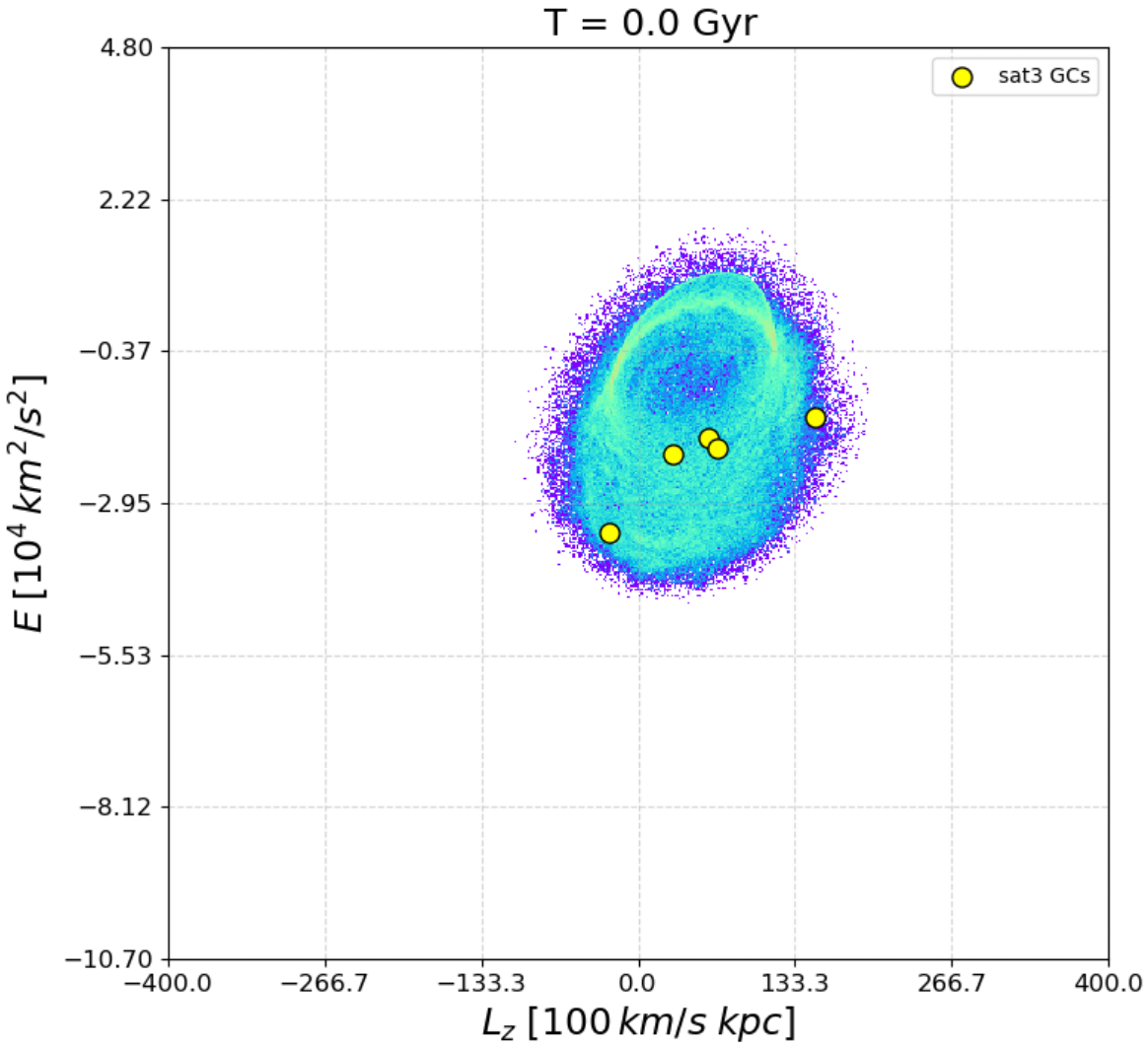}\par
\includegraphics[width=.67\linewidth]{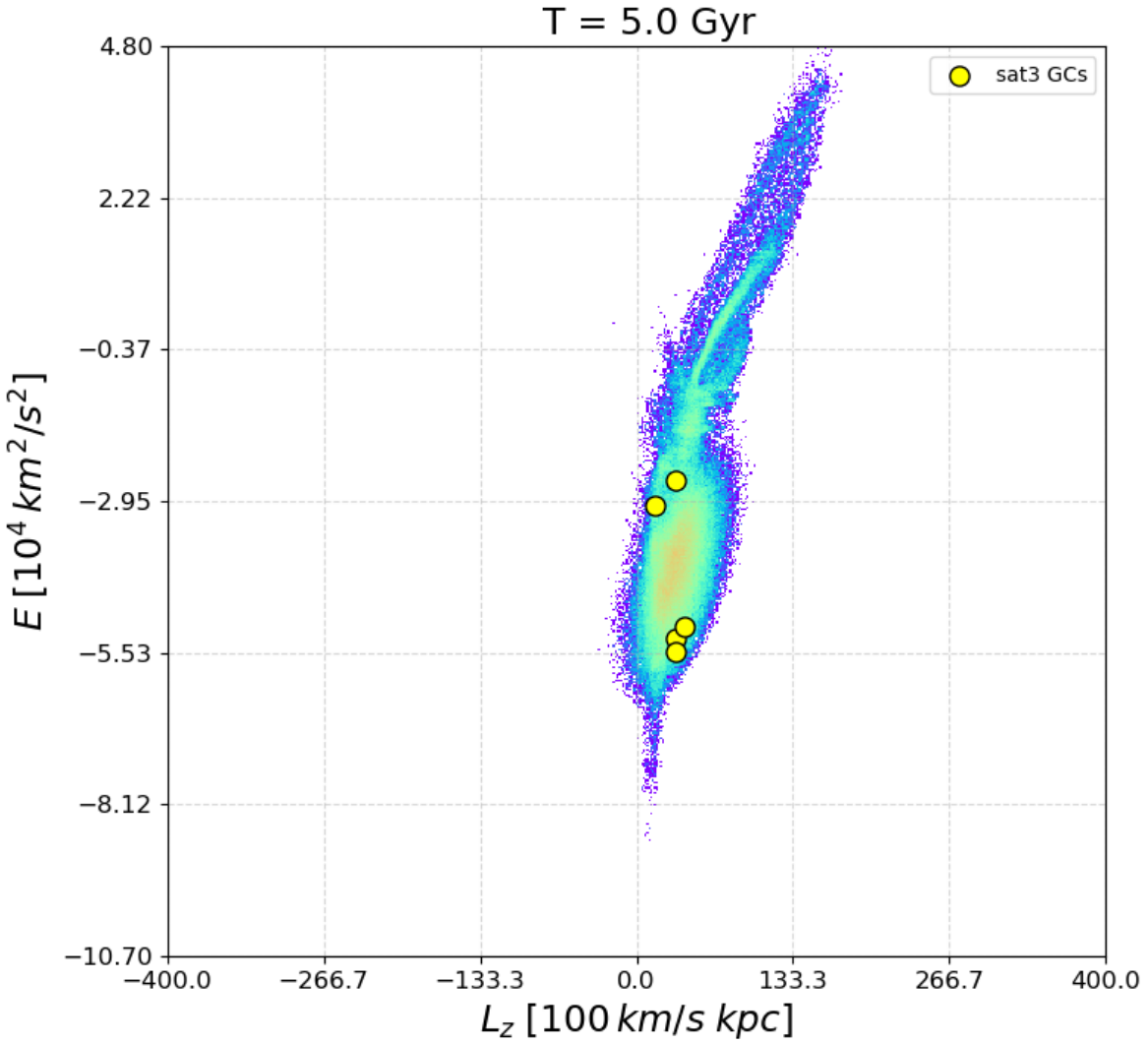}
    \end{multicols}
    \vspace{-20pt}
        \begin{multicols}{2}
\includegraphics[width=.67\linewidth]{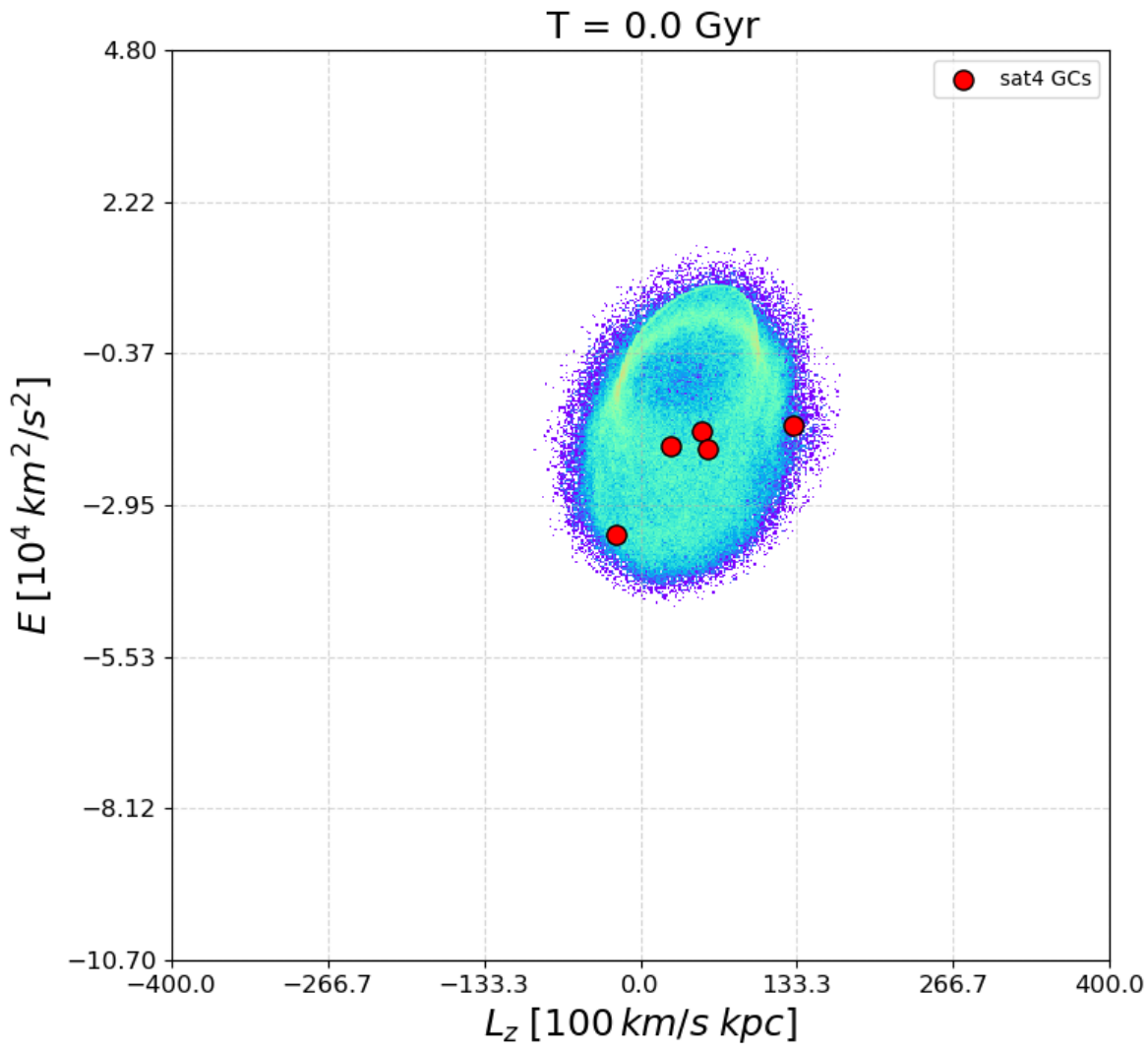}\par
\includegraphics[width=.67\linewidth]{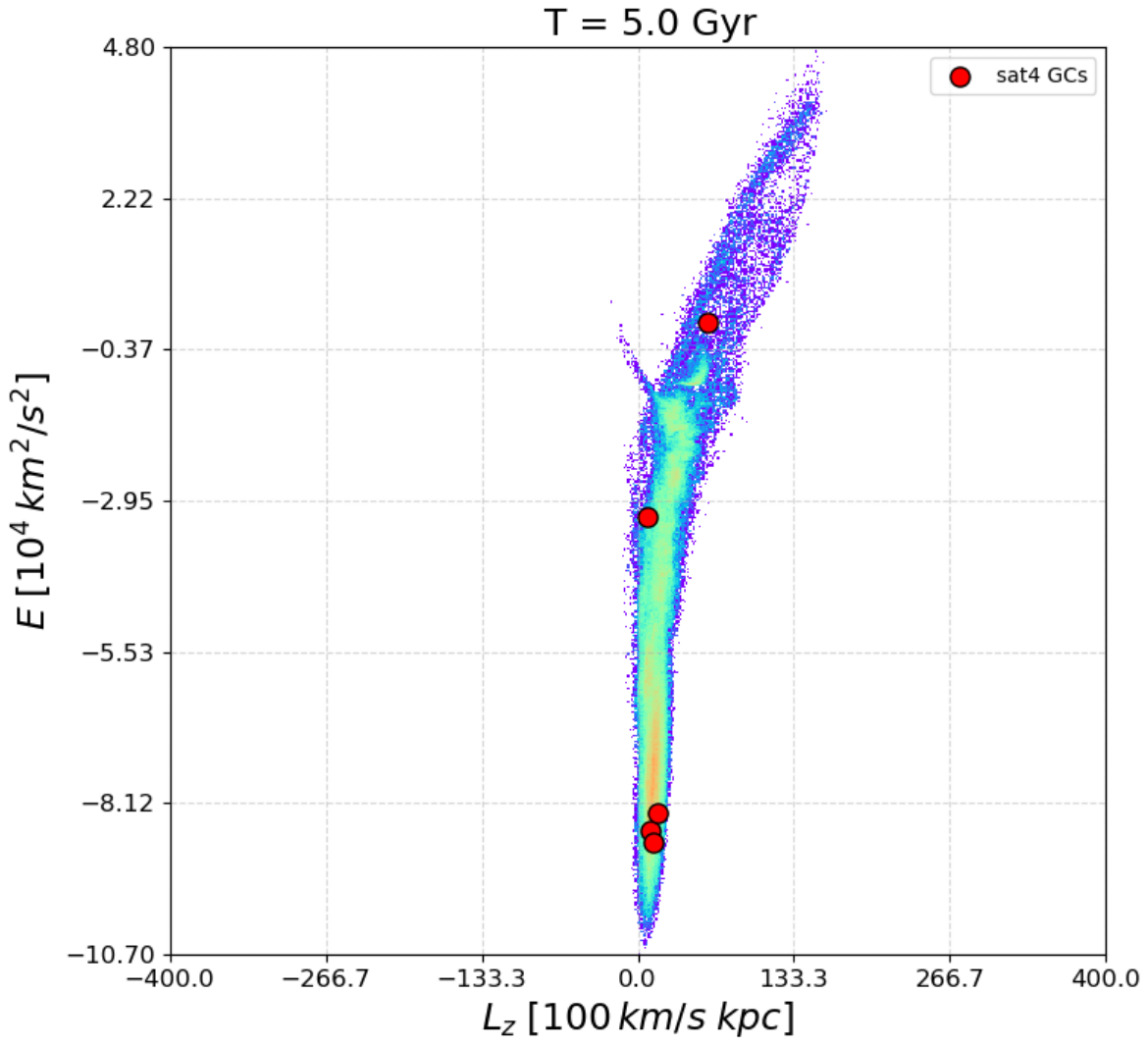} 
    \end{multicols}
\end{centering}
\caption{Initial (\textit{left panel}) and final (\textit{right panel}) distributions of accreted globular clusters and stars of the same satellite in the $E - L_{z}$ space for the four accreted satellites having 1:100 mass ratio with respect to the MW-type galaxy (ID = MWsat\_n4\_$\Phi$150-60-0-30). Note that the ranges on the $x$ and $y$ axes are the same for all panels.}
\label{fig:e_lz_4sat_1}
\end{figure*}

\begin{figure*}
\begin{centering}
\begin{multicols}{2}
     \includegraphics[width=.67\linewidth]{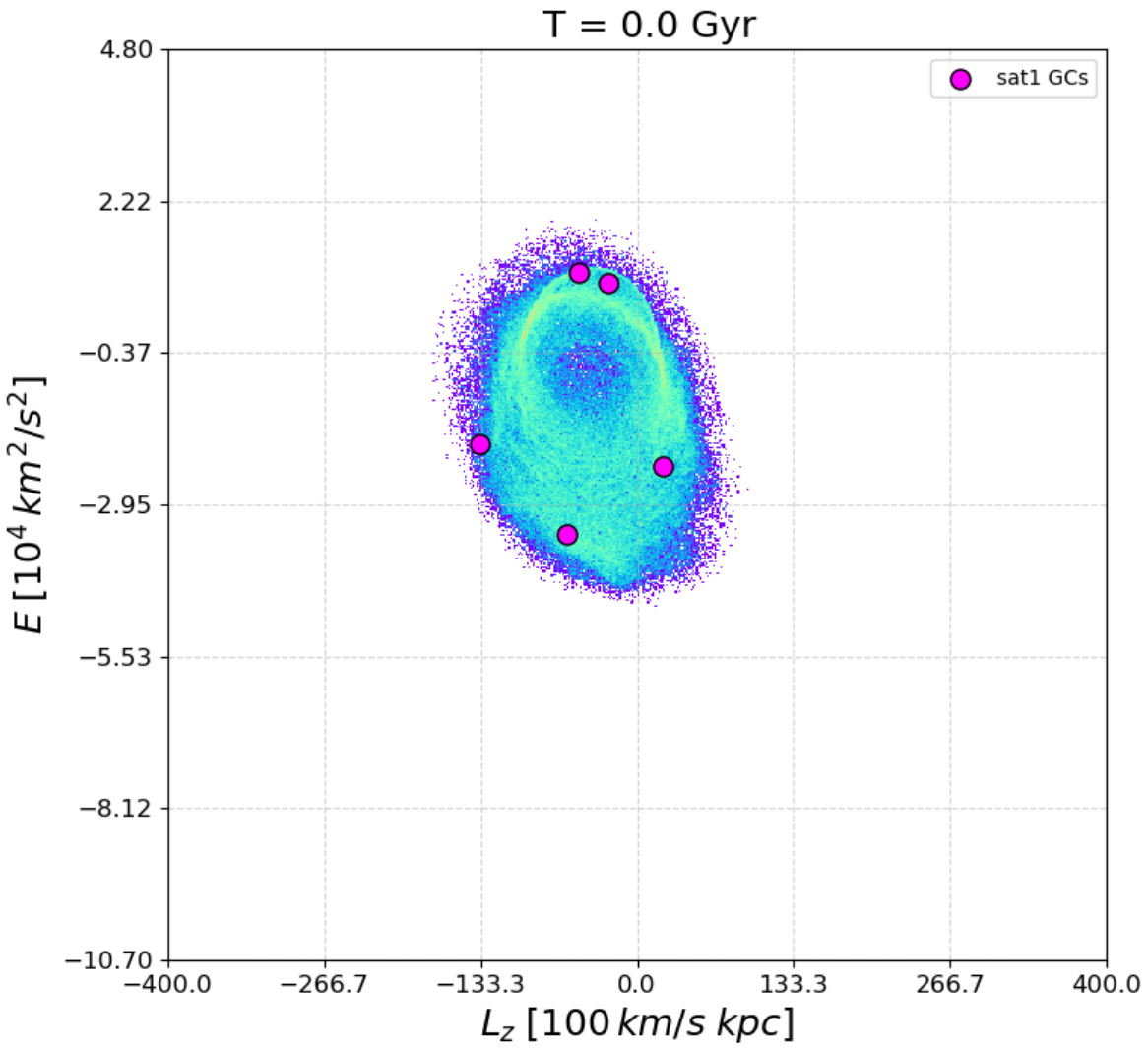}\par
         \includegraphics[width=.67\linewidth]{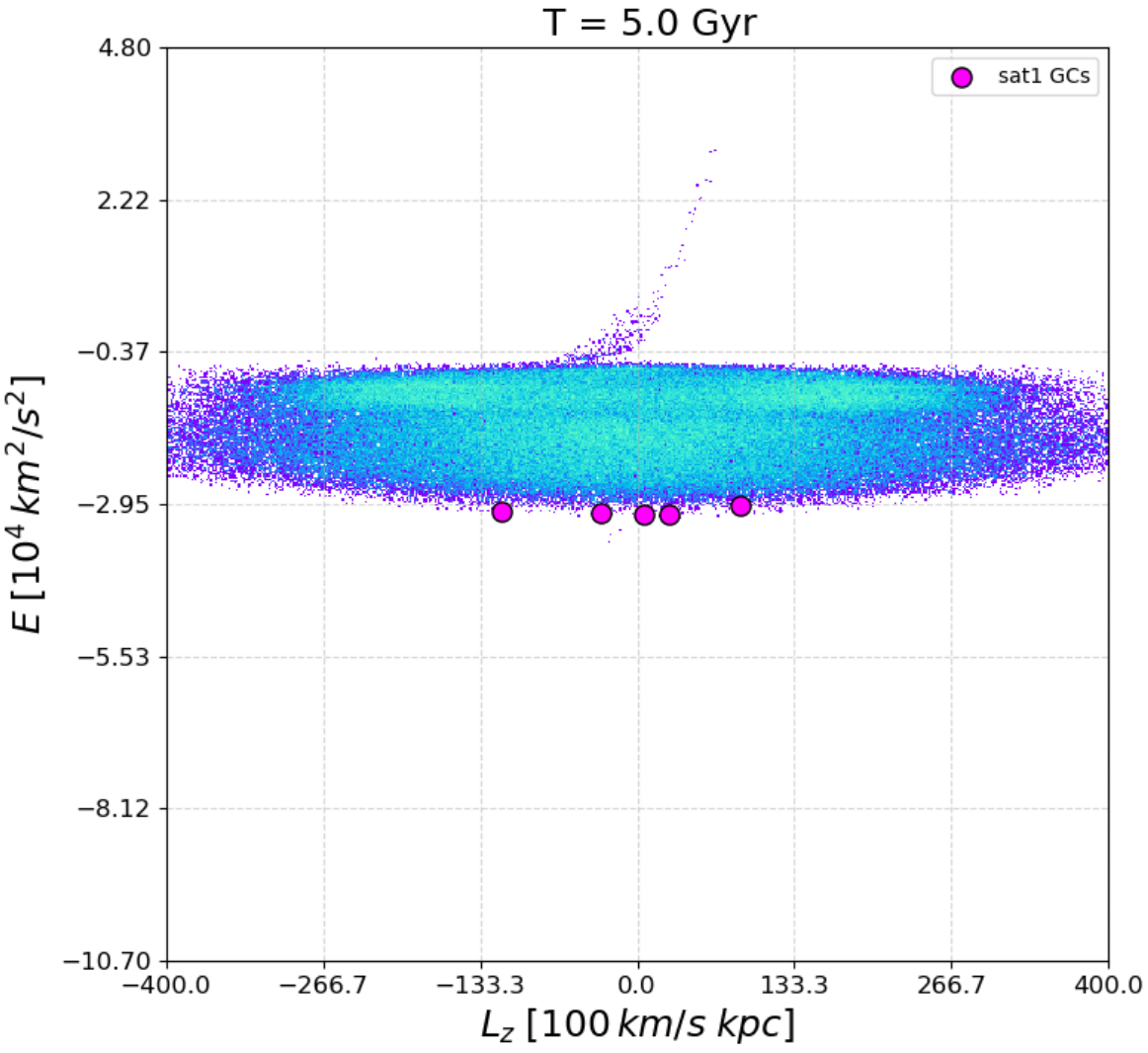} 
    \end{multicols}
    \begin{multicols}{2}
\includegraphics[width=.67\linewidth]{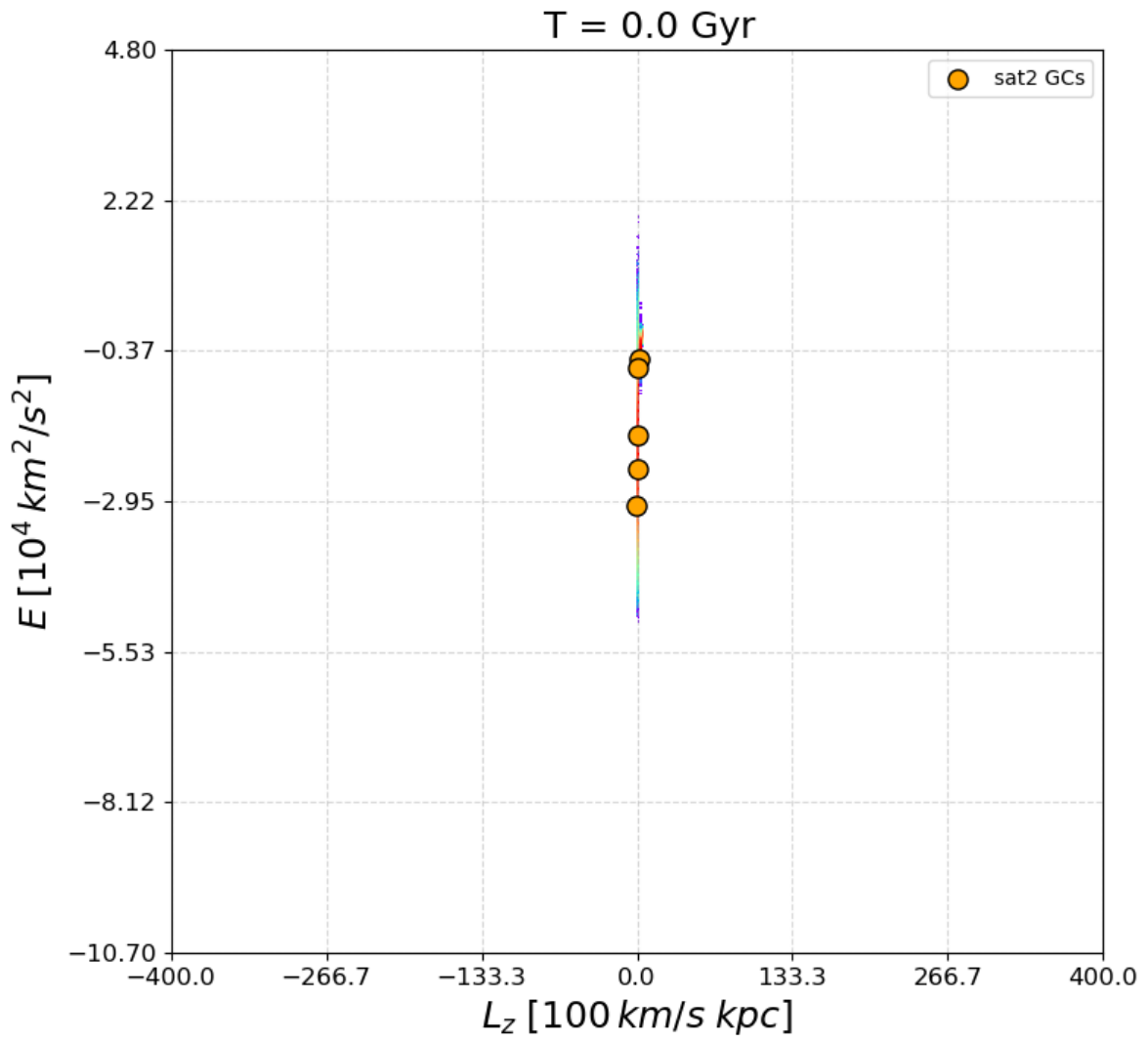}\par
\includegraphics[width=.67\linewidth]{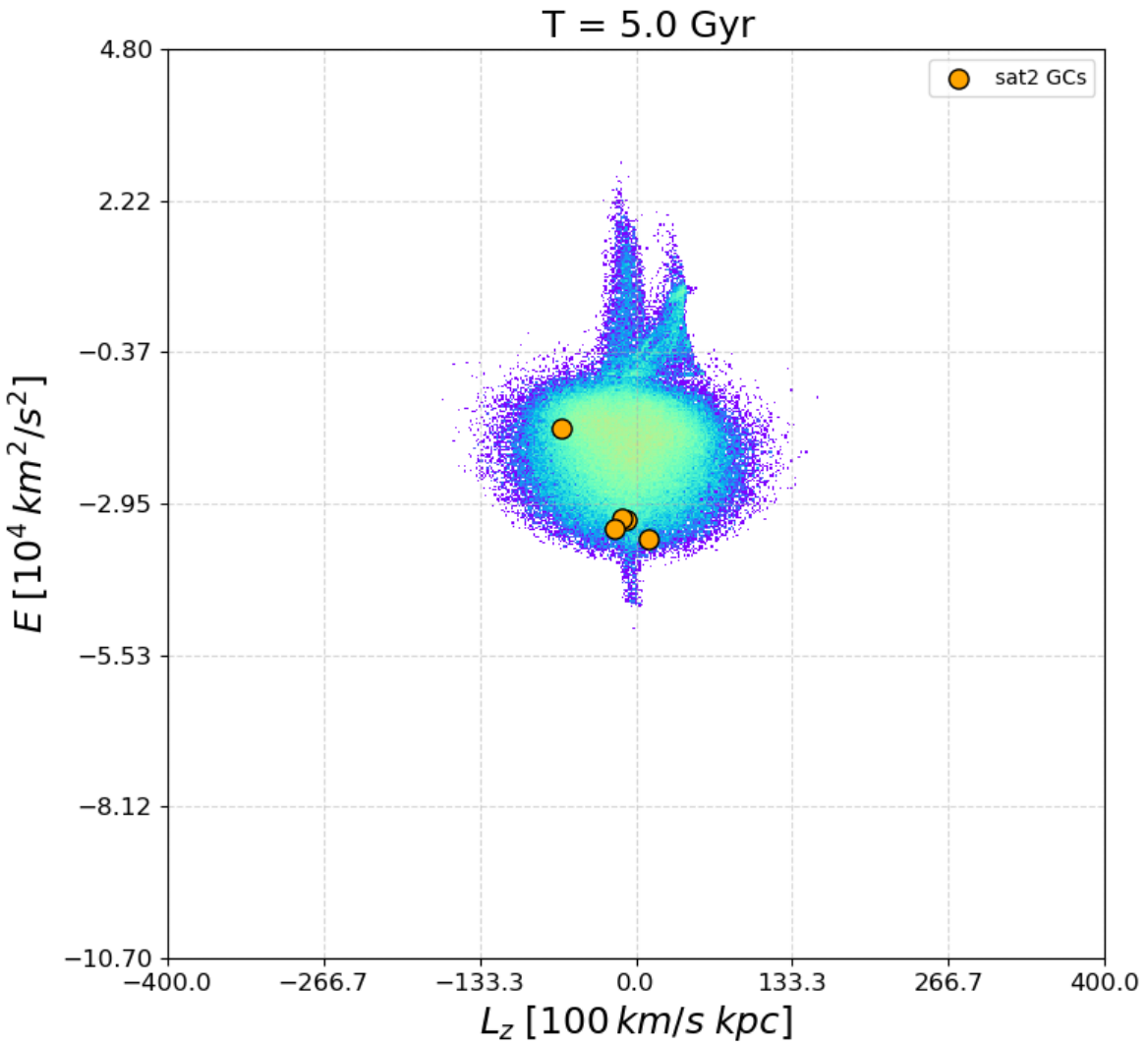}
    \end{multicols}
    \begin{multicols}{2}
\includegraphics[width=.67\linewidth]{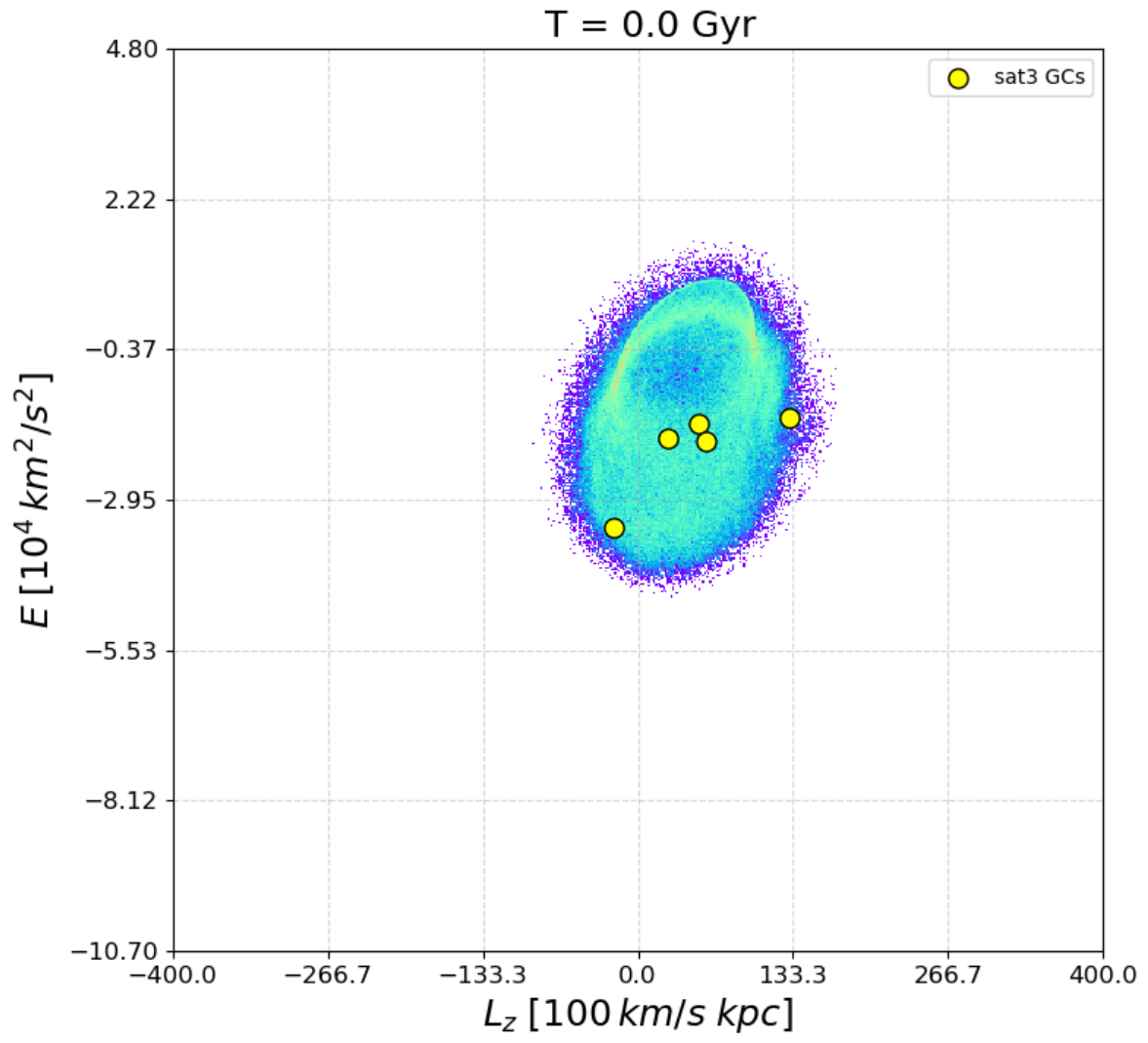}\par
\includegraphics[width=.67\linewidth]{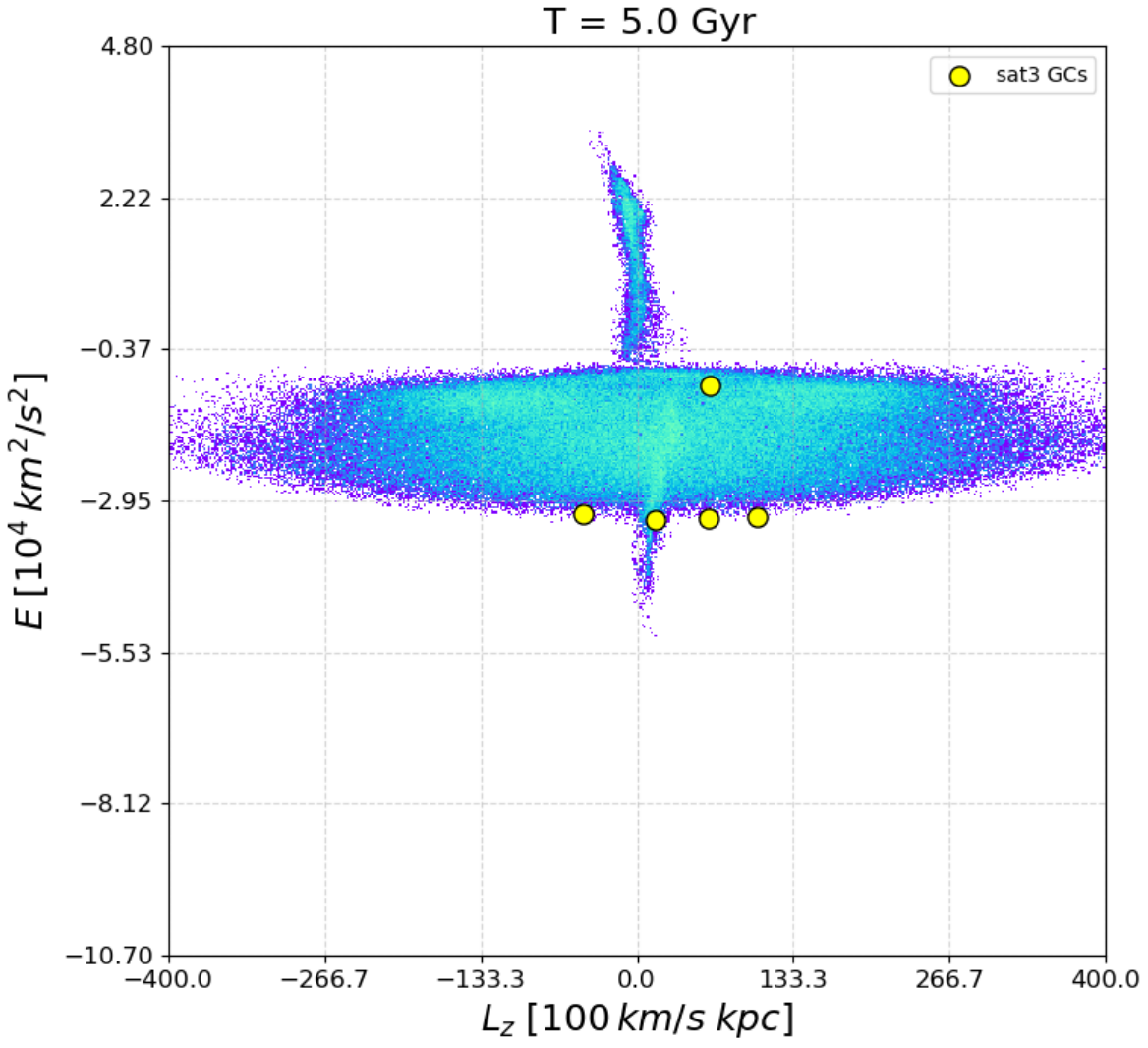}
    \end{multicols}
        \begin{multicols}{2}
\includegraphics[width=.67\linewidth]{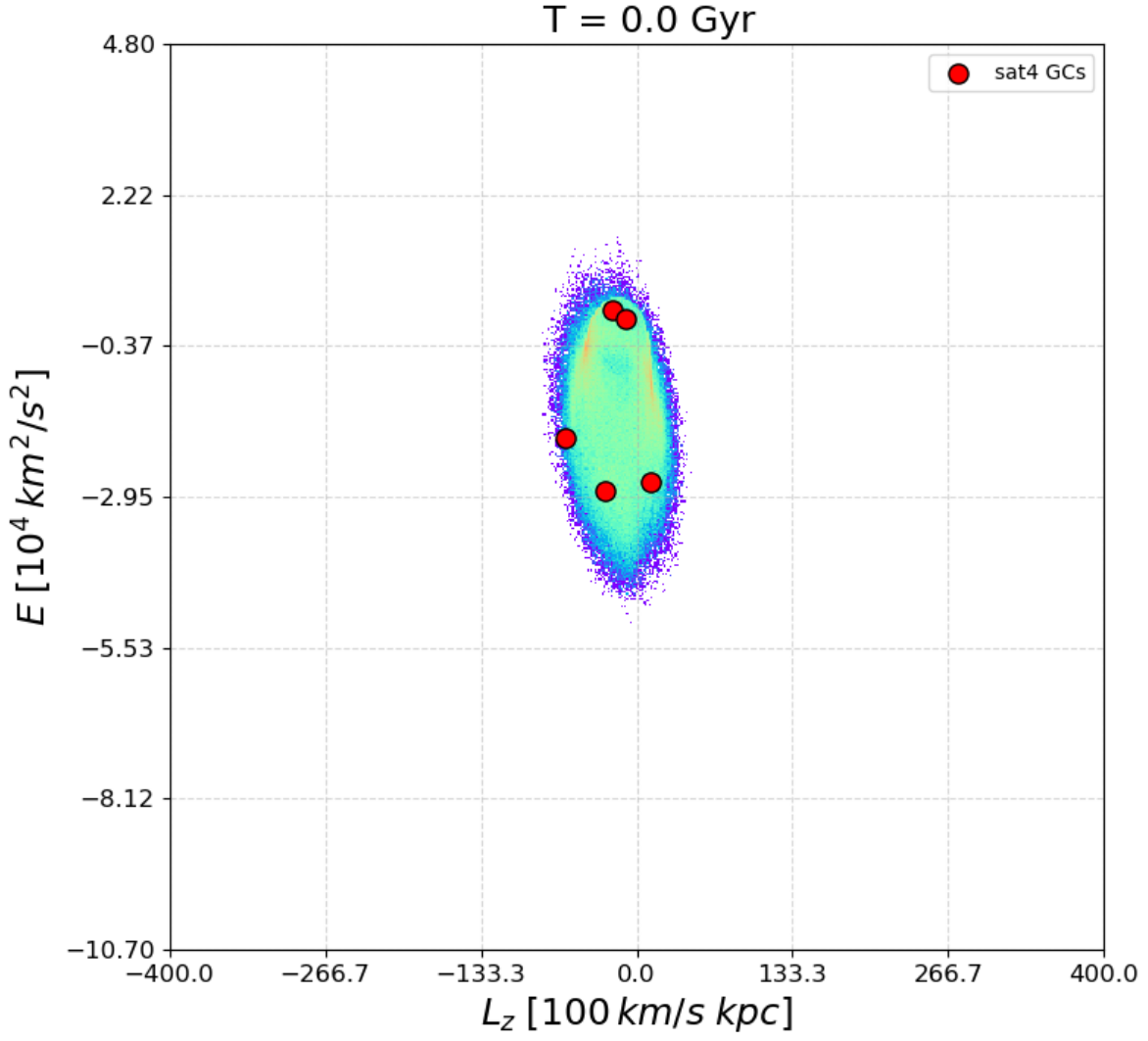}\par
\includegraphics[width=.67\linewidth]{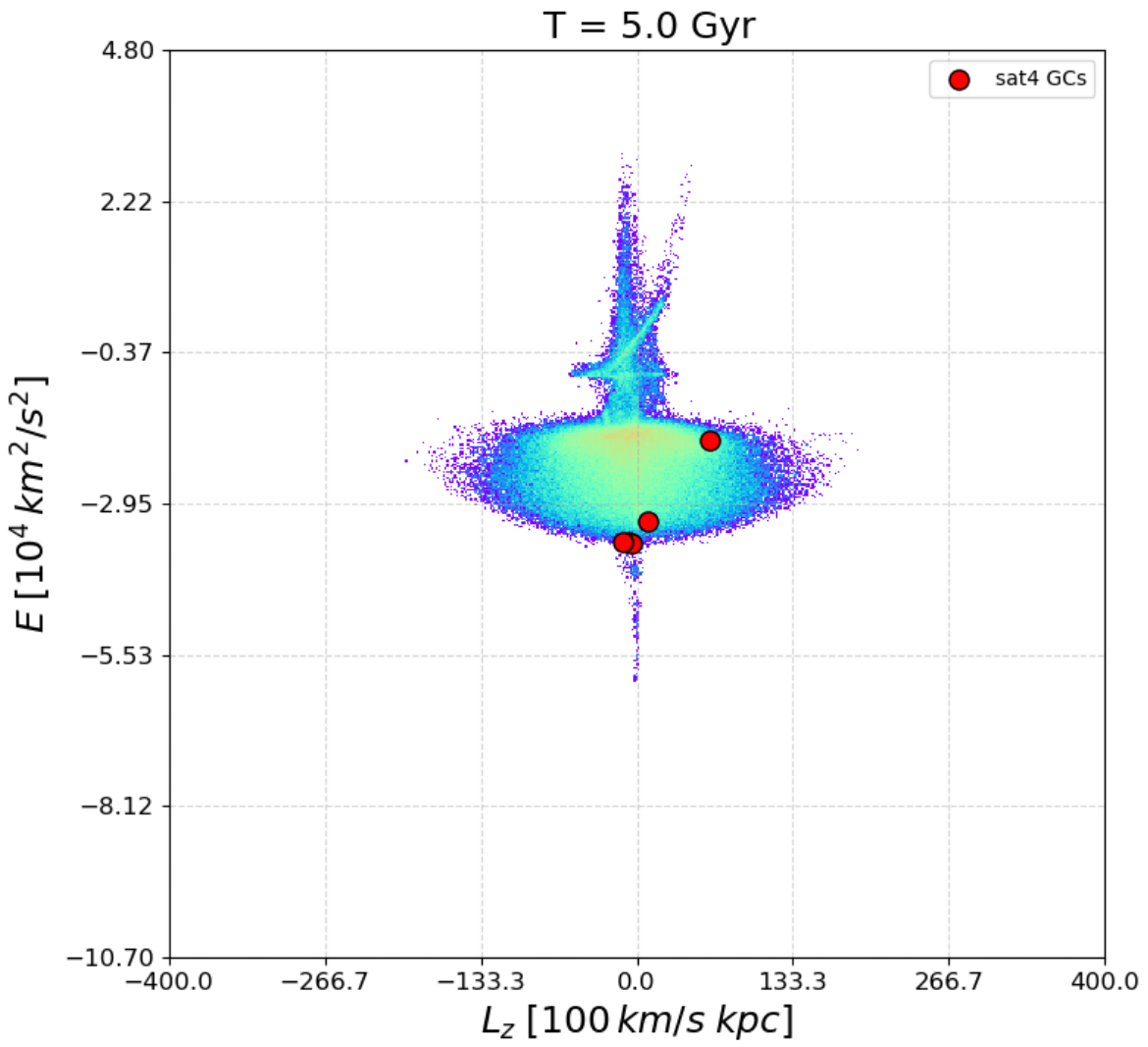} 
    \end{multicols}
\end{centering}
\caption{Same as Fig.~\ref{fig:e_lz_4sat_1} but for the simulation with ID = MWsat\_n4\_$\Phi$180-90-30-120.}
\label{fig:e_lz_4sat_2}
\end{figure*}

Figs.~\ref{fig:e_lz_4sat_1} and \ref{fig:e_lz_4sat_2} show the initial (left panel) and final (right panel) distributions of GCs and stars belonging to each accreted satellite for the two simulations. As we can see, all the satellites in both the simulations start with a clumpy distribution of stars and clusters.
On the other hand, the final stellar distributions are different as we can identify three types of shapes: high-energy and angular-momentum spread distributions (see satellites 1,2 in Fig.~\ref{fig:e_lz_4sat_1} and satellites 1,3 in Fig.~\ref{fig:e_lz_4sat_2}), which correspond to those satellites that do not end up merging but move away from the MW after an initial approach, more clumped distributions in energy-angular momentum (see satellite 2,4 in Fig.~\ref{fig:e_lz_4sat_2}) corresponding to satellites that are at the beginning of the merging, and more elongated distributions in energy (see satellite 3,4 in Fig.~\ref{fig:e_lz_4sat_1}) corresponding to satellites that have orbited the MW for a while and are towards the end of the merging. For some satellites (see for example satellite 2,3,4 in Fig.~\ref{fig:e_lz_4sat_2}), we can also see structures that start from the main distribution and elongate towards higher and lower energies that correspond to the trailing and leading tails, respectively. If we look at how the clusters are distributed, we can see that, in all the accreted systems, most of them are positioned at the lower extremes in energy of the corresponding stellar distributions. 

\begin{figure*}
\centering
\includegraphics[width=.48\textwidth]{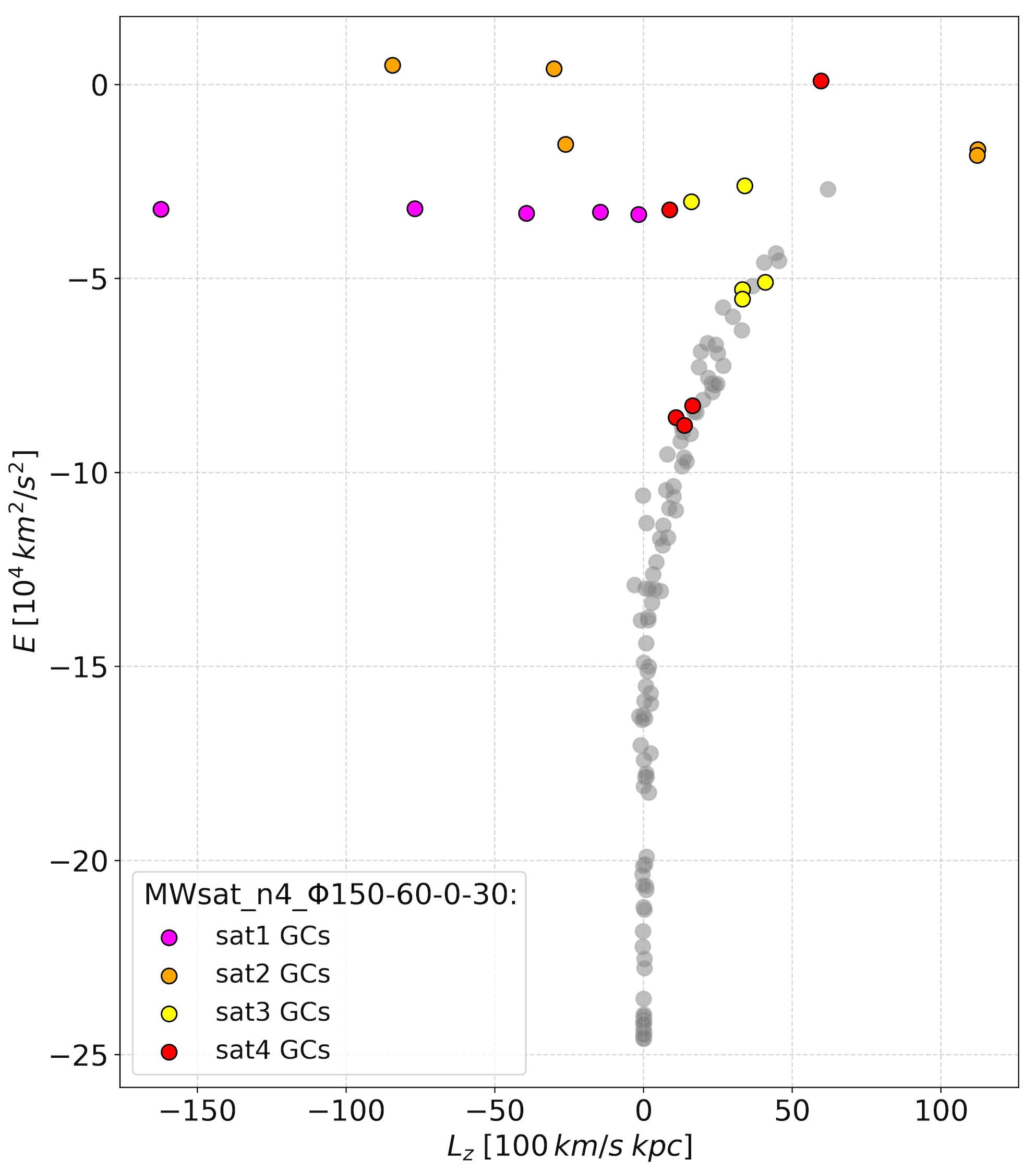}
\includegraphics[width=.48\textwidth]{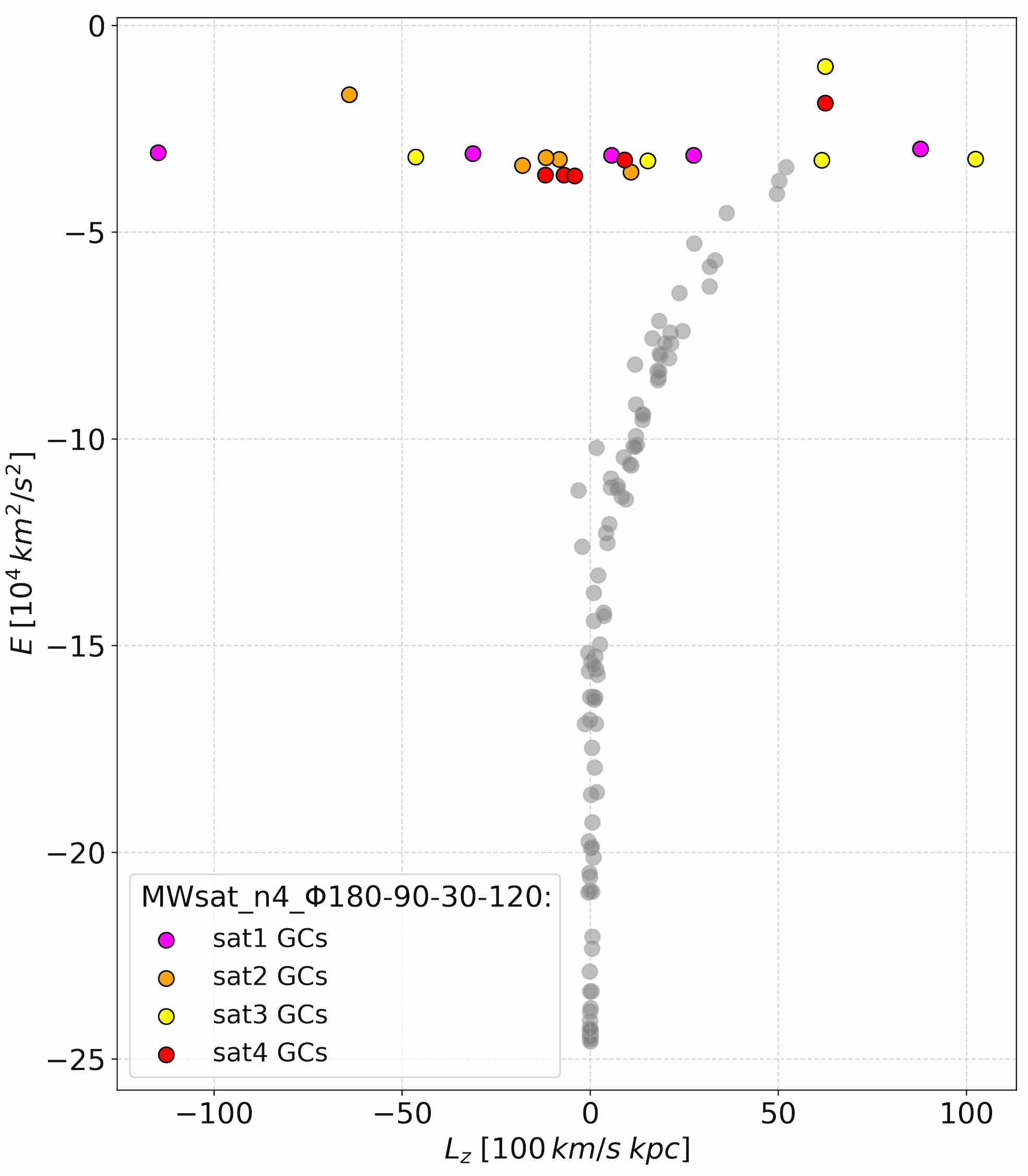} \\
\caption{Final $E - L_{z}$ distribution of globular clusters belonging to all the four accreted satellites (colour coded circles) and to the MW-type galaxy (grey circles) for the two simulations MWsat\_n4\_$\Phi$150-60-0-30 (left panel) and MWsat\_n4\_$\Phi$180-90-30-120 (right panel).}
\label{fig:mw4_e_lz}
\end{figure*}

Fig.~\ref{fig:mw4_e_lz} shows the final $E - L_{z}$ distributions of globular clusters belonging to all the four accreted satellites (colour coded circles) and to the MW-type galaxy (grey circles) respectively for the MWsat\_n4\_$\Phi$150-60-0-30 and MWsat\_n4\_$\Phi$180-90-30-120 simulations.
As a general trend, we can see that accreted GCs remain at high energies compared to the cases involving mergers with satellites of 1:10 mass ratio. This happens because the progenitors to which they belong, being less massive, cannot penetrate deep enough into the MW-type galaxy potential. The only satellites that are able to lose more energy and carry a part of their clusters at energies lower than -5 (see satellites 3,4 in Fig.~\ref{fig:e_lz_4sat_1}), deposit some GCs also at high energies thus resulting in a final distribution that is not clustered as expected. In fact, even in this case we cannot identify clumps of clusters belonging to the same progenitor, being overall mixed and spread over an extended range of angular momentum.

\end{appendices}
\end{document}